\documentclass[fleqn,usenatbib]{mnras}
\usepackage{fix-cm}

\usepackage{newtxtext,newtxmath}

\usepackage[T1]{fontenc}

\usepackage{verbatim}

\DeclareRobustCommand{\VAN}[3]{#2}
\let\VANthebibliography\thebibliography
\def\thebibliography{\DeclareRobustCommand{\VAN}[3]{##3}\VANthebibliography}

\usepackage{graphicx}	
\usepackage{amsmath}	
\usepackage[left]{lineno}
\usepackage{xspace}
\usepackage{tabularx}
\usepackage{array}
\usepackage{caption}
\usepackage{subcaption}
\graphicspath{{./}{figures/}}
\usepackage{geometry}
\usepackage{natbib}
\usepackage{tabu}
\usepackage{appendix}
\usepackage[utf8]{inputenc}
\usepackage[flushleft]{threeparttable}
\usepackage{url}
\usepackage{lipsum} 
\usepackage[bottom]{footmisc} 
\usepackage{xcolor}
\usepackage{fancyhdr}
\usepackage{rotating}
\usepackage{lscape}
\usepackage{bm}
\usepackage{tabularx}
\definecolor{code}{HTML}{990033}
\definecolor{link}{HTML}{006699}
\usepackage{float}
\usepackage{morefloats}  
\usepackage{caption}
\usepackage{longtable}
\usepackage{graphicx} 
\usepackage{booktabs} 
\usepackage{pdflscape} 
\usepackage{longtable} 
\usepackage{caption} 
\usepackage{adjustbox}
\usepackage{float}
\usepackage{placeins}
\usepackage{hyperref}
\usepackage{orcidlink}
\newcommand{\fermi}{{\em Fermi}\xspace}

\title[Joint spectral analysis of LAT and GBM GRBs]{%
\begin{center}
Implications of Joint Spectral Analysis of Gamma-Ray Bursts detected by \textit{Fermi} Large Area Telescope and Gamma-ray Burst Monitor on Phenomenological Correlations
\end{center}}

\author[Aldowma, Razzaque, Martinelli et al.]{
T.~K.~M.~Aldowma\orcidlink{0000-0003-3479-1341},$^{1, 2}$\thanks{E-mail: tamador-khalil@oiu.edu.sd}
S.~Razzaque\orcidlink{0000-0002-0130-2460},$^{1, 3, 4}$\thanks{E-mail: srazzaque@uj.ac.za}
R.~Martinelli\orcidlink{0009-0004-0133-7227},$^{5, 8}$\thanks{E-mail: riccardo.martinelli@phd.units.it}
R.~Gupta\orcidlink{0000-0003-4905-7801},$^{6, 7}$
F.~Longo\orcidlink{0000-0003-2501-2270},$^{5, 8}$ 
N.~Omodei\orcidlink{0000-0002-5448-7577},$^{9}$ 
\newauthor
N.~Di~Lalla\orcidlink{0000-0002-7574-1298},$^{9}$ 
J.~L.~Racusin\orcidlink{0000-0002-4744-9898},$^{6}$
A.~H~.~Airasca\orcidlink{0009-0007-8169-4719},$^{10,11}$
\\  
$^{1}$ Centre for Astro-Particle Physics (CAPP) and Department of Physics, University of Johannesburg, PO Box 524, Auckland Park 2006, South Africa\\
$^{2}$ Department of Astronomy and Meteorology, Faculty of Science and Technology, Omdurman Islamic University, PO Box 382, Omdurman, 14415, Sudan\\
$^{3}$ Department of Physics, The George Washington University, Washington, DC 20052, USA\\
$^{4}$ National Institute for Theoretical and Computational Sciences (NITheCS), Private Bag X1, Matieland, South Africa\\
$^{5}$ Department of Physics, University of Trieste, I-34127 Trieste, Italy\\
$^{6}$ Astrophysics Science Division, NASA Goddard Space Flight Center, Mail Code 661, Greenbelt, MD 20771, USA\\
$^{7}$ NASA Postdoctoral Program Fellow\\
$^{8}$ Istituto Nazionale di Fisica Nucleare, Sezione di Trieste, I-34127 Trieste, Italy\\
$^{9}$ W.\ W.\ Hansen Experimental Physics Laboratory, Kavli Institute for Particle Astrophysics and Cosmology,\\
Department of Physics and SLAC National Accelerator Laboratory, Stanford University, Stanford, CA 94305, USA\\
$^{10}$ Istituto Nazionale di Fisica Nucleare, Sezione di Bari, I-70126 Bari, Italy\\
$^{11}$ Università degli studi di Trento, via Calepina 14, 38122 Trento, Italy
}

\date{Accepted XXX. Received YYY; in original form ZZZ}

\pubyear{\the\year{}}

\begin{document}
\label{firstpage}

\pagerange{\pageref{firstpage}--\pageref{lastpage}}
\maketitle

\begin{abstract}
Gamma-ray bursts (GRBs) have emerged as powerful cosmological probes for exploring the distant Universe, owing to their immense luminosities and detectability at high redshifts. Several empirical correlations have been established, particularly involving their energy properties. This work aims to enhance the precision of these correlations through joint spectral analysis, focusing on reducing uncertainties in both the spectral indices and the peak energy ($E_{\rm p}$) derived from spectral fitting. We extend previous studies using both traditional and novel spectral models, utilizing a sample of 37 GRBs observed by the \textit{Fermi} Gamma-ray Burst Monitor (GBM) and Large Area Telescope (LAT), incorporating the LAT Low-Energy (LLE) technique, over the period 2008--2024. Our analysis compares results from joint fits (GBM--LAT--LLE) against those from GBM-only fits. The study focuses on fitting time-integrated ${\rm T}_{90}$ and peak flux in the rest frame. Among the observable phenomenological correlations, we revisit the Amati and Yonetoku relations: the Amati relation links the intrinsic peak energy ($E_{i,\rm p}$) to the total isotropic energy ($E_{\rm iso}$) emitted during ${\rm T}_{90}$, while the Yonetoku relation connects $E_{i,\rm p}$ to the isotropic luminosity ($L_{\rm iso}$). Refining these correlations aims to deepen our understanding of GRB energetics and improve the precision of cosmological parameter estimates derived from GRB observations.
\end{abstract}




\section{Introduction}
\label{sec:sec_1} 

Gamma-ray bursts (GRBs) are the most luminous explosions known, releasing enormous energy in $\gamma$ rays and occurring at cosmological distances. Data from the Burst Alert and Transient Source Experiment (\textit{BATSE}) onboard the {\it Compton Gamma-Ray Observatory}, which operated in the 1990s and detected $\sim 3000$ GRBs, showed that they were uniformly distributed on the sky \citep{Meegan1992Natur.355..143M}, indicating an extragalactic origin. GRBs emit promptly in $\sim 0.1$–$1$ MeV $\gamma$ rays and are generally classified into two categories based on their duration: short GRBs (SGRBs) and long GRBs (LGRBs). Historically, the dividing line has been ${\rm T}_{90} \approx 2$~s \citep[e.g.,][]{Kouveliotou1993}, though ${\rm T}_{90}$ is instrument dependent; in the fourth \textit{Fermi}-GBM GRB catalog, the separation occurs at ${\rm T}_{90} \approx 4.2$~s \citep{VonKienlin2020}.

This classification relies on the time duration ${\rm T}_{90}$, defined as the interval during which the photon flux accumulates between 5\% and 95\% of the total fluence \citep{Paciesas2012_199}. Because of their cosmological distances, GRBs are among the most powerful explosions observed in the universe, releasing $\gtrsim 10^{51}$ ergs in $\gamma$ rays. Prompt $\gamma$-ray emission from GRBs continues to raise questions regarding their energy source and underlying emission mechanisms, which originate from either the death of massive stars or the merger of two neutron stars \citep[for reviews, see e.g.,][]{Fishman1995, 1999PhR...314..575P, Meszaros+06, kumar2015PhR...561}.
 
Observations of GRB prompt emissions are conducted using space telescopes, each covering different energy ranges (e.g.,  {\textit Fermi} Gamma-ray Burst Monitor (GBM)  8 keV -- 40 MeV \citep{Meegan2009_702}, {\it Fermi} Large Area Telescope (LAT) 20 MeV -- {$>$300~GeV} \citep{Atwood2009_697}, \textit{Konus-}Wind $\sim$ 10 keV -- 10 MeV \citep{Aptekar1995_71}, \textit{Swift} Burst Alert Telescope (BAT) 15 -- 150 keV \citep{Gehrels2004E_552}, etc.) 
A combined energy range, from $\sim 8$~keV to $\gtrsim 300$~GeV, by these instruments has greatly improved our understanding of GRB prompt emission, including multiple emission components, delayed onset and extended duration of $\gtrsim 100$~MeV emission beyond ${\rm T}_{90}$. This extended emission is now known as afterglow emission \citep[see for review,][]{2013FrPhy...8..661G}. The spectrum of prompt emission is non-thermal, and synchrotron emission is likely responsible for it \citep{2018A&A...613A..16R, 2020NatAs...4..174B, 2020NatAs...4..210Z}. {Recent studies further suggest that GRBs may arise from more than a single progenitor, with metallicity and the surrounding environment playing a role in shaping their observed rates} \citep{2024ApJ...967L..30D, PetrosianDainotti2024}.
However, synchrotron emission is inconsistent with the observed spectra for many GRBs \citep{2014IJMPD..2330002Z, 2015AdAst2015E..22P, 2024ApJ...972..166G}.
Alternatively, the spectrum is typically fitted by the so-called GRB spectrum or Band function, the {empirical} function, which connects two power laws with an exponential that peaks in the $\sim 0.1-1$~MeV range \citep{Band1993_413}. Other {empirical} functions, such as a smoothly-broken power law or a power-law with exponential cutoff, are also used \citep[see, e.g.,][]{Kaneko:06}. A quasi-thermal blackbody component has been observed in some GRBs  \citep[see, e.g.,][]{Ryde2010, 2011ApJ...727L..33G, 2014IJMPD..2330002Z}, as well as an additional power-law component extending to higher and lower energies (i.e., 10 keV–GeV band) is observed in the joint spectral analysis of GRBs from the LAT and GBM instruments \citep{2013ApJS..209...11A}. Broadband, multi-component spectral analysis with LAT and GBM data has shown a more accurate estimate of the total $\gamma$-ray energy release from GRBs and better constrains the peak energy of the fitted spectrum \citep{Dirirsa2019}, which is the main focus of this paper. By applying joint spectral fitting to a sample of LAT and GBM-detected GRBs, we aim to reduce uncertainties in energy parameters, ultimately improving the reliability of GRB-based cosmological constraints.

Several phenomenological relations have been developed over the past two decades connecting the spectral peak energy to the total $\gamma$-ray energy or power release from a GRB. Since the latter depends on an assumed cosmological model for luminosity distance calculation, GRBs have been used to better constrain cosmological parameters and study the distant Universe, rather than directly constraining cosmological models. In contrast to the optical radiation from Type Ia supernovae (SNe Ia; \citealp{Riess1998}), which have been detected to redshifts up to $z$ = 2.6 and 2.9 \citep{2022ApJ...934L...7R, 2025A&A...701A..70V}, the high-energy radiation from GRBs is not significantly affected by dust extinction as it travels to us. Consequently, the observed $\gamma$-ray flux directly measures the prompt emission of energy \citep{2008MNRAS.391..577A}. Furthermore, GRBs have been detected at high redshifts up to \( z = 8.2 \) and even \( z \sim 9.4 \) \citep{2009Natur.461.1254T, Cucchiara2011_736}, thus they can probe a more distant universe than SNe Ia. 
An example of GRB phenomenological relations is the Amati relation, expressed as $E_{\rm iso}/({10}^{52}\, \text{erg}) = k (E_{\rm i, p}/\text{keV})^m$ \citep{Amati2002_81A, Amati2006_372}, which links the intrinsic peak energy $E_{\rm i, p}$, the energy at which the prompt $\gamma$-ray spectral energy distribution $\nu{f}_{\nu}$ peaks in the cosmological rest frame to the isotropic-equivalent $\gamma$-ray energy release $E_{\rm iso}$ in the source frame. Another correlation, the Yonetoku relation, describes the connection between $E_{\rm i, p}$ and the isotropic-equivalent luminosity $L_{\rm iso}$ during the peak flux interval with a similar mathematical relation \citep{2004ApJ...609..935Y, 2014Ap&SS.351..267Z, 2019NatCo..10.1504I, 2020ApJ...900...33P}. When the parameters are tightly constrained, these correlations offer the potential for GRBs to use as cosmological standard candles \citep[see, e.g.,][]{2005ApJ...627....1F, 2006NJPh....8..123G, 2008MNRAS.391..577A, Basilakos2008_391, 2011MNRAS.415.3580D, Wang2016_585, 2016ApJ...831L...8G, Demianski2017_693, Dirirsa2019, Khadka2021, Demianski2021, Kumar2023}. 

In this paper, we fit broadband spectra of GRBs jointly detected by LAT and GBM with measured redshift 
since the launch of the \textit{Fermi} Gamma-ray Space Telescope in 2008. We use single- or multi-component spectral models to fit data in the $10$~keV -- $100$~GeV energy range and calculate the fluence over ${\rm T}_{90}$, {defined as the time interval containing 90\% of the fluence}, and during the peak flux interval. We also compare joint LAT$+$GBM fits to GBM-only fits.  We compute $E_{\rm iso}$ and $L_{\rm iso}$ using a flat $\Lambda$CDM cosmology to derive the Amati ($E_{\rm iso} - E_{i,\rm p}$) and Yonetoku ($L_{\rm iso} - E_{i,\rm  p}$) correlations. We also explore the potential to constrain cosmological parameters 
by jointly fitting with the parameters of the phenomenological correlations using GRB and SNe Ia data. We describe our GRB sample in Section~\ref{sec:sec_2}, the spectral analysis method in Section~\ref{sec:sec_3}, and fitting the Amati and Yonetoku relations in Section~\ref{sec:sec_4}. In Section~\ref{sec:sec_5}, we present results from our analyses and discuss them in Section~\ref{sec:sec_6}. We provide conclusions in Section~\ref{sec:conclusion}.

\section{GRB Samples}
\label{sec:sec_2} 
Our sample covers {sixteen} years of \textit{Fermi} observations from 2008 to 2024, using data from its two instruments, GBM and LAT. We include GRBs with known redshifts to explore the Amati \citep{Amati2002_81A} and Yonetoku \citep{2004ApJ...609..935Y} relations through joint spectral analysis. Redshift information, obtained via spectroscopic or photometric follow-up observations, was collected from the General Coordinates Network (GCN\footnote{\url{https://gcn.nasa.gov/}}) notices and associated publications.

\subsection{\fermi Gamma-ray Burst Monitor (GBM) Data}
The \textit{Fermi}-GBM is equipped with 12 sodium iodide (NaI) scintillation detectors and 2 bismuth germanate (BGO) detectors. The NaI detectors are sensitive to photons in the energy range of approximately 8~keV to 1~MeV; each detector has an effective area of around 100~cm$^2$ {near 60~keV} \citep{Meegan2009_702}. The BGO detectors extend this coverage to higher energies, operating between 250~keV and 30~MeV \citep{2009ExA....24...47B}. With its wide field of view, GBM continuously monitors a large fraction of the sky for transient events.

Data is collected in three formats: Time-Tagged Events (TTE), continuous spectrum (CSPEC), and continuous time (CTIME). TTE data, which provides the highest time and energy resolutions by recording individual photon events across 128 Pulse Height Analysis (PHA) channels, was utilized for the analysis of GRBs in this study, while CSPEC and CTIME are binned in smaller versions of the raw TTE data, the availability of continuous TTE has improved significantly, particularly for observations post-2011, allowing for more detailed analyses. We obtained data from the \textit{Fermi}-GBM catalog\footnote{\url{https://heasarc.gsfc.nasa.gov/W3Browse/fermi/fermigbrst.html}} \citep{VonKienlin2020}. For the NaI detectors, we select the energy range excluding the K-edge, using 10--25 and 45--900~keV, while the BGO detectors cover the range from 250~keV to 30~MeV.
 
The fourth \textit{Fermi}-GBM GRB catalog \citep{VonKienlin2020} spans events from 2008 to 2018; it lists 2,356 GRBs triggered by GBM during this decade. The comprehensive spectral information of these events can be found in \citet{2021ApJ...913...60P}. Of these, 135 GRBs have measured redshifts, including 16 classified as SGRBs and 122 as LGRBs. In this analysis, however, we use data through 2024.

\subsection{\fermi Large Area Telescope (LAT) Data}\label{sec:2_2}
The LAT is a pair-conversion telescope with silicon tracking detectors, a cesium-iodide calorimeter, and an Anti-Coincidence Detector (ACD). It detects photons from $\sim$ 20 MeV to $\gtrsim$ 300 GeV and observes 20\% of the sky at any given time \citep{Atwood2009_697}. Regular LAT analysis is done in the energy range 100~MeV–300~GeV \citep{2013ApJS..209...11A, 2019ApJ...878...52A}, while the LAT Low Energy (LLE\footnote{\url{https://heasarc.gsfc.nasa.gov/FTP/fermi/data/lat/triggers}}; \citealp{2010arXiv1002.2617P, 2014ApJ...789...20A}) analysis uses a looser event selection to increase effective area in the 30–100~MeV band, at the cost of higher background rates. For bright transients, background levels are estimated by fitting pre- and post-burst intervals with a polynomial, similar to GBM analysis, enabling spectral extraction in this range.

The second \textit{Fermi}-LAT catalog\footnote{\url{https://heasarc.gsfc.nasa.gov/W3Browse/fermi/fermifhl.html}} of GRBs covers a period from 2008 until 2018 August 4, and provides results for a total of 186 GRBs, of which 91 show emissions in the range 30–100 MeV (17 of which are seen only in this band) and 169 are detected above 100 MeV \citep{2019ApJ...878...52A}. The diffuse background models used in this work (\texttt{gll\_iem\_v07.fits}\footnote{\url{https://fermi.gsfc.nasa.gov/ssc/data/analysis/software/aux/4fgl/Galactic_Diffuse_Emission_Model_for_the_4FGL_Catalog_Analysis.pdf}} and the corresponding isotropic templates) are described in the LAT 4FGL analyses \citep{Abdollahi2020, Abdollahi2022, Ballet2023}. Overall, 40 GRBs detected by LAT between 2008 and 2024 have known redshifts.

\subsection{Joint GBM and LAT Data with known Redshift}
A total of 40 GRBs with identified redshifts were detected by \textit{Fermi} GBM and LAT. However, \textit{Fermi}-GBM did not trigger on GRB\,081203A and GRB\,130907A, and thus they were excluded from our sample. Furthermore, GRB\,120711A, GRB\,130702A, and GRB\,160623A were also excluded due to insufficient LAT photon statistics within the GBM \({\rm T}_{90}\) duration to perform a time-integrated joint spectral analysis combining LAT and GBM data. This insufficiency is primarily due to the GRBs being located at large off-axis angles, resulting in low LAT effective area, or partially outside the LAT field of view (FOV) during the prompt emission phase.  
Therefore, after these exclusions, we analyzed a final sample of 37 GRBs detected by \textit{Fermi} between 2008 and 2024 as listed in Table~\ref{tab1}. In certain cases, the \textit{Fermi}-LAT data lacked photons during the time-integrated or peak intervals, or no corresponding trigger was observed in the \textit{Fermi}-GBM data. 
 
Our dataset features some of the most luminous GRBs ever recorded, including GRB\,130427A \citep{2014Sci...343...42A}, GRB\,190114C \citep{2020ApJ...890....9A}, and GRB\,221009A \citep{axelsson2024grb}, which has been dubbed the ``brightest of all time'' \citep[BOAT;][]{2023ApJ...946L..31B}. For GRB\,221009A, the presence of Bad Time Intervals (BTIs), as detailed in \citet{axelsson2024grb, 2024HEAD...2120402D}, necessitated determining the \({\rm T}_{90}\) duration using the methodology described in \citet{edvige2023bright}. This particular burst has also been the focus of extensive studies due to its high-energy emissions, as highlighted in many publications \citep[see, e.g.,][]{2023ApJ...952L..42L, Das2023A&A...670L..12D, Abdalla:2024qtv, Finke_2024, 2024ApJ...971..163R}. Additionally, we have integrated findings from prior research, such as \citet{Dirirsa2019}, to enable comparisons of spectral analysis results for selected GRBs. For instance, the analysis of GRB\,220101A references the results from \citet{Scotton_2023}. Furthermore, our study includes one exceptionally bright SGRB, GRB\,090510, as investigated in \citet{2010ApJ...716.1178A, Muccino_2013}.

\section{Joint spectral analysis}
\label{sec:sec_3}
We have extracted \fermi GBM and LAT data to perform joint spectral analysis
using the Multi-Mission Maximum Likelihood framework \citep[\texttt{3ML} (v2.4.3)\footnote{\url{https://threeml.readthedocs.io/en/latest/}};][]{2015arXiv150708343V}, which enabled the selection and download of GBM, LAT, and LLE data. For NaI and BGO detectors, background intervals were taken from the GBM catalog when available or manually selected as pre- and post-burst emission. The same procedure was applied for LLE data, as described in Sec~\ref{sec:2_2}, and unbinned likelihood analysis for LAT data. With \texttt{3ML}, we have performed joint spectral analyses of the GRBs in our sample, including time-integrated and peak flux intervals. The GBM and LLE data were imported into \texttt{3ML} using the {\tt TimeSeriesBuilder} class; GBM detectors were selected based on an angle of less than ${50}^\circ$, with the background either automatically fitted using \texttt{3ML} or manually selected using two-time intervals before and after the GRB prompt emission for each energy channel. For LAT data \citep{2013arXiv1303.3514A, 2018arXiv181011394B}, we employed the {\tt FermiLATLike} plugin for unbinned analysis, interfacing with {\tt gtburst\footnote{\url{https://fermi.gsfc.nasa.gov/ssc/data/analysis/scitools/gtburst.html}}}, a tool within the {\tt Fermitools} (v2.2.0)\footnote{\url{https://fermi.gsfc.nasa.gov/ssc/data/analysis/documentation/}} suite.

The main spectral models we used in our analysis are Band \citep{Band1993_413}, Comptonized (CPL, \citealt{Kaneko:06, 2009PASP..121.1279S}), Internal Shock Synchrotron Model (ISSM, \citealt{2020A&A...640A..91Y}), Smoothly Broken Power-law (SBPL, \citealt{1999ApL&C..39..281R}), and Double SBPL (DSBPL, \citealt{2019A&A...625A..60R}). Additionally, we used thermal Black Body (BB) and power-law (PL) models to improve fits. In Appendix \ref{apx1}, we outline the functional forms of these models. 
We applied these models to data collected during the whole ${\rm T}_{90}$ duration of the burst and during a $1\text{\,s}$ interval chosen around the time when the GRB’s prompt‐emission spectrum reaches its maximum—i.e., the time of peak spectrum accumulation measured from the trigger at which the count rate is the highest. The former is relevant for the Amati relation \citep{Amati2006_372} while the latter is used in the Yonetoku relation \citep{2004ApJ...609..935Y}. 
The same time intervals were used for both LAT and GBM data, during which LAT photons may influence the co-temporal spectral fits. By incorporating LLE data to bridge the energy gap between GBM and LAT, particularly for LAT events with low {Test Statistic} (TS) values, the joint fit is better constrained.

\subsection{Selecting the Best-fit Spectral Model}\label{sec:3_1}
One challenge in performing spectral analysis with \texttt{3ML} is that different models often yield similarly good residuals for the same GRBs. We initially start with a single-component spectral model, such as Band, SBPL, CPL, and ISSM, as base models to fit the fluence and peak flux spectra.
We employed the Bayesian Information Criterion (BIC; \citealt{kass1995bayes}) to distinguish between these models and select the one giving the lowest BIC value. The BIC automatically incorporates a penalty term proportional to the number of free parameters, ensuring that models with greater complexity are selected only if they provide a substantially improved fit. An illustrative example of using the BIC to differentiate spectral models can be found in the analysis of GRB\,160625B by \citet{2017ApJ...836...81W}. See also \citet{2018ApJ...862..154C} for the methodology that we follow here. In particular, we calculate \(\Delta\)BIC, where a value of \(\Delta\)BIC \(\geq 6\) indicates a significant improvement in model performance. For single-model fits, we adopt the Band model as the default, and $\Delta{\rm BIC}$ would be:
\[
\Delta{\rm BIC} = {\rm BIC}_{\rm Band} - {\rm BIC}_{\rm Comp},
\]
if, e.g., the alternate model is the Compton. For bright GRBs, however, more complex models are often necessary. For instance, if the model with the lowest BIC is Band+BB+PL, we calculate \(\Delta\)BIC as: 
\[
\Delta{\rm BIC} = {\rm BIC}_{\rm Band+BB+PL} - {\rm BIC}_{\rm Band+BB},
\]
and follow a similar approach for other model comparisons. 

Although the $\Delta\mathrm{BIC}$ criterion offers a quantitative framework for selecting the best-fit spectral model, it does not inherently capture the physical validity of the chosen model, particularly when $\Delta\mathrm{BIC}$ values are marginal (e.g., $<6$). In such cases, the adoption of more complex models, such as Band+CPL+PL, must be supported by physical reasoning in addition to statistical metrics. Spectral analysis alone is often insufficient to break model degeneracies; therefore, joint spectral and polarization measurements are essential to provide complementary constraints on the emission mechanisms \citep[see, e.g.,][]{2022MNRAS.511.1694G, 2025arXiv250405038G}.

Accordingly, we highlight the importance of evaluating not only the statistical preference through $\Delta\mathrm{BIC}$ but also the model's parameter stability, associated uncertainties, and overall physical plausibility, especially when dealing with models that offer only marginal improvements in fit quality.

\subsection{Time Selection: Fluence and Flux}\label{sec:3_2}
\subsubsection{Bolometric fluence}

The bolometric fluence, $S_{\rm bol}$, is defined as the total energy per unit area integrated over the burst duration and energy range. For the purpose of this analysis, where spectra are derived from counts integrated over the ${\rm T}_{90}$ duration, the bolometric fluence is given by:
\begin{eqnarray}
S_{\rm {bol}} = \int_{E_{\text{min}}/1+z}^{E_{\text{max}}/1+z} E \frac{dN}{dE} \, dE\,.
\label{Eq1}
\end{eqnarray}
Here, the photon energy ranges from a minimum of \(E_{\text{min}} = 1\,\text{keV}\) to a maximum of \(E_{\text{max}} = 10^4\,\text{keV}\).

\subsubsection{Bolometric peak flux}
We define the bolometric energy flux of a burst during a one-second peak flux time interval as: 
\begin{eqnarray} 
P_{\rm {bol}} = \int_{E_{\text{min}}/1+z}^{E_{\text{max}}/1+z} E \frac{dN}{dE} \, dE\,.
\label{Eq2}
\end{eqnarray}

We calculate both flux and fluence using the \texttt{get\_flux} method
in \texttt{3ML}, which can provide individual or total flux values. 
For models such as Band+BB, we determine the energy peak from the Band component, and similar logic is applied to other models.

\section{Phenomenological Correlations}
\label{sec:sec_4} 

We proceed to investigate the Amati \citep{Amati2002_81A, Amati2006_372} and Yonetoku \citep{2004ApJ...609..935Y, 2014Ap&SS.351..267Z} correlations in this paper using the broadband spectral fits to the joint GBM and LAT data. We calculate the isotropic-equivalent $\gamma$-ray energy release, $E_{\rm iso}$, using the bolometric fluence, ${S}_{\rm bol}$, as:
\begin{eqnarray}\label{Eq3}
E_{\rm iso} = \frac{4\pi d_{\rm L}^2}{1+z} {S}_{\rm bol},
\end{eqnarray}
where ${S}_{\rm bol}$ is derived from the spectral fitting, see equation~(\ref{Eq1}). The intrinsic peak energy of the ${\rm \nu}f_{\rm \nu}$ spectrum is given by $E_{\rm i,p} = E_{\rm p}(1+z)$. The luminosity distance $d_L$ in equation~(\ref{Eq3}) is calculated using a flat $\Lambda$CDM cosmological model as follows:
\begin{eqnarray}\label{Eq4}
d_{\rm L} = (1+z)\frac{c}{H_{\rm 0}}\int_0^z \frac{dz'}{\sqrt{(1-\Omega_{\Lambda})(1+z')^3+\Omega_{\Lambda}}}.
\end{eqnarray} 
We use nominal values of the cosmological parameters from Planck data as $H_0 = 70 \, \text{km s}^{-1} \, \text{Mpc}^{-1}$ and $\Omega_{\Lambda} = 0.73$ \citep{2011ApJS..192...14J}. We compute the isotropic-equivalent peak luminosity $L_{\rm iso}$ using the peak bolometric flux $P_{\rm bol}$ as follows:
\begin{eqnarray}\label{Eq5}
L_{\rm iso} = 4\pi d_{\rm L}^2\ {P}_{\rm bol},
\end{eqnarray} 
where ${P}_{\rm bol}$ is obtained from spectral fitting in equation~(\ref{Eq2}). It is important to note that $E_{\rm i,p}$ in this context is calculated from the peak of the ${\rm \nu}f_{\rm \nu}$ spectrum at the time of peak flux.

The Amati and Yonetoku correlations are formulated as follows:
\begin{eqnarray}
\nonumber
&& \frac{E_{\rm i,p}}{100\;  \text{keV}} = 10^{k} \left( \frac{E_{\rm iso}}{10^{52}\; \text{erg}}\right)^{m} \,\, {\rm and} \\ 
&& \frac{E_{\rm i,p}}{100\; \text{keV}} = 10^{k} \left( \frac{L_{\rm iso}}{10^{52}\; \text{erg/s}}\right)^{m},
\label{Eq6}
\end{eqnarray}
where $100$~keV serves as a reference energy, while parameters $m$ and $k$ denote the index and intercept, respectively. We compute the combinations of $E_{\rm iso} - E_{\rm i,p}$ and $L_{\rm iso} - E_{\rm i,p}$ for each GRB in our sample, utilizing the cosmological parameters outlined earlier.

For data fitting, we employ the linearized form of these correlations, expressed as $y = mx + k$, where:
\begin{eqnarray}
\nonumber
&& x = \log_{10}\left(\frac{E_{\rm iso}}{10^{52}\;\text{erg}}\right) \,\,{\rm or}\,\, 
\log_{10} \left( \frac{L_{\rm iso}}{10^{52}\;\text{erg/s}}\right)
\\
&& y = \log_{10}\left(\frac{E_{\rm i,p}}{100\;\text{keV}}\right)\,  .
\label{Eq7}
\end{eqnarray}
We calculate the uncertainty on $y$ using the following formula \citep{Wang2016_585, Demianski2017_693}:
\begin{eqnarray}\label{Eq8}
\sigma_{y} = \sqrt{\sigma_{k}^2+m^2\sigma_{x}^2+\sigma_{m}^2x^2+\sigma_{\rm ext}^2}\,,
\end{eqnarray}
where $\sigma_{\rm ext}$ is the extrinsic uncertainty on $y$, which is treated as an unknown parameter. The terms $\sigma_{x}$ and $\sigma_{y}$ represent the uncertainties associated with the measurements of the $x$ and $y$ data points, respectively. To determine the parameters \(k\), \(m\), and \(\sigma_{\text{ext}}\) we minimize the likelihood function \citep{2005physics..11182D}
\begin{eqnarray}\label{Eq9}
\begin{aligned}
{L}(m,k,\sigma_{\rm ext}) = {} & \displaystyle\frac{1}{2}\sum_{{i}}^N \ln (\sigma^2_{\rm ext}+\sigma^2_{y_{\rm i}}+m^2\sigma^2_{x_{\rm i}}) 
 \\
 &  + \displaystyle\frac{1}{2} \sum_{{i}}^N{\displaystyle\frac{(y_{\rm i}-mx_{\rm i}-k)^2}{(\sigma^2_{\rm ext}+\sigma^2_{\rm y_{i}}+m^2\sigma^2_{x_{\rm i}})}}.
\end{aligned}
\end{eqnarray}
We compute errors on the fitting parameters $k$, $m$, and $\sigma_{\rm ext}$ following \citet{2005physics..11182D}.

\section{Results}
\label{sec:sec_5}
Here, we present results from our joint GBM, LAT, and LLE spectral analyses, which fit Amati and Yonetoku relations and have implications for cosmological parameters.

\subsection{Joint Spectral Analysis}

\begin{figure*}
    \centering

    \hfill
        \begin{subfigure}[b]{0.464\textwidth}
        \centering
        \includegraphics[width=\textwidth]{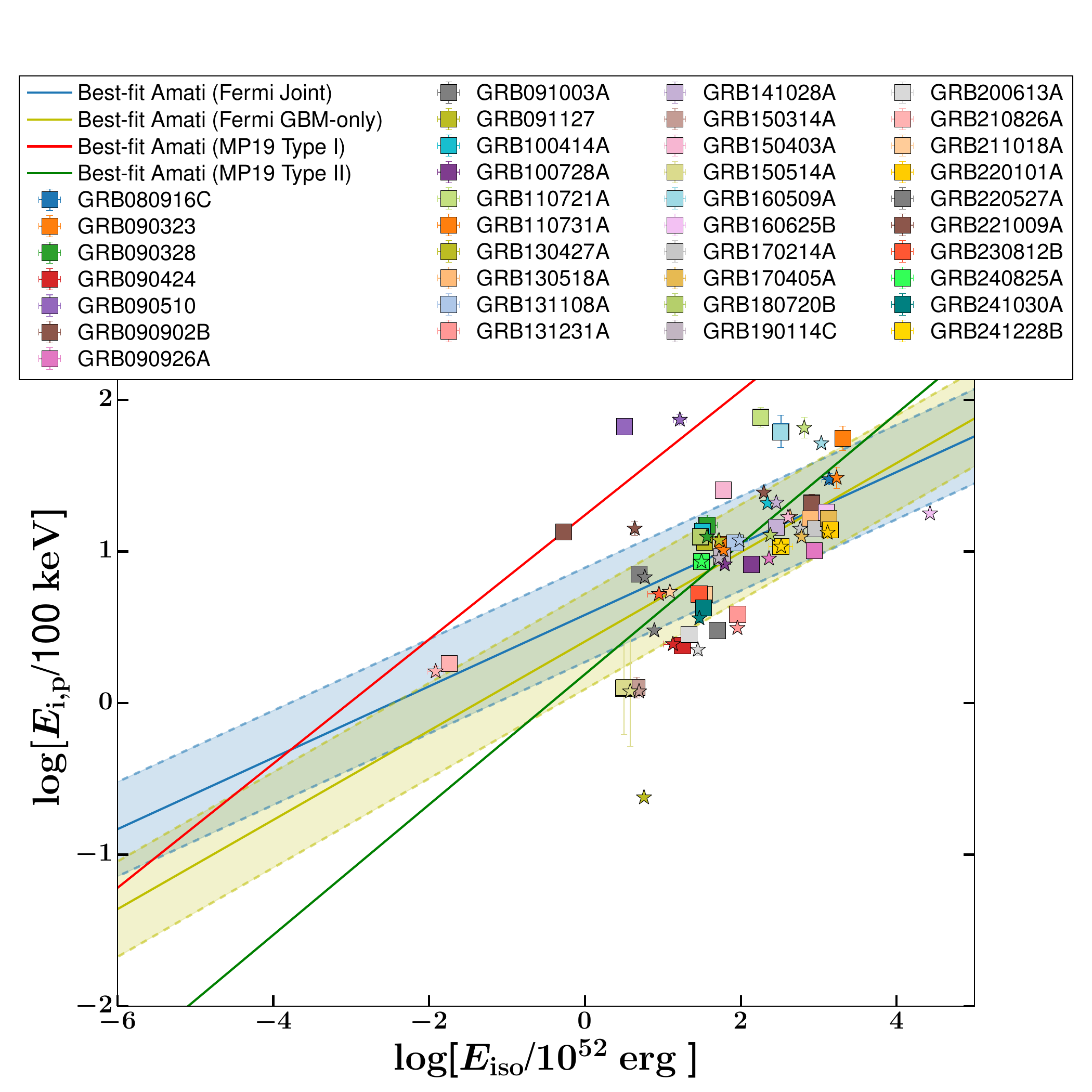}
    \end{subfigure}
    \hfill
        \begin{subfigure}[b]{0.50\textwidth}
        \centering
        \includegraphics[width=\textwidth]{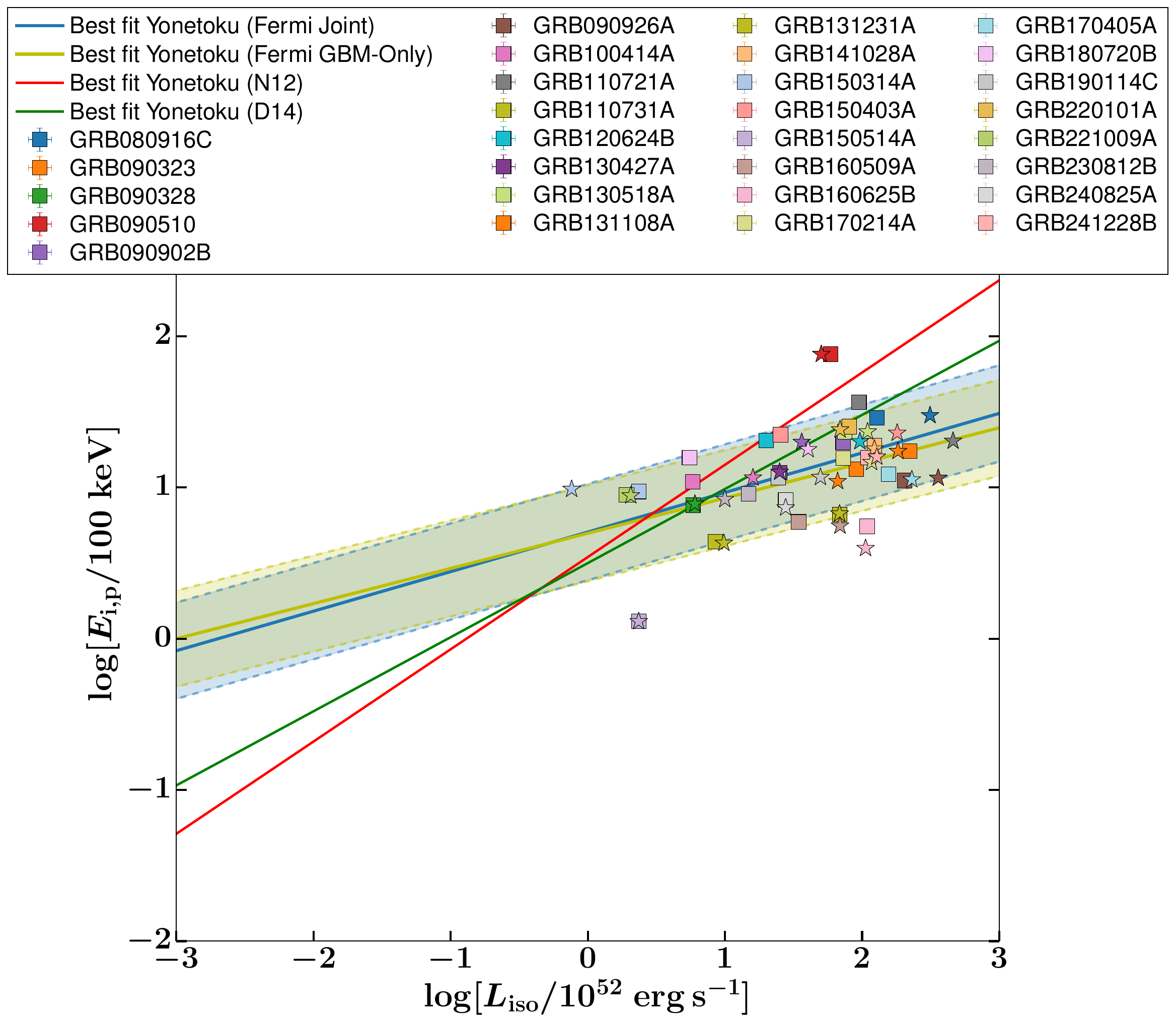}
    \end{subfigure}
    \caption{
        The \({E}_{i,\mathrm{p}} - {E}_{\mathrm{iso}}\) data is shown (left panel) for both LGRBs and SGRBs from our analysis (labeled as ``GRBs'' in the legend) and reported in Table~\ref{tab2}. Data points from the GBM-only fits are represented by stars, while those from the joint GBM+LAT+LLE fits are shown as squares. The \({E}_{i,\mathrm{p}} - {L}_{\mathrm{iso}}\) data (right panel) is also plotted similarly and reported in Table~\ref{tab5}. The Amati and Yonetoku relation fits are shown separately for the joint spectral analyses (blue line) and GBM-only spectral analyses (yellow line). The shaded regions correspond to $1\sigma$ scatter in the fits to data. The fit parameters for the Amati (Yonetoku) relation are reported in Table~\ref{tab3} (Table~\ref{tab6}). The red and green lines in the left (right) panels correspond to fits reported in MP19 (N12 and D14).
}
    
    \label{fig:fig1}
\end{figure*}

We present in Table \ref{tab1} the spectral fits for the GRB sample using different spectral models, with the best-fit model determined as outlined in Sec~\ref{sec:3_1} for the fluence over ${\rm T}_{90}$. The corresponding \(E_{\rm i,p}\)–\(E_{\rm iso}\) values and fitting parameters are listed in Tables~\ref{tab2} and \ref{tab3}. All count spectra and the corresponding best-fit model plots are provided in Appendix~\ref{apx2}. Similarly, we present results for the spectra during a one-second peak flux time interval in Table~\ref{tab4}, with the corresponding \(E_{\rm i,p}\)–\(L_{\rm iso}\) values and fitting parameters listed in Tables~\ref{tab5} and \ref{tab6}. In a similar manner, in Appendix~\ref{apx3} we present the spectral fitting results for the peak flux interval of 29 GRBs. The reduction in the sample size occurred because the one-second time-binning failed to capture photons in the LAT or LAT-LLE data for some GRBs. As a result, the final sample consists of the remaining GRBs for which spectral data were available.

Note that, in both of these tables, each GRB was fitted twice: first, using a joint analysis (GBM+LAT+LLE) and then, an analysis of GBM data alone. A similar comparison was also carried out in \cite{2025A&A...700A..88M}. This approach was employed specifically to evaluate the impact of including LAT and LLE data, allowing us to examine potential systematic differences in key spectral parameters, such as the peak energy ($E_{\rm{p}}$) and spectral indices, as well as to assess the fit quality and whether the joint analysis provides any significant improvement over the GBM-only fit. The fit quality is quantified using the BIC and $\Delta$BIC values (calculated with respect to the Band model). Finally, we report the bolometric flux during ${\rm T}_{90}$ in Table~\ref{tab1} and during the time interval of peak flux in Table~\ref{tab4}.

Several GRBs, including GRB\,080916C, GRB\,090510, GRB\,110721A, and GRB\,160509A, as well as others listed in Tables~\ref{tab1} and~\ref{tab4}, show clear differences between the two fitting approaches.
In these cases, the inclusion of LAT data often results in a higher $E_{\rm{p}}$ or a preference for more complex models (e.g., Band+PL instead of Band alone). This improvement is likely attributed to LAT's sensitivity to high-energy photons, which provides tighter constraints on the high-energy tail of the spectrum and enables the identification of additional spectral components not apparent in GBM-only fits. Conversely, for GRBs with low LAT significance (e.g., low TS values), the results from the joint and GBM-only fits remain largely consistent, indicating that the benefits of joint fitting are closely tied to the strength of LAT detections. 

Furthermore, based on the analysis of the Amati and Yonetoku correlations, approximately 44.7\% and 27.6\% of the GRBs, respectively, show a reduction in ${E}_{i, \mathrm{p}}$ uncertainty when using the joint fits {(GBM+LAT+LLE)}. These findings emphasize the importance of incorporating LAT data to enhance the precision of spectral parameter estimates and reduce model degeneracies.

\subsection{Amati Relation}\label{sec:results_Amati}
\begin{figure*}
    \centering

    \hfill
    \begin{subfigure}[b]{0.47\textwidth}
        \centering
    \includegraphics[width=\textwidth]{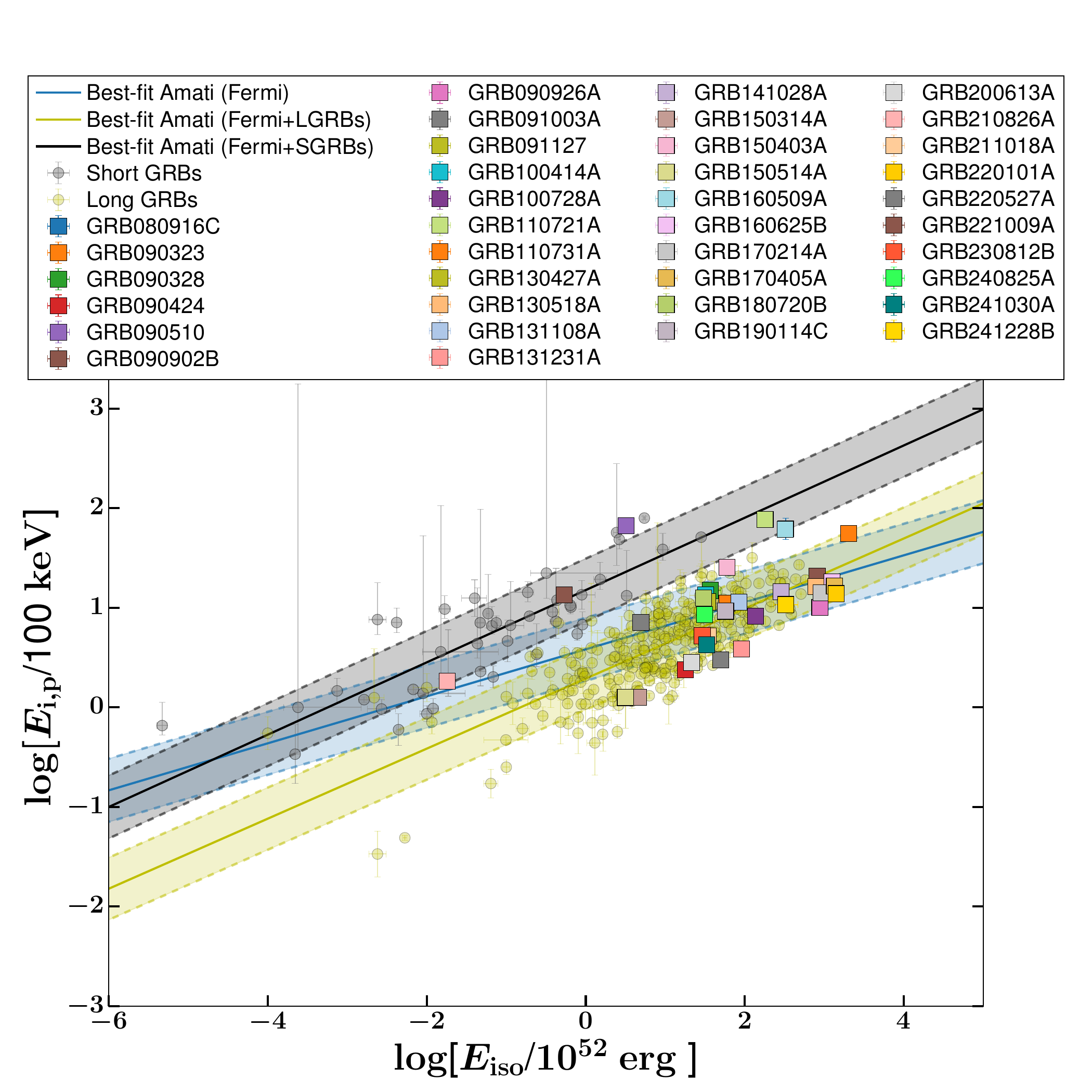}

    \end{subfigure}
    \hfill
    \begin{subfigure}[b]{0.47\textwidth}
        \centering
        \includegraphics[width=\textwidth]{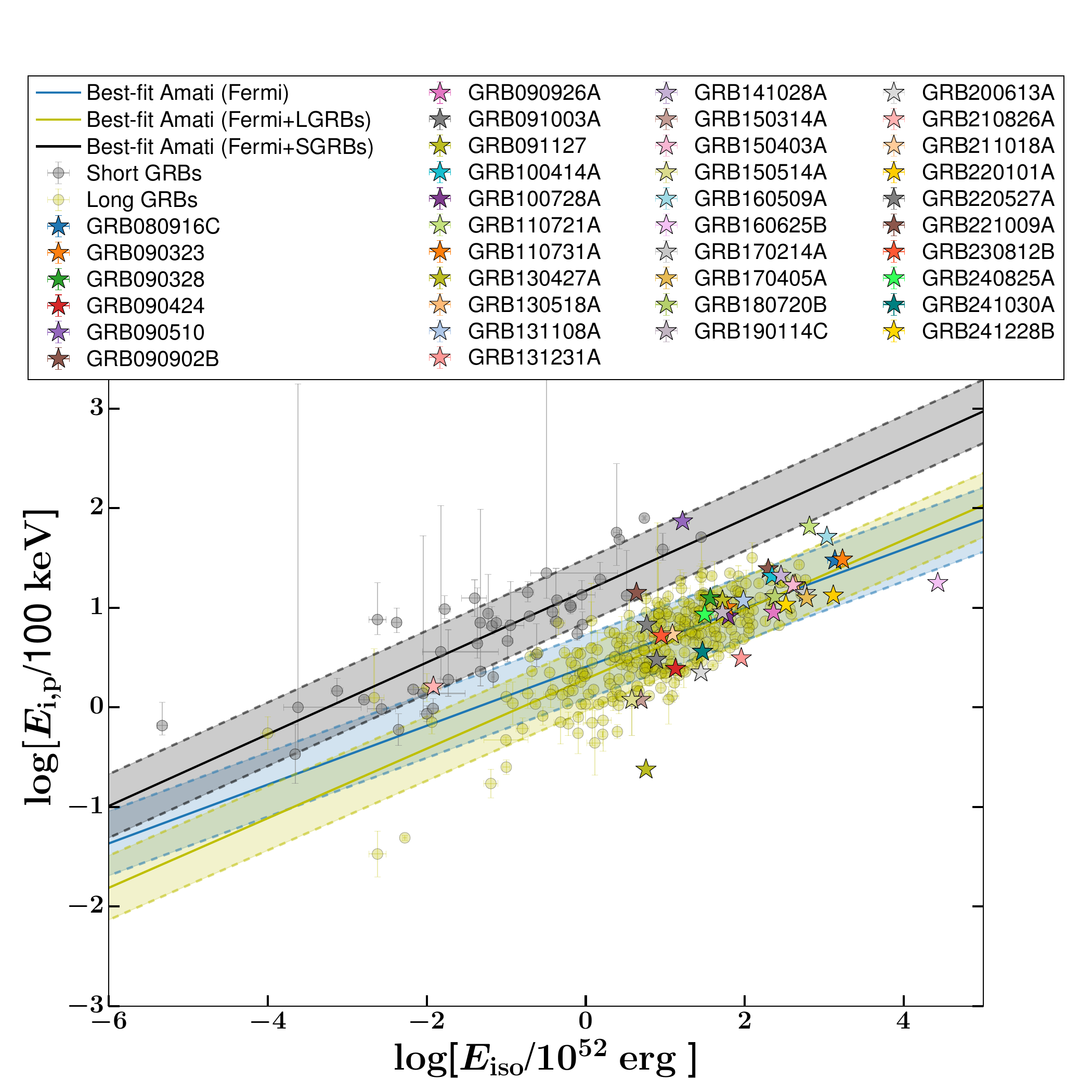}
    \end{subfigure}
    \caption{
    The \({E}_{i,\mathrm{p}} - {E}_{\mathrm{iso}}\) data for 36 LGRBs and 1 SGRB from our analyses as shown in the left panel of Figure~\ref{fig:fig1}. Here we combine our data with 275 LGRBs (yellow markers) and 45 SGRBs (gray markers) from MP19. The left (right) panel uses results from our joint GBM+LAT+LLE (GBM-only) fits to the spectra. The Amati relation fits are for the LGRBs from our analysis (blue line), the combined LGRBs (yellow line), and the combined SGRBs (gray line). The fit parameters are reported in Table~\ref{tab3}. 
}
    
    \label{fig:fig2}
\end{figure*}

We use the spectral analysis results reported in Table~\ref{tab1} and redshift information to calculate \({E}_{i,\mathrm{p}}\) and \({E}_{\mathrm{iso}}\) using equation~(\ref{Eq3}). 
These values, together with their corresponding errors, are shown in Tables~\ref{tab2} for both the joint GBM+LAT+LLE fits and GBM-only fits, and are plotted in Figure~\ref{fig:fig1} (left panel), which shows a significant difference between the two spectral analysis results. In general there is substantial scatter in the \({E}_{i,\mathrm{p}}\) -- \({E}_{\mathrm{iso}}\) plane for both the joint and GBM-only analyses results.  This is likely due to the limited number of GRBs in our sample. 
Figure~\ref{fig:fig2} shows a combination of \fermi-GRBs with a joint (GBM+LAT+LLE) fit (left panel) and a GBM-only fit (right panel) together with data for 45 short and 275 LGRBs reported in \citet{short_long_grbs}, MP19 from hereafter. MP19 used a large sample of GRBs with known redshifts from 1997 to 2019, collected from various telescopes (Konus-Wind, BeppoSAX, BATSE/CGRO, HETE-2, {\it Swift}, and \fermi), to investigate the Amati correlation.

We run a Markov Chain Monte Carlo (MCMC) method with \texttt{emcee} \citep{Foreman_Mackey_2013} using the likelihood function as described in 
equation~(\ref{Eq9}) to the \({E}_{i,\mathrm{p}} - {E}_{\mathrm{iso}}\) data. We adopt uninformative flat priors for the parameters. The prior range are defined as follow: $m,k \in [0.0,0.1]$ and $\sigma_{\rm ext} \in [0.0,2.0]$.

The results of the Amati correlation fits for different combinations of the GRB sample analyzed in this work and in MP19 are reported in Table~\ref{tab3}. Note that there is a significant difference between the slope $m$ for the joint fit and GBM-only fit cases for the 36 \fermi-LGRBs in our sample. This difference, however, disappears when our sample is combined with the 275 LGRBs in MP19. The fits to the combined SGRBs and LGRBs give very similar values for $m$.

\begin{figure*}
    \centering
    \begin{subfigure}[b]{0.45\textwidth}
        \centering
        \includegraphics[width=\textwidth]{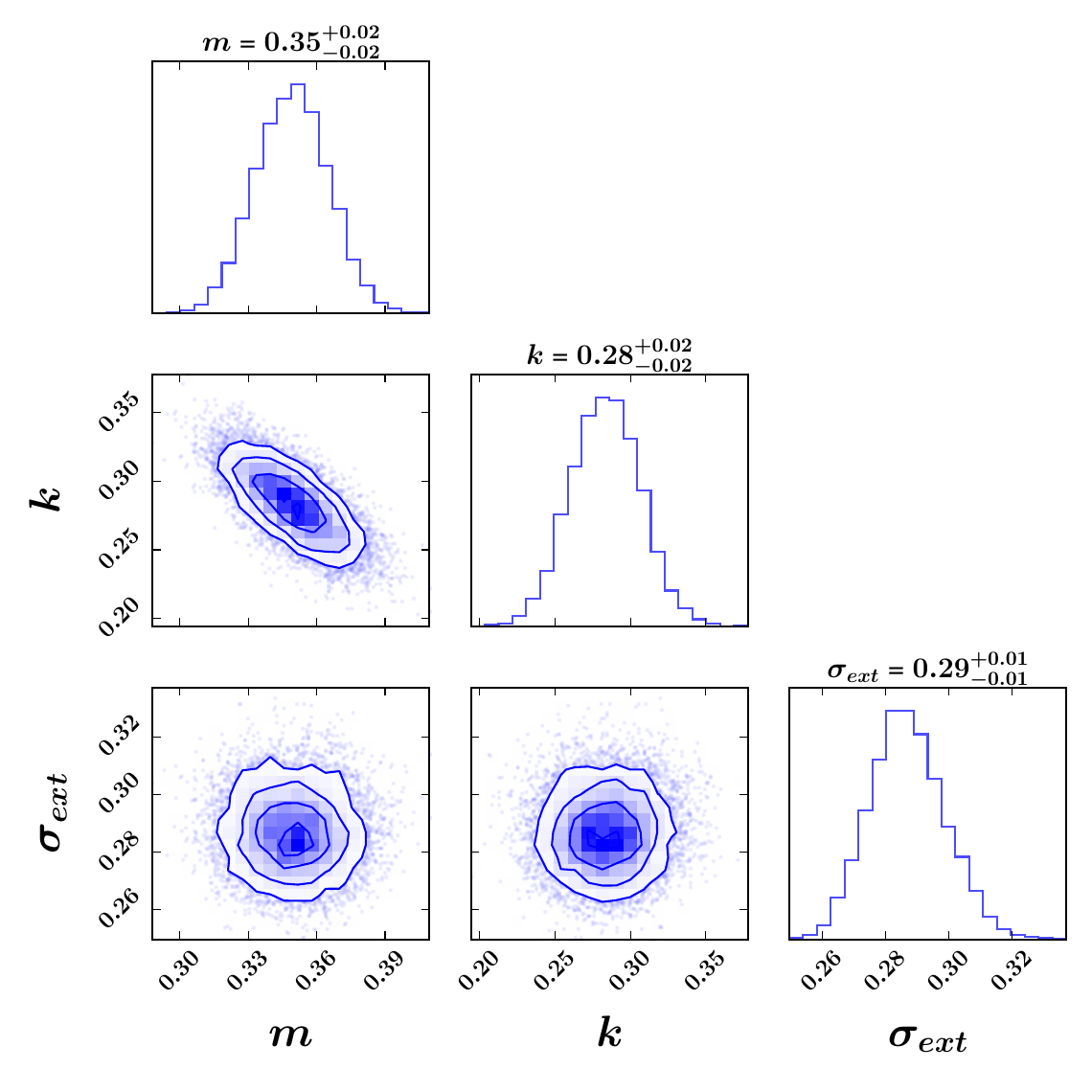}
    \end{subfigure}
    \hfill
    \begin{subfigure}[b]{0.45\textwidth}
        \centering
        \includegraphics[width=\textwidth]{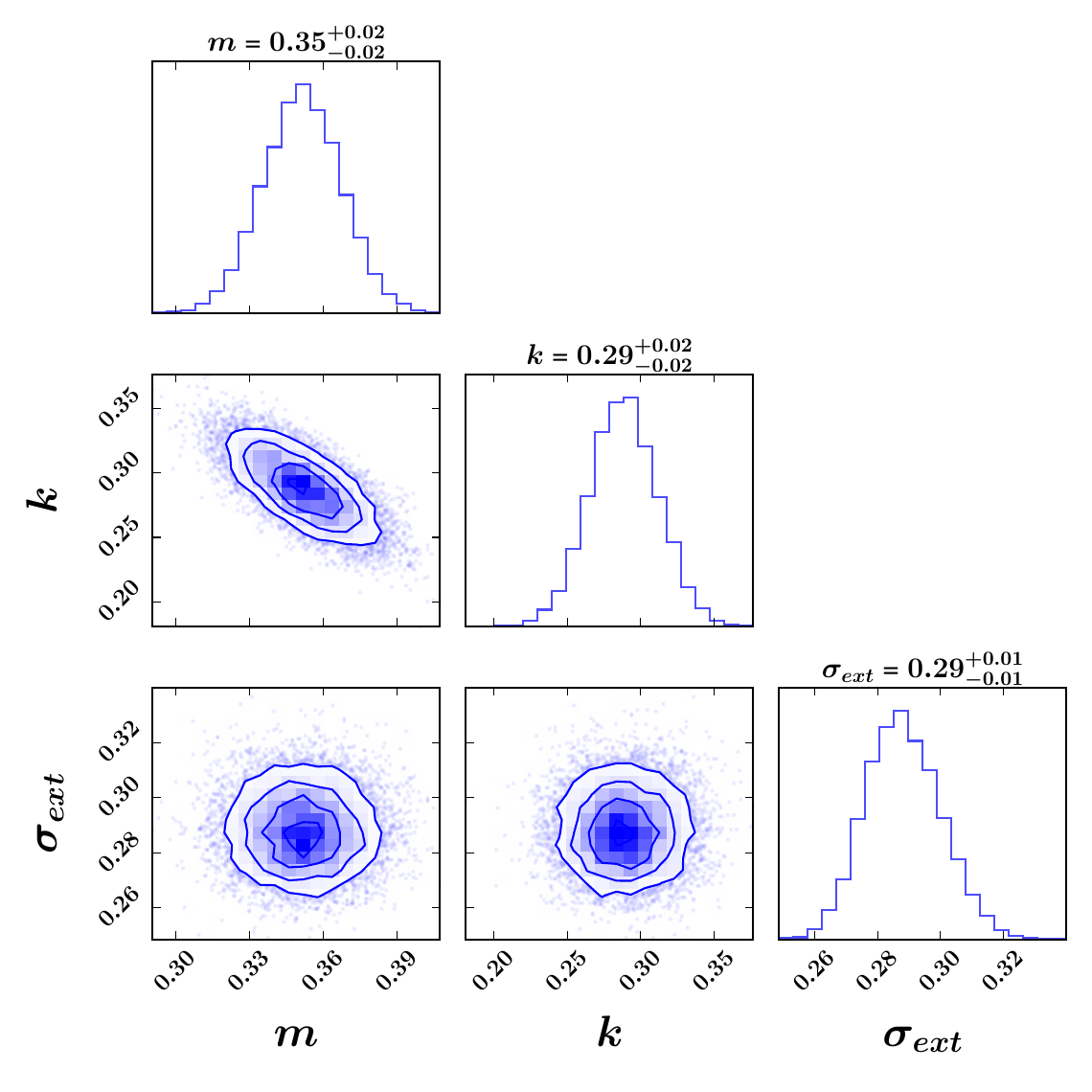}
    \end{subfigure}
    \caption{Corner plots of the fitting parameters of the Amati relation, namely \(k\), \(m\), and \(\sigma_{\rm ext}\), were determined using the MCMC method for 36 \fermi-LGRBs and 275 MP19-LGRBs. These are also reported in Table~\ref{tab3}. 
    The left panel displays results from the fits of the GBM-only dataset, while the right panel corresponds to fits derived from GBM+LAT+LLE joint analysis.} 
    \label{fig:fig3}
\end{figure*}

In Figure~\ref{fig:fig1} (left panel), we also show the Amati relation fits along with 1$\sigma$ scatter for the case of 36 \fermi-LGRBs using GBM-only fit data (yellow shaded region) and joint fit data (blue shaded region). In Figure~\ref{fig:fig2} (both panels), we show 36 \fermi-LGRBs and 275 MP19-LGRBs (yellow shaded region), and 1 \fermi-SGRB and 45 MP19-SGRBs (gray shaded region). The left panel shows the joint (GBM+LAT+LLE) fits, and the right panel is for the GBM-only fits. In Figure~\ref{fig:fig3} we show corner plots of the Amati correlation parameters in the case of 36 \fermi-LGRBs and 275 MP19-LGRBs. Again, the left (right) panel is for GBM-only (GBM+LAT+LLE) fit results.

\subsection{Yonetoku Relation}\label{sec:5_3}
\begin{figure*}
    \centering
    \begin{subfigure}[b]{0.47\textwidth}
        \centering
        \includegraphics[width=\textwidth]{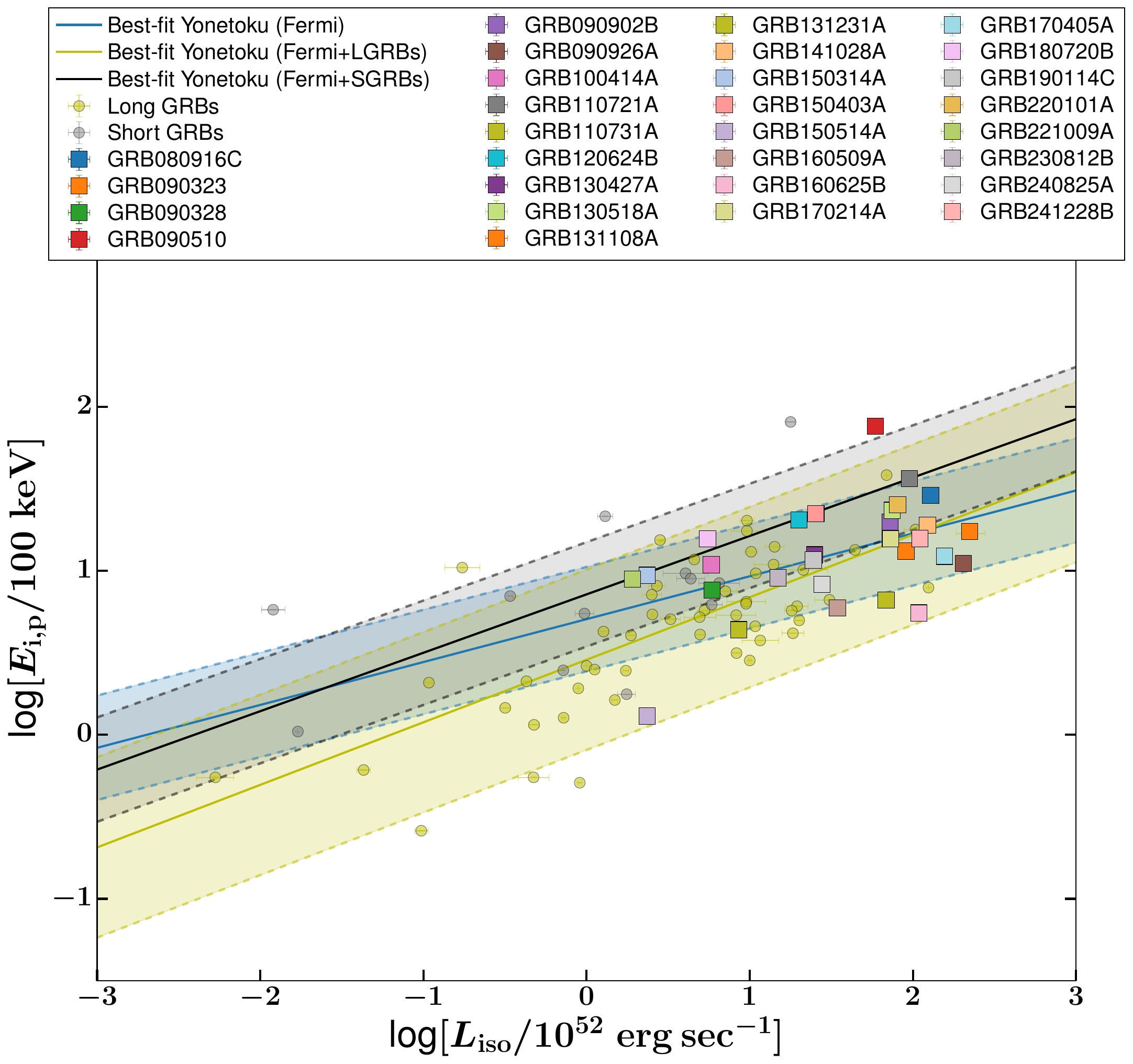}
    \end{subfigure}
    \hfill
        \begin{subfigure}[b]{0.47\textwidth}
        \centering
        \includegraphics[width=\textwidth]{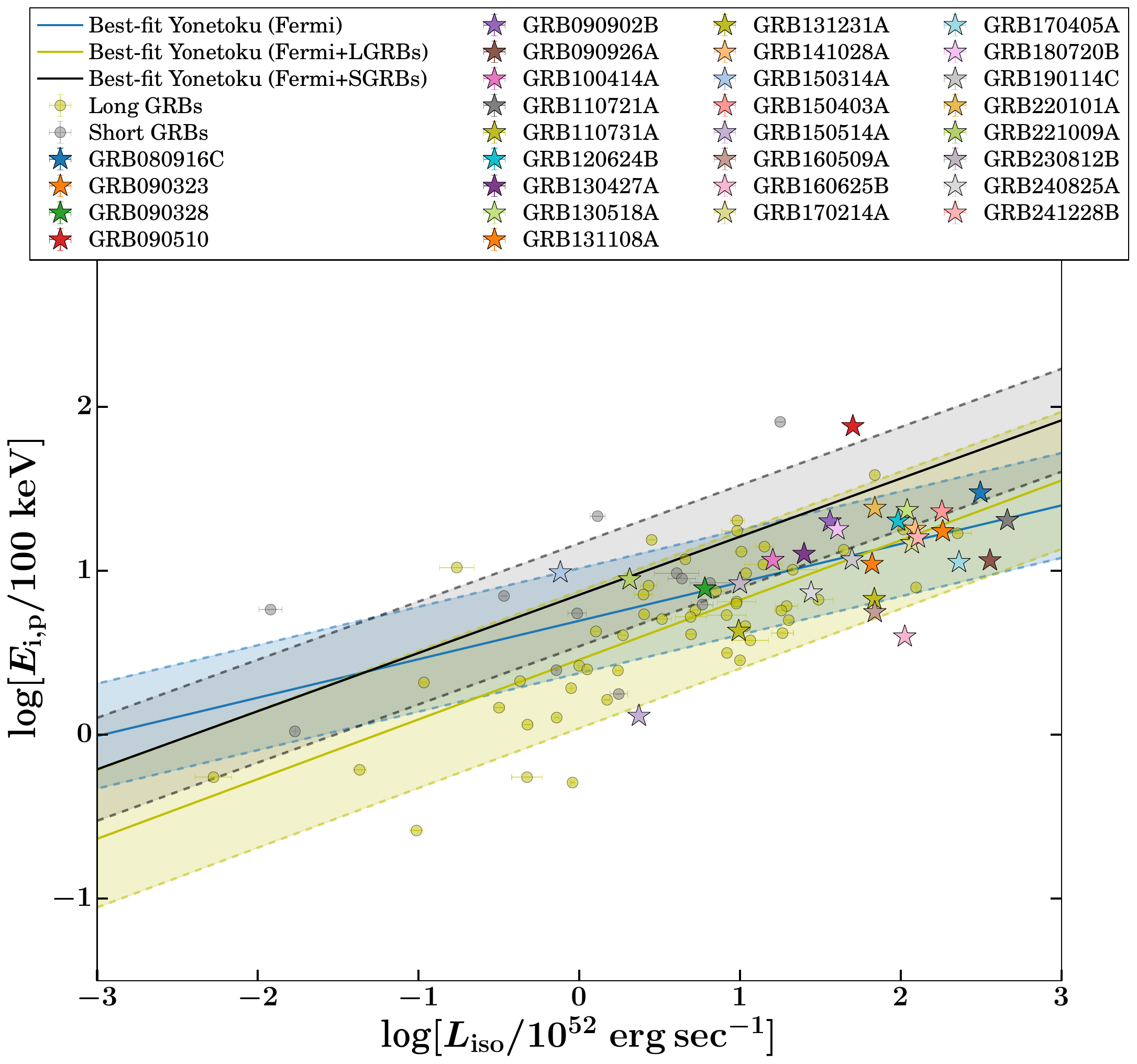}
    \end{subfigure}%
    \caption{The \({E}_{i,\mathrm{p}} - {L}_{\mathrm{iso}}\) data for 28 LGRBs and 1 SGRB from our analyses as shown in the right panel of Figure~\ref{fig:fig1}. Here we combine our data with 58 LGRBs (yellow markers) from N12 and 12 SGRBs (gray markers) from D14. The left (right) panel uses results from our joint GBM+LAT+LLE (GBM-only) fits to the spectra. The Yonetoku relation fits are for the LGRBs from our analysis (blue line), the combined LGRBs (yellow line), and the combined SGRBs (gray line). The fit parameters are reported in Table~\ref{tab6}. 
}
    \label{fig:fig4}
\end{figure*}

Similar to the Amati relation discussed in Sec.~\ref{sec:results_Amati}, we use the spectral analysis results for the peak flux reported in Table~\ref{tab4} to calculate \(E_{i,\rm p}\) and \(L_{\rm iso}\) for the Yonetoku relation. The calculated values, along with their respective uncertainties, are presented in Table~\ref{tab5} for both the joint GBM+LAT+LLE fits and the GBM-only fits. These results are also shown in Figure~\ref{fig:fig1} (right panel) for 28 \fermi-LGRBs, which again shows a significant difference between 
\(L_{\rm iso}\) for some GRBs. 
Figure~\ref{fig:fig4} shows the combination of \fermi-GRBs with the GBM+LAT+LLE joint fit data (left panel) and GBM-only fit data (right panel) together with data for 58 LGRBs reported in \citet{2012MNRAS.421.1256N}, referred to as N12 hereafter, and 12 SGRBs reported in \citet{2014MNRAS.442.2342D}, referred to as D14 hereafter.

In Table~\ref{tab6} we report results of the Yonetoku correlation fits for \fermi-LGRBs and show them with $1\sigma$ scatter in Figure~\ref{fig:fig1} (right panel) for the GBM-only fit data (yellow shaded region) and joint fit data (blue shaded region). Note that, unlike in the Amati correlation case, the values of $m$ in the Yonetoku correlation for the GBM-only and joint fits are very similar for the 26 \fermi-LGRBs.

\begin{figure*}
    \centering
        \begin{subfigure}[b]{0.45\textwidth}
        \centering
        \includegraphics[width=\textwidth]{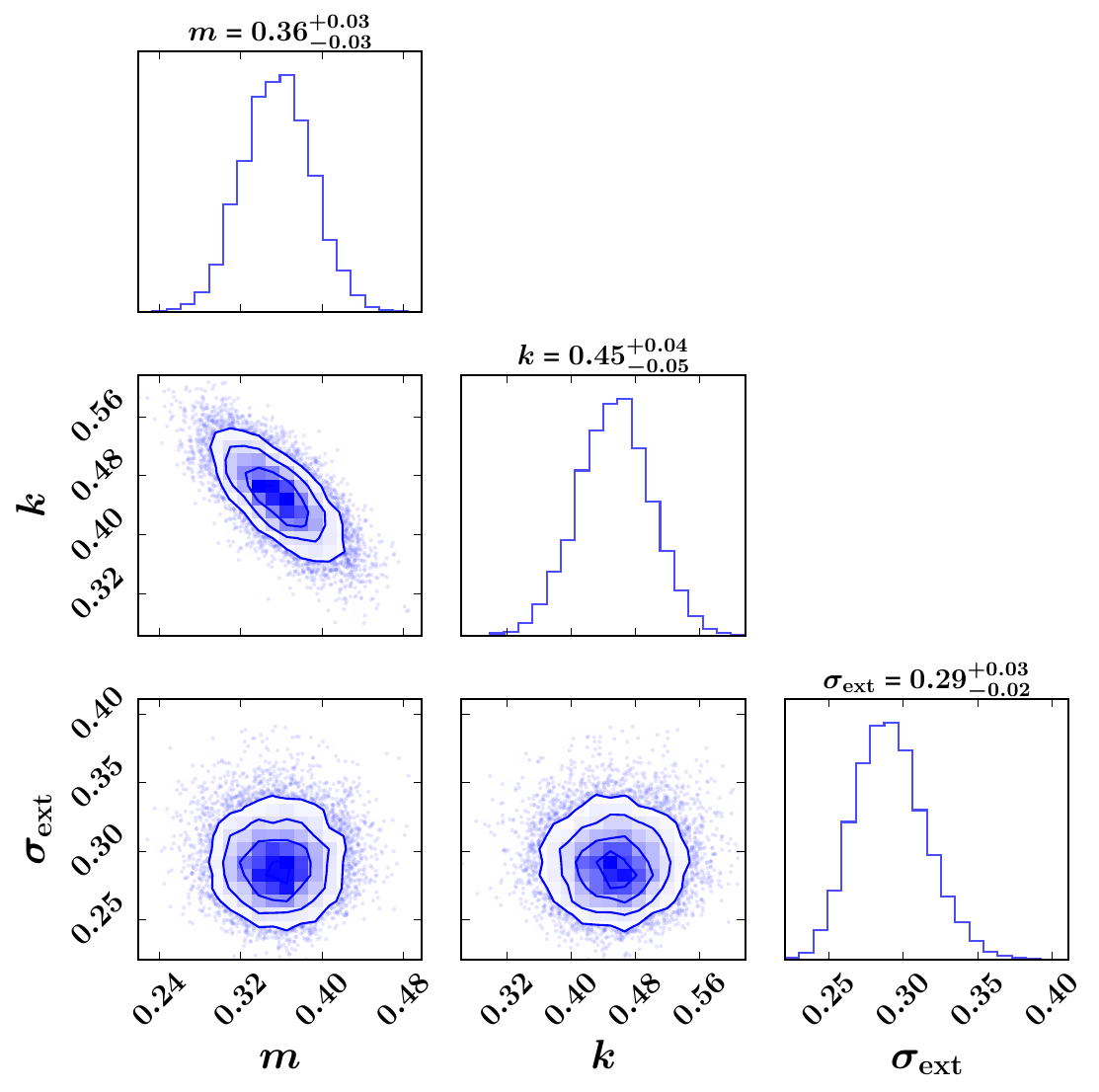}
    \end{subfigure}
    \hfill
     \begin{subfigure}[b]{0.45\textwidth}
        \centering
        \includegraphics[width=\textwidth]{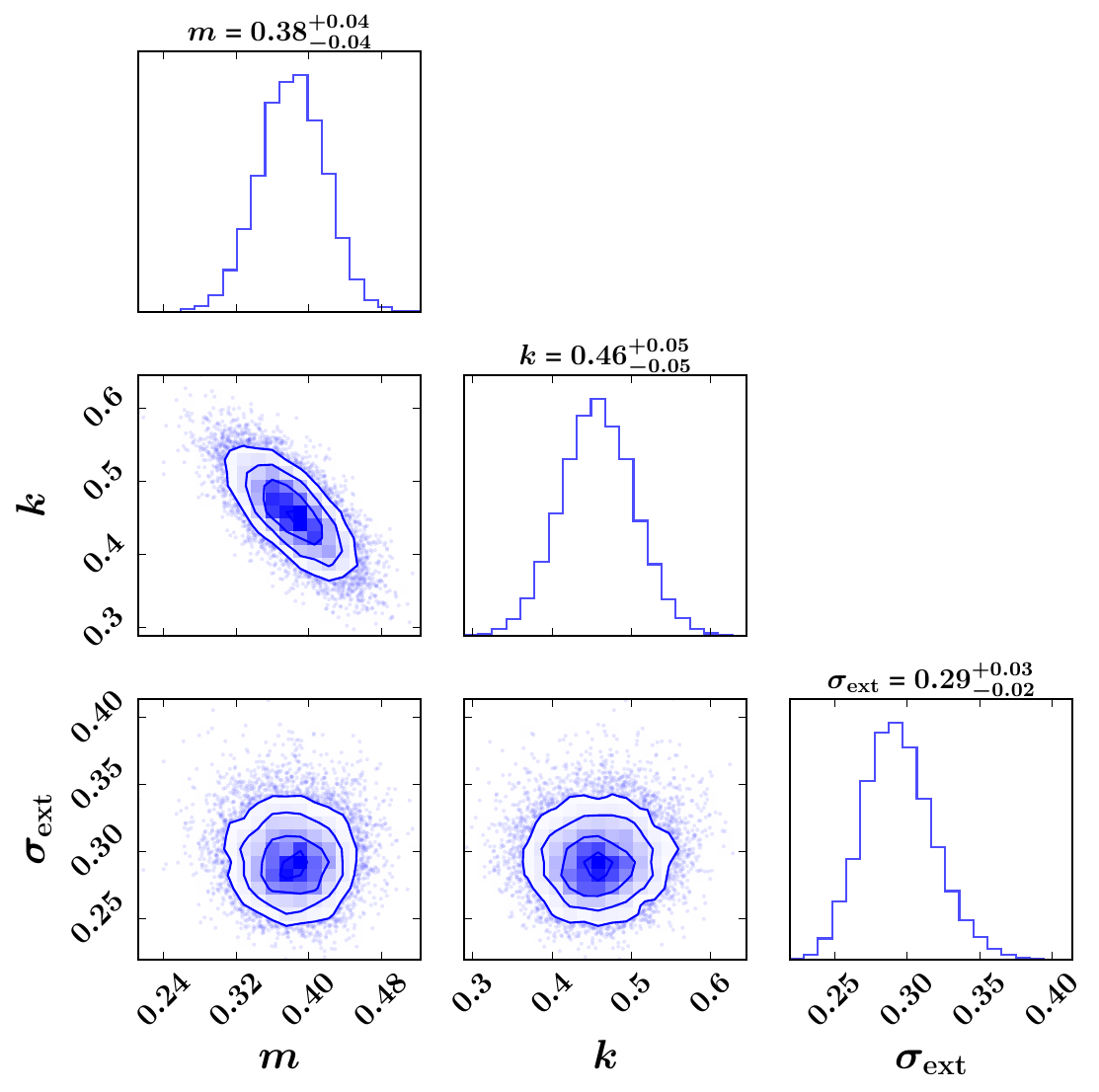}
    \end{subfigure}
    \caption{
    Corner plots of the fitting parameters of the Yonetoku relation, namely \(k\), \(m\), and \(\sigma_{\rm ext}\), which were determined using the MCMC method for 28 \fermi-LGRBs and 58 N12-LGRBs. These are also reported in Table~\ref{tab6}. The left panel displays results from the  GBM-only fits, while the right panel corresponds to fits derived from the GBM+LAT+LLE fits.} 
    \label{fig:fig5}
\end{figure*}

In Figure~\ref{fig:fig4} we show the Yonetoku correlation fits with 28 \fermi-LGRBs and 58 N12-LGRBs (yellow shaded region), and 12 D14-SGRBs (gray shaded region). The left (right) panel is for the joint LAT+GBM+LLE (GBM-only) fit results for the \fermi-GRBs. Figure~\ref{fig:fig5} shows corner plots from MCMC analysis for the case of 26 \fermi-LGRBs and 58 N12-LGRBs. The left (right) panel is for GBM-only (join LAT+GBM+LLE) fit results for the \fermi-GRBs. The best-fit parameter values are reported in Table~\ref{tab6}.

\subsection{Constraints on cosmological parameters}\label{sec:5_4}
Once the parameters of the Amati and Yonetoku correlation are obtained, we can use GRBs as {standardizable} candles to constrain cosmological parameters such as $H_0$ and $\Omega_\Lambda$.

\subsubsection{Amati correlation}
\label{Amati_cosmo}
In the case of the Amati relation, we can invert equation~(\ref{Eq3}), together with equation~(\ref{Eq6}), to obtain an experimental calculation of the luminosity distance as
\begin{align}\label{Eq12}
d_L &= \left[ \frac{1+z}{4\pi S_{\rm bol}} \, E_{\rm iso} \right]^{1/2} \,,\notag \\
\Rightarrow \log_{10}(d_L) &= \tfrac{1}{2} \log_{10}\!\left(\frac{1+z}{4\pi S_{\rm bol}}\right) + \tfrac{1}{2} \log_{10}(E_{\rm iso})\,.
\end{align}

\begin{figure}
    \centering
    \begin{subfigure}[b]{0.48\textwidth}
        \centering
        \includegraphics[width=\textwidth]{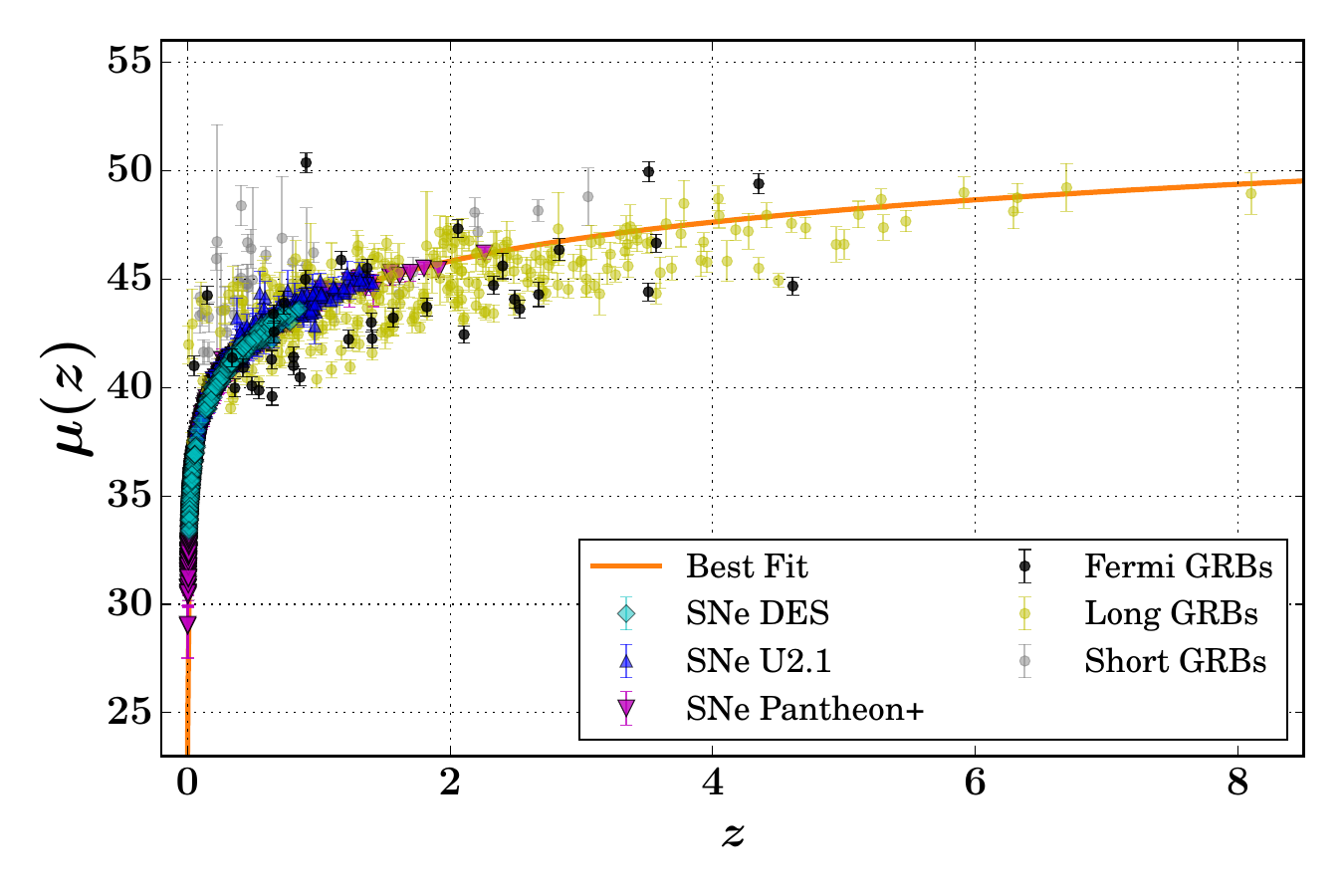}
    \end{subfigure}
    \hskip -0.5cm
    \caption{Distance modulus from the Amati correlation for 
    The \fermi-GRBs (dark gray points) with MP19-LGRBs (yellow points) and MP19-SGRBs (light gray points). Also shown are the SNe U2.1 data \citep[blue points;][]{Suzuki_2012}, DES SNe sample \citep[cyan color points;][]{DES_sne}, and Pantheon+ \citep[purple points;][]{2022ApJ...938..113S}. The orange curve corresponds to $\mu^{\rm th}$ with the best-fit $H_0$ and $\Omega_\Lambda$ values we obtained from the MCMC fit to the combined GRB and SNe data, as shown in Fig.~\ref{fig:fig7}.}
    \label{fig:fig6}
\end{figure}

On the other hand, using the Amati relation, we can write 
\begin{align}
   \log_{10}(E_{\rm iso}) &= 52 + \frac{1}{m}\,\log_{10}\!\left(\frac{E_{i,\rm p}}{100~\rm keV}\right) - \frac{k}{m} \,,
\end{align}
where $E_{\rm iso}$ is measur{e}d in ergs. This expression can be inserted in equation~(\ref{Eq12}) to calculate $d_L$. We can therefore use this experimental $d_{\rm L}$ to calculate the distance modulus as a function of redshift as $\mu_{\rm GRB}(z) = 5\log_{10}(d_{\rm L}(z)/ 1\; \text{Mpc}) + 25$. In Figure~\ref{fig:fig6}, we show these calculated distance modulus for our sample of \fermi-GRBs, combined with MP19-GRBs  from \citet{short_long_grbs}.
Also shown in Figure~\ref{fig:fig6} are the distance moduli for Type-Ia supernovae from the SNe U2.1 sample \citep{Suzuki_2012}, from the Dark Energy Survey \citep[DES\footnote{\url{https://www.darkenergysurvey.org/the-des-project/data-access/}};][]{DES_sne} and the latest Pantheon+\footnote{\url{https://github.com/PantheonPlusSH0ES/DataRelease}} sample \citep{2022ApJ...938..113S}, which comprises 1701 data entries corresponding to 1550 unique SNe Ia data.

Next, {we simultaneously constrain the cosmological parameters ($H_{0}$ and $\Omega_{\Lambda}$ ) of a flat $\Lambda$CDM model along with the correlation parameters of the Amati relation (\(k, m, \, \text{and } \sigma_{\rm ext}\))}  using joint spectral analysis results of \fermi-GRBs together with MP19-GRBs, as well as SNe U2.1, DES SNe, and {Pantheon+ SNe} data. We use the MCMC method with flat priors, and  with a Log-Likelihood given by
\begin{eqnarray}\label{Eq13}
\ln{L} = \ln{L_{\rm GRBs}} + \ln{L_{\rm SNe}}\,
\end{eqnarray} 
where
\begin{align}\label{Eq14}
\ln{L_{\rm GRBs}} = 
& - \frac{1}{2} \sum_{i}^{N} \bigg[ \ln{\left(\sigma_{\rm \mu_{\rm GRB}(z_{i})}^{2}\right)} \notag \\
& + \frac{\left(\mu_{\rm GRB}(z_{i}; m, k) - \mu^{\rm th}(z_{i};H_{0},\Omega_{\Lambda})\right)^{2}}{\sigma_{\rm \mu_{\rm GRB}(z_{i})\, }^{2}}  \bigg]\,.
\end{align}
And the error on $\mu_{\rm GRB}$ is given by
\begin{eqnarray}\label{Eq15}
\sigma_{\rm \mu_{\rm GRB}}^{{2}} = \left( \frac{5}{2}\,\frac{\sigma_{S_{\rm bol}}}{S_{\rm bol} \ln 10} \right)^2
+ \left( \frac{5}{2m}\,\frac{\sigma_{E_{\rm i,p}}}{E_{\rm i,p} \ln 10} \right)^2
+ \sigma^2_{{\rm ext}}.
\end{eqnarray}
We calculate $\ln{L_{\rm SNe}}$ from the published values of $\mu$ and $\sigma_{\mu}$ by U2.1 SNe \citet{Suzuki_2012}, DES SNe \citet{DES_sne}, {and Pantheon+ SNe \citep{2022ApJ...938..113S}} data.

\begin{figure*}
    \centering
    \begin{subfigure}[b]{0.80\textwidth}
        \centering
        \includegraphics[width=\textwidth]{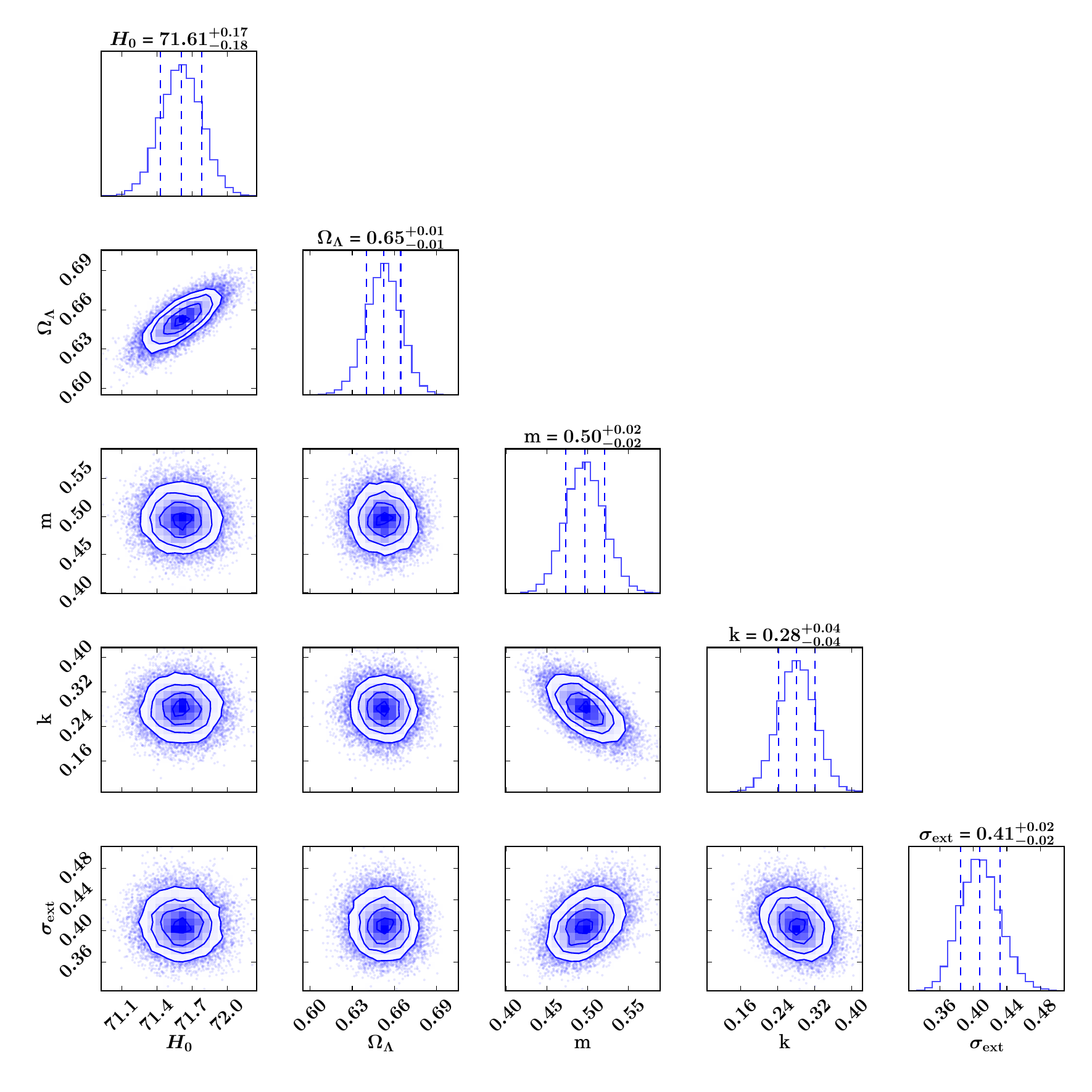}
    \end{subfigure}
    \caption{Corner plots of the estimated cosmological parameters \( H_0 \), \( \Omega_\Lambda \), \(k\), \(m\), and \(\sigma_{\rm ext}\) using MCMC based on the Amati correlation obtained for \fermi-LGRBs (joint fits) and MP19-LGRBs.}
    \label{fig:fig7}
\end{figure*}

In Figure~\ref{fig:fig7} we show the corner plots for the parameters estimated by MCMC, resulting in $H_0 = 71.61 \pm 0.17$~km~s$^{-1}$~Mpc$^{-1}$ and $\Omega_\Lambda = 0.65 \pm 0.01$. These values are then used to calculate $\mu^{\rm th}$ (orange curve), in Figure~\ref{fig:fig7}. We also estimated these parameters in the case of using GBM-only fits to \fermi-LGRBs, resulting in $H_0$ and $\Omega_\Lambda$  the same as in the joint-fit case. 


\subsubsection{Yonetoku correlation}
\label{Yonetoku_Cosmo}
\begin{figure}
    \centering
    \begin{subfigure}[b]{0.48\textwidth}
        \centering
        \includegraphics[width=\textwidth]{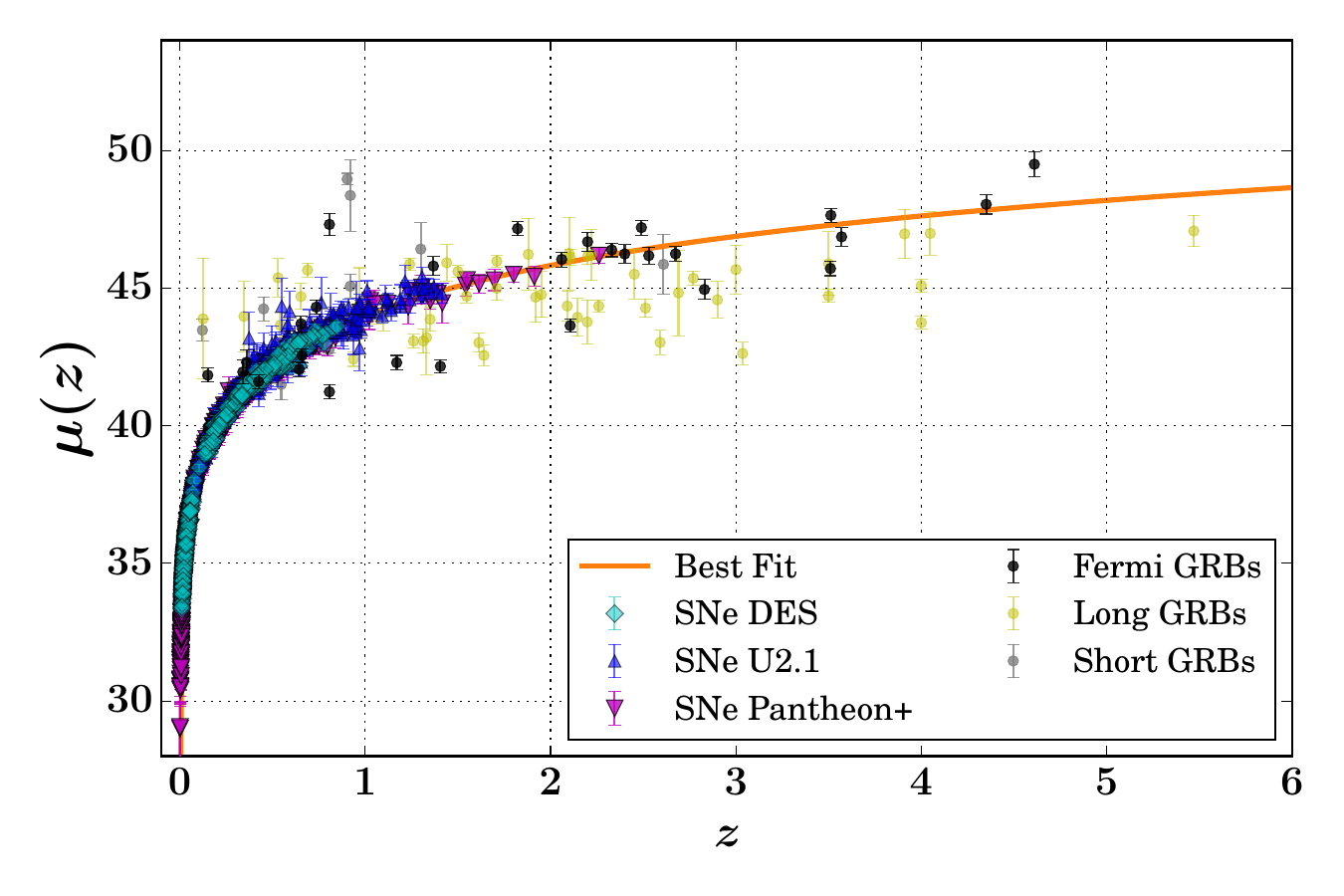}
    \end{subfigure}

    \caption{Distance modulus from the Yonetoku correlation for the \fermi-GRBs (dark gray points) with N12-LGRBs (yellow points) and D14-SGRBs (light gray points). Also shown are the SNe U2.1 (blue points; \citet{Suzuki_2012}), DES SNe sample \citep[cyan color points;][]{DES_sne}, and Pantheon+ \citep[purple points;][]{2022ApJ...938..113S}. The orange line
    corresponds to $\mu^{\rm th}$ with the best-fit $H_0$ and $\Omega_\Lambda$ values we obtained from the MCMC fit to the combined GRB and SNe data, as shown in Figure~\ref{fig:fig8}.}
    \label{fig:fig9}
\end{figure}

\begin{figure*}
    \centering
    \begin{subfigure}[b]{0.80\textwidth}
        \centering
        \includegraphics[width=\textwidth]{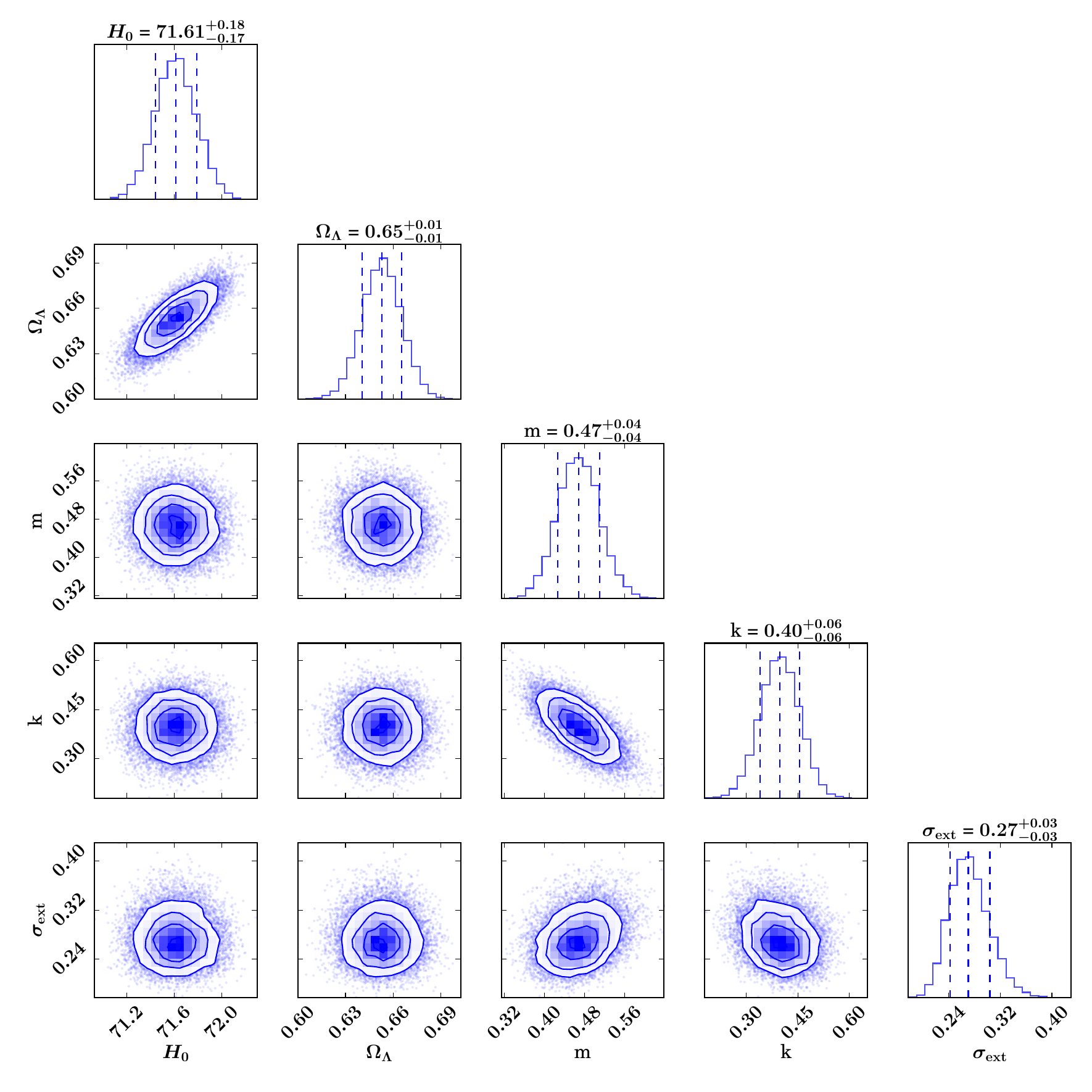}

    \end{subfigure}

    \caption{Corner plots of the estimated cosmological parameters \( H_0 \), \( \Omega_\Lambda \), \(k\), \(m\), and \(\sigma_{\rm ext}\) using MCMC based on the Yonetoku correlation obtained for \fermi-LGRBs (joint fits) and N12-LGRBs.
}
    \label{fig:fig8}
\end{figure*}

In the case of the Yonetoku correlation, we use Eqs.~(\ref{Eq5}) and (\ref{Eq6}) to compute experimental $d_{\rm L}$ as
\begin{eqnarray}\label{Eq18}
d_L &= \left[ \frac{{L_{\rm iso}}}{4\pi P_{\rm bol}} \right]^{1/2}.
\end{eqnarray} 
We calculate the distance modulus, \(\mu(z)\), as before, and calculate its uncertainty using the error propagation method as
\begin{eqnarray}\label{Eq19}
\sigma_{\rm \mu_{\rm GRB}}^{{2}} = \left( \frac{5}{2}\,\frac{\sigma_{P_{\rm bol}}}{P_{\rm bol} \ln 10} \right)^2
+ \left( \frac{5}{2m}\,\frac{\sigma_{E_{i,p}}}{E_{i,p} \ln 10} \right)^2
+ \sigma^2_{{\rm ext}}.
\end{eqnarray}
We show in Figure~\ref{fig:fig9} the distance modulus derived from the Yonetoku correlation results with joint spectral fits to \fermi-GRBs along with 58 N12-LGRBs and 12 D14-SGRBs. We also show distance modulus of SNe data from \citet{Suzuki_2012}, DES \citet{DES_sne},
and Pantheon+ SNe \citep{2022ApJ...938..113S} data. 
In Figure~\ref{fig:fig8} we show the corner plots for the parameters estimated by MCMC with LGRBs (joint spectral fits for \fermi-LGRBs) and SNe data, resulting in $H_0 = 71.61\pm0.17$~km~s$^{-1}$~Mpc$^{-1}$ and $\Omega_\Lambda = 0.65\pm0.01$. These values are then used to calculate $\mu^{\rm th}$ (orange curve) in Figure~\ref{fig:fig9}. We also estimated these parameters using GBM-only fits to \fermi-GRBs, and the values for $H_0$ and $\Omega_\Lambda$ remained consistent with those from the joint fit.


\section{Discussion}\label{sec:sec_6}

Out of the 40 GRBs detected by \textit{Fermi} (LAT-GBM), we performed spectral analysis on 37 GRBs with known redshifts, extending previous studies \citep{Dirirsa2019} by incorporating data up to 2024. This allowed for a joint spectral analysis of \fermi-GRBs detected by LAT and GBM, aiming for improved spectral index and peak energy (\(E_{\rm p} \)) determination, and to better estimate energetics. The selection criteria for this subset were discussed earlier in Section~\ref{sec:sec_2}.  

\begin{itemize}
    \item \textbf{Spectral models and fit comparisons:}  
    All spectral models used in this study are detailed in Appendix~\ref{apx1}. The Band model \citep{Band1993_413} serves as the primary model and provides the best fit among Band+BB models, while the ISSM+BB \citep{2020A&A...640A..91Y} yields the best fit for four GRBs in the time-integrated analysis (see Table~\ref{tab1}). The second-best fit model is Band+PL. For certain GRBs, additional spectral components such as BB, PL, and/or (CPL or Comp) \citep{Kaneko:06, 2009PASP..121.1279S} are required to achieve an optimal fit, e.g., Band+CPL+PL for GRB\,090902B.  
    The spectral plots in Appendices~\ref{apx2} and \ref{apx3} illustrate the improvement in fit quality between the joint and GBM-only fits.

    Notably, the reduced uncertainties (\(\sigma_1\)) for most GRBs indicate a more constrained spectral fit when comparing the two approaches. Although the residuals between the two fits may sometimes appear similar, the best-fit spectral models are determined based on the \(\Delta\)BIC values, with the Band function set as the default model. The BIC and \(\Delta\)BIC values from both time-integrated and peak flux fits are provided in Tables~\ref{tab1} and \ref{tab4}, where more than 24 GRBs have \(\Delta\)BIC greater than 6 in the time-integrated analysis, and more than 14 GRBs in the peak time analysis. 

    \item \textbf{Multi-component models for bright GRBs:}  
    For multi-component models of bright GRBs, we adopt the model with the lowest BIC as the best fit. In such cases, more complex models are selected as the default, given the prevalence of bright GRBs in our \fermi-GRB sample. 

    \item \textbf{Bolometric flux calculations:}  
    We calculated the bolometric flux for the time-integrated spectra using equations~(\ref{Eq1}) and (\ref{Eq2}), and cross-checked with {\tt 3ML}, which conveniently provides these values directly. The bolometric flux is essential for calculating \(E_{\rm iso}\) and \(L_{\rm iso}\), which are reported in Tables~\ref{tab2} and \ref{tab5}, respectively. For \(E_{\rm i,p}\), we followed the standard approach of scaling \(E_{\rm p}\) by \((1 + z)\). 

    \item \textbf{Energy range exclusions:}  
    For GRB\,221009A, only LLE data were usable at the peak, yielding peak luminosity of \({L}_{\rm iso} = (2.10 \pm 0.03) \times 10^{52}\,{\rm erg\,s^{-1}}\). This value is {not directly comparable} to the keV-band and bolometric peak luminosities reported by \citet{2023A&A...677L...2R}, who found \({L}_{\rm iso} = 3.7^{+2.5}_{-0.5} \times 10^{52}\,{\rm erg\,s^{-1}}\) in the 92–920\,keV band, and  \({L}_{{\rm iso,bol}} = 8.4^{+2.5}_{-1.5} \times 10^{52}\,{\rm erg\,s^{-1}}\) in the 1–10$^4$\,keV band; both in the rest frame at 4\,s. They obtained these results using data from GRBAlpha. Differences are expected due to the distinct energy bands, integration timescales, and instruments. GRBAlpha is a small satellite mission that provided the observational data for the GRB episode analyzed in \citet{2023A&A...677L...2R}.

    \item \textbf{Spectral model performance in peak-time analysis:} The Band \citep{Band1993_413} spectral model often provided the best fit for peak-time analysis in both joint and GBM-only cases, followed by SBPL, especially in comparison to time-integrated fits. These findings were recently discussed by \citet{mei2025gamma}.

    \item \textbf{Likelihood analysis and correlation fitting:}  
    The likelihood function described in equation~(\ref{Eq9}), along with its associated uncertainties, accounts for extrinsic systematics (\(\sigma_{\text{ext}}\)) in \(y\) values caused by unknown physical parameters. By maximizing the likelihood function in equation~(\ref{Eq9}), we determine the best-fit parameters (\(k\), \(m\), and \(\sigma_{\text{ext}}\)) for the Amati (Table~\ref{tab3}) and Yonetoku (Table~\ref{tab6}) correlations. The fit parameters are derived during the MCMC analysis.

    \item \textbf{Amati correlation results:}  
    Figure~\ref{fig:fig1} (left panel) illustrates the Amati correlation \citep{Amati2002_81A}, based on the best spectral fitting results presented in Table~\ref{tab1}. The resulting intrinsic peak energy \( E_{\rm i,p}\) of the \(\nu f_{\nu}\) spectra ranges from \( 125 \pm 11 \) keV (GRB\,091127) to $6.7 \pm 0.8~\mathrm{MeV}$ (GRB\,090510) for the joint spectral fits, and from \( 24 \pm 0.74 \) keV to \( 7.40 \pm 0.59  \) MeV for the GBM-only spectral fits.  The isotropic energy \( E_{\rm iso} \) ranges from \( (3.2 \pm 0.59) \times 10^{52} \) erg (GRB\,150514A) to $(2.04 \pm 0.07) \times 10^{55}~\mathrm{erg}$ (GRB\,090323) for the joint fit, and from \( (4.0 \pm 0.76) \times 10^{52} \) erg to \((1.7 \pm 0.1) \times 10^{55}\) erg for GBM-only bursts of the same GRB. The remaining \fermi-GRBs with their corresponding \( E_{\rm i,p}\)-\( E_{\rm iso} \) values are provided in Table~\ref{tab2}.

    The Amati relation slope parameter for LGRBs, as reported by \citet{short_long_grbs}, is \( m = 0.43 \pm 0.03 \) and the values of \(m\) from our combined fit for the \fermi-GRBs sample with MP19, where we find \( m = 0.25 \pm 0.07 \) for the joint spectral fits and \( m = 0.30 \pm 0.06 \) for the GBM-only spectral fits (see Figure~\ref{fig:fig2}). Other parameter values, including \( k \), \( m \), and \( \sigma_{\rm ext} \), are reported in Table~\ref{tab3}.  
    For SGRBs, MP19 reports \( m = 0.41 \pm 0.05 \), while our results yield \( m = 0.36 \pm 0.04 \)  for the joint spectral fits and the same for the GBM-only spectral fits. The Amati relation slope obtained from our joint \fermi and MP19 GRB fits is lower than the canonical value of \(m \sim 0.4\). However, this result remains consistent with MP19 within the stated uncertainty. The deviation may be attributed to several factors. First, a potential selection bias arises from LAT detectability: GRBs detected by LAT tend to be brighter and possess higher peak energies, which can affect the distribution and slope of the $E_{\mathrm{i,p}}$–$E_{\mathrm{iso}}$ correlation. Second, differences in the spectral modeling and parameter estimation techniques, particularly in the determination of $E_{\mathrm{p}}$, may influence the resulting fit. In our analysis, $E_{\mathrm{p}}$ values are derived from joint spectral fits across multiple instruments, which likely provides a more accurate but systematically different estimate compared to the GBM-only or time-integrated methods used in earlier works. Together, these factors could explain the observed slope discrepancy.

  \item \textbf{Yonetoku correlation results:}  
  Figure~\ref{fig:fig1}  (right panel) shows the Yonetoku correlation \citep{2004ApJ...609..935Y} based on the best spectral fit parameters listed in Table~\ref{tab4}. For the joint GBM+LAT+LLE analysis, the intrinsic peak energy \(E_{\rm i,p}\) of the \(\nu f_{\nu}\) spectra ranges from \(130 \pm 90\)~keV (GRB\,150314A) to \(7.6 \pm 1.0\)~MeV (GRB\,090510). {For the SGRB\,090510, we follow the GBM catalog convention and use a 64\, ms peak interval, here corresponding to \(T_{0}+(0.512\)–\(0.576)\)~s. } In the GBM-only analysis, \(E_{\rm i,p}\) ranges from \(125 \pm 99\)~keV to \(7.61 \pm 1.045\)~MeV. The isotropic luminosity \(L_{\rm iso}\) spans from \((2.0 \pm 0.1) \times 10^{52}\)~erg\,s\(^{-1}\) (GRB\,221009A) to \((313 \pm 72) \times 10^{52}\)~erg\,s\(^{-1}\) (GRB\,080916C) in the joint sample, and from \((2.0 \pm 0.03) \times 10^{52}\)~erg\,s\(^{-1}\) to \((3.0 \pm 0.48) \times 10^{55}\)~erg\,s\(^{-1}\) in the GBM-only sample (see Table~\ref{tab5}). The \(E_{\rm i,p}\)–\(L_{\rm iso}\) values for all the \fermi GRBs are listed in Table~\ref{tab5}.
  
  The slope parameter for LGRBs, as reported by \citet{2012MNRAS.421.1256N}, is \( m = 0.49 \pm 0.03 \). This value is consistent within uncertainties with our combined fit for the \fermi-GRB sample with N12, where we find \( m = 0.38 \pm 0.04 \) for the joint spectral fits and \( m = 0.36 \pm 0.03 \) for the GBM-only spectral fits (Figure~\ref{fig:fig5}). Other parameter values, including \( k \), \( m \), and \( \sigma_{\rm ext} \), are reported in Table~\ref{tab6}. For SGRBs, D14 reports \( m = 0.89 \pm 0.20 \), while our fit without the \fermi-GRB sample yields \( m = 0.36 \pm 0.13 \).
  
  \item \textbf{Biases in the Amati and Yonetoku Correlations:}  The observed Amati \citep{Amati+02} and Yonetoku \citep{2004ApJ...609..935Y} correlations may be influenced by selection effects. In particular, bursts that produce bright high-energy emission tend to be more luminous and spectrally harder, which biases the sample toward such events. Additionally, LAT provides precise localizations that enable follow-up observations, increasing the likelihood of redshift measurements for these bursts. Moreover, redshift measurements are typically obtained for GRBs with bright prompt emission or afterglows, introducing another bias toward intrinsically luminous events. As a result, fainter or high-redshift GRBs may be underrepresented in the sample. These factors can reduce the statistical robustness and limit the cosmological interpretation of the correlations. Such selection biases should be carefully considered when interpreting the best-fit results and comparing them with findings from other studies. These selection effects likely contribute to the differences in the slopes we obtain for both the Amati and Yonetoku correlations compared to those reported in previous studies. Interestingly, \citet{Xu2023} shows that viewing angle can also play an important role in determining the slope of these correlations.
  
  \item \textbf{Cosmological parameter estimation:} In Section~\ref{sec:5_4}, we derived the cosmological parameters using the Amati and Yonetoku correlations. In both cases, the luminosity distance was calculated using equations~(\ref{Eq12}) and (\ref{Eq18}). We incorporated SNe data from three sources: the U2.1 sample from \citet{Suzuki_2012}, DES sample from \citet{DES_sne}, and the Pantheon+ sample from  \citet{2022ApJ...938..113S}, along with our \fermi-GRB sample, including LGRBs and SGRBs from \fermi, MP19, N12, and D14 see Figure~\ref{fig:fig7} for the Amati correlation, and Figure~\ref{fig:fig8} for the Yonetoku correlation. It is clear that GRBs alone cannot provide tight constraints on the cosmological parameters because of an insufficient sample size, especially at low redshifts. Therefore, when jointly fitting GRB and SNe Ia samples, the statistical weight of the SNe Ia is significantly larger than that of the GRBs, so the resulting cosmological parameters are primarily dominated by the SNe Ia data.
  
  \item \textbf{Constraints on cosmology:}  To constrain the cosmological parameters \(H_0\) and \(\Omega_\Lambda\) within a flat \(\Lambda\)CDM model, we employed a maximum likelihood approach. This analysis compares the theoretical distance modulus \(\mu\) with its observational estimate derived from either the Amati or Yonetoku correlations.  Within this framework, GRBs are treated as standardizable cosmological candles, with their luminosity distances inferred from the Amati and Yonetoku relation fits. This methodology has been widely applied in previous studies \citep[e.g.,][]{2008MNRAS.391..577A, Wang2016_585, Demianski2017_693, Dirirsa2019}. In our analysis, we performed a joint likelihood fit of the cosmological parameters (\(H_0\) and \(\Omega_\Lambda\)) together with the correlation parameters (\(k\), \(m\), and \(\sigma_{\rm ext}\)) to mitigate the circularity problem. Table~\ref{tab7} presents a comparison of the inferred \(H_0\) and \(\Omega_\Lambda\) values using different datasets (GRBs+SNe and GRBs only), alongside results from previous studies. 

\end{itemize}

\section{Conclusion}
\label{sec:conclusion}
In this work, we performed a comprehensive joint spectral analysis of GRBs detected by \textit{Fermi} GBM and LAT instruments over a 16-year period (2008–2024). Focusing on a sample of GRBs with known redshifts, we investigated the Amati and Yonetoku correlations by incorporating spectral data from both the ${\rm T}_{90}$ duration and the peak flux intervals. Our results demonstrate that joint fits provide more precise spectral parameter estimates compared to GBM-only fits, thereby improving the reliability of phenomenological correlations. This underscores the advantages of multi-instrument datasets, as the combination of GBM and LAT data enables systematic cross-validation of results and enhances the accuracy of key parameters such as the spectral peak energy ($E_{\rm p}$).  

A major challenge arose in the treatment of exceptionally bright GRBs, such as GRB\,221009A, where standard spectral techniques required modifications. In several cases, energy ranges had to be excluded from the time-integrated fits to ensure stable solutions. Similarly, for some peak flux intervals, we encountered zero TS values due to the absence of LAT-detected photons. Nevertheless, the availability of LAT-LLE data mitigated this issue, enabling reliable spectral fits even when LAT photons were not detected.  

By leveraging these refined analyses, we successfully derived the Amati and Yonetoku correlations and compared our results with previous datasets. Our GRB sample aligns well with distributions reported in earlier works, reinforcing the validity of these relations. {Furthermore, without imposing any prior assumptions on either the cosmological parameters or the correlation parameters, we jointly constrained them under the assumption of a flat universe, incorporating Type Ia SNe data spanning a redshift range of $0.08 \leq z \leq 2.3$, and thus avoided the circularity problem.} When combined with datasets from \citet{short_long_grbs, 2012MNRAS.421.1256N, 2014MNRAS.442.2342D}, our analysis produced cosmological constraints that differ from those of \citet{Dirirsa2019}, who used the \fermi-GRB sample alongside \citet{Wang2016_585}. Despite these differences, our inferred $H_0$ remain broadly consistent within uncertainties. Our $\Omega_\Lambda$, however, is smaller than the values found by \citet{Dirirsa2019}, in Cepheid + SNe analysis by \citet{Riess2022} and in Galaxy Cluster analysis by \citet{eROSITA}, but consistent with the Pantheon+ SNe analysis by \citet{Pantheon} and with the CMB analysis by \citet{Planck_Collaboration2018}.
Interestingly, our best-fit value of $H_0$ is in between the measurements using CMB data~\citep{Planck_Collaboration2018}, and using Cepheids and SNe data \citep{Riess2022}. See Table~\ref{tab7} for detailed comparisons.

In our analysis, we used a total of 311 GRBs for the Amati correlation. However, achieving a precision on \(\Omega_m\) comparable to the current SNe Ia data might require approximately 390 GRBs~\citep{DainottiNielson2022}. Nevertheless, significant scatter persists in both the $E_{\mathrm{i,p}}$–$E_{\rm iso}$ and $L_{\rm iso}$–$E_{\mathrm{i,p}}$ relations. This suggests that statistical improvements alone cannot fully explain the observed dispersion, which may arise from the intrinsic diversity of GRB emission mechanisms. For instance, varying contributions from thermal and non-thermal components could systematically shift bursts within the correlation plane. Reducing such intrinsic scatter remains essential for improving GRB-based cosmological constraints.  

Our findings highlight the necessity of enlarging GRB samples with well-constrained spectra to enable robust cosmological applications. Future observations with next-generation gamma-ray observatories will play a central role in addressing these challenges. For example, the StarBurst Multi-messenger Pioneer (StarBurst; \citealt{2024HEAD...2150003B}), while not capable of fully characterizing GRBs on its own, could, together with the Compton Spectrometer and Imager (COSI; \citealt{2023AAS...24114606T}), provide complementary energy coverage and improved prompt emission spectral measurements for joint analyses. Likewise, missions such as the Space-based Multi-band Astronomical Variable Objects Monitor (SVOM; \citealt{2021Galax...9..113B}) and ground-based facilities like the Cherenkov Telescope Array Observatory (CTAO; \citealt{2022SPIE12182E..0PW}) will extend coverage from keV to TeV energies, enabling prompt follow-up and improved redshift determination. Coordinated multi-instrument efforts of this kind will enhance our understanding of GRB physics and strengthen their use as cosmological probes.

\onecolumn
\setlength{\tabcolsep}{0.136cm}
\begin{landscape}
\begingroup
\footnotesize 
\setlength{\LTcapwidth}{\textwidth} 
\small
\centering
\renewcommand{\arraystretch}{1.3}
\begin{longtable}{lccccccccccc}

\caption{Best-fit parameters from time-integrated spectral analysis of joint GBM-LLE+LAT data and GBM-only data. $\alpha_1$, and $\beta$ are the lower and higher photon indices for the Band and SBPL models, respectively. $\gamma$ is the photon index of the ExpCut model, while $\alpha_2$ is that of the PL model. $E_{\rm p}$ is the Band or ExpCut peak energy. $kT$ is the temperature of the BB model.}\label{tab1} \\
\hline
Names & Instruments    & ${\rm T}_{90}$ interval & Model & ${\alpha}_1, \gamma$ & $\beta$ & $E_{\rm p}\, (\rm keV)$ & ${\alpha}_2$  &  $kT\,(\rm keV)$   &  {Fluence} ($\rm erg$ ${\rm cm}^{-2}$) & BIC  & $\Delta$ BIC \\
\hline
\endfirsthead
\caption[]{(Continued)}\\
\hline
Names & Instruments    & ${\rm T}_{90}$ interval & Model & ${\alpha}_1, \gamma$ & $\beta$ & $E_{\rm p}$ (keV)& ${\alpha}_2$  &  $kT$ (\rm keV)   &  Fluence ($\rm erg$ ${\rm cm}^{-2}$) & BIC  & $\Delta$ BIC \\
\hline
\endhead
\hline
\endfoot
GRB\,080916C  & GBM+LAT+LLE   &0.0--66  & Band+BB & $-1.20 \pm 0.04 $ & $-2.27 \pm 0.03$ & $1160 \pm 240$ &  - &  $45.5 \pm 3.3$ &    $1.32 \pm 0.27 \times {10}^{-6}$ & 9393 & 5 \\

GRB\,080916C  & GBM  & ...  & SBPL & $-1.02 \pm 0.02$ & $-2.21 \pm 0.12$ & $560 \pm 50$ &  - &  - &   $5.5 \pm 1.2 \times {10}^{-6}$  & 5739 & 0\\

\hline

GRB\,090323  & GBM+LAT+LLE  & 8.704--142.594 & Band+BB & $-1.39\pm 0.03$ &  $-3.1  \pm 0.10$  & $1220 \pm 230$ &  - &  $40 \pm 2.0 $ & $6.0\pm 1.9 \times {10}^{-6}$ & 8902 & 33\\

GRB\,090323  & GBM  & ... & Band+BB & $-1.31\pm0.04$ &  $-5.0 \pm 0.03$  & $670\pm 110$ & -  & $41 \pm 1.8$ & $4.2 \pm 0.4 \times {10}^{-6}$ & 8403 & 17\\

\hline

GRB\,090328 & GBM+LAT+LLE   & 4.352–-66.049 & ISSM & $-1.06\pm0.03$ & $-3.02\pm 0.19 $&  $860\pm130$  & - &  - & $4.1 \pm 1.5 \times 10^{-6}$ & 7450 & 3\\

GRB\,090328 & GBM   &... & ISSM & $-1.09\pm0.02$ & $-10.06\pm2.3$ &  $720 \pm 60$  & - & -  & $4.1 \pm 1.5 \times 10^{-6}$  & 7040 & 2\\
\hline

GRB\,090424 & GBM+LAT  & 0.512–-14.656 & Band+BB & $-0.55  \pm 0.08 $ &  $-2.71 \pm  0.07 	$ & $155 \pm 4.0$ & - &  $ 7.5 \pm 0.4 $ & $1.6\pm 0.1 \times {10}^{-5} $ & 6091 & 32\\

GRB\,090424 & GBM  & ... & Band+BB & $-0.58 \pm 0.05 $ &  $-2.82  \pm 0.11$  & $159\pm 4.0$ & -  & $7.67 \pm 0.33$  & $1.22 \pm 0.07 \times {10}^{-5}$ &  6006 & 30\\

\hline
GRB\,090510 & GBM+LAT+LLE  & 0.0–-1.74  & Band+PL & $ -0.50 \pm 0.14$ &  $-2.45 \pm 0.15$  &  $3500\pm400$ & $-1.6\pm0.03$ & - & $8.4\pm 1.5 \times {10}^{-6}$ & 6945 & 8 \\

GRB\,090510 & GBM   & ... & Band+BB & $-0.60 \pm 0.09$  & $-2.7 \pm 0.25$ &   $3890\pm 310$ & -  &  $12.5 \pm 2.6$ &  $4.3\pm 0.8 \times {10}^{-5}$ & 5864 & 7  \\
\hline

GRB\,090902B & GBM+LAT+LLE    & 2.816--22.144 & Band+CPL+PL & $-0.72 \pm 0.01$ & $-4.21 \pm 0.26$ & $738 \pm 100 $ & $-1.1 \pm 0.02$ & - & $5.0\pm 1.01 \times {10}^{-5}$ & 9950 & 384\\

GRB\,090902B & GBM  & ... & Band & $-0.96\pm0.04$ & $-4.30 \pm 0.33$ & $867 \pm 11$ & - & - &  $1.2 \pm 0.07 \times {10}^{-5}$  & 8999 & 0\\

\hline

\multicolumn{12}{l}{\footnotesize\textit{{Note. See the online supplementary material for the full table.}}}\\  
\end{longtable}

\setlength{\tabcolsep}{0.158cm}
\renewcommand{\arraystretch}{1.2}

\begin{longtable}{@{\extracolsep{\fill}}lcccccl@{}}
\caption{\(E_{i,\rm p} - {E}_{\rm iso}\) values for 37 \fermi GRBs based on Joint and GBM-Only spectral fits. \label{tab2}} \\

\hline
GRBs\,Names & Redshift &  $E_{i,\rm p}$ (keV) -- Joint &  $E_{\rm iso}$ (${10}^{52}$ -- Joint erg) & 
$E_{i,\rm p}$ (keV) -- GBM-only & $E_{\rm iso}$ (${10}^{52}$ erg) -- GBM-only  & Reference ($z$) \\

\hline
\endfirsthead

\multicolumn{7}{c}%
{{ \tablename\ \thetable{} -- (Continued)}} \\
\hline
GRBs\,Names & Redshift &  $E_{i,\rm p}$ (keV) -- Joint  &  $E_{\rm iso}$ (${10}^{52}$ erg) -- Joint  & 
$E_{i,\rm p}$ (keV) -- GBM-only  & $E_{\rm iso}$ (${10}^{52}$ erg) -- GBM-only  & Reference ($z$) \\
\hline
\endhead

\hline
\endlastfoot

GRB\,080916C & 4.35 & 6206 $\pm $1525 & 327 $\pm$ 67 & 2996 $\pm$ 268 & 1361 $\pm$ 285 & \cite{greiner2009redshift}\\
GRB\,090323  & 3.57 & 5575 $\pm$ 1051  & 2035 $\pm$ 68 & 3062 $\pm$ 502 & 1697 $\pm$ 104 & \cite{2009GCN.10100....1C}\\
GRB\,090328  & 0.736 & 1493 $\pm$ 226 & 37 $\pm$ 11 & 1250 $\pm$ 104 & 37 $\pm$ 511 & \cite{2009GCN..9053....1C} \\
GRB\,090424  & 0.544 & 239 $\pm$ 6.2 & 18 $\pm$ 1.0 & 245 $\pm$ 6.0 & 14 $\pm$ 4.0 & \cite{2009GCN..9243....1C} \\
GRB\,090510  & 0.903 & 6661 $\pm$ 761 & 3.23 $\pm$ 1.0 & 7402 $\pm$ 590 & 17 $\pm$ 3.1 & \cite{2009GCN..9353....1R} \\
GRB\,090902B & 1.822 & 2083 $\pm$ 282& 815 $\pm$ 2.12 & 2447 $\pm$ 31 & 198 $\pm$ 1.2  & \cite{2009GCN..9873....1C}\\

\hline
\multicolumn{7}{l}{\footnotesize\textit{{Note. See the online supplementary material for the full table.}}}\\
\end{longtable}

\setlength{\LTcapwidth}{\textwidth} 
\begin{longtable}{@{\extracolsep{\fill}}lccc@{}}
\caption{The Amati \citep{Amati2006_372} correlation fit results are presented for the \fermi GRB sample from our analysis, using both joint fits and GBM-only fits, and the GRB samples from \citet{short_long_grbs}.}\label{tab3}\\
\hline
\bf Samples & \(k\) & \(m\) & \(\sigma_{\rm ext}\) \\
\hline
37 \fermi (Joint fit)             & 0.56$\pm$0.13 & 0.25$\pm$0.07 & 0.42$\pm$0.06 \\
37 \fermi (GBM-only)              & 0.39$\pm$0.14 & 0.30$\pm$0.06 & 0.41$\pm$0.05 \\ 
33 \fermi LGRBs (Joint fit) + 275 LGRBs & 0.29$\pm$0.02 & 0.35$\pm$0.02 & 0.29$\pm$0.01 \\
37 \fermi LGRBs (GBM-only) + 275 LGRBs  & 0.28$\pm$0.02 & 0.35$\pm$0.02 & 0.29$\pm$0.01 \\
1 \fermi SGRBs (Joint fit) + 45 SGRBs  & 1.18$\pm$0.06 & 0.36$\pm$0.04 & 0.33$\pm$0.04 \\
1 \fermi SGRBs (GBM-only) + 45 SGRBs   & 1.17$\pm$0.07 & 0.36$\pm$0.04 & 0.32$\pm$0.04 \\
\hline
\end{longtable}
\endgroup

\setlength{\tabcolsep}{0.06cm}
\begingroup
\footnotesize 
\setlength{\LTcapwidth}{\textwidth} 
\small
\centering
\renewcommand{\arraystretch}{1.3}
\begin{longtable}{lccccccccccc}

\caption{Best-fit parameters from spectral analysis during the peak flux interval of joint GBM-LLE+LAT data and GBM-only data. $\alpha_1$, and $\beta$ are the lower and higher photon indices for the Band and SBPL models, respectively. $\gamma$ is the photon index of the ExpCut model, while $\alpha_2$ is that of the PL model. $E_{\rm p}$ is the Band or ExpCut peak energy. $kT$ is the temperature of the BB model.}\label{tab4} \\
\hline
Names & Instruments    & Peak flux interval & Model & ${\alpha}_1, \gamma$ & $\beta$ & $E_{\rm p}$ (keV) & ${\alpha}_2$  &  $kT$ (keV) &   Peak flux ($\rm erg$ ${\rm cm}^{-2}$  ${\rm s}^{-1}$)  & BIC  & $\Delta$BIC \\
\hline
\endfirsthead
\caption[]{(Continued)}\\
\hline
Names & Instruments    & Peak flux interval & Model & ${\alpha}_1, \gamma$ & $\beta$ & $E_{\rm p}$ (keV)& ${\alpha}_2$  &  $kT$ (keV)&   Peak flux ($\rm erg$ ${\rm cm}^{-2}$  ${\rm s}^{-1}$) & BIC  & $\Delta$BIC \\
\hline
\endhead
\hline
\endfoot
GRB\,080916C  & GBM+LAT+LLE  & 2.048--3.072 & Band & $-0.77 \pm 0.07 $ & $-2.8 \pm 0.16$ & $540 \pm 90$ &  - &  - &   $6.4 \pm 2.0 \times 10^{-6}$ & 1692 & 0 \\

GRB\,080916C  & GBM  & ... & Band & $-0.78 \pm 0.06$ & $-4 \pm 5$ & $560 \pm 90$ &  - &  - & $1.56 \pm 0.36 \times {10}^{-5} $   & 1692 & 0\\

\hline
    
GRB\,090323  & GBM+LLE   & 65.536--66.56 & Band & $-0.52 \pm 0.02$ &  $-3.00 \pm 0.04$  & $380 \pm 34 $ &  - & - & $1.78 \pm 0.10 \times {10}^{-5}$ & 2185 & 0\\

GRB\,090323 &  GBM  &  ... & Band & $-0.51 \pm 0.06$ & $-4.3 \pm 2.6$  & $380 \pm 30 $ & -  & - & $1.46 \pm 0.32 \times {10}^{-5}$ & 2019 & 0\\

\hline

GRB\,090328 & GBM+LLE   & 23.552--24.576 & Band & $-0.75 \pm 0.05$ & $-2.82 \pm 0.10$ &  $440 \pm 40$ & - &  - & $2.39 \pm 0.13 \times {10}^{-5} $ &  1796 & 0 \\ 

GRB\,090328 & GBM   & ... & Band & $-0.75 \pm 0.04$ & $-5.0 \pm 0.05$ &  $448 \pm 35 $  & - & -  & $2.39 \pm 0.13 \times {10}^{-5}$ & 1809 & 0 \\

\hline

GRB\,090510 & GBM+LLE   & 0.512--0.576 & SBPL & $-0.79 \pm 0.07$ & $-5.0 \pm 0.002$ &  $4000 \pm 550$ & - &  - & $1.4 \pm 0.5 \times {10}^{-4} $ &  404 & 1 \\

GRB\,090510 & GBM   &  ... & SBPL & $-0.79 \pm 0.07$ & $-5.0 \pm 0.002$ &  $4000 \pm 550 $  & - & -  & $1.2 \pm 0.9 \times {10}^{-4}$ & 395 & 0 \\

\hline

GRB\,090902B & GBM+LAT+LLE & 9.216--10.24 & ISSM+PL & $0.58 \pm 0.6$ & $-7.8 \pm 1.1$ & $703 \pm 29$ & $-1.93 \pm 0.01$ & - & $2.15 \pm 0.16 \times {10}^{-5}$ & 2889 & 1061\\

GRB\,090902B & GBM  & ... & ISSM & $0.40 \pm 0.13$ & $-9.4 \pm 3.1$ & $707\pm 22$ & - & - & $1.13\pm0.11 \times {10}^{-4}$ & 2689 & 0\\

\hline

\multicolumn{12}{l}{\footnotesize\textit{{Note. See the online supplementary material for the full table.}}}\\
\end{longtable}

\setlength{\tabcolsep}{0.158cm}
\renewcommand{\arraystretch}{1.2}

\begin{longtable}{@{\extracolsep{\fill}}lcccccl@{}}
\caption{\(E_{i,\rm p} - {L}_{\rm iso}\) values for 29 \fermi GRBs based on Joint and GBM-Only spectral fits. \label{tab5}} \\

\hline
GRBs\,Names & Redshift &  $E_{i,\rm p}$ (keV) -- Joint  &  $L_{\rm iso}$  (${10}^{52}$ erg $s^{-1}$) -- Joint  & 
$E_{i,\rm p}$ (keV) -- GBM-only & $L_{\rm iso}$ (${10}^{52}$ erg $s^{-1}$) -- GBM-only & Reference ($z$) \\
\hline
\endfirsthead

\multicolumn{7}{c}%
{{ \tablename\ \thetable{} -- (Continued)}} \\
\hline
GRBs\,Names & Redshift &  $E_{i,\rm p}$ (keV) -- Joint &  $L_{\rm iso}$  (${10}^{52}$ erg $s^{-1}$) -- Joint  & 
$E_{i,\rm p}$ (keV) -- GBM-only & $L_{\rm iso}$ (${10}^{52}$ erg $s^{-1}$) -- GBM-only & Reference ($z$) \\
\hline
\endhead

\hline
\endlastfoot

GRB\,080916C        & 4.35   & 2889 $\pm$ 455 & 313 $\pm$ 72  & 1926 $\pm$ 295 & 2996 $\pm$ 481 & \cite{greiner2009redshift} \\ 
GRB\,090323         & 3.57       & 1736 $\pm$ 153  & 222 $\pm$ 13   & 1737 $\pm$ 151  & 183 $\pm$ 38  & \cite{2009GCN.10100....1C}  \\ 
GRB\,090328         & 0.736      & 764 $\pm$ 69 & 6.0 $\pm$ 0.30 & 777 $\pm$ 64 & 6.0 $\pm$ 0.33      & \cite{2009GCN..9053....1C}   \\ 
GRB\,090510         & 0.903  & 7612 $\pm$ 1047 & 59 $\pm$ 12  & 7612 $\pm$ 1047 & 50 $\pm$ 18     & \cite{2009GCN..9353....1R}  \\ 
GRB\,090902B        & 1.822      & 1984 $\pm$ 268 & 73 $\pm$ 23  & 1995 $\pm$ 63 & 36 $\pm$ 3.0   & \cite{2009GCN..9873....1C}  \\ 

\hline
\multicolumn{7}{l}{\footnotesize\textit{{Note. See the online supplementary material for the full table.}}}\\
\end{longtable}

\setlength{\LTcapwidth}{\textwidth} 
\begin{longtable}{@{\extracolsep{\fill}}lccc@{}}
\caption{The Yonetoku \citep{2004ApJ...609..935Y} correlation fit results are presented for the \fermi GRB sample and GRB samples from \citet{2012MNRAS.421.1256N, 2014MNRAS.442.2342D}, analyzed using both joint fits and GBM-only fits.}\label{tab6}\\
\hline
\bf Samples & \(k\) & \(m\) & \(\sigma_{\rm ext}\) \\
\hline
28 \fermi LGRBs (Joint fit)            & 0.71$\pm$0.14 & 0.25$\pm$0.08 & 0.27$\pm$0.04\\
28 \fermi LGRBs (GBM-only)             & 0.67$\pm$0.13 & 0.23$\pm$0.07 & 0.26$\pm$0.04 \\ 
28 \fermi LGRBs (Joint fit) + 58 LGRBs & 0.46$\pm$0.05 & 0.38$\pm$0.04 & 0.29$\pm$0.03 \\
28 \fermi LGRBs (GBM-only) + 58 LGRBs  & 0.44$\pm$0.05 & 0.36$\pm$0.03 & 0.27$\pm$0.03 \\
1 \fermi SGRBs + 12 SGRBs  & 0.86 $\pm$ 0.13 & 0.36 $\pm $0.13 & 0.41 $\pm$ 0.1\\
\hline
\end{longtable}

\setlength{\LTcapwidth}{\textwidth} 
\begin{longtable}{@{\extracolsep{\fill}}lccccc@{}}
\caption{Best-fit cosmological and correlation parameters from GRBs combined with different SNe Ia compilations, using both the Amati and Yonetoku relations.}
\label{tab7} \\

\hline
Dataset & $H_0$~(km~s$^{-1}$~Mpc$^{-1}$) & $\Omega_\Lambda$ & $m$ & $k$ & $\sigma_{\rm ext}$ \\
\hline
\endfirsthead

\multicolumn{6}{c}%
{{\bfseries \tablename\ \thetable{} -- continued from previous page}} \\
\hline
Dataset & $H_0$ & $\Omega_\Lambda$ & $m$ & $k$ & $\sigma_{\rm ext}$ \\
\hline
\endhead

\hline \multicolumn{6}{r}{{Continued on next page}} \\
\endfoot

\hline
\endlastfoot

\multicolumn{6}{c}{\textbf{Amati Relation}} \\
\hline
GRBs only                          & $69.64 \pm 13.56$ & $0.58 \pm 0.06$ & $0.54 \pm 0.02$ & $0.66 \pm 0.10$ & $0.31 \pm 0.02$ \\
GRBs + SNe (U2.1, DES, Pantheon+) & $71.60 \pm 0.17$ & $0.65 \pm 0.01$ & $0.50 \pm 0.03$ & $0.28 \pm 0.04$ & $0.41 \pm 0.02$ \\
GRBs + SNe (U2.1, DES)            & $70.32 \pm 0.23$ & $0.65 \pm 0.02$ & $0.54 \pm 0.02$ & $0.65 \pm 0.03$ & $0.31 \pm 0.02$ \\

\hline
\multicolumn{6}{c}{\textbf{Yonetoku Relation}} \\
\hline
GRBs only                          & $70.00 \pm 13.60$ & $0.64 \pm 0.10$ & $0.44 \pm 0.04$ & $0.36 \pm 0.09$ & $0.25 \pm 0.03$ \\
GRBs + SNe (U2.1, DES, Pantheon+) & $71.61 \pm 0.18$ & $0.65 \pm 0.01$ & $0.47 \pm 0.04$ & $0.40 \pm 0.06$ & $0.27 \pm 0.03$ \\
GRBs + SNe (U2.1, DES)            & $70.34 \pm 0.24$ & $0.66 \pm 0.02$ & $0.44 \pm 0.04$ & $0.36 \pm 0.06$ & $0.25 \pm 0.03$ \\
\hline
\multicolumn{6}{c}{\textbf{Previous Studies}} \\
\hline
GRB+W2016 + SNe \citep{Dirirsa2019}        & $70.0 \pm 0.50$   & $0.72 \pm 0.03$   & $1.18 \pm 0.18$ & $1.34 \pm 0.07$ & $0.38 \pm 0.06$ \\
CMB  \citep{WMAP9}                        & $69.30 \pm 0.80$   & $0.72 \pm 0.02$ & – & – & – \\
CMB \citep{Planck_Collaboration2018}             & $67.40 \pm 0.50$   & $0.68 \pm 0.01$ & – & – & – \\

Cepheids + SNe \citep{Riess2022}               & $73.04 \pm 1.04$ & $0.70 \pm 0.02$        & – & – & – \\
Pantheon+ SNe \citep{Pantheon} & $73.5 \pm 1.10$ & $0.67 \pm 0.02$& – & – & – \\

\end{longtable}
\endgroup    
\end{landscape}

\twocolumn
\section*{Acknowledgments}
The \textit{Fermi} LAT Collaboration acknowledges generous ongoing support from a number of agencies and institutes that have supported both the development and the operation of the LAT as well as scientific data analysis. These include the National Aeronautics and Space Administration and the Department of Energy in the United States, the Commissariat \`a l'Energie Atomique and the Centre National de la Recherche Scientifique / Institut National de Physique Nucl\'eaire et de Physique des Particules in France, the Agenzia Spaziale Italiana and the Istituto Nazionale di Fisica Nucleare in Italy, the Ministry of Education, Culture, Sports, Science and Technology (MEXT), High Energy Accelerator Research Organization (KEK) and Japan Aerospace Exploration Agency (JAXA) in Japan, and the K.~A.~Wallenberg Foundation, the Swedish Research Council and the Swedish National Space Board in Sweden.

Additional support for science analysis during the operations phase is gratefully acknowledged from the Istituto Nazionale di Astrofisica in Italy and the Centre National d'\'Etudes Spatiales in France. This work performed in part under DOE Contract DE-AC02-76SF00515.

T.K.M.A.\ acknowledges receiving a PhD fellowship from the Organization for Women in Science for the Developing World (OWSD) and SIDA (Swedish International Development Cooperation Agency). T.K.M.A.\ and S.R.\ were partially supported by a BRICS STI grant (No. 150504) from the National Research Foundation (NRF) of South Africa and additional support from the South African Gamma-ray Astronomy Programme (SA-GAMMA).

R.G.\ was sponsored by the National Aeronautics and Space Administration (NASA) through a contract with ORAU. The views and conclusions contained in this document are those of the authors and should not be interpreted as representing the official policies, either expressed or implied, of the National Aeronautics and Space Administration (NASA) or the U.S. Government. The U.S. Government is authorized to reproduce and distribute reprints for Government purposes, notwithstanding any copyright notation herein.

\section*{Data Availability}
All data used in this analysis are publicly available from the official archives of the \fermi Gamma-ray Burst Monitor (GBM) and \fermi Large Area Telescope (LAT) missions, which provide comprehensive datasets including spectral information, light curves, and other relevant data products.


\bibliographystyle{mnras}
\bibliography{refs, Fermi_GRB_2}

\appendix

\section{Typical Spectral Models} 
\label{apx1}
To conduct the spectral analysis of GRBs, we utilize various empirical functions, as detailed below:
\subsection{Power Law (PL):}  
The power law (PL) model represents one of the simplest forms of spectral analysis. It characterizes the spectrum using two parameters defined as:  
\begin{equation}
N_{\text{PL}}(E) = A_{\text{PL}} \left( \frac{E}{100 \, \text{keV}} \right)^{\alpha_1},
\label{eq:PL}
\end{equation}  
where \(A_{\text{PL}}\) is the normalization constant, and \(\alpha_1\) denotes the photon index that governs the spectral slope.

\subsection{Band Function:}  
The Band function \citep{1993ApJ...413..281B} is widely used to describe GRB spectra. It combines two power-law components with a smooth exponential transition, expressed as:
\begin{equation}
\resizebox{0.85\linewidth}{!}{$
N_{\text{Band}}(E) = A_{\text{Band}} \left\{
\begin{array}{l}
\left(\frac{E}{100 \text{ keV}}\right)^\alpha \exp\left[- \frac{E(2+\alpha)}{E_p}\right], \\ 
\text{if } E \leq E_b; \\
\\
\left(\frac{E}{100 \text{ keV}}\right)^\beta \exp(\beta - \alpha) \left[- \frac{E_p}{100 \text{ keV}} \frac{\alpha - \beta}{2 + \alpha}\right]^{\alpha - \beta}. \\ 
\text{if } E > E_b
\end{array}
\right.$}
\label{eq:Band}
\end{equation}
Here, \(E_b = E_p (\alpha - \beta)/(2 + \alpha)\), and \(A_{\text{Band}}\) serves as the normalization parameter.

\subsection{Comptonized Model (Comp):}  
The Comptonized model \citep{Kaneko:06, 2009PASP..121.1279S} provides an alternative spectral description, combining a power law with an exponential cutoff:  
\begin{equation}
N_{\text{Comp}}(E) = A_{\text{Comp}} \left( \frac{E}{100 \, \text{keV}} \right)^\alpha \exp\left[-(2 + \alpha)\frac{E}{E_p}\right].
\label{eq:Comp}
\end{equation}
where \(A_{\text{Comp}}\) is the normalization, \(\alpha\) is the photon index, and \(E_p\) is the peak energy.

\subsection{Blackbody (BB):}  
Thermal contributions to GRB spectra are represented using a blackbody function \citep{2003A&A...406..879G, 2004ApJ...614..827R}:  
\begin{equation}
N_{\text{BB}}(E) = A_{\text{BB}} \frac{E^2}{\exp(E/kT) - 1}.
\end{equation}
where \(A_{\text{BB}}\) is the normalization constant, and \(kT\) represents the temperature.

\subsection{Internal Shock Synchrotron Model (ISSM):}  
The ISSM \citep{2020A&A...640A..91Y} describes the synchrotron radiation of internal shocks. It is given by:  
\begin{align}
\frac{dN_{\text{ISSM}}}{dE}(E) = 
\frac{A_{\text{ISSM}}}{
\left[ 
1 - \frac{E_p}{E_r} \left(\frac{2 + \beta}{2 + \alpha}\right)
\right]^{\beta - \alpha}}
\left( \frac{E}{E_r} \right)^\alpha \notag \\ 
\quad \times 
\left[ 
\frac{E}{E_r} - \frac{E_p}{E_r} \left(\frac{2 + \beta}{2 + \alpha}\right)
\right]^{\beta - \alpha}\,,
\label{eq:ISSM}
\end{align}
where \(E_r\) is the reference energy and \(A_{\text{ISSM}}\) is the normalization factor.

\subsection{Smooth Broken Power Law (SBPL):}  
The SBPL \citep{1999ApL&C..39..281R} models spectra with curvature using:  
\begin{equation}
N_{\text{SBPL}}(E) = A E_j^{\alpha} 
\left[ 
\left( \frac{E}{E_j} \right)^{-\alpha n} + 
\left( \frac{E}{E_j} \right)^{-\beta n} 
\right]^{-\frac{1}{n}}\,,
\label{eq:SBPL}
\end{equation}
where \(n\) determines the smoothness of the curvature.

\subsection{Double Smooth Broken Power Law (DSBPL):}  
An extension of the SBPL, the DSBPL \citep{2018A&A...613A..16R} provides a more flexible representation:  
\begin{eqnarray}
N^{\text{2SBPL}}_{E} = A \, E_{\text{break}}^{\alpha_1} 
\left[ 
\left[ \left( \frac{E}{E_{\text{break}}} \right)^{-\alpha_1 n_1} + 
\left( \frac{E}{E_{\text{break}}} \right)^{-\alpha_2/n_1} \right]^{n_2/n_1} 
\right. \nonumber \\
\left.
+ \left( \frac{E}{E_j} \right)^{-\beta n_2} 
\cdot \left[ \left( \frac{E_j}{E_{\text{break}}} \right)^{-\alpha_1 n_1} + 
\left( \frac{E_j}{E_{\text{break}}} \right)^{-\alpha_2 n_1} \right]^{n_2/n_1} 
\right]^{-1/n_2}.
\label{eq:2SBPL}
\end{eqnarray}
where \(E_j = E_{\text{peak}} \cdot \left( - \frac{\alpha_2 + 2}{\beta + 2} \right)^{1/[(\beta - \alpha_2) n_2]}\).

\section{Count and Spectra Plots: Time-Integrated}
\label{apx2}
In these figures\footnote{See the online supplementary material for the full plots.}, we present the results of the spectral analysis for a sample of 37 GRBs, which were analyzed using joint spectral fitting and GBM-only fitting. Each GRB is represented by a series of plots that illustrate the differences between the two fitting approaches.

On the right side of the figure, we display the count spectra for each GRB. The upper panel corresponds to the results from the joint spectral fit, while the lower panel shows the spectra derived from the GBM-only fit. These spectra offer a comparison of the model fits applied to the GRB data, highlighting the impact of using the full spectrum.

On the right side, we show the ${\rm \nu}f_{\rm \nu}$ spectra, which represent the energy distribution of the GRB emission for the sample of \fermi-GBM-LAT-LLE GRBs with known redshifts. These spectra were derived from time-integrated spectral analysis over the duration ${T}_{90}$. The ${\rm \nu}f_{\rm \nu}$ spectra allow us to visualize the spectral energy distribution (SED) of the GRBs, providing valuable information on their emission properties.

The models used to fit the spectra are displayed as solid lines, with the confidence regions $1\sigma$ indicated by shaded areas of the same color as the model lines. These shaded regions provide an indication of the uncertainty associated with each model, illustrating the range of possible spectral fits within the given confidence interval.

This analysis enables a comprehensive comparison of the joint spectral fits and GBM-only spectral fits, highlighting the effects of different spectral modeling techniques.

It is important to note that for some GRBs, we had to limit the energy range due to poor fitting results within the ${T}_{90}$ time interval. This restriction was primarily applied to the GBM detectors, where the fits were inadequate. In such cases, we retained only the detector that provided the best spectral fit, ensuring the most reliable results for the analysis. This selective approach helps mitigate the influence of poorly constrained data and improves the overall accuracy of the spectral modeling.


\section{Count and Spectra Plots: Peak Flux}
\label{apx3}

The plots\footnote{See the online supplementary material for the full plots.} provided are consistent with those described in Appendix \ref{apx2}, following the same format and analysis method.

\setcounter{table}{0}
\renewcommand{\thetable}{\arabic{table}}
\onecolumn
\setlength{\tabcolsep}{0.136cm}
\begin{landscape}
\section*{Online Supplementary Material}
\begingroup
\footnotesize 
\setlength{\tabcolsep}{0.136cm} 

\small
\centering
\renewcommand{\arraystretch}{1.3}
\begin{longtable}{lccccccccccc}

\caption{Best-fit parameters from time-integrated spectral analysis of joint GBM-LLE+LAT data and GBM-only data. $\alpha_1$, and $\beta$ are the lower and higher photon indices for the Band and SBPL models, respectively. $\gamma$ is the photon index of the ExpCut model, while $\alpha_2$ is that of the PL model. $E_{\rm p}$ is the Band or ExpCut peak energy. $kT$ is the temperature of the BB model.}\\
\hline
Names & Instruments    & ${\rm T}_{90}$ interval & Model & ${\alpha}_1, \gamma$ & $\beta$ & $E_{\rm p}\, (\rm keV)$ & ${\alpha}_2$  &  $kT\,(\rm keV)$   &  {Fluence} ($\rm erg$ ${\rm cm}^{-2}$) & BIC  & $\Delta$ BIC \\
\hline
\endfirsthead
\caption[]{(Continued)}\\
\hline
Names & Instruments    & ${\rm T}_{90}$ interval & Model & ${\alpha}_1, \gamma$ & $\beta$ & $E_{\rm p}$ (keV)& ${\alpha}_2$  &  $kT$ (\rm keV)   &  Fluence ($\rm erg$ ${\rm cm}^{-2}$) & BIC  & $\Delta$ BIC \\
\hline
\endhead
\hline
\endfoot
GRB\,080916C  & GBM+LAT+LLE   &0.0--66  & Band+BB & $-1.20 \pm 0.04 $ & $-2.27 \pm 0.03$ & $1160 \pm 240$ &  - &  $45.5 \pm 3.3$ &    $1.32 \pm 0.27 \times {10}^{-6}$ & 9393 & 5 \\

GRB\,080916C  & GBM  & ...  & SBPL & $-1.02 \pm 0.02$ & $-2.21 \pm 0.12$ & $560 \pm 50$ &  - &  - &   $5.5 \pm 1.2 \times {10}^{-6}$  & 5739 & 0\\

\hline

GRB\,090323  & GBM+LAT+LLE  & 8.704--142.594 & Band+BB & $-1.39\pm 0.03$ &  $-3.1  \pm 0.10$  & $1220 \pm 230$ &  - &  $40 \pm 2.0 $ & $6.0\pm 1.9 \times {10}^{-6}$ & 8902 & 33\\

GRB\,090323  & GBM  & ... & Band+BB & $-1.31\pm0.04$ &  $-5.0 \pm 0.03$  & $670\pm 110$ & -  & $41 \pm 1.8$ & $4.2 \pm 0.4 \times {10}^{-6}$ & 8403 & 17\\

\hline

GRB\,090328 & GBM+LAT+LLE   & 4.352–-66.049 & ISSM & $-1.06\pm0.03$ & $-3.02\pm 0.19 $&  $860\pm130$  & - &  - & $4.1 \pm 1.5 \times 10^{-6}$ & 7450 & 3\\

GRB\,090328 & GBM   &... & ISSM & $-1.09\pm0.02$ & $-10.06\pm2.3$ &  $720 \pm 60$  & - & -  & $4.1 \pm 1.5 \times 10^{-6}$  & 7040 & 2\\
\hline

GRB\,090424 & GBM+LAT  & 0.512–-14.656 & Band+BB & $-0.55  \pm 0.08 $ &  $-2.71 \pm  0.07 	$ & $155 \pm 4.0$ & - &  $ 7.5 \pm 0.4 $ & $1.6\pm 0.1 \times {10}^{-5} $ & 6091 & 32\\

GRB\,090424 & GBM  & ... & Band+BB & $-0.58 \pm 0.05 $ &  $-2.82  \pm 0.11$  & $159\pm 4.0$ & -  & $7.67 \pm 0.33$  & $1.22 \pm 0.07 \times {10}^{-5}$ &  6006 & 30\\

\hline
GRB\,090510 & GBM+LAT+LLE  & 0.0–-1.74  & Band+PL & $ -0.50 \pm 0.14$ &  $-2.45 \pm 0.15$  &  $3500\pm400$ & $-1.6\pm0.03$ & - & $8.4\pm 1.5 \times {10}^{-6}$ & 6945 & 8 \\

GRB\,090510 & GBM   & ... & Band+BB & $-0.60 \pm 0.09$  & $-2.7 \pm 0.25$ &   $3890\pm 310$ & -  &  $12.5 \pm 2.6$ &  $4.3\pm 0.8 \times {10}^{-5}$ & 5864 & 7  \\
\hline

GRB\,090902B & GBM+LAT+LLE    & 2.816--22.144 & Band+CPL+PL & $-0.72 \pm 0.01$ & $-4.21 \pm 0.26$ & $738 \pm 100 $ & $-1.1 \pm 0.02$ & - & $5.0\pm 1.01 \times {10}^{-5}$ & 9950 & 384\\

GRB\,090902B & GBM  & ... & Band & $-0.96\pm0.04$ & $-4.30 \pm 0.33$ & $867 \pm 11$ & - & - &  $1.2 \pm 0.07 \times {10}^{-5}$  & 8999 & 0\\
\hline
GRB\,090926A & GBM+LAT   & 2.176--15.936 & Band+BB & $-0.82 \pm 0.012$ & $-2.51 \pm 0.01$ & $326 \pm 12$  &  -  & $38 \pm 1.5$  & $5.7 \pm 0.05 \times {10}^{-5} $ & 8949 & 14\\

GRB\,090926A &  GBM   &... & Band & $-0.69 \pm 0.08$ & $-2.6 \pm 0.05$ & $289 \pm 4.0$ & - & - & $1.5 \pm 0.5 \times {10}^{-5}$  & 7920 & 0\\
\hline

GRB\,091003A & LAT+GBM  & 0.832--21.056 & Band & $-0.95\pm 0.03$ & $-2.58\pm 0.06$ & $374\pm 21$	& - & - & $1.12 \pm 0.34 \times 10^{-6}$  & 9487 & 0\\

GRB\,091003A & GBM  & ... & Band & $-0.93\pm0.03$& $-2.39\pm0.13$ & $356 \pm 25$& - & - & $1.33 \pm 0.11 \times 10^{-6}$  & 9412 & 0\\
\hline

GRB\,091127 & LAT+GBM   & 0.0--8.7 & SBPL+BB & $-1.5 \pm 0.05	$ & $-2.7 \pm 0.07$ &  $84 \pm 7.5 $   & - & $ 4.93\pm0.22 $  & $5.72\pm0.36 \times {10}^{-6} $ & 5322 & 1 \\

GRB\,091127 & GBM   & ... &  SBPL & $-0.84 \pm 0.05$ & $-2.29 \pm 0.01$ & $16\pm0.50$   & - & -  & $1.14\pm 0.04\times {10}^{-5} $  & 6386 & 10\\
\hline
GRB\,100414A & LAT+GBM  & 1.856--28.352 & Band+PL & $-0.36\pm0.04$ & $-4.6\pm1.4$ & $570 \pm 14 $ & - & -  & $2.42\pm 0.12 \times 10^{-6}$  & 7735 & 22 \\

GRB\,100414A & GBM  & ... & SBPL & $-0.63\pm0.02$ & $-5.0\pm0.01$ & $879\pm22$ & - &  & $1.66\pm 0.17 \times 10^{-5}$ & 7631 & 15 \\

\hline
GRB\,100728A & LAT+GBM  & 13.312--178.69 & Band & $-0.61\pm 0.04$ & $-2.93 \pm 0.10$ & $320 \pm 15 $ & - & - & $1.28 \pm 0.17 \times 10^{-6}$ & 11399 & 0\\

GRB\,100728A & GBM  & ... & Band & $-0.61\pm0.05$ & $-2.93\pm0.30$ & $320\pm 18$& - & - &  $5.8 \pm 0.7\times 10^{-7}$ & 11205 & 0\\
\hline

GRB\,110721A & GBM+LAT+LLE  & 0.003--21.825 & ISSM+BB & $-1.20\pm 0.03$ & $-3.0 \pm 0.1$ &  $1700\pm260$  &  - & $35 \pm 2.7$  & $3.1\pm 0.65 \times {10}^{-6}$  & 6499 & 42\\

GRB\,110721A & GBM   & ... & ISSM & $-9.78 \pm 0.32$ & $-2.21 \pm 0.05$ &  $1450 \pm 230$  &  - & - & $1.12\pm 0.07 \times {10}^{-5}$  & 5963 & 21\\
\hline
GRB\,110731A   & GBM+LAT+LLE  & 0.003--7.488  & Band+ExpCut  & $0.08 \pm 0.02$  &  $-2.31 \pm 0.06$ & $290 \pm 25$ & - & - & $4.3 \pm 1.2\times {10}^{-6} $ & 3645 & 33\\

GRB\,110731A   & GBM   & ...  & Band  & $0.31\pm 0.17$  &  $-1.84 \pm 0.15$ & $266 \pm 36$ & - & -  & $4.3 \pm1.2 \times {10}^{-6}$  & 3116 & 24\\
\hline
GRB\,130427A  & GBM+LAT+LLE  & 11.23--142.34  & Band+PL & $-0.95 \pm0.01$ &  $-3.0\pm0.04$ & $886\pm11$ & $-1.83\pm0.03$ &  -  & $8.92 \pm 1.3 \times {10}^{-6}$ & 12757 & 201\\ 

GRB\,130427A & GBM & ... & Band & $-0.98 \pm 0.03$ & $-3.07 \pm 0.09$ & $884 \pm11$ & - & - & $1.37 \pm 0.06 \times {10}^{-5}$  & 7230 & 0\\

\hline
GRB\,130518A & LAT+LLE+GBM  & 9.92--58.497 & ISSM & $-0.73\pm0.02$ & $-3.26\pm0.04$ & $468\pm12$ & - & - & $1.08\pm0.05\times 10^{-5}$  & 12534 &  15 \\

GRB\,130518A & GBM  & ... & ISSM & $-0.71\pm0.03$ & $-3.01\pm0.14$ & $486\pm 18$ & - & - & $5.96\pm0.32\times 10^{-6}$ & 12235 & 21 \\

\hline
GRB\,131108A & GBM+LAT+LLE   &  0.32--18.496 &  Band & $-0.874 \pm0.026$  & $-2.27 \pm 0.11$ & $336 \pm 7.0$  &  - & - & $3.3 \pm 1.3 \times {10}^{-6} $  & 7244 & 0\\

GRB\,131108A & GBM   &  ... &  Band & $-1.27 \pm 0.06 $  & $-2.48 \pm 0.15$ & $348 \pm 7.0$&  - & - & $ 3.8 \pm 0.36 \times {10}^{-5}$  & 5598 & 0\\
\hline

GRB\,131231A & GBM+LAT+LLE   &  13.312--44.544 &  SBPL  & $-1.266 \pm 0.06$  &  $-3.01 \pm 0.024$ & $234 \pm 0.62$  & - & -  & $2.64 \pm 0.03 \times {10}^{-5}$  & 7039 & 36\\

GRB\,131231A & GBM   & ... & SBPL & $-1.25\pm0.01$  &  $-2.73 \pm 0.04$ & $190\pm0.7$  & - & -  & $2.62 \pm 0.6 \times {10}^{-5}$  & 6787 & 3\\

\hline
GRB\,141028A & GBM+LAT+LLE  & 6.656--38.145  & ISSM & $-0.70 \pm 0.03$  &  $-2.86 \pm 0.03$ & $433 \pm 18$ & -  & -  & $6.80 \pm 0.45 \times {10}^{-6} $  & 6367 & 55\\

GRB\,141028A & GBM  & ...  & ISSM &$-0.53 \pm 0.06$  &  $-2.30\pm0.04$ & $630 \pm 50$ & -  & -   & $6.8 \pm 0.45 \times {10}^{-6}  $  & 5949 & 8 \\

\hline
GRB\,150314A  & GBM+LAT+LLE  &  0.003--10.816 &  Band & $-1.27 \pm 0.11$ & $-2.53 \pm 0.06$  & $70\pm11$ & - & - & $2.1 \pm 0.4 \times {10}^{-6}$  & 5410 & 0\\

GRB\,150314A  & GBM &  ... &  Band & $-1.23 \pm 0.08$ & $-2.38 \pm 0.14$  & $ 66 \pm 6.0$ & - & -  & $2.6 \pm 0.4 \times {10}^{-6}$  & 5250 & 0\\
\hline

GRB\,150403A    & LLE+GBM   & 3.328--25.6 & ISSM+BB & $-0.99 \pm 0.04$ & $-3.9 \pm 0.40$  & $830 \pm 70$ &  - &$35 \pm 1.5$ & $ 5.2 \pm 0.04 \times {10}^{-6} $ & 4564 & 29\\

GRB\,150403A    & GBM   & ... & ISSM+BB &  $-0.09 \pm 0.25$ & $-2.37 \pm 0.06$ & $550 \pm 40$ & -  &$5.2 \pm 0.4$ & $1.69 \pm 0.90 \times {10}^{-5}$ & 4539 & 4\\
\hline

GRB\,150514A & GBM+LAT+LLE   & 0.0--10.816 &  Band & $-1.27\pm 0.08 $  & $-2.55 \pm0.07$ & $70 \pm 50$  & - & -  &  $1.67  \pm 0.31 \times {10}^{-6}$ & 5368 & 0\\

GRB\,150514A & GBM   & ... &  Band & $-1.23 \pm0.09 $  & $-2.38 \pm 0.14 $ & $66 \pm 55 $  & - & -  & $2.0 \pm 0.4 \times {10}^{-6} $  & 5250 & 0\\
\hline

GRB\,160509A & GBM+LAT+LLE  &  8.192--377.862 & Band+PL & $-0.94\pm0.02$ & $-3.25 \pm 0.06$ & $2820 \pm 210$ & $-1.35 \pm 0.26$ & - & $2.04 \pm 0.08 \times 10^{-6}$ & 14150 & 27 \\

GRB\,160509A & GBM  &  ... & Band & $-0.98\pm0.02$ & $-5.0\pm0.02$ & $4600 \pm 900$ & - & - &  $3.6 \pm 0.5 \times 10^{-6}$ & 11296 & 0\\
\hline

GRB\,160625B & GBM+LAT+LLE  & 188.451--214 & ISSM+PL & $-0.68\pm0.01$ & $-3.154 \pm0.03$ & $755 \pm 8.0 $ & $-1.6 \pm 0.02$ & - & $5.22 \pm 0.26 \times 10^{-5}$& 7338 & 4 \\

GRB\,160625B & GBM  & ... & ISSM & $-0.67\pm0.08$ & $-3.12 \pm 0.04$ & $740\pm 8.0$ & - & - & $1.12\pm 0.02 \times 10^{-4}$ & 7409 & 0 \\

\hline

GRB\,170214A &  GBM+LAT+LLE  & 12.544--135.426 & ISSM+BB  & $-1.04 \pm 0.05$ & $-2.81 \pm 0.05$  & $760 \pm 80$ & - & $50 \pm 2.1$  & $4.82\pm 0.34 \times {10}^{-6}$ & 11284 & 17\\

GRB\,170214A &  GBM  & ... & ISSM+BB  & $-0.02 \pm 0.20$ & $-2.78 \pm 0.13$  & $400 \pm 19$ & - & $3.42 \pm 0.23$ &  $6.40\pm 0.35 \times {10}^{-6}$  & 8664 & 35\\
\hline

GRB\,170405A    & GBM+LAT+LLE   & 7.36--86.08  & ISSM &  $-0.64 \pm 0.4$ &  $-3.01 \pm 0.03$ &  $367 \pm 14$ &  - & - & $6.40 \pm 0.40 \times {10}^{-6}$  & 8664 & 13\\

GRB\,170405A    & GBM   & ...  & Band &  $-0.77 \pm 0.03$ &  $-2.34 \pm 0.11$ &  $277 \pm 12 $ &  - & - & $ 2.87 \pm 0.24 \times {10}^{-6}$  & 7638 & 0\\

\hline
GRB\,180720B & LAT+LLE+GBM  & 4.352--53.249 & ISSM & $-1.05\pm0.01$ & $-2.99\pm0.04$ & $755 \pm 17$ & - & - & $5.46\pm 0.06 \times 10^{-6}$ & 10286 & 126 \\

GRB\,180720B & GBM  & ... & ISSM & $-1.04\pm0.01$ & $-2.85 \pm 0.07$ & $755\pm 21$ &  & & $4.29 \pm 0.8 \times 10^{-5}$ & 9764 &79 \\

\hline
GRB\,190114C & GBM+LAT+LLE  & 2.3--116.4  & Band+PL &  $-0.54 \pm 0.03$ & $-3.01 \pm 0.09$  & $648 \pm 18$ &$-2.01 \pm 0.01$ & - & $1.06 \pm 0.04 \times {10}^{-5} $ & 10288 &  1142 \\

GRB\,190114C & GBM  & ...  & Band+PL &  $-0.41 \pm 0.04$ & $-4.4 \pm 1.4$  & $632\pm 17$ &$-1.89 \pm 0.02$  & -  &  $9.66 \pm 0.34 \times {10}^{-5}$  & 6063 & 1009\\
\hline

GRB\,200613A & LAT+GBM  & 3.648--470 & Band+CPL& $-1.16 \pm 0.14$ & $-2.52 \pm 0.06$ & $127 \pm 6.0$ & - & - & $1.11 \pm0.1 \times 10^{-7}$& 10175 & 51\\

GRB\,200613A & GBM  & ... & Band+CPL & $-0.5 \pm 0.19 $ & $-2.25 \pm 0.16$ & $101 \pm 9$ & - & - & $1.45 \pm 0.22 \times 10^{-7}$& 9843 & 7\\

\hline
GRB\,210826A & GBM+LAT & 0.512--20.224 & ISSM & $-0.48\pm0.15$ & $-3.3 \pm 0.4$ & $174 \pm 13$ & - & - & $1.63 \pm 0.30 \times {10}^{-6}$  & 4343 &  7\\

GRB\,210826A & GBM  & ...  & Band & $-0.75 \pm 0.06$  & $-2.40 \pm 0.21$ &  $154 \pm 12$  & - & - & $1.09 \pm 0.25 \times {10}^{-7}$  & 4294 & 0\\
\hline

GRB\,211018A & LAT+GBM & 4.384--128.290 & Band & $-0.60\pm0.03$ & $-2.52\pm0.03$ & $318 \pm 10$ & - & - & $2.55\pm0.16 \times 10^{-6}$ & 10887 & 0 \\

GRB\,211018A & GBM  & ... & Band & $-0.62\pm0.03$ & $-2.89\pm0.31$ & $330 \pm 13 $ & - & - & $9.1 \pm 0.7 \times 10^{-7}$ & 10451 & 0\\

\hline

GRB\,220101A    & GBM+LAT+LLE  & 17.472--145.731  & SBPL & $-0.99 \pm 0.03$  &  $-2.80 \pm 0.08$ & $247 \pm 22 $ & - & - &  $2.67 \pm 0.08 \times {10}^{-6}$ & 8928 & 3\\
 
 GRB\,220101A    & GBM  & ...  & SBPL & $-0.98 \pm 0.03$  &  $-2.75 \pm 0.15$ & $237 \pm 22$ & - & - &  $2.47 \pm 0.21 \times {10}^{-6}$  & 8261 &  1\\

\hline
GRB\,220527A & LAT+GBM &  2.624--13.120 & Band & $-0.714\pm0.015	$ & $-2.76\pm0.04$ & $162 \pm 2.3 $ &  -&  - & $2.40 \pm 0.8 \times 10^{-5}$ & 8904 & 0 \\

GRB\,220527A & GBM  &  ... & Band & $-0.714\pm0.015$ & $-2.76\pm0.06$ & $162 \pm 2.5$ &  - &   - & $3.73\pm0.16 \times 10^{-6}$ & 8788 & 0 \\

\hline
GRB\,221009A & GBM+LAT+LLE  &  290--295 & ISSM+BB & $-1.57 \pm 0.02$ & $-2.51 \pm 0.07$ & $1170 \pm 100$ &  - & $78 \pm10$ & $1.98 \pm 0.21 \times{10}^{-5}$& 3914 & 141 \\

GRB\,221009A & GBM  &  ... & ISSM+BB & $-1.57 \pm 0.02$ & $-2.49 \pm 0.13$ & $1230\pm 130$ & - & $78 \pm 13$ & $1.61 \pm 0.06 \times {10}^{-4}$ & 3732 & 41 \\

\hline

GRB\,230812B  & LAT+GBM   & 0.384--3.648  & Band+ExpCut & $-0.85 \pm 0.03$  & $-3.05 \pm 0.04$  & $385 \pm 4.0$  & - & - &  $2.72 \pm 0.06 \times {10}^{-4}$   & 4531 & 186\\

GRB\,230812B  & GBM  & ...  &  Band & $-0.57 \pm 0.01$  & $-2.63 \pm 0.016$  & $207\pm 1.5$  & - & - & $1.53 \pm 0.06 \times {10}^{-4}$  & 5818 & 0\\
\hline

GRB\,240825A  & LAT+GBM   &  1.152--5.12 & Band+ExpCut & $-0.82 \pm 0.04$  & $-2.33 \pm 0.04$  & $516 \pm 38$  & - & - &  $6.8 \pm 1.7 \times {10}^{-5}$   & 8995 & 5\\

GRB\,240825A  & GBM  & ...  &  Band+ExpCut &$-0.83 \pm 0.02$ & $-2.27 \pm 0.03$  &$516 \pm 21$  & - & - & $6.8 \pm 1.7 \times {10}^{-5}$   & 7049 & 1\\
\hline

GRB\,241030A  & LAT+GBM   & 20.9--165 & Band & $-1.30 \pm 0.03$  & $-2.76 \pm 0.11$  & $177 \pm 13$  & - & - &  $4.4 \pm 0.6 \times {10}^{-7}$   & 10188 & 0\\

GRB\,241030A  & GBM  & ...  & Band & $-1.3 \pm 0.03$  & $-2.42 \pm 0.18$  & $151 \pm 11$  & - & - &  $3.9 \pm 0.5 \times {10}^{-7}$   & 10140 & 0\\

\hline
GRB\,241228B  & LAT+GBM   & 1.60--19.456 & SBPL+BB & $-0.54 \pm 0.07$  & $-3.12 \pm 0.10$  & $293   \pm 23$  & - & $7.4 \pm 0.5$&  $1.2 \pm 0.1 \times {10}^{-5}$   & 7734 & 16\\

GRB\,241228B  & GBM  & ...  & Band & $-0.81 \pm 0.07$  & $-5.0\pm 0.006$  & $426 \pm 12$  & - & - &  $1.1 \pm 0.7 \times {10}^{-5}$   & 7478 & 0\\

\hline

\end{longtable}

\begin{longtable}{@{\extracolsep{\fill}}lcccccl@{}}
\caption{\(E_{i,\rm p} - {E}_{\rm iso}\) values for 37 Fermi GRBs based on Joint and GBM-Only spectral fits. } \\

\hline
GRBs\,Names & Redshift &  $E_{i,\rm p}$ (keV) -- Joint &  $E_{\rm iso}$ (${10}^{52}$ -- Joint erg) & 
$E_{i,\rm p}$ (keV) -- GBM-only & $E_{\rm iso}$ (${10}^{52}$ erg) -- GBM-only  & Reference ($z$) \\

\hline
\endfirsthead

\multicolumn{7}{c}%
{{ \tablename\ \thetable{} -- (Continued)}} \\
\hline
GRBs\,Names & Redshift &  $E_{i,\rm p}$ (keV) -- Joint  &  $E_{\rm iso}$ (${10}^{52}$ erg) -- Joint  & 
$E_{i,\rm p}$ (keV) -- GBM-only  & $E_{\rm iso}$ (${10}^{52}$ erg) -- GBM-only  & Reference ($z$) \\
\hline
\endhead

\hline
\endlastfoot

GRB\,080916C & 4.35 & 6206 $\pm $1525 & 327 $\pm$ 67 & 2996 $\pm$ 268 & 1361 $\pm$ 285 & \cite{greiner2009redshift}\\
GRB\,090323  & 3.57 & 5575 $\pm$ 1051  & 2035 $\pm$ 68 & 3062 $\pm$ 502 & 1697 $\pm$ 104 & \cite{2009GCN.10100....1C}\\
GRB\,090328  & 0.736 & 1493 $\pm$ 226 & 37 $\pm$ 11 & 1250 $\pm$ 104 & 37 $\pm$ 511 & \cite{2009GCN..9053....1C} \\
GRB\,090424  & 0.544 & 239 $\pm$ 6.2 & 18 $\pm$ 1.0 & 245 $\pm$ 6.0 & 14 $\pm$ 4.0 & \cite{2009GCN..9243....1C} \\
GRB\,090510  & 0.903 & 6661 $\pm$ 761 & 3.23 $\pm$ 1.0 & 7402 $\pm$ 590 & 17 $\pm$ 3.1 & \cite{2009GCN..9353....1R} \\
GRB\,090902B & 1.822 & 2083 $\pm$ 282& 815 $\pm$ 2.12 & 2447 $\pm$ 31 & 198 $\pm$ 1.2  & \cite{2009GCN..9873....1C}\\
GRB\,090926A & 2.106 & 1013 $\pm$ 37 & 880 $\pm$ 7.7 & 898 $\pm$ 12 & 230 $\pm$ 8.0 & \cite{2009GCN..9948....1O}\\
GRB\,091003A & 0.897 & 710 $\pm$ 40 & 4.9 $\pm$ 0.2 & 675 $\pm$ 47 & 5.9 $\pm$ 0.5 & \cite{2009GCN.10031....1C}\\
GRB\,091127  & 0.49  & 125 $\pm$ 11 & 3.12 $\pm$ 0.10 & 24 $\pm$ 0.74 & 5.73 $\pm$ 0.27 & \cite{2009GCN.10233....1T} \\
GRB\,100414A & 1.368 & 1350 $\pm$ 33 & 32 $\pm$ 1.6 & 2082 $\pm$ 52& 220 $\pm$ 22 & \cite{2010GCN.10606....1C}\\
GRB\,100728A & 1.567 & 821 $\pm$ 39 & 137 $\pm$ 18 & 821 $\pm$ 46 & 62 $\pm$ 8.0 & \cite{2013GCN.14500....1K}\\
GRB\,110721A & 3.512 & 7670 $\pm$ 1173 & 180 $\pm$ 38 & 6542 $\pm$ 1038 & 652 $\pm$ 41 & \cite{2011GCN.12192....1G} \\
GRB\,110731A & 2.83  & 1111 $\pm$ 96 & 60 $\pm$ 17 & 1019 $\pm$ 138 & 60 $\pm$ 17 & \cite{2011GCN.12225....1T} \\
GRB\,130427A & 0.34  & 1146 $\pm$ 15 & 34 $\pm$ 1.0 & 1185 $\pm$ 15 & 53 $\pm$ 1.1 & \cite{2013GCN.14455....1L}\\
GRB\,130518A & 2.49  & 1633 $\pm$ 42 & 782 $\pm$ 31 & 1696 $\pm$ 63& 429 $\pm$ 23 & \cite{2013GCN.14687....1C} \\
GRB\,131108A & 2.4   & 1142 $\pm$ 58 & 84 $\pm$ 33 & 1183 $\pm$ 56 & 96 $\pm$ 10 & \cite{2013GCN.15470....1D}\\
GRB\,131231A & 0.6439 & 385 $\pm$ 1.1 & 91 $\pm$ 0.5 & 312 $\pm$ 1.2 & 90 $\pm$ 2.1 & \cite{2014GCN.15645....1X}\\
GRB\,141028A & 2.33  & 1442 $\pm$ 60 & 286 $\pm$ 19 & 2098 $\pm$ 167 & 286 $\pm$ 19 & \cite{2014GCN.16982....1C}\\
GRB\,150314A & 0.807 & 127 $\pm$ 20 & 4.6 $\pm$ 0.15 & 119 $\pm$ 10 & 4.94 $\pm$ 0.1 & \cite{2015GCN.17583....1D}\\
GRB\,150403A & 2.06  & 2540 $\pm$ 214 & 60 $\pm$ 1.0 & 1683 $\pm$ 122 & 403 $\pm$ 22 & \cite{2015GCN.17672....1P}\\
GRB\,150514A & 0.807 & 127 $\pm$ 90 & 3.2 $\pm$ 0.59 & 119 $\pm$ 99 & 4.0 $\pm$ 0.76 & \cite{2015GCN.17822....1D} \\
GRB\,160509A & 1.17  & 6119 $\pm$ 434 & 323 $\pm$ 17 & 5165 $\pm$ 271 & 1080 $\pm$ 130 & \cite{2016GCN.19419....1T}\\
GRB\,160625B & 1.406 & 390 $\pm$ 5.0 & 505 $\pm$ 7.0 & 404 $\pm$ 5.0 & 49 $\pm$ 5.0 & \cite{2016GCN.19600....1X}\\
GRB\,170214A & 2.53  & 1412 $\pm$ 57 & 991 $\pm$ 42 & 1412 $\pm$ 67 & 579 $\pm$ 37 & \cite{2017GCN.20686....1K} \\
GRB\,170405A & 3.51  & 1655 $\pm$ 63 & 1342 $\pm$ 73 & 1249 $\pm$ 54 & 602 $\pm$ 50 & \cite{2017GCN.20990....1D}\\
GRB\,180720B & 0.654 & 1249 $\pm$ 28 & 31  $\pm$ 0.33 & 1282 $\pm$ 35 & 240 $\pm$ 5.0 & \cite{2018GCN.22996....1V}\\
GRB\,190114C & 0.425 & 923 $\pm$ 26 & 57 $\pm$ 2.2 & 901 $\pm$ 24 & 52 $\pm$ 1.84 & \cite{2019GCN.23708....1C}\\
GRB\,200613A & 1.227 & 138 $\pm$ 11 & 29 $\pm$ 1.3 & 225 $\pm$ 25 & 18 $\pm$ 3.49 & \cite{2020GCN.27937....1P}\\
GRB\,210826A & 0.05  & 181 $\pm$ 14 & 0.02 $\pm$ 0.01 & 162 $\pm$ 13 & 0.01 $\pm$ 0.001 & \cite{2021GCN.30725....1D}\\
GRB\,211018A & 0.64  & 522 $\pm$ 16 & 35 $\pm$ 2.5 & 541 $\pm$ 21 & 12 $\pm$ 1.0 & \cite{2021GCN.31010....1D}\\
GRB\,220101A & 4.61  & 1386 $\pm$ 123 & 1406 $\pm$ 42 & 1330 $\pm$ 124 & 1300 $\pm$ 111 & \cite{2022GCN.31357....1P}\\
GRB\,220527A & 0.857 & 301 $\pm$ 4.3 & 50 $\pm$ 1.6 & 301 $\pm$ 5.0 & 7.8 $\pm$ 0.27 & \cite{2022GCN.32144....1S} \\
GRB\,221009A & 0.151 & 1347 $\pm$ 115 & 0.53 $\pm$ 0.01 & 1416 $\pm$ 150 & 4.4 $\pm$ 0.16 & \cite{2022GCN.32648....1D}\\
GRB\,230812B & 0.36  & 524 $\pm$ 5.4 & 29 $\pm$ 1.0 & 524 $\pm$ 5.44 & 8.9 $\pm$ 2.22 & \cite{2023GCN.34409....1D}\\
GRB\,240825A & 0.659 & 856 $\pm$ 63 & 31 $\pm$ 8.0 & 856 $\pm$ 34   & 31 $\pm$ 7.8 &   \cite{2024GCN.37293....1M}\\
GRB\,241030A & 1.4   & 425 $\pm$ 31 & 33 $\pm$ 4.2 & 362 $\pm$ 24  & 29 $\pm$ 3.8 &   \cite{2024GCN.38027....1L}\\
GRB\,241228B & 2.674 & 1077 $\pm$ 84 & 326 $\pm$ 166 & 1077 $\pm$ 84   & 329 $\pm$ 20 &   \cite{2024GCN.38704....1A}\\
\hline
\end{longtable}
\setcounter{table}{3}
\small
\centering
\begin{longtable}{lccccccccccc}
\caption{Best-fit parameters from spectral analysis during the peak flux interval of joint GBM-LLE+LAT data and GBM-only data. $\alpha_1$, and $\beta$ are the lower and higher photon indices for the Band and SBPL models, respectively. $\gamma$ is the photon index of the ExpCut model, while $\alpha_2$ is that of the PL model. $E_{\rm p}$ is the Band or ExpCut peak energy. $kT$ is the temperature of the BB model.} \\
\hline
Names & Instruments    & Peak flux interval & Model & ${\alpha}_1, \gamma$ & $\beta$ & $E_{\rm p}$ (keV) & ${\alpha}_2$  &  $kT$ (keV) &   Peak flux ($\rm erg$ ${\rm cm}^{-2}$  ${\rm s}^{-1}$)  & BIC  & $\Delta$BIC \\
\hline
\endfirsthead
\caption[]{(Continued)}\\
\hline
Names & Instruments    & Peak flux interval & Model & ${\alpha}_1, \gamma$ & $\beta$ & $E_{\rm p}$ (keV)& ${\alpha}_2$  &  $kT$ (keV)&   Peak flux ($\rm erg$ ${\rm cm}^{-2}$  ${\rm s}^{-1}$) & BIC  & $\Delta$BIC \\
\hline
\endhead
\hline
\endfoot
GRB\,080916C  & GBM+LAT+LLE  & 2.048--3.072 & Band & $-0.77 \pm 0.07 $ & $-2.8 \pm 0.16$ & $540 \pm 90$ &  - &  - &   $6.4 \pm 2.0 \times {10^-6}$ & 1692 & 0 \\

GRB\,080916C  & GBM  & ... & Band & $-0.78 \pm 0.06$ & $-4 \pm 5$ & $560 \pm 90$ &  - &  - & $1.56 \pm 0.36 \times {10}^{-5} $   & 1692 & 0\\

\hline
    
GRB\,090323  & GBM+LLE   & 65.536--66.56 & Band & $-0.52 \pm 0.02$ &  $-3.00 \pm 0.04$  & $380 \pm 34 $ &  - & - & $1.78 \pm 0.10 \times {10}^{-5}$ & 2185 & 0\\

GRB\,090323 &  GBM  &  ... & Band & $-0.51 \pm 0.06$ & $-4.3 \pm 2.6$  & $380 \pm 30 $ & -  & - & $1.46 \pm 0.32 \times {10}^{-5}$ & 2019 & 0\\

\hline

GRB\,090328 & GBM+LLE   & 23.552--24.576 & Band & $-0.75 \pm 0.05$ & $-2.82 \pm 0.10$ &  $440 \pm 40$ & - &  - & $2.39 \pm 0.13 \times {10}^{-5} $ &  1796 & 0 \\ 

GRB\,090328 & GBM   & ... & Band & $-0.75 \pm 0.04$ & $-5.0 \pm 0.05$ &  $448 \pm 35 $  & - & -  & $2.39 \pm 0.13 \times {10}^{-5}$ & 1809 & 0 \\

\hline

GRB\,090510 & GBM+LLE   & 0.512--0.576 & SBPL & $-0.79 \pm 0.07$ & $-5.0 \pm 0.002$ &  $4000 \pm 550$ & - &  - & $1.4 \pm 0.5 \times {10}^{-4} $ &  404 & 1 \\

GRB\,090510 & GBM   &  ... & SBPL & $-0.79 \pm 0.07$ & $-5.0 \pm 0.002$ &  $4000 \pm 550 $  & - & -  & $1.2 \pm 0.9 \times {10}^{-4}$ & 395 & 0 \\

\hline

GRB\,090902B & GBM+LAT+LLE & 9.216--10.24 & ISSM+PL & $0.58 \pm 0.6$ & $-7.8 \pm 1.1$ & $703 \pm 29$ & $-1.93 \pm 0.01$ & - & $2.15 \pm 0.16 \times {10}^{-5}$ & 2889 & 1061\\

GRB\,090902B & GBM  & ... & ISSM & $0.40 \pm 0.13$ & $-9.4 \pm 3.1$ & $707\pm 22$ & - & - & $1.13\pm0.11 \times {10}^{-4}$ & 2689 & 0\\

\hline

GRB\,090926A & GBM+LAT   & 3.072--4.096 & SBPL & $-0.47 \pm 0.028$ & $-3.00 \pm 0.09$ & $253 \pm  15$  &  -  &-& $5.9 \pm 0.5 \times {10}^{-5}$  & 2826 & 14\\

GRB\,090926A &  GBM   & ... & SBPL & $-0.47 \pm 0.02$ & $-3.05 \pm 0.11$ & $ 255\pm 18$ & - & - &  $7.0 \pm 0.65 \times {10}^{-4} $  & 2797 & 14\\

\hline
GRB\,100414A & LAT+GBM  & 23.552--24.576 & Band & $-0.67\pm0.05$ & $-2.50\pm0.16$ & $460 \pm 40$ & - & - & $4.9 \pm 0.6 \times 10^{-6}$ & 2421 & 0 \\

GRB\,100414A & GBM  & ... & Band & $-0.69\pm0.05$ & $-2.9\pm0.7$ & $490\pm50$ & - & - & $1.35\pm0.3 \times 10^{-5}$ & 2399 & 0 \\
\hline

GRB\,110721A & GBM+LAT+LLE   & 2.048--3.072 & Band+ExpCut & $-1.11 \pm 0.05$ & $-2.85 \pm 0.14$ &  $810\pm70$  &  -  & - & $7.9 \pm 1.2 \times {10}^{-6} $  & 2248 & 30\\

GRB\,110721A & GBM   & ... & Band+ExpCut & $-1.11 \pm 0.05$ & $-2.74 \pm 0.35$ &  $450\pm40 $  & - & - & $3.83\pm 0.18 \times {10}^{-5}$ & 2223 & 1\\

\hline

GRB\,110731A   & GBM+LLE   & 0.0--1.024  & Band  & $-0.97 \pm 0.12$  &  $-3.07 \pm 0.18$ & $ 173 \pm 23$ & - & - & $9.6 \pm 2.3\times {10}^{-6}$ & 1506 & 0\\

GRB\,110731A   & GBM   & ...  & Band  & $-0.97 \pm 0.08$  &  $-4.1 \pm 2.5$ & $175\pm17$ & - & - & $9.6\pm 2.3 \times {10}^{-6}$ & 1516 & 0\\

\hline

GRB\,120624B & LLE+GBM  & 11.264--12.288& Band & $-0.77\pm0.06$ & $-2.19\pm0.16$ & $640\pm 110$ & - & - & $5.4\pm 1.15 \times 10^{-6}$ & 3046 & 0 \\

GRB\,120624B & GBM  & ... & Band & $-0.76 \pm 0.06$ & $-2.17 \pm 0.15$ & $630\pm100$ & - & - &$5.2\pm 1.15 \times 10^{-6}$ & 3042 & 0 \\
\hline

GRB\,130427A  & GBM+LAT+LLE  & 8.192--9.216  &Band+BB & $-0.49 \pm 0.01$ &  $-3.88 \pm 0.16$ & $934 \pm 16$ & -  &  $43 \pm 16$& $ 6.4\pm 2.3 \times {10}^{-4} $ & 2588 & 399\\

GRB\,130427A & GBM  & ...  & Band+BB & $-0.49 \pm 0.011$ & $-4.05 \pm 0.15$ & $944\pm 13$ & - & $43.5 \pm 1.4$ & $6.4\pm 2.2 \times {10}^{-4}$   & 2474 & 406\\

\hline

GRB\,130518A & LLE+GBM  & 25.6--26.624& ISSM & $-0.57\pm0.04$ & $-2.82\pm0.13$ & $670\pm 40$ & - & - & $1.43 \pm 0.19\times 10^{-5}$ & 3747 & 24 \\

GRB\,130518A & GBM  & ... & ISSM & $-0.57 \pm 0.04$ & $-2.82 \pm 0.13$ & $670 \pm 40$ & - & - & $2.1 \pm 0.21 \times 10^{-5}$ & 3728 & 24 \\
\hline

GRB\,131108A & GBM+LAT+LLE   &  0.0--1.0 &  ISSM  & $-0.64 \pm 0.07$  &  $-2.3 \pm 0.027$ & $388 \pm 16$  & - & -  & $1.9 \pm 0.4 \times {10}^{-5}$  & 2298 & 15\\

GRB\,131108A & GBM   & ... & Band  & $-0.63 \pm 0.57$  &  $-2.0 \pm0.07$ & $323 \pm 35 $  & - & - & $1.39 \pm 0.30 \times {10}^{-5}$ & 1979 & 0\\

\hline
GRB\,131231A & GBM+LAT+LLE   & 22.528--23.552 & SBPL  & $-0.81 \pm 0.02$  &  $-3.11 \pm 0.07$ & $266 \pm 16 $  & - & -  & $ 4.7 \pm 0.45 \times {10}^{-5} $  & 2321 & 3 \\

GRB\,131231A & GBM   & ... & SBPL  & $-0.82 \pm 0.03$  &  $-3.08 \pm 0.14$ & $261\pm 23$  & - & - & $5.4 \pm 0.5 \times {10}^{-5} $ & 2282 & 2\\

\hline

GRB\,141028A & GBM+LAT+LLE & 12.288--13.312 & ISSM & $-0.47 \pm 0.09$  &  $-3.16 \pm  0.09$ & $570 \pm 50$ & -  & -  & $2.77 \pm 0.11 \times {10}^{-5}$  & 1686 & 4\\

GRB\,141028A & GBM  &... & ISSM & $-0.43 \pm 0.03$  &  $-3.13 \pm 0.30$ & $540 \pm 40$ & -  & -  & $2.77 \pm 0.16 \times {10}^{-5} $  & 1646 &  3\\

  \hline
GRB\,150314A  & GBM+LLE  &  1.024--2.048 & Band & $-0.69 \pm 0.23$ & $-2.57 \pm 0.04$  & $520 \pm 40$ & - & - & $7.4 \pm 0.4 \times {10}^{-6}$  & 2242 & 0\\

GRB\,150314A  & GBM  & ... &  Band & $-0.76 \pm 0.12$ & $-2.73 \pm 0.18$  & $540 \pm 34$ & - & - & $2.4 \pm 0.7 \times {10}^{-6}$  & 2229 & 0\\

\hline

GRB\,150403A    & GBM+LLE  & 11.264--12.288 & ISSM & $-0.50 \pm 0.06$ & $-2.71 \pm 0.09$  & $730 \pm 60$ &  - &  - & $7.76 \pm 0.23 \times {10}^{-6}$ & 1533 & 21\\

GRB\,150403A & GBM  & ... & ISSM &  $-0.48 \pm 0.07$ & $-2.63 \pm 0.15$ & $750 \pm 70$ & -  & - & $5.5 \pm 0.6\times {10}^{-5}$ & 1642 & 9\\
\hline

GRB\,150514A & GBM+LLE   & 1.024--2.048 &  Band & $-0.69 \pm 0.23 $  & $-2.57 \pm 0.09$ & $53\pm40$  & - & -  & $7.4 \pm 0.4 \times {10}^{-6}$  & 2242 & 0\\

GRB\,150514A & GBM   & ... &  Band & $-0.76 \pm 0.80 $  & $-2.73 \pm 0.16$ & $54 \pm 32$  & - & -  & $7.4\pm 0.57 \times {10}^{-6}$ & 2228 & 0\\
\hline

GRB\,160509A & GBM+LAT+LLE  & 16.384--17.408 & SBPL & $-0.83\pm0.02$ & $-2.65 \pm 0.10$ & $273 \pm 24$ & - & - & $4.3\pm 1.45 \times 10^{-5}$ & 3822 & 16 \\

GRB\,160509A & GBM  &  ... & SBPL & $-0.82\pm0.02$ & $-2.57 \pm 0.09$ & $257 \pm 22$ & - & - &  $8.6\pm 0.7 \times 10^{-5}$ & 3728 & 14\\
\hline

GRB\,160625B & GBM+LLE  & 188.45--189.47 & DSBPL+BB & $-1.57\pm0.03$ & $-2.66 \pm 0.07$ & $2000 \pm 140 $ & $-0.66 \pm 0.02$ & $20 \pm 1.1$ & $ 8.57 \pm 0.14 \times 10^{-5}$& 5375 & 72 \\

GRB\,160625B & GBM  & ... & DSBPL+BB & $-1.562 \pm 0.04$ & $-2.49 \pm 0.06$ & $1680\pm140 $ & $-0.67 \pm 0.02$ & $20 \pm 1.1$ & $8.35\pm0.14 \times 10^{-5}$ & 5329 & 76 \\
\hline

GRB\,170214A &  GBM+LAT+LLE  &  61.44--62.464 & Band+ExpCut  & $-1.21 \pm 0.05$ & $-3.25 \pm 0.16$  & $4800\pm750$ & - & - & $1.34 \pm 0.29 \times {10}^{-5}$ & 2575 & 28\\

GRB\,170214A &  GBM  & ... & Band & $-0.036 \pm 0.20$ & $-1.92 \pm 0.06$  & $300 \pm 0.6 $ & - & - & $2.16 \pm 0.36 \times {10}^{-5}$  & 2448 & 0 \\

\hline

GRB\,170405A    & GBM+LAT+LLE   & 29.696--30.72  & SBPL &  $-0.73 \pm 0.07$ & $-2.88 \pm 0.08$ & $272 \pm 40$&  - & - & $1.3 \pm 0.4 \pm {10}^{-5} $ & 2045 &  1\\

GRB\,170405A    & GBM  &  ... & SBPL & $-0.63 \pm 0.07$  &  $-2.39 \pm 0.23$ & $250 \pm 60$  &  - & - & $1.92\pm 0.16 \times {10}^{-5} $  & 2018 & 0\\

\hline

GRB\,180720B & GBM+LAT+LLE  & 16.384--17.408 & ISSM& $-0.83\pm0.021$ & $-2.70 \pm 0.04$ & $790 \pm 60 $ &  - & - & $3.00\pm0.07 \times 10^{-5}$& 5070 & 90 \\

GRB\,180720B & GBM  & ... & ISSM & $-0.78\pm0.03$ & $-2.47 \pm 0.07$ & $1080\pm 70$ & - & - & $1.78\pm 0.35 \times 10^{-4}$& 3578 &37 \\
\hline

GRB\,190114C & GBM+LAT+LLE  & 3.0--4.0 & Band  &  $-0.28 \pm 0.012$ & $-2.77 \pm 0.04$  & $881\pm13$ & - & -  & $4.1 \pm 0.5 \times {10}^{-4}$  & 2813 & 0 \\

GRB\,190114C & GBM &  ... & Band&  $-0.28 \pm 0.011$ & $-2.81 \pm 0.04$  & $885\pm13$ & -  & -  & $7.72\pm0.25 \times {10}^{-4}$ & 2592 & 0\\
\hline

 GRB\,220101A    & GBM+LLE & 104--105  & Band & $-0.74 \pm 0.09$  &  $-2.50 \pm 0.13$ & $450 \pm 75$ & - & - &  $3.5\pm 1.3 \times {10}^{-6}$& 1911 & 0\\
 
 GRB\,220101A    & GBM  & ... & Band & $-0.74 \pm 0.09$  &  $-2.09 \pm 0.23$ & $430 \pm 80$ & - & - & $3.0\pm1.1 \times {10}^{-6} $ & 1934 &  0\\

\hline
 GRB\,221009A    & GBM+LLE  & 219--220  & DSBPL & $-0.90 \pm0.01$  &  $-3.54 \pm 0.11$ & $775 \pm 23$ & $-1.2 \pm 0.12$ & - &  $3.08 \pm 0.04 \times {10}^{-4}$& 2769 & 166\\
 
 GRB\,221009A    & GBM  & ... & DSBPL & $-0.90\pm0.007$  &  $-3.99 \pm 0.17$ & $768\pm 22$ & $-0.90 \pm 0.05$ & - & $3.31 \pm 0.13 \times {10}^{-4}  $ & 2812 &  123\\

\hline

GRB\,230812B  & LAT+GBM   &  0.0--1.0 & SBPL+BB  & $-0.571 \pm 0.18$  & $-3.29 \pm 0.05$  & $667\pm 295$  & - & $47 \pm 0.9$ & $3.33 \pm 0.13 \times {10}^{-4} $  & 2849 & 150\\

GRB\,230812B  & GBM  &  ... & SBPL+BB  & $-0.54 \pm 0.17$  & $-3.21 \pm 0.06$  & $614 \pm 285$  & - & $47 \pm0.9$ & $2.23 \pm  0.08 \times {10}^{-4}$  & 3672 & 117\\

\hline
GRB\,240825A  & GBM+LAT   & 1.024--2.048  & DSBPL & $-1.32 \pm 0.01	$  & $-2.4 \pm 0.02$  & $498 \pm 20$  & - & - &  $1.44 \pm 0.07 \times {10}^{-4}$   & 5426 & 36\\

GRB\,240825A  & GBM  & ...  &  DSBPL &$-0.34 \pm 0.01$ & $-1.21 \pm 0.03$  &$443 \pm 14$  & - & - & $1.44 \pm 0.07 \times {10}^{-4}$   & 4893 & 31\\

\hline
GRB\,241228B  & GBM+LAT   &4.0--5.0 & Band & $-0.49 \pm 0.05$  & $-3.09 \pm 0.28$  & $428   \pm 26$  & - & - &  $2.06 \pm 0.25 \times {10}^{-5}$   & 2785 & 0\\

GRB\,241228B  & GBM  & ...  & Band & $-0.50 \pm 0.09$  & $-4.4 \pm 2.5$  & $434 \pm 24$  & - & - &  $1.78 \pm 0.2 \times {10}^{-5}$   & 2759 & 0\\
\end{longtable}
\begin{longtable}{@{\extracolsep{\fill}}lcccccl@{}}
\caption{\(E_{i,\rm p} - {L}_{\rm iso}\) values for 29 Fermi GRBs based on Joint and GBM-Only spectral fits. } \\

\hline
GRBs\,Names & Redshift &  $E_{i,\rm p}$ (keV) -- Joint  &  $L_{\rm iso}$  (${10}^{52}$ erg $s^{-1}$) -- Joint  & 
$E_{i,\rm p}$ (keV) -- GBM-only & $L_{\rm iso}$ (${10}^{52}$ erg $s^{-1}$) -- GBM-only & Reference ($z$) \\
\hline
\endfirsthead

\multicolumn{7}{c}%
{{ \tablename\ \thetable{} -- (Continued)}} \\
\hline
GRBs\,Names & Redshift &  $E_{i,\rm p}$ (keV) -- Joint &  $L_{\rm iso}$  (${10}^{52}$ erg $s^{-1}$) -- Joint  & 
$E_{i,\rm p}$ (keV) -- GBM-only & $L_{\rm iso}$ (${10}^{52}$ erg $s^{-1}$) -- GBM-only & Reference ($z$) \\
\hline
\endhead

\hline
\endlastfoot

GRB\,080916C        & 4.35   & 2889 $\pm$ 455 & 313 $\pm$ 72  & 1926 $\pm$ 295 & 2996 $\pm$ 481 & \cite{greiner2009redshift} \\ 
GRB\,090323         & 3.57       & 1736 $\pm$ 153  & 222 $\pm$ 13   & 1737 $\pm$ 151  & 183 $\pm$ 38  & \cite{2009GCN.10100....1C}  \\ 
GRB\,090328         & 0.736      & 764 $\pm$ 69 & 6.0 $\pm$ 0.30 & 777 $\pm$ 64 & 6.0 $\pm$ 0.33      & \cite{2009GCN..9053....1C}   \\ 
GRB\,090510         & 0.903  & 7612 $\pm$ 1047 & 59 $\pm$ 12  & 7612 $\pm$ 1047 & 50 $\pm$ 18     & \cite{2009GCN..9353....1R}  \\ 
GRB\,090902B        & 1.822      & 1984 $\pm$ 268 & 73 $\pm$ 23  & 1995 $\pm$ 63 & 36 $\pm$ 3.0   & \cite{2009GCN..9873....1C}  \\ 

GRB\,090926A    & 2.106    & 1112 $\pm$ 37 & 204 $\pm$ 3.0  & 1155 $\pm$ 12 & 359 $\pm$ 9.2  & \cite{2009GCN..9948....1O}  \\ 

GRB\,100414A        & 1.368      & 202 $\pm$ 9.0 & 31 $\pm$ 45  & 1376 $\pm$ 28  & 6.4 $\pm$ 1.0 & \cite{2010GCN.10606....1C}  \\ 

GRB\,110721A        & 3.512      & 3655 $\pm$ 316 & 89 $\pm$ 13  & 2030 $\pm$ 180  & 461 $\pm$ 23 &   \cite{2011GCN.12192....1G} \\ 

GRB\,110731A        & 2.83       & 663 $\pm$ 83 & 69 $\pm$ 16  & 670 $\pm$ 63 & 68 $\pm$ 16  & \cite{2011GCN.12225....1T}  \\

GRB\,120624B        & 2.2        & 3936 $\pm$ 576   & 101 $\pm$ 10   & 2848 $\pm$ 48   & 97 $\pm$ 12   & \cite{2012GCN.13389....1D} \\ 

GRB\,130427A        & 0.34       & 1252 $\pm$  21 & 25 $\pm$ 9.0  & 1265 $\pm$ 17  & 25 $\pm$ 8.0   & \cite{2013GCN.14455....1L} \\ 

GRB\,130518A        & 2.49       & 2125 $\pm$ 80  & 113 $\pm$ 10  & 1644 $\pm$ 80      & 347 $\pm$ 22  & \cite{2013GCN.14687....1C}  \\

GRB\,131108A        & 2.4        & 1319 $\pm$ 54  & 91 $\pm$ 19  & 1098 $\pm$ 122 & 133 $\pm$ 17 & \cite{2013GCN.15470....1D} \\ 

GRB\,131231A        & 0.6439     & 437 $\pm$ 25  & 9.0 $\pm$ 1.0 & 429 $\pm$ 37 & 10 $\pm$ 1.0 &  \cite{2014GCN.15645....1X}    \\ 

GRB\,141028A        & 2.33       & 1898 $\pm$ 167  & 123$\pm$ 5.0  & 1798 $\pm$ 133      & 123 $\pm$ 5.0 &    \cite{2014GCN.16982....1C}   \\

GRB\,150314A        & 0.807      & 940 $\pm$ 73 & 2.4 $\pm$ 0.13  & 976 $\pm$ 61 & 1.0 $\pm$ 0.22   &   \cite{2015GCN.17583....1D}  \\ 

GRB\,150403A        & 2.06       & 2234 $\pm$   184  & 25 $\pm$ 1.0  & 2295 $\pm$ 214  & 180 $\pm$ 19  & \cite{2015GCN.17672....1P}   \\ 

GRB\,150514A        & 0.807      & 130 $\pm$ 90  & 2.5 $\pm$ 0.12 & 125 $\pm$ 99 & 1.0 $\pm$ 0.21 & \cite{2015GCN.17822....1D}      \\ 

GRB\,160509A        & 1.17       & 6119 $\pm$ 434 & 68 $\pm$ 5.0 &  5164 $\pm$ 271 & 21 $\pm$ 2.0  & \cite{2016GCN.19419....1T}   \\ 

GRB\,160625B        & 1.406      & 553 $\pm$ 11   & 109 $\pm$ 2.0  & 397 $\pm$ 6.0   & 106 $\pm$ 2.0  & \cite{2016GCN.19600....1X}    \\ 

GRB\,170214A        & 2.53       & 1564 $\pm$ 56  & 73 $\pm$ 15 & 1469 $\pm$ 63  & 117 $\pm$ 19  & \cite{2017GCN.20686....1K}   \\ 

GRB\,170405A        & 3.51       & 1226 $\pm$ 180 & 156 $\pm$ 48   & 1128 $\pm$  271 & 231 $\pm$ 19   & \cite{2017GCN.20990....1D}  \\ 

GRB\,180720B        & 0.654      & 1470 $\pm$ 42 & 6.0 $\pm$ 1.3   & 1179 $\pm$ 40 & 40 $\pm$ 2.2  & \cite{2018GCN.22996....1V}   \\ 

GRB\,190114C        & 0.425      & 1197 $\pm$  20 & 7.0 $\pm$ 1.0  & 1198 $\pm$  18 & 50 $\pm$ 5.0 &    \cite{2019GCN.23708....1C}   \\ 

GRB\,220101A        & 4.61       & 2524 $\pm$ 420 & 81 $\pm$ 29 & 2412 $\pm$ 449 & 62 $\pm$ 24     & \cite{2022GCN.31357....1P}  \\ 

GRB\,221009A        & 0.151      & 892 $\pm$ 26  & 2.1 $\pm$ 0.03   & 884 $\pm$ 25 & 2.1 $\pm$ 0.1   &     \cite{2022GCN.32648....1D}          \\

GRB\,230812B        & 0.36       & 907 $\pm$ 408 & 15 $\pm$ 1.0 & 835 $\pm$ 383   & 10 $\pm$ 4.0 &   \cite{2023GCN.34409....1D}               \\

GRB\,240825A        & 0.659       & 826 $\pm$ 33.18 & 27.7 $\pm$ 1.34 & 937 $\pm$ 23.23   & 27.7 $\pm$ 1.35 &   \cite{2024GCN.37293....1M}  \\

GRB\,241228B        & 2.674       & 1572 $\pm$ 96 & 110 $\pm$ 12 & 1594 $\pm$ 88.18   & 128 $\pm$ 9.3 &   \cite{2024GCN.38704....1A}             \\
\hline
\end{longtable}
\endgroup    
\end{landscape}

\appendix
\setcounter{section}{2}


\begin{figure*}
    \centering
    \begin{subfigure}[b]{0.40\textwidth}
        \centering
        \includegraphics[width=\textwidth]{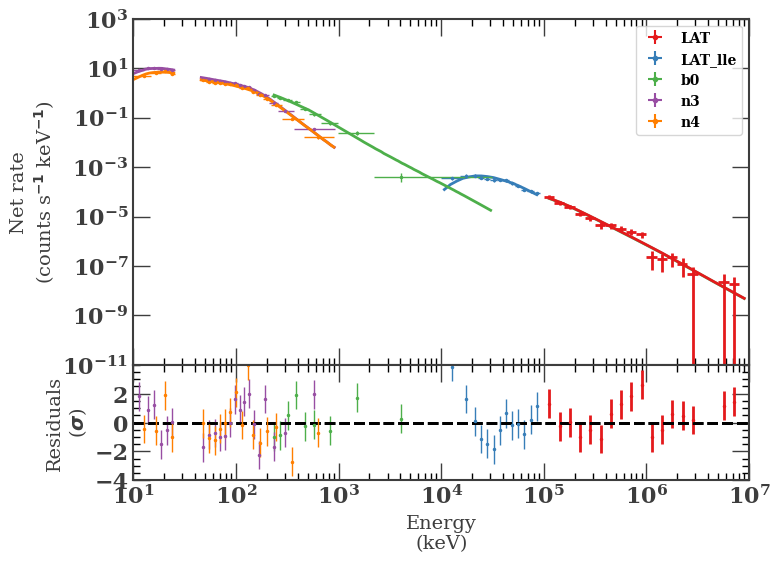}
            
    \end{subfigure}
    \hfill
    \begin{subfigure}[b]{0.40\textwidth}
        \centering
        \includegraphics[width=\textwidth]{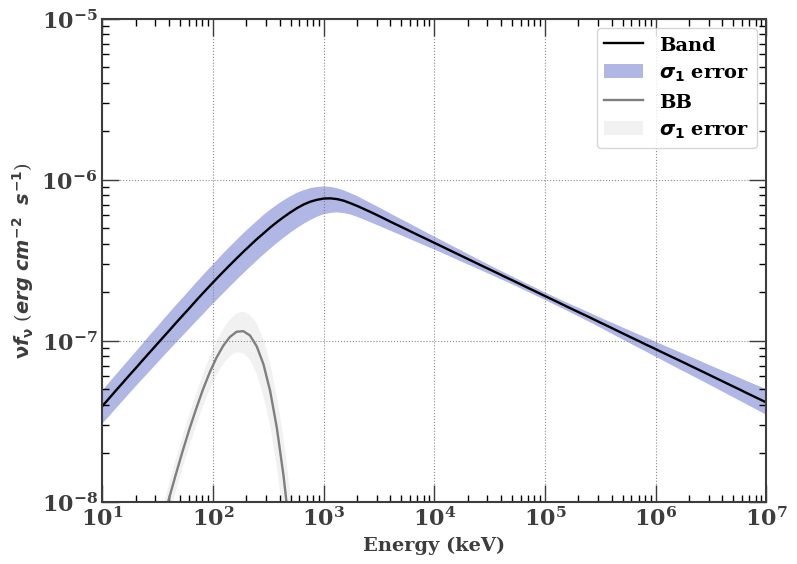}
          
    \end{subfigure}
        \hfill
    \begin{subfigure}[b]{0.40\textwidth}
        \centering
        \includegraphics[width=\textwidth]{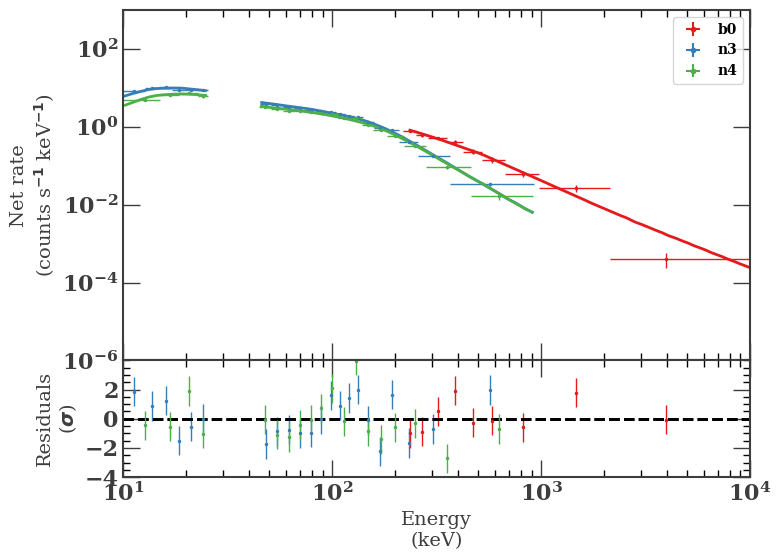}
          
    \end{subfigure}
        \hfill
    \begin{subfigure}[b]{0.40\textwidth}
        \centering
        \includegraphics[width=\textwidth]{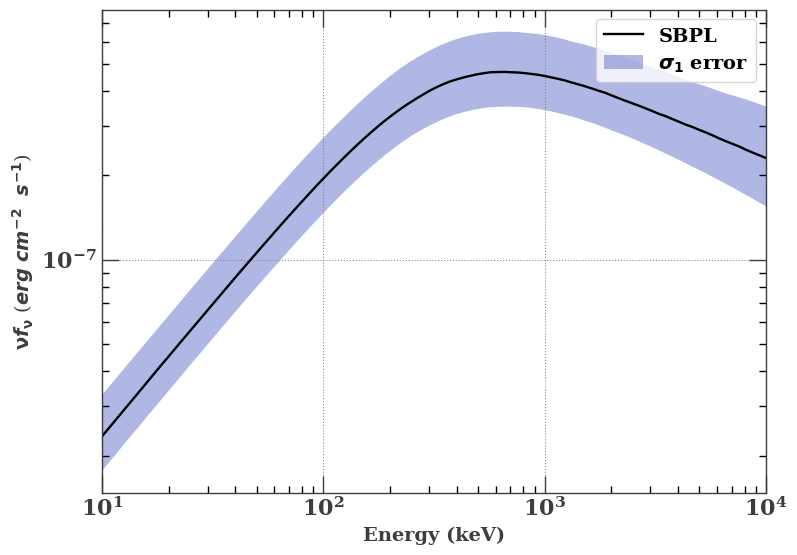}
          
    \end{subfigure}
    \caption{GRB 080916C: The count spectra (left panels) and ${\rm \nu}f_{\rm \nu}$ spectra (right panels).  The top (bottom) panels are for the joint (GBM-only) fits.}
    \label{fig_a1}
\end{figure*}

\begin{figure*}
    \centering
    \begin{subfigure}[b]{0.40\textwidth}
        \centering
        \includegraphics[width=\textwidth]{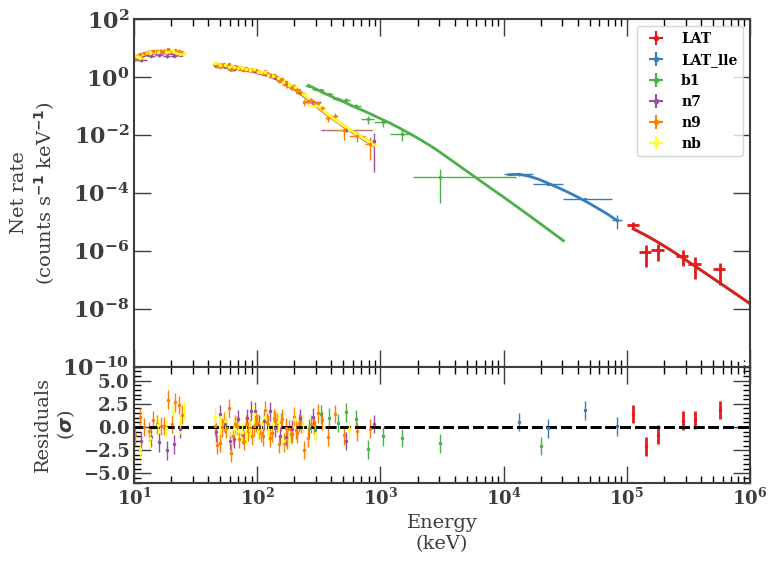}
            
    \end{subfigure}
    \hfill
    \begin{subfigure}[b]{0.40\textwidth}
        \centering
        \includegraphics[width=\textwidth]{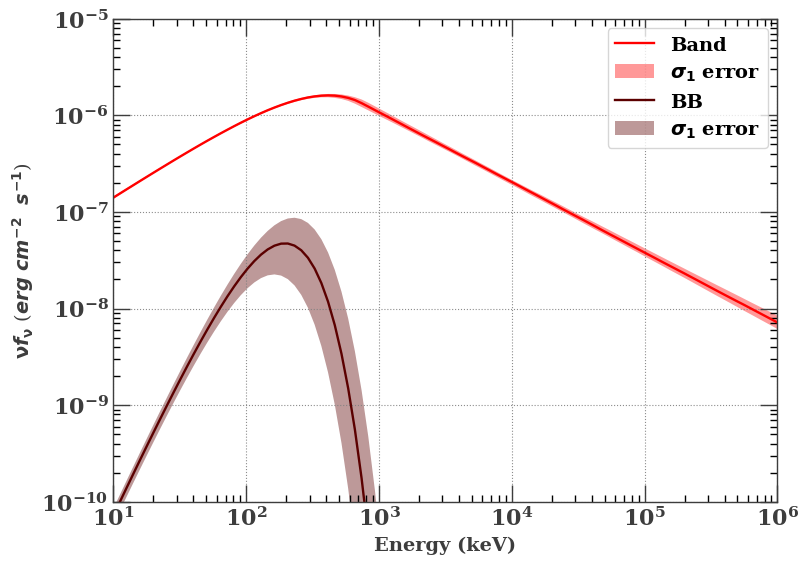}
          
    \end{subfigure}
        \hfill
    \begin{subfigure}[b]{0.40\textwidth}
        \centering
        \includegraphics[width=\textwidth]{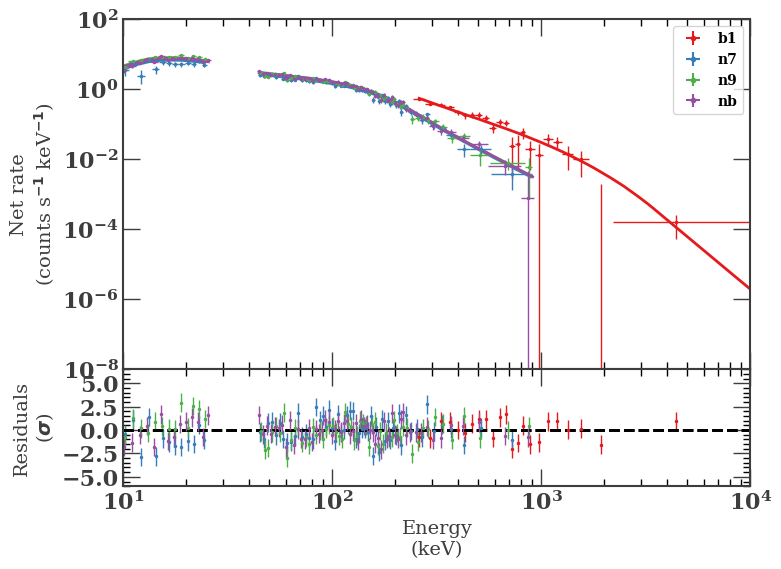}
        \caption{Count Rate plot: GBM}
          
    \end{subfigure}
        \hfill
    \begin{subfigure}[b]{0.40\textwidth}
        \centering
        \includegraphics[width=\textwidth]{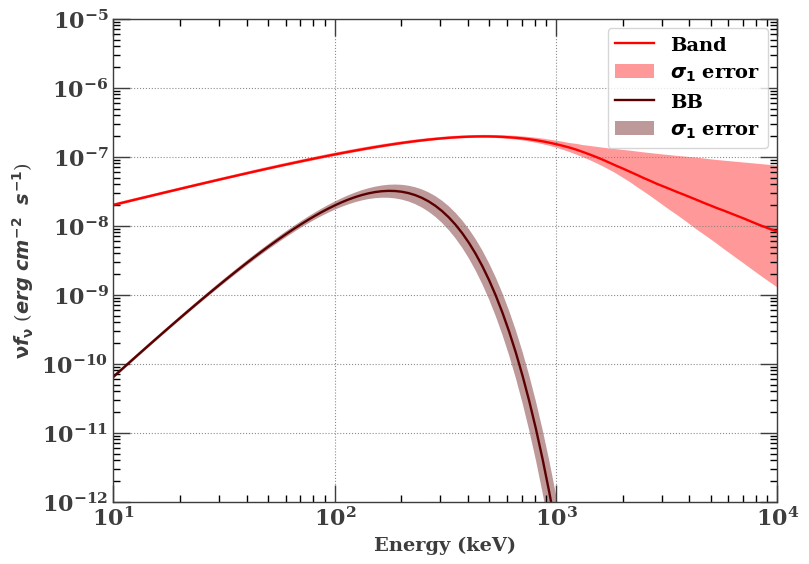}
        \caption{${\rm \nu}f_{\rm \nu}$ plot: GBM}
          
    \end{subfigure}
    \caption{GRB 090323: The count spectra (left panels) and ${\rm \nu}f_{\rm \nu}$ spectra (right panels).  The top (bottom) panels are for the joint (GBM-only) fits.}
    \label{fig_a2}
\end{figure*}

\newpage

\begin{figure*}
    \centering
    \begin{subfigure}[b]{0.40\textwidth}
        \centering
        \includegraphics[width=\textwidth]{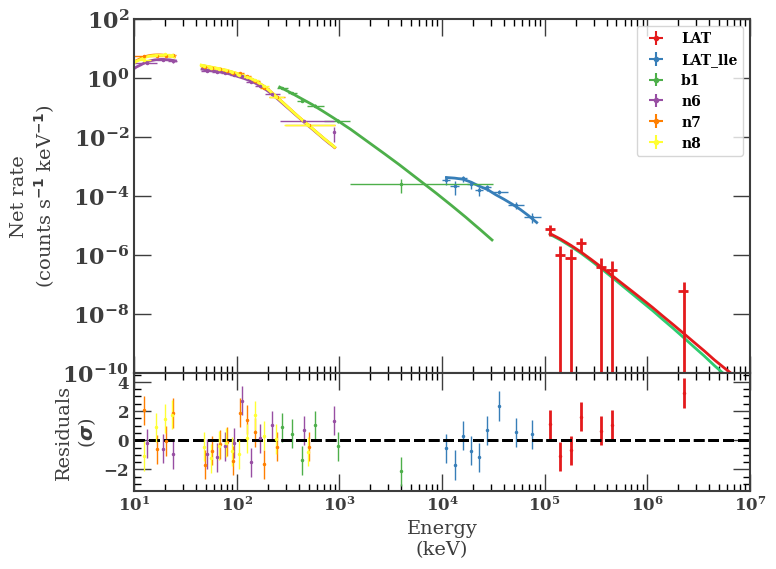}
            
    \end{subfigure}
    \hfill
    \begin{subfigure}[b]{0.40\textwidth}
        \centering
        \includegraphics[width=\textwidth]{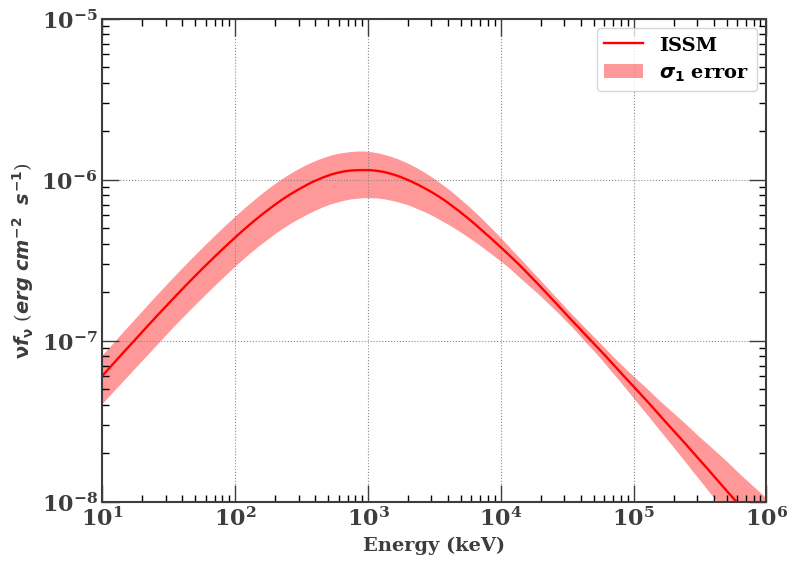}
          
    \end{subfigure}
        \hfill
    \begin{subfigure}[b]{0.40\textwidth}
        \centering
        \includegraphics[width=\textwidth]{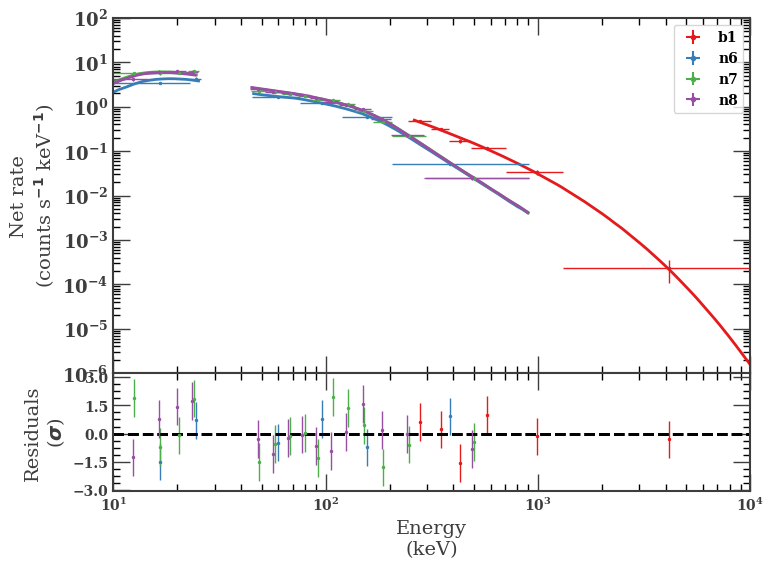}
          
    \end{subfigure}
        \hfill
    \begin{subfigure}[b]{0.40\textwidth}
        \centering
        \includegraphics[width=\textwidth]{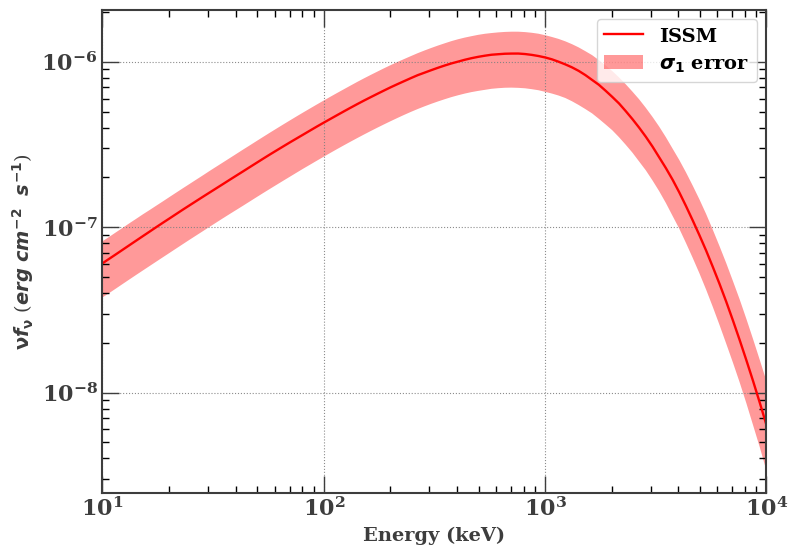}
          
    \end{subfigure}
    \caption{GRB 090328: The count spectra (left panels) and ${\rm \nu}f_{\rm \nu}$ spectra (right panels).  The top (bottom) panels are for the joint (GBM-only) fits.}
    \label{fig_a3}
\end{figure*}

\newpage

\begin{figure*}
    \centering
    \begin{subfigure}[b]{0.40\textwidth}
        \centering
        \includegraphics[width=\textwidth]{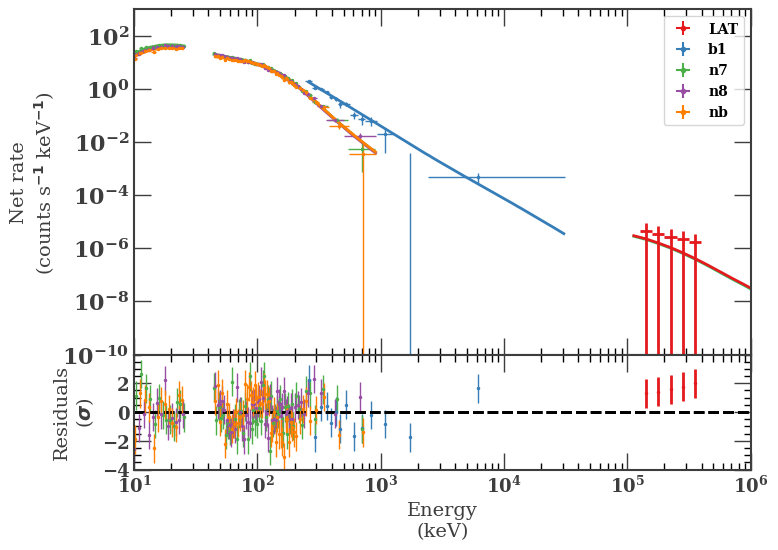}
            
    \end{subfigure}
    \hfill
    \begin{subfigure}[b]{0.40\textwidth}
        \centering
        \includegraphics[width=\textwidth]{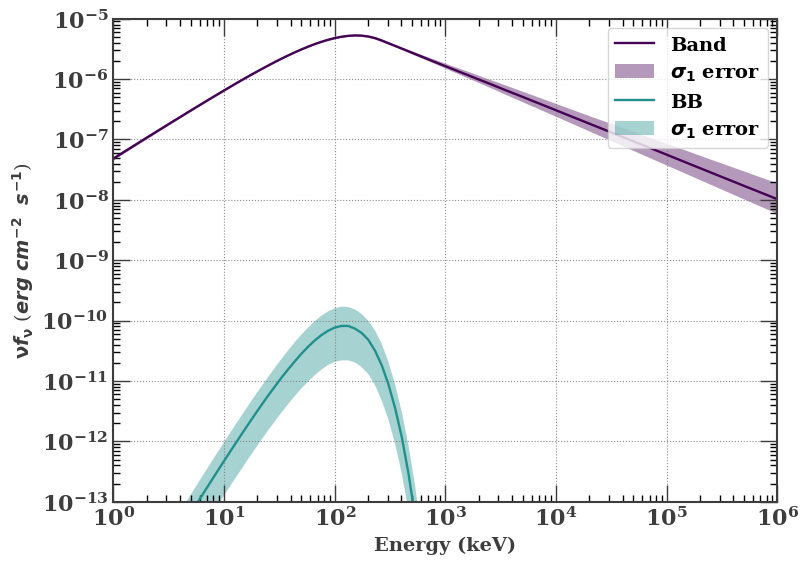}
          
    \end{subfigure}
        \hfill
    \begin{subfigure}[b]{0.40\textwidth}
        \centering
        \includegraphics[width=\textwidth]{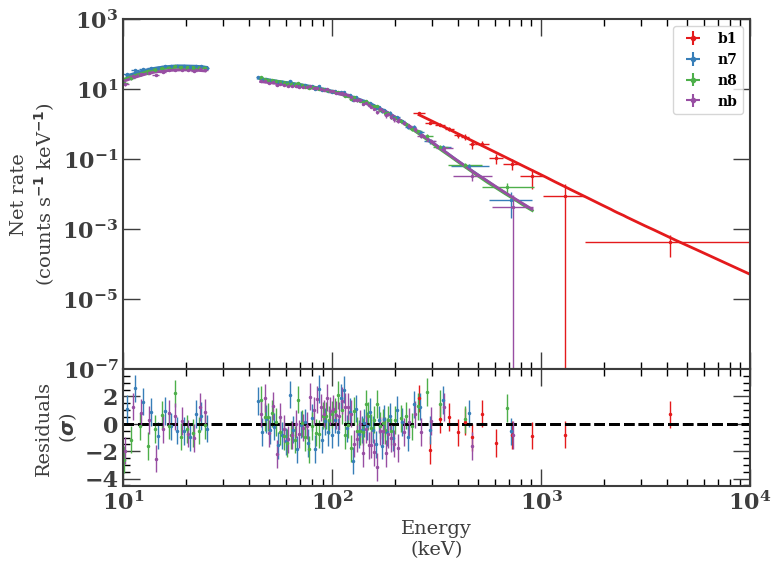}
          
    \end{subfigure}
        \hfill
    \begin{subfigure}[b]{0.40\textwidth}
        \centering
        \includegraphics[width=\textwidth]{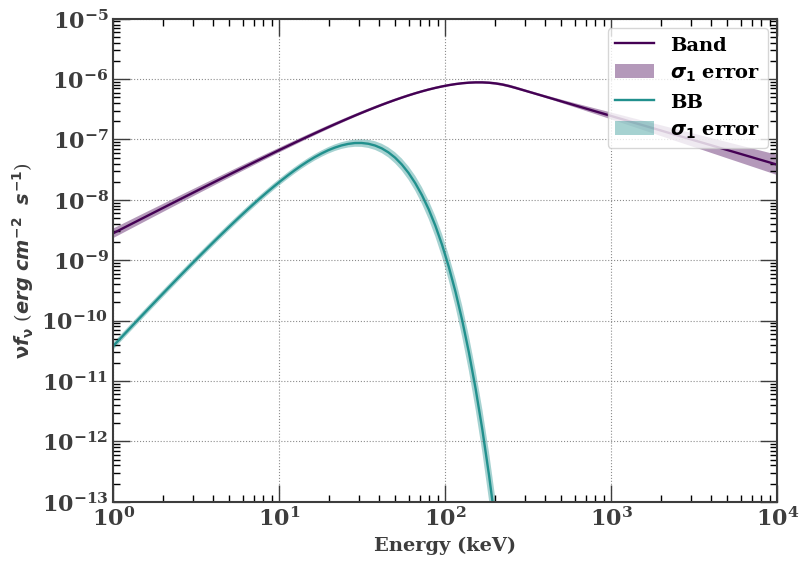}
          
    \end{subfigure}
    \caption{GRB 090424: The count spectra (left panels) and ${\rm \nu}f_{\rm \nu}$ spectra (right panels).  The top (bottom) panels are for the joint (GBM-only) fits.}
    \label{fig_a4}
\end{figure*}

\begin{figure*}
    \centering
    \begin{subfigure}[b]{0.40\textwidth}
        \centering
        \includegraphics[width=\textwidth]{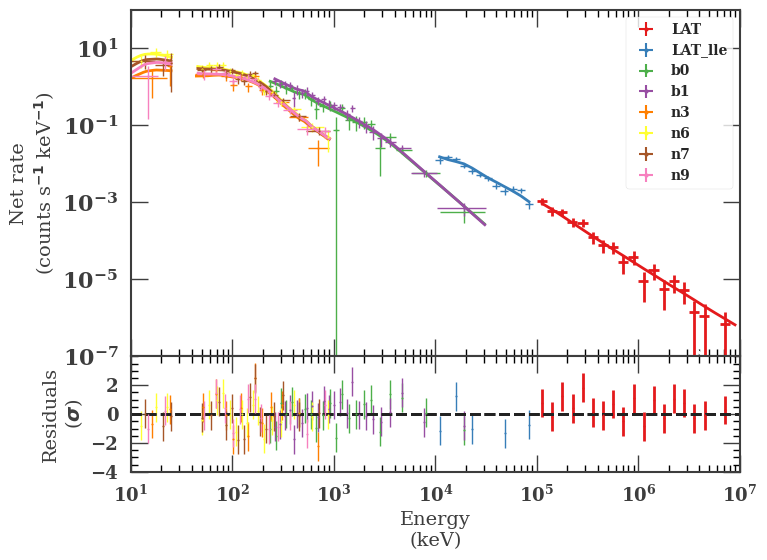}
            
    \end{subfigure}
    \hfill
    \begin{subfigure}[b]{0.40\textwidth}
        \centering
        \includegraphics[width=\textwidth]{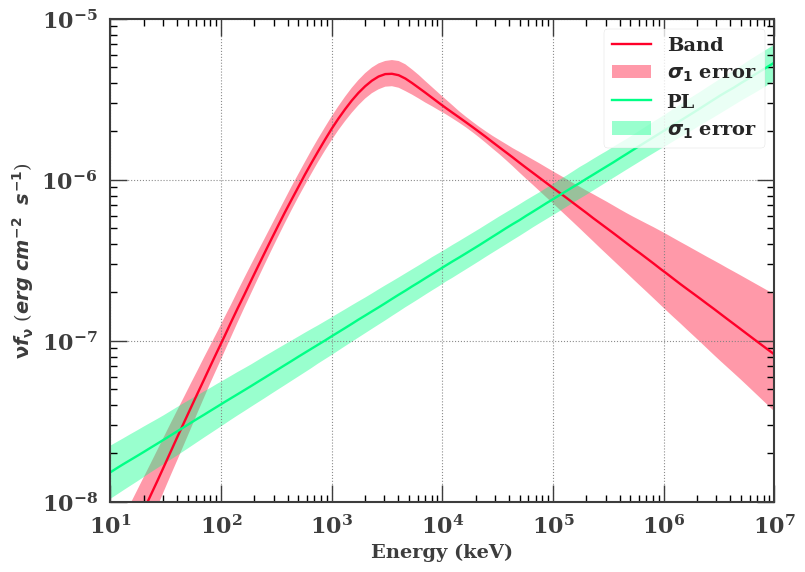}
          
    \end{subfigure}
        \hfill
    \begin{subfigure}[b]{0.40\textwidth}
        \centering
        \includegraphics[width=\textwidth]{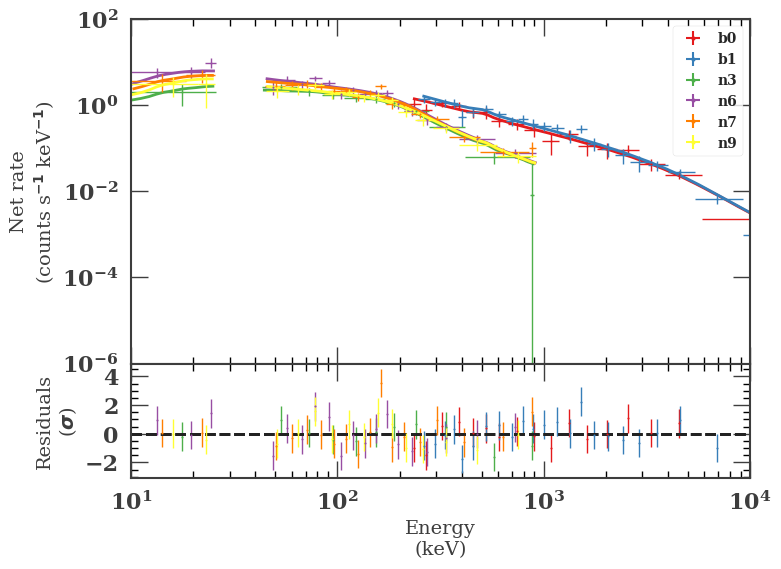}
          
    \end{subfigure}
        \hfill
    \begin{subfigure}[b]{0.40\textwidth}
        \centering
        \includegraphics[width=\textwidth]{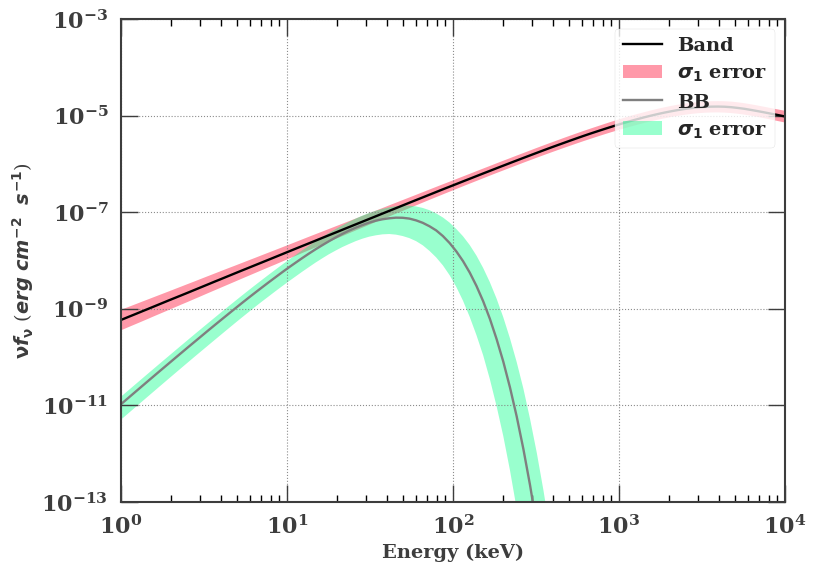}
          
    \end{subfigure}
    \caption{GRB 090510: \textbf{SGRB} - The count spectra (left panels) and ${\rm \nu}f_{\rm \nu}$ spectra (right panels).  The top (bottom) panels are for the joint (GBM-only) fits.}
    \label{fig_a5}
\end{figure*}
\newpage

\begin{figure*}
    \centering
    \begin{subfigure}[b]{0.40\textwidth}
        \centering
        \includegraphics[width=\textwidth]{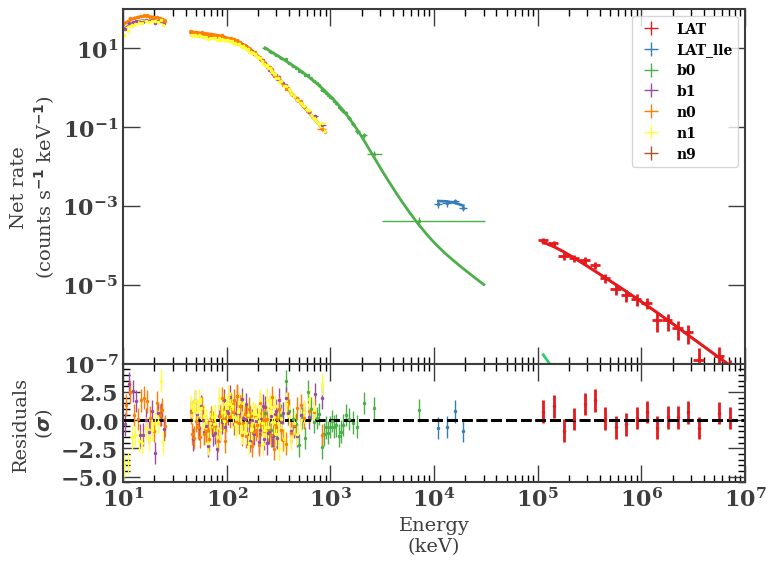}
            
    \end{subfigure}
    \hfill
    \begin{subfigure}[b]{0.40\textwidth}
        \centering
        \includegraphics[width=\textwidth]{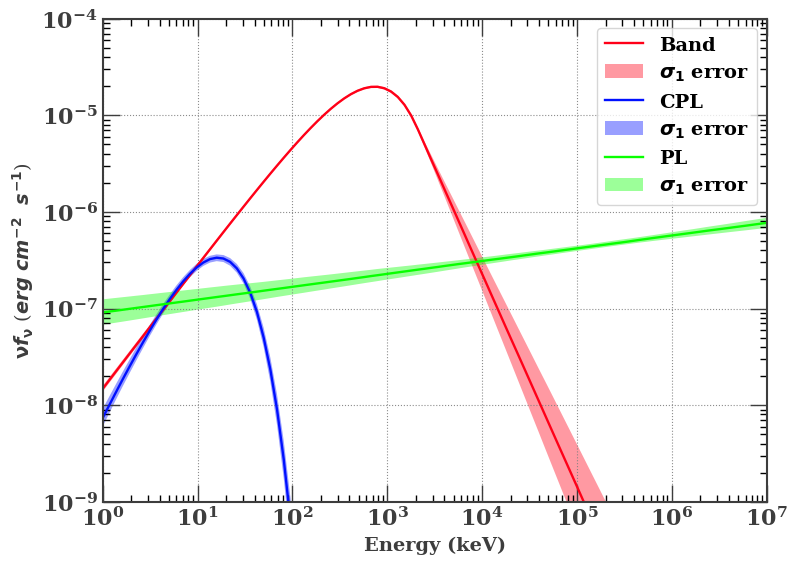}
          
    \end{subfigure}
        \hfill
    \begin{subfigure}[b]{0.40\textwidth}
        \centering
        \includegraphics[width=\textwidth]{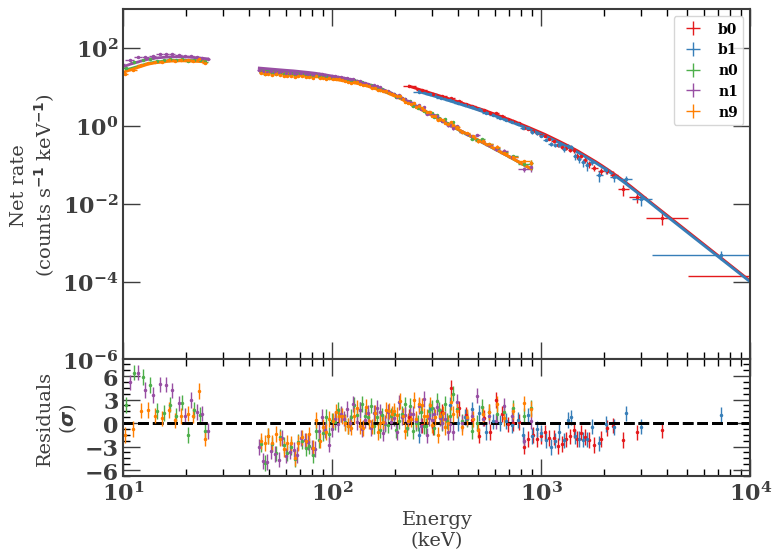}
          
    \end{subfigure}
        \hfill
    \begin{subfigure}[b]{0.40\textwidth}
        \centering
        \includegraphics[width=\textwidth]{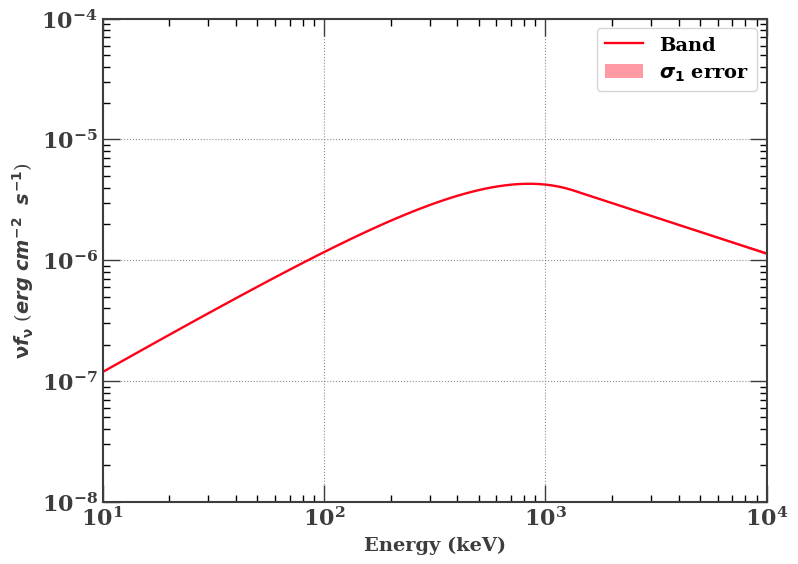}
          
    \end{subfigure}
    \caption{GRB 090902B: The count spectra (left panels) and ${\rm \nu}f_{\rm \nu}$ spectra (right panels).  The top (bottom) panels are for the joint (GBM-only) fits.}
    \label{fig_a6}
\end{figure*}


\begin{figure*}
    \centering
    \begin{subfigure}[b]{0.40\textwidth}
        \centering
        \includegraphics[width=\textwidth]{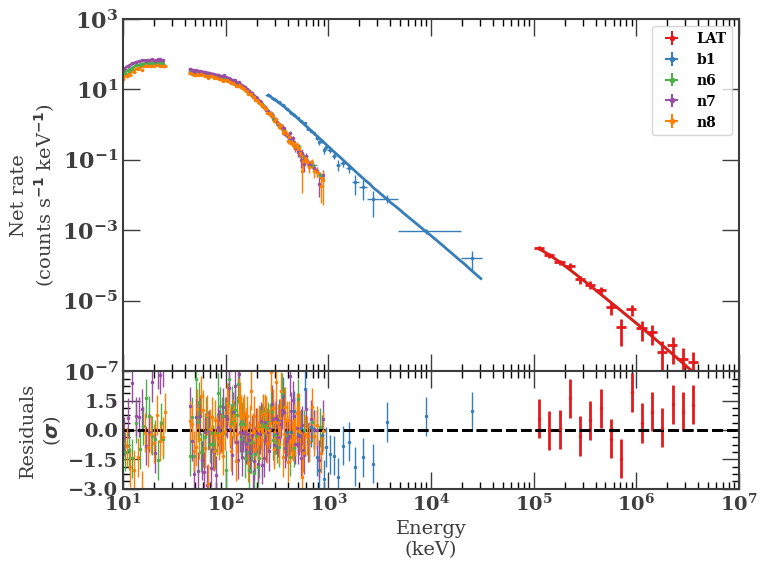}
            
    \end{subfigure}
    \hfill
    \begin{subfigure}[b]{0.40\textwidth}
        \centering
        \includegraphics[width=\textwidth]{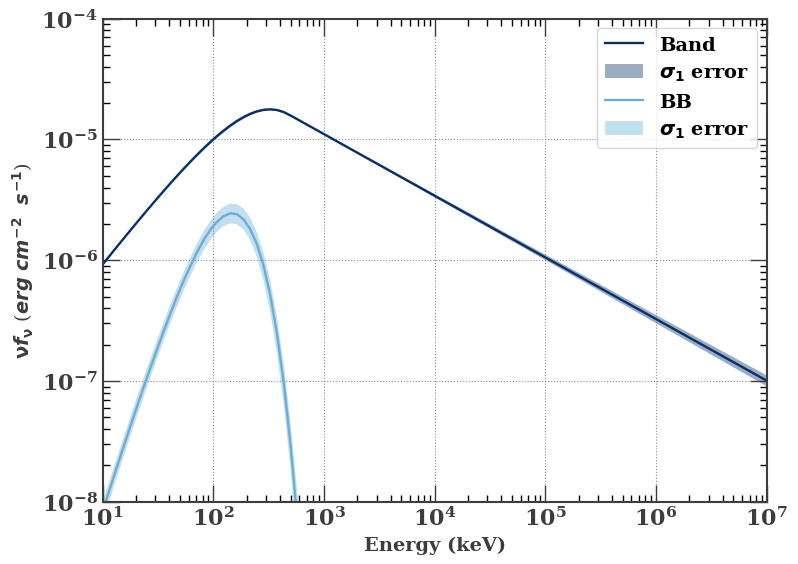}
          
    \end{subfigure}
        \hfill
    \begin{subfigure}[b]{0.40\textwidth}
        \centering
        \includegraphics[width=\textwidth]{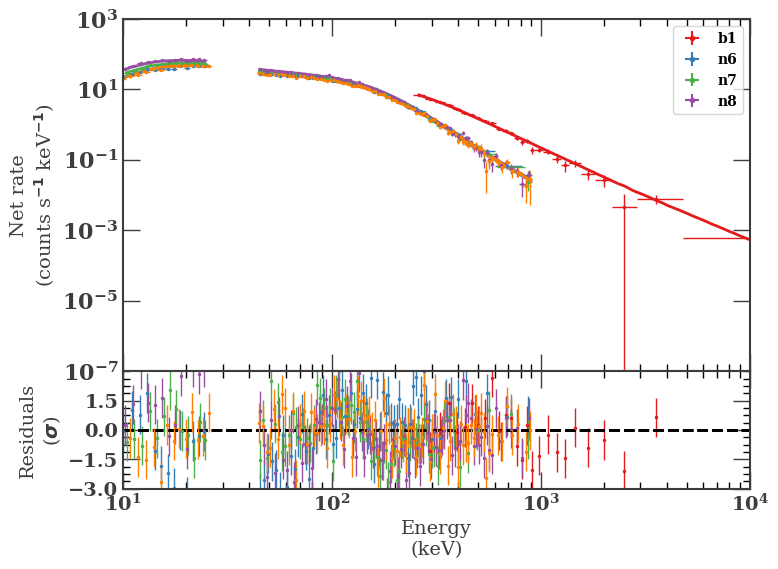}
          
    \end{subfigure}
        \hfill
    \begin{subfigure}[b]{0.40\textwidth}
        \centering
        \includegraphics[width=\textwidth]{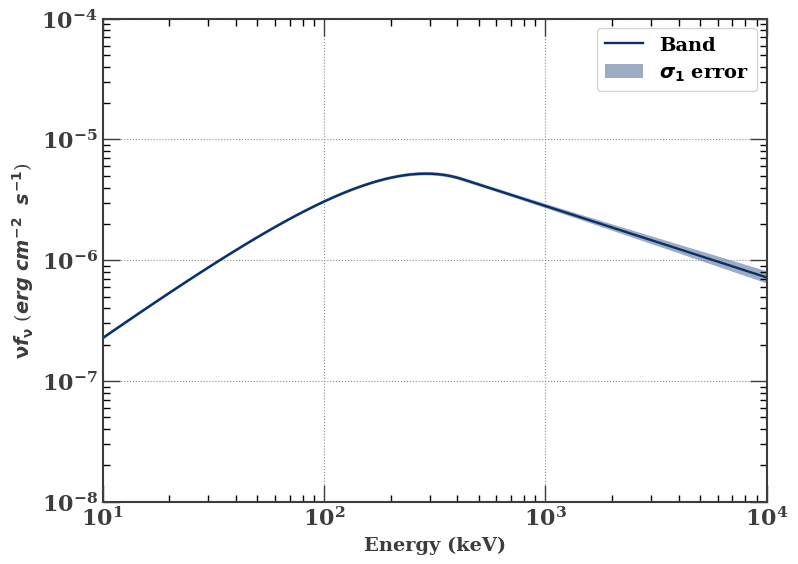}
          
    \end{subfigure}
    \caption{GRB 090926A: The count spectra (left panels) and ${\rm \nu}f_{\rm \nu}$ spectra (right panels).  The top (bottom) panels are for the joint (GBM-only) fits.}
    \label{fig_a7}
\end{figure*}


\begin{figure*}
    \centering
    \begin{subfigure}[b]{0.40\textwidth}
        \centering
        \includegraphics[width=\textwidth]{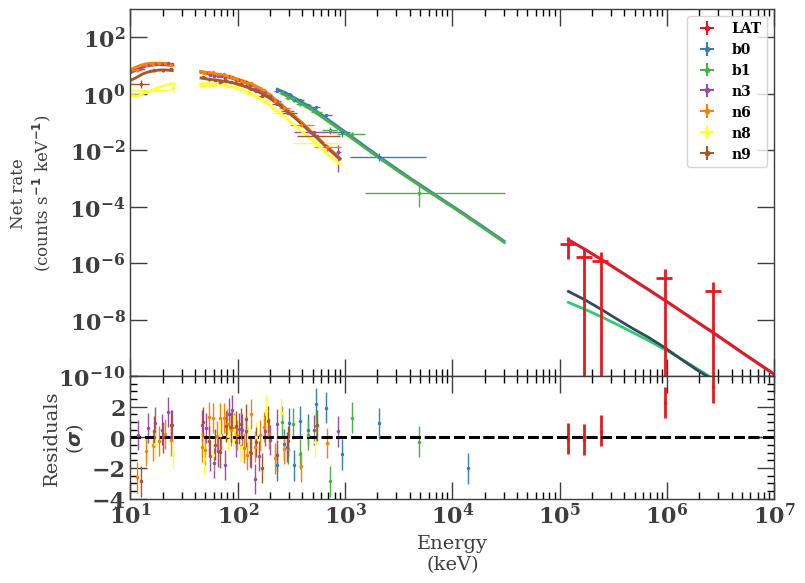}
            
    \end{subfigure}
    \hfill
    \begin{subfigure}[b]{0.40\textwidth}
        \centering
        \includegraphics[width=\textwidth]{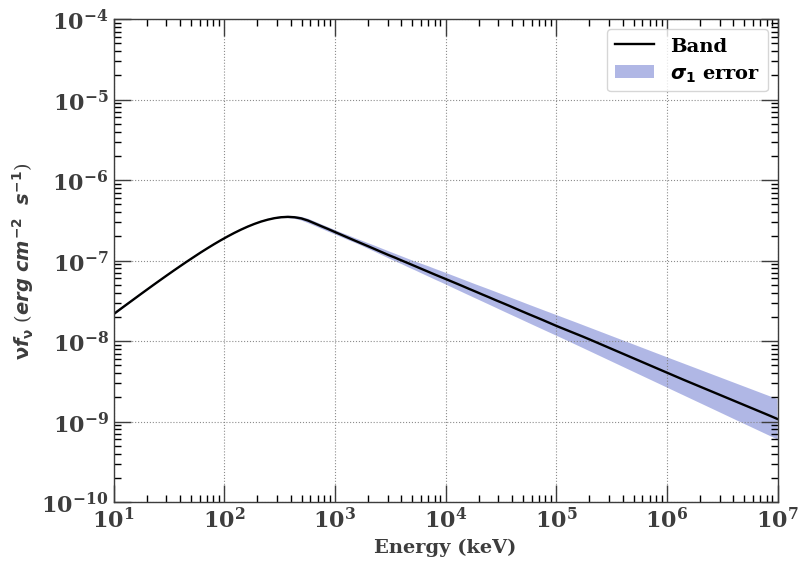}
          
    \end{subfigure}
        \hfill
    \begin{subfigure}[b]{0.40\textwidth}
        \centering
        \includegraphics[width=\textwidth]{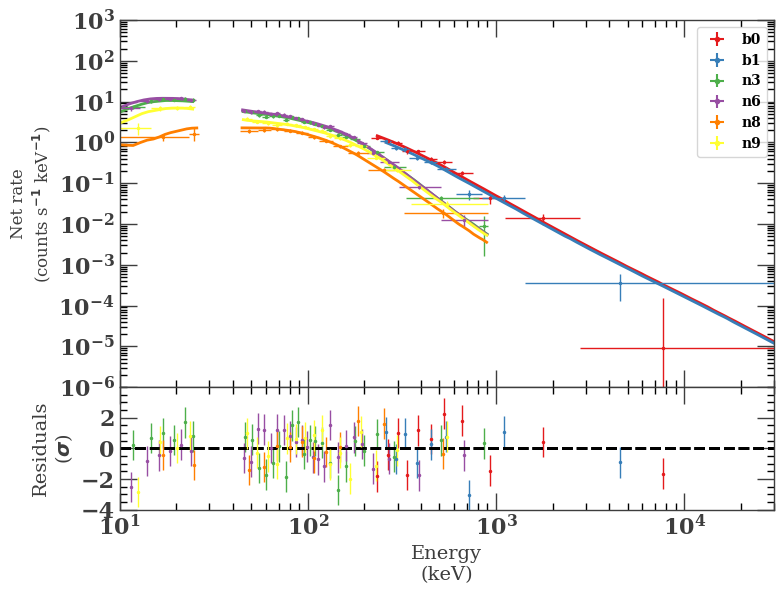}
          
    \end{subfigure}
        \hfill
    \begin{subfigure}[b]{0.40\textwidth}
        \centering
        \includegraphics[width=\textwidth]{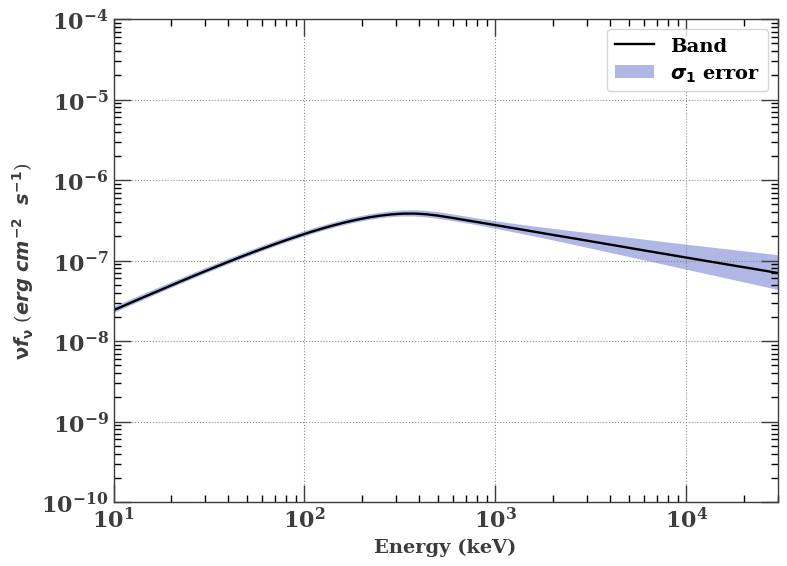}
          
    \end{subfigure}
    \caption{GRB 091003: The count spectra (left panels) and ${\rm \nu}f_{\rm \nu}$ spectra (right panels).  The top (bottom) panels are for the joint (GBM-only) fits.}
    \label{fig_a8}
\end{figure*}


\begin{figure*}
    \centering
    \begin{subfigure}[b]{0.40\textwidth}
        \centering
        \includegraphics[width=\textwidth]{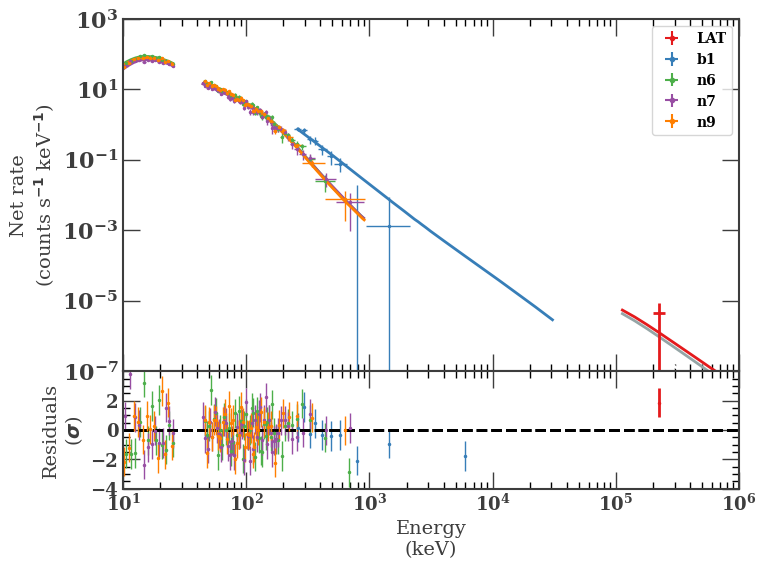}
            
    \end{subfigure}
    \hfill
    \begin{subfigure}[b]{0.40\textwidth}
        \centering
        \includegraphics[width=\textwidth]{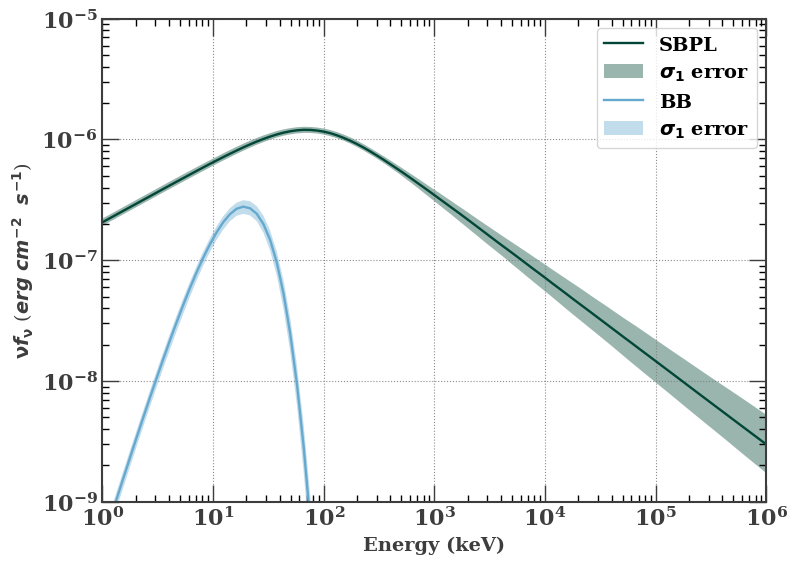}
          
    \end{subfigure}
        \hfill
    \begin{subfigure}[b]{0.40\textwidth}
        \centering
        \includegraphics[width=\textwidth]{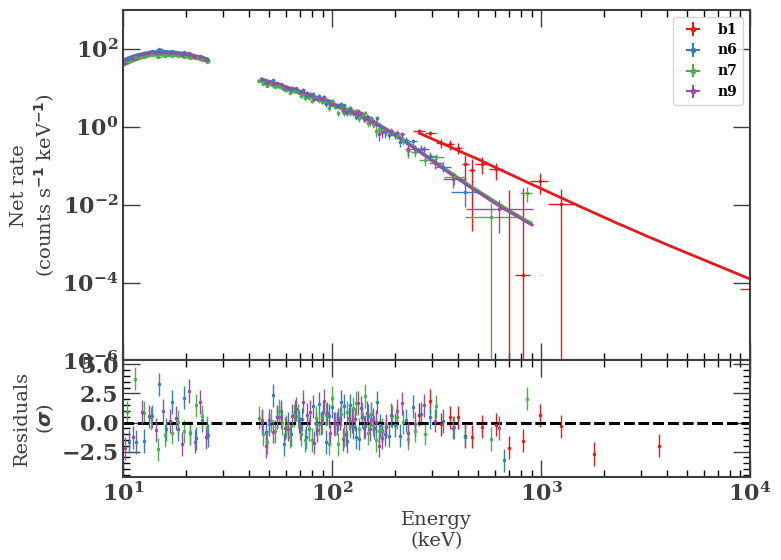}
          
    \end{subfigure}
        \hfill
    \begin{subfigure}[b]{0.40\textwidth}
        \centering
        \includegraphics[width=\textwidth]{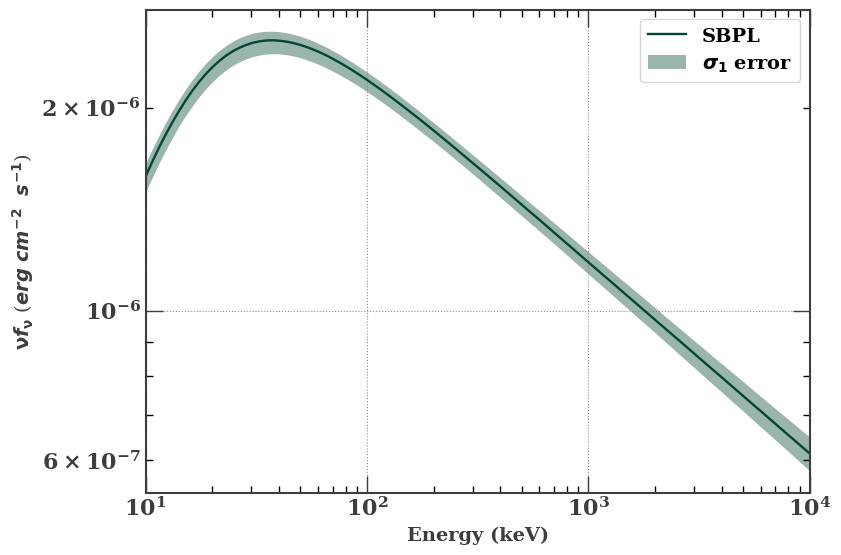}
          
    \end{subfigure}
    \caption{GRB 091127: The count spectra (left panels) and ${\rm \nu}f_{\rm \nu}$ spectra (right panels).  The top (bottom) panels are for the joint (GBM-only) fits.}
    \label{fig_a9}
\end{figure*}


\begin{figure*}
    \centering
    \begin{subfigure}[b]{0.40\textwidth}
        \centering
        \includegraphics[width=\textwidth]{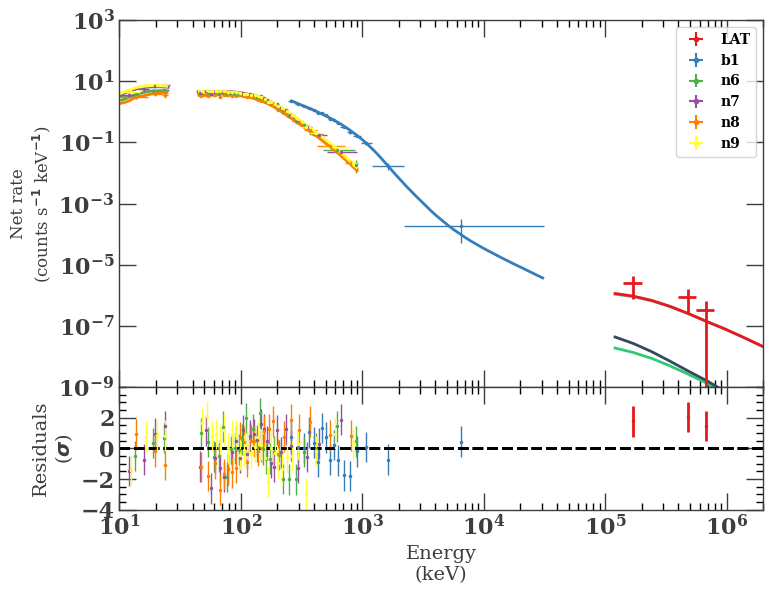}
            
    \end{subfigure}
    \hfill
    \begin{subfigure}[b]{0.40\textwidth}
        \centering
        \includegraphics[width=\textwidth]{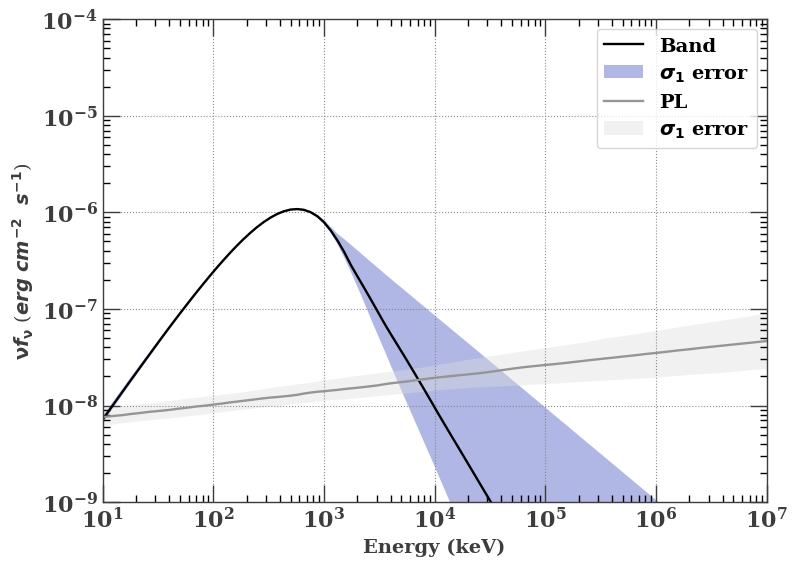}
          
    \end{subfigure}
        \hfill
    \begin{subfigure}[b]{0.40\textwidth}
        \centering
        \includegraphics[width=\textwidth]{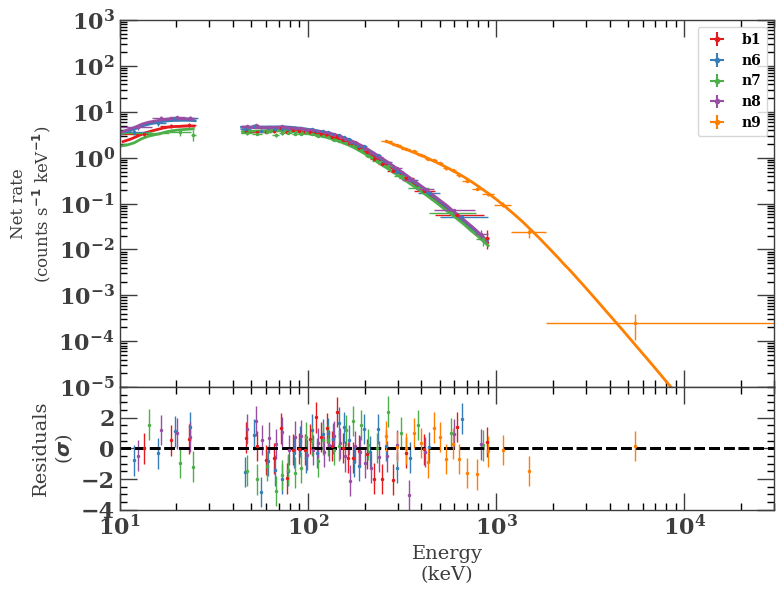}
          
    \end{subfigure}
        \hfill
    \begin{subfigure}[b]{0.40\textwidth}
        \centering
        \includegraphics[width=\textwidth]{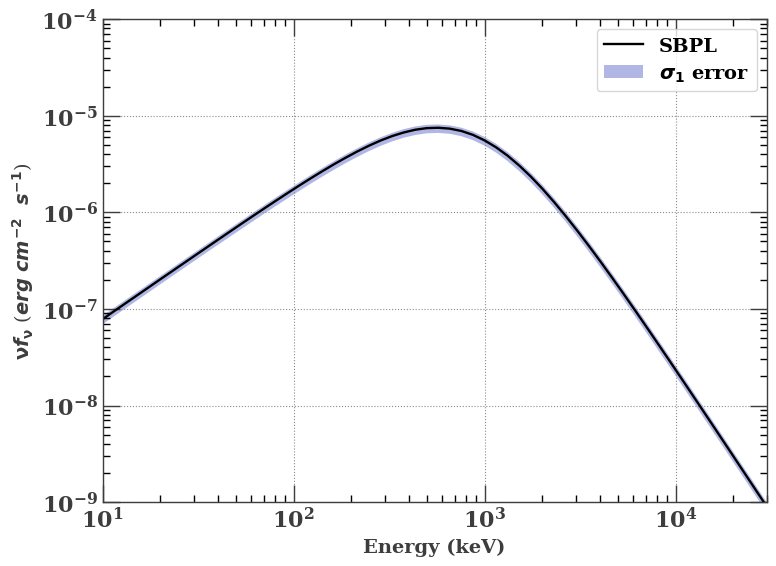}
          
    \end{subfigure}
    \caption{GRB 100414A: The count spectra (left panels) and ${\rm \nu}f_{\rm \nu}$ spectra (right panels).  The top (bottom) panels are for the joint (GBM-only) fits.}
    \label{fig_a10}
\end{figure*}


\begin{figure*}
    \centering
    \begin{subfigure}[b]{0.40\textwidth}
        \centering
        \includegraphics[width=\textwidth]{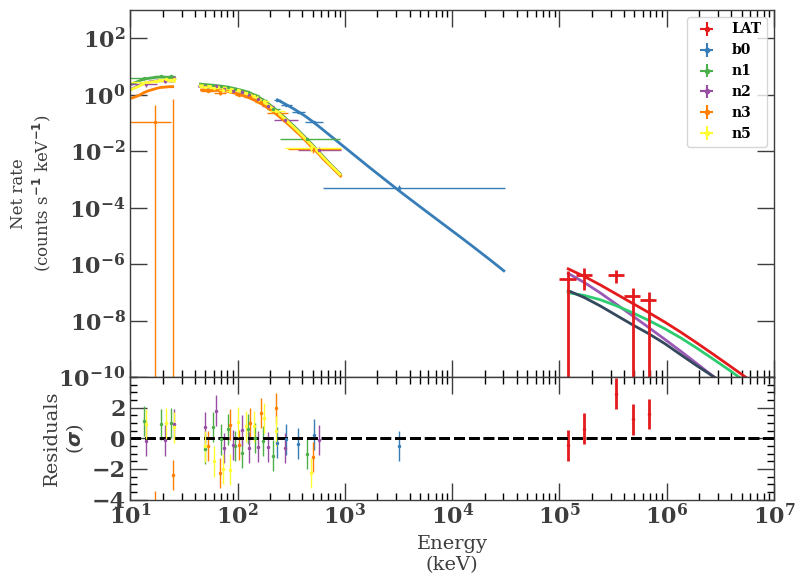}
            
    \end{subfigure}
    \hfill
    \begin{subfigure}[b]{0.40\textwidth}
        \centering
        \includegraphics[width=\textwidth]{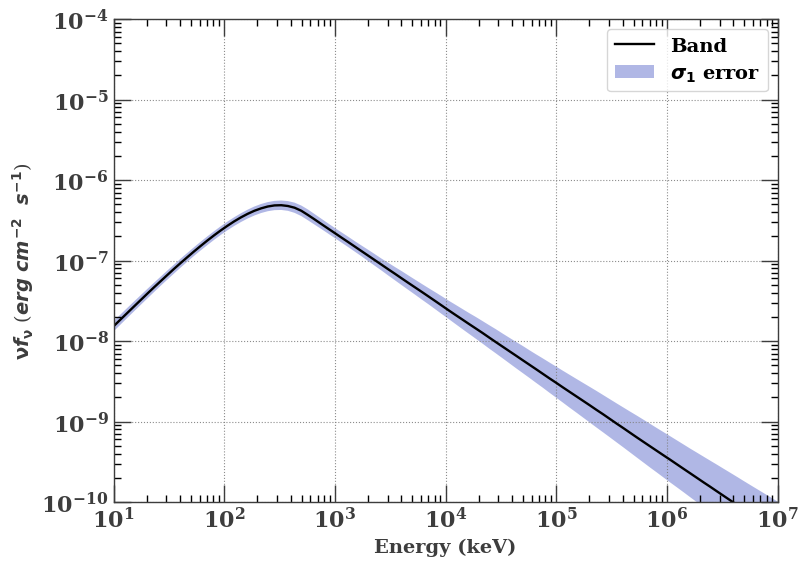}
          
    \end{subfigure}
        \hfill
    \begin{subfigure}[b]{0.40\textwidth}
        \centering
        \includegraphics[width=\textwidth]{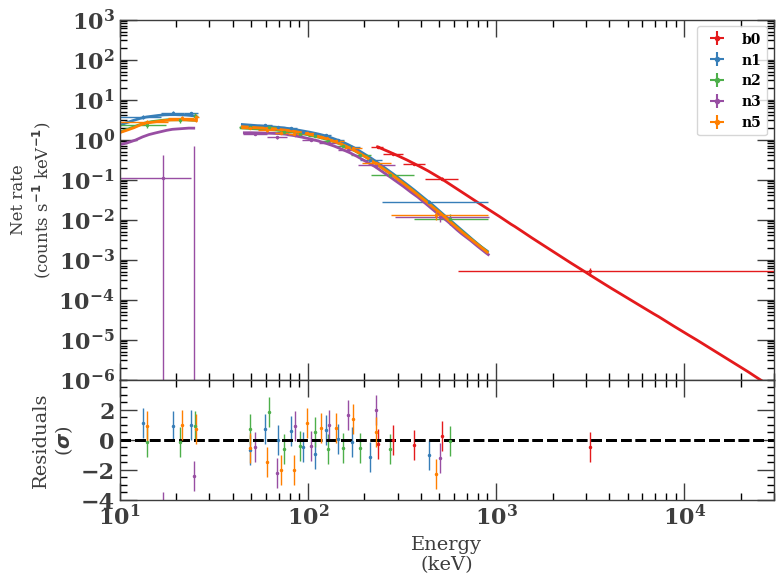}
          
    \end{subfigure}
        \hfill
    \begin{subfigure}[b]{0.40\textwidth}
        \centering
        \includegraphics[width=\textwidth]{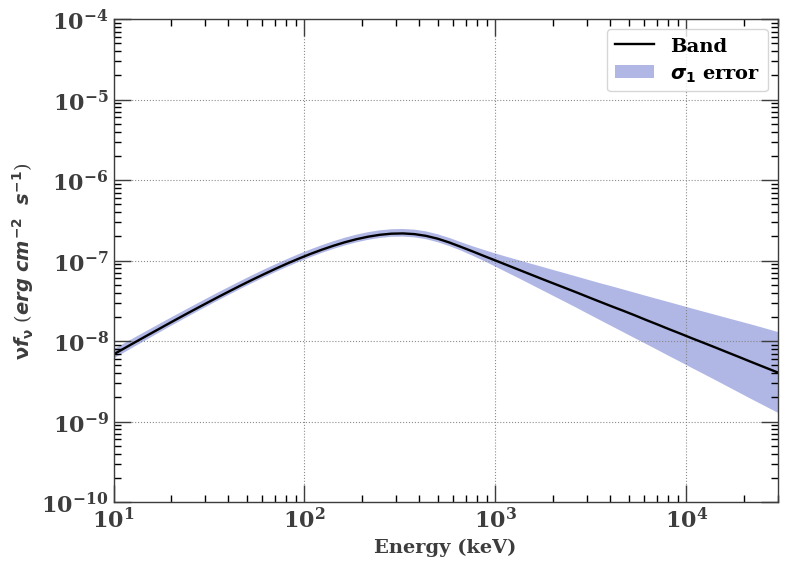}
          
    \end{subfigure}
    \caption{GRB 100728A: The count spectra (left panels) and ${\rm \nu}f_{\rm \nu}$ spectra (right panels).  The top (bottom) panels are for the joint (GBM-only) fits..}
    \label{fig_a11}
\end{figure*}


\begin{figure*}
    \centering
    \begin{subfigure}[b]{0.40\textwidth}
        \centering
        \includegraphics[width=\textwidth]{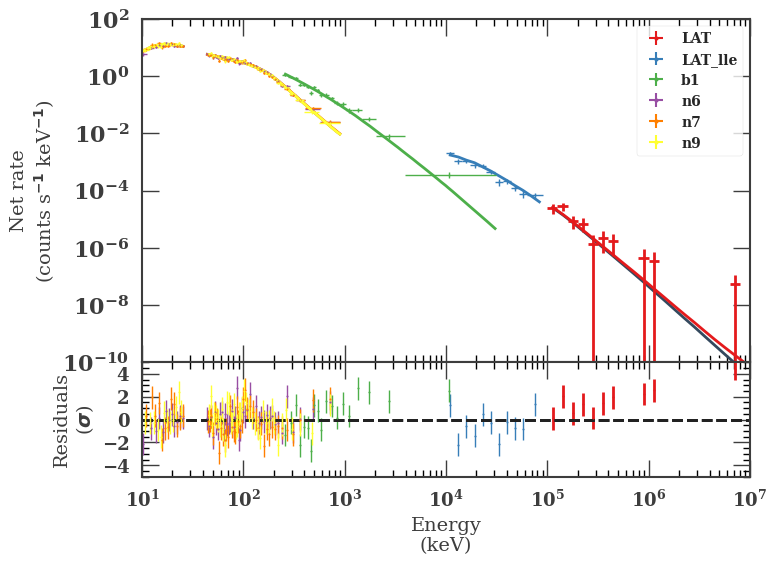}
            
    \end{subfigure}
    \hfill
    \begin{subfigure}[b]{0.40\textwidth}
        \centering
        \includegraphics[width=\textwidth]{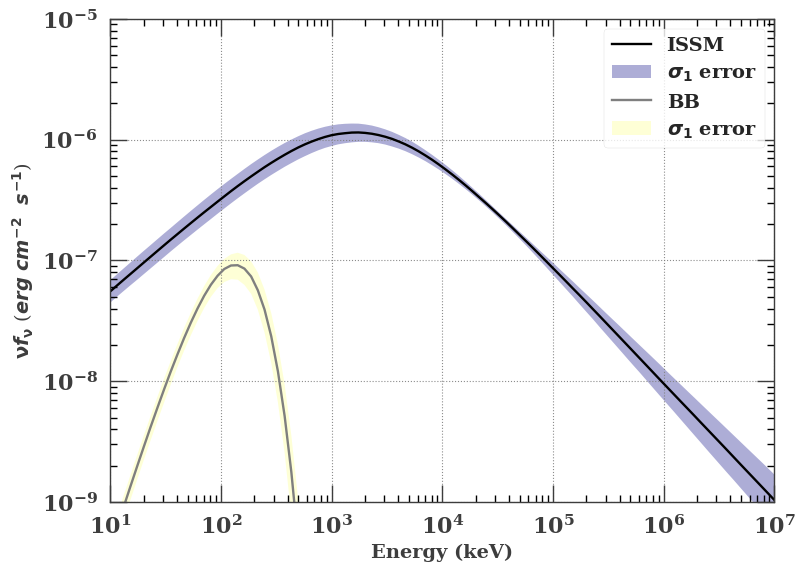}
          
    \end{subfigure}
        \hfill
    \begin{subfigure}[b]{0.40\textwidth}
        \centering
        \includegraphics[width=\textwidth]{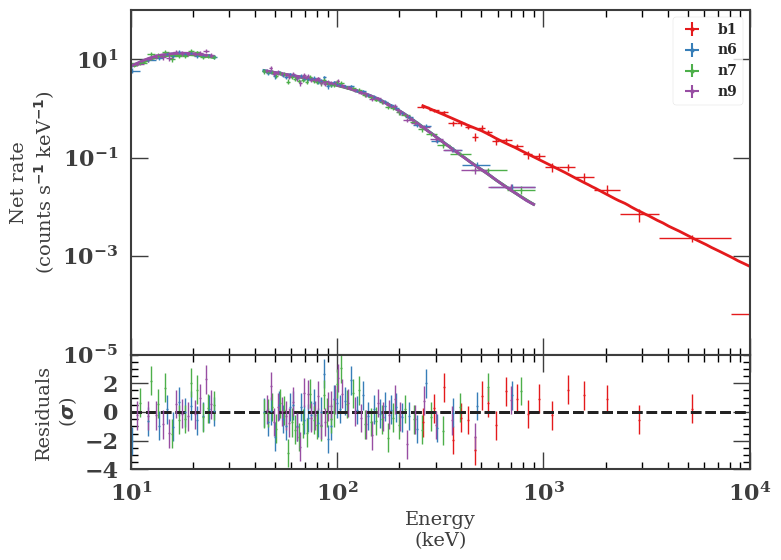}
          
    \end{subfigure}
        \hfill
    \begin{subfigure}[b]{0.40\textwidth}
        \centering
        \includegraphics[width=\textwidth]{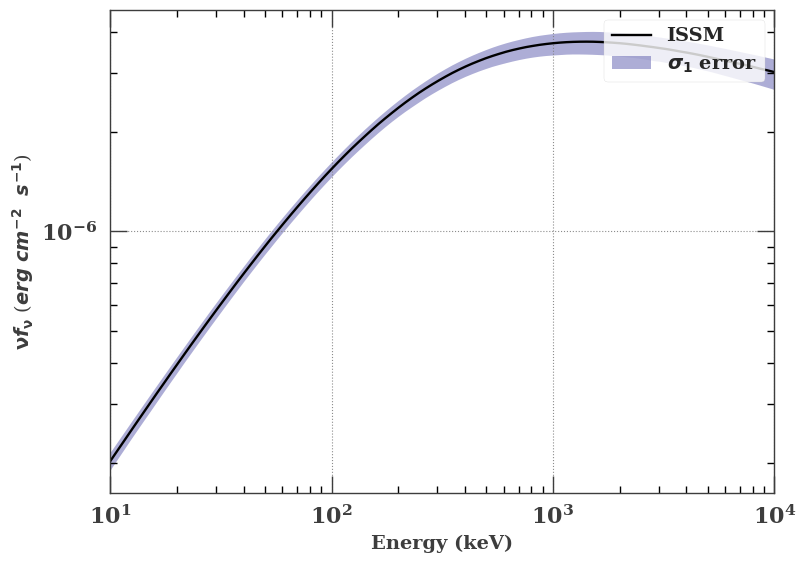}
          
    \end{subfigure}
    \caption{GRB 110721A: The count spectra (left panels) and ${\rm \nu}f_{\rm \nu}$ spectra (right panels).  The top (bottom) panels are for the joint (GBM-only) fits.}
    \label{fig_a_a12}
\end{figure*}


\begin{figure*}
    \centering
    \begin{subfigure}[b]{0.40\textwidth}
        \centering
        \includegraphics[width=\textwidth]{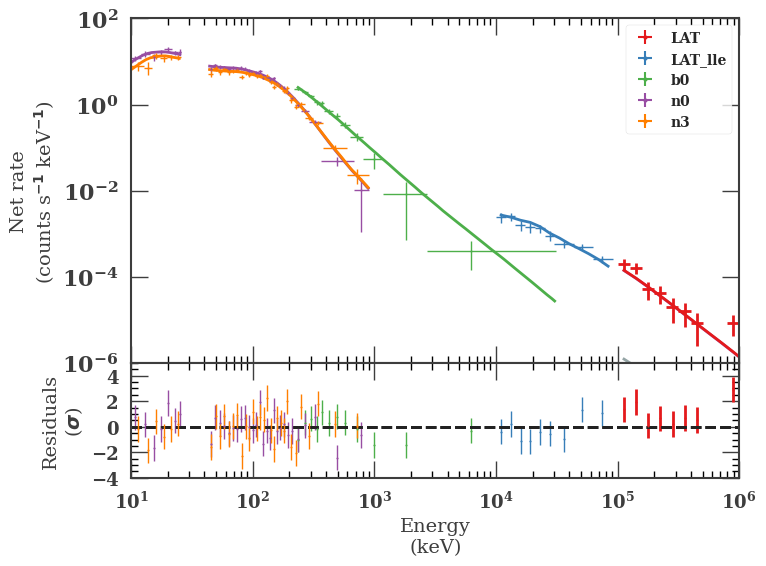}
            
    \end{subfigure}
    \hfill
    \begin{subfigure}[b]{0.40\textwidth}
        \centering
        \includegraphics[width=\textwidth]{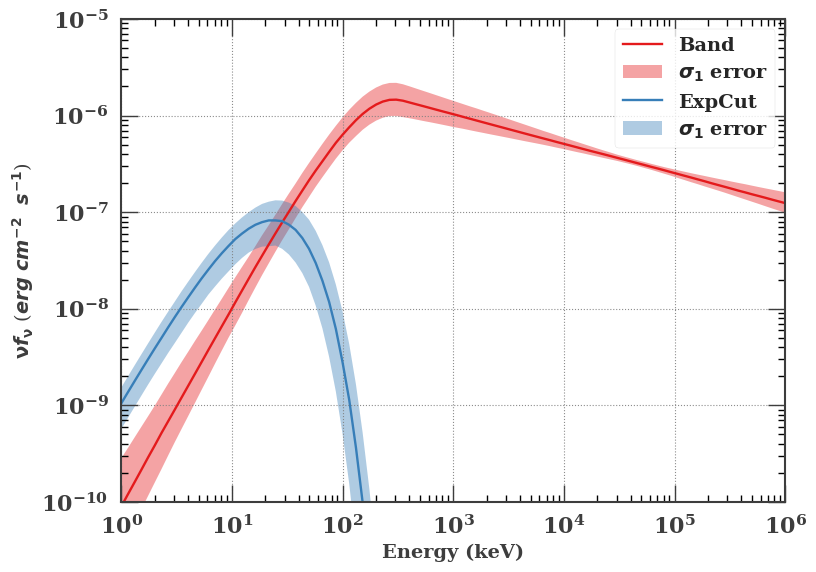}
          
    \end{subfigure}
        \hfill
    \begin{subfigure}[b]{0.40\textwidth}
        \centering
        \includegraphics[width=\textwidth]{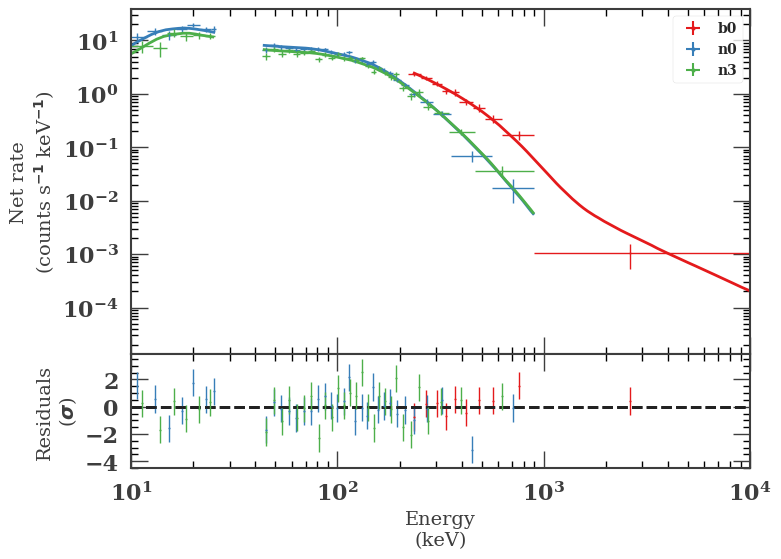}
          
    \end{subfigure}
        \hfill
    \begin{subfigure}[b]{0.40\textwidth}
        \centering
        \includegraphics[width=\textwidth]{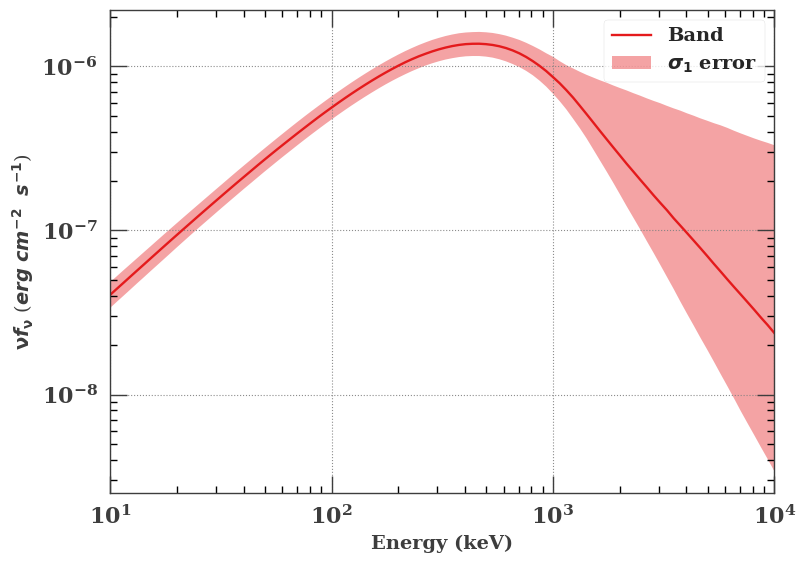}
          
    \end{subfigure}
    \caption{GRB 110731A: The count spectra (left panels) and ${\rm \nu}f_{\rm \nu}$ spectra (right panels).  The top (bottom) panels are for the joint (GBM-only) fits.}
    \label{fig_a13}
\end{figure*}


\begin{figure*}
    \centering
    \begin{subfigure}[b]{0.40\textwidth}
        \centering
        \includegraphics[width=\textwidth]{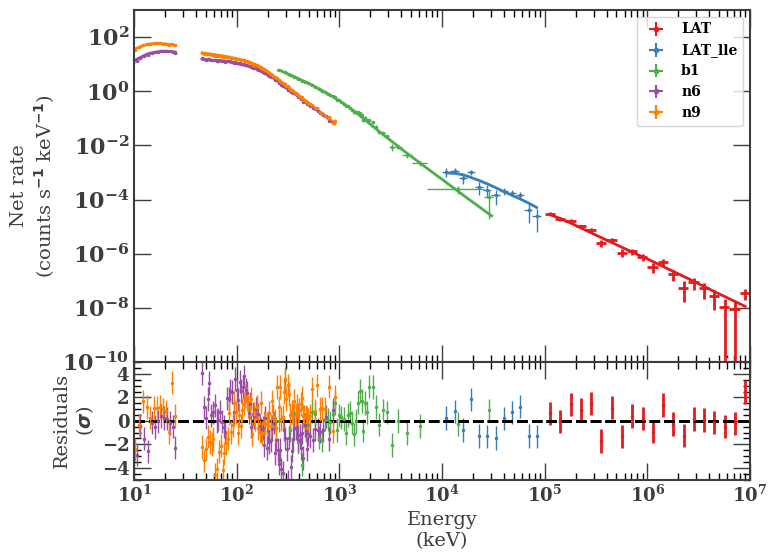}
            
    \end{subfigure}
    \hfill
    \begin{subfigure}[b]{0.40\textwidth}
        \centering
        \includegraphics[width=\textwidth]{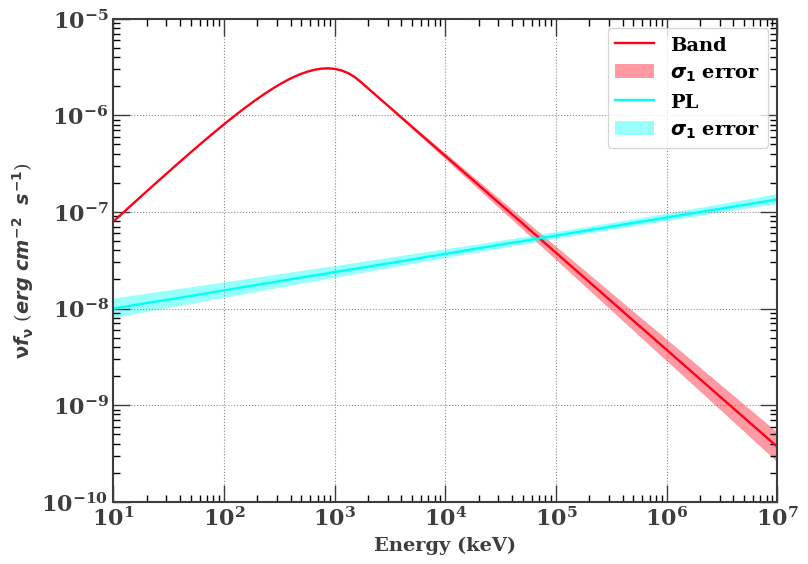}
          
    \end{subfigure}
        \hfill
    \begin{subfigure}[b]{0.40\textwidth}
        \centering
        \includegraphics[width=\textwidth]{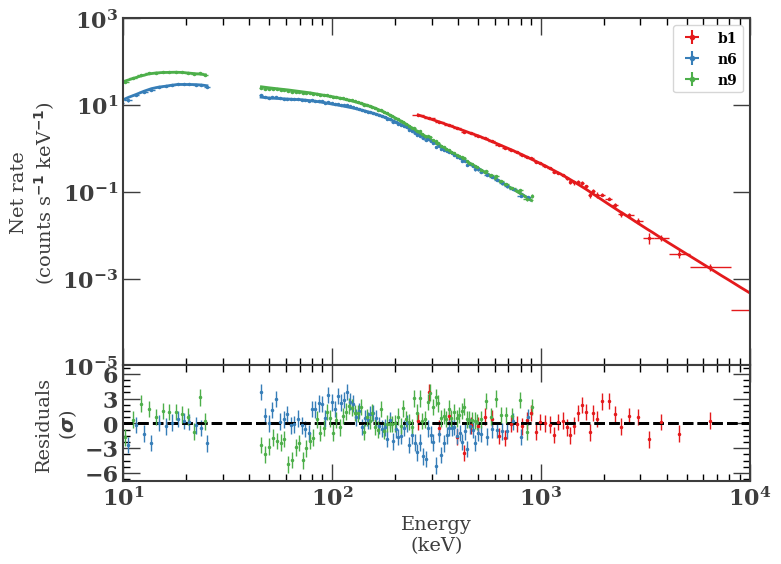}
          
    \end{subfigure}
        \hfill
    \begin{subfigure}[b]{0.40\textwidth}
        \centering
        \includegraphics[width=\textwidth]{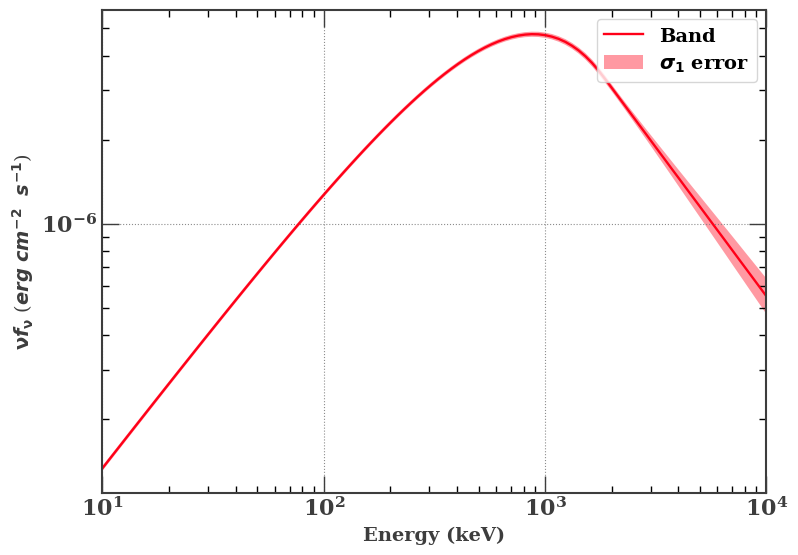}
          
    \end{subfigure}
    \caption{GRB 130427A: The count spectra (left panels) and ${\rm \nu}f_{\rm \nu}$ spectra (right panels).  The top (bottom) panels are for the joint (GBM-only) fits.}
    \label{fig_a14}
\end{figure*}


\begin{figure*}
    \centering
    \begin{subfigure}[b]{0.40\textwidth}
        \centering
        \includegraphics[width=\textwidth]{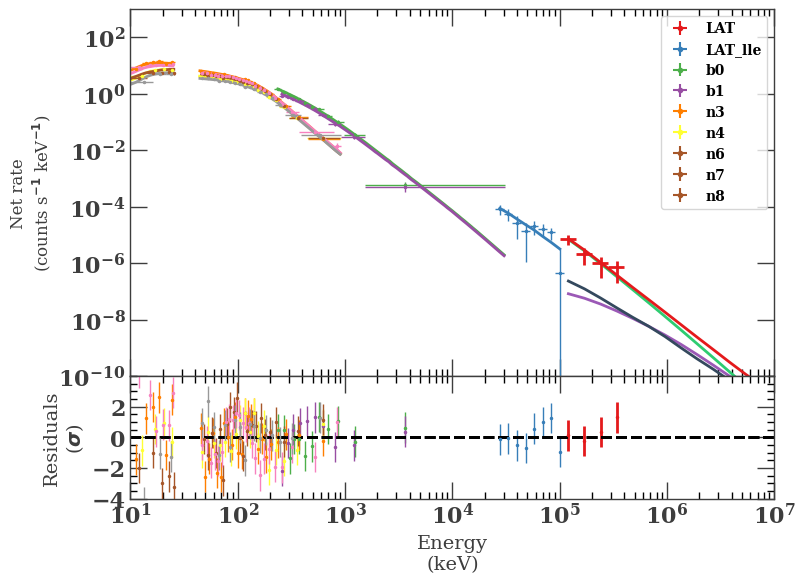}
            
    \end{subfigure}
    \hfill
    \begin{subfigure}[b]{0.40\textwidth}
        \centering
        \includegraphics[width=\textwidth]{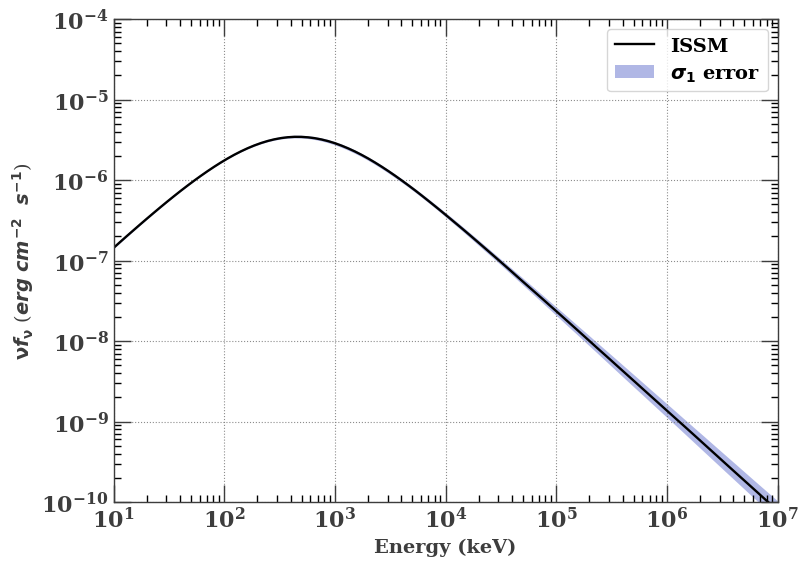}
          
    \end{subfigure}
        \hfill
    \begin{subfigure}[b]{0.40\textwidth}
        \centering
        \includegraphics[width=\textwidth]{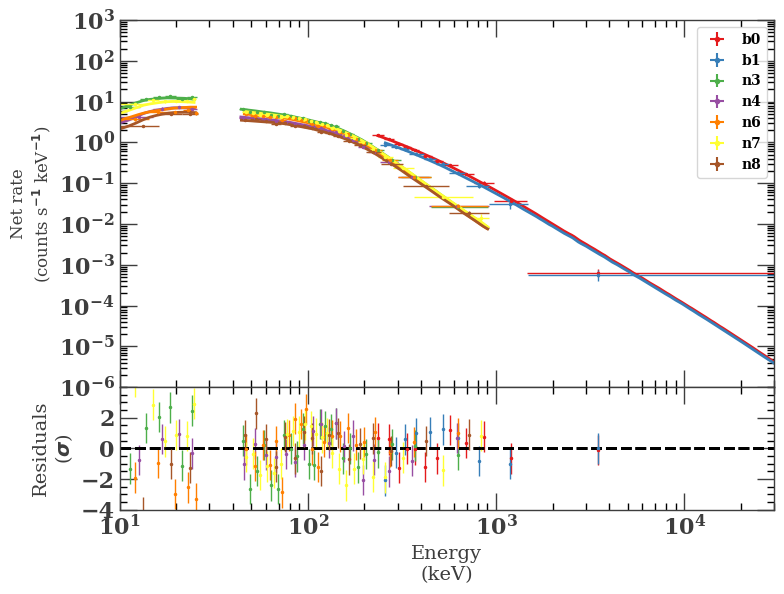}
          
    \end{subfigure}
        \hfill
    \begin{subfigure}[b]{0.40\textwidth}
        \centering
        \includegraphics[width=\textwidth]{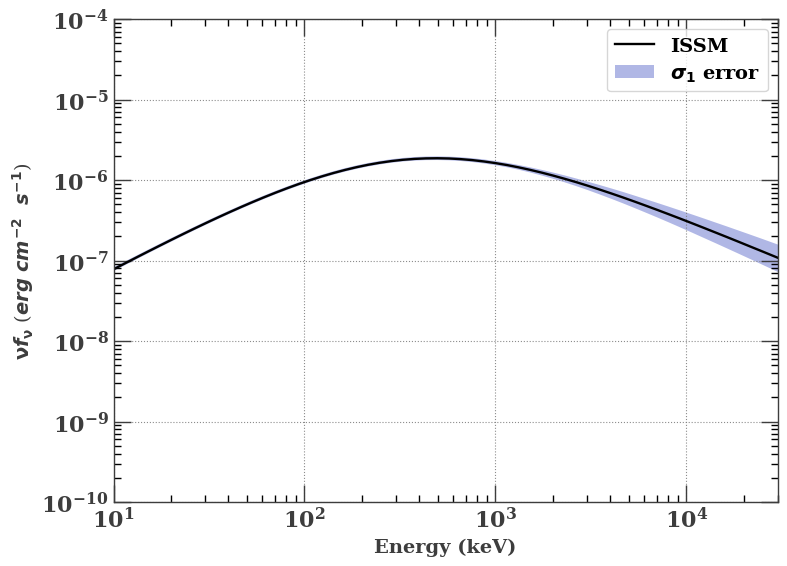}
          
    \end{subfigure}
    \caption{GRB 130518A: The count spectra (left panels) and ${\rm \nu}f_{\rm \nu}$ spectra (right panels).  The top (bottom) panels are for the joint (GBM-only) fits.}
    \label{fig_a15}
\end{figure*}


\begin{figure*}
    \centering
    \begin{subfigure}[b]{0.40\textwidth}
        \centering
        \includegraphics[width=\textwidth]{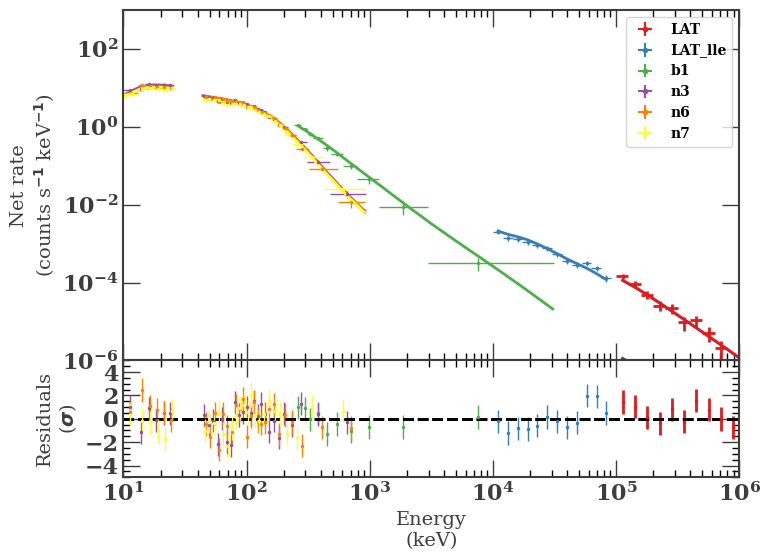}
            
    \end{subfigure}
    \hfill
    \begin{subfigure}[b]{0.40\textwidth}
        \centering
        \includegraphics[width=\textwidth]{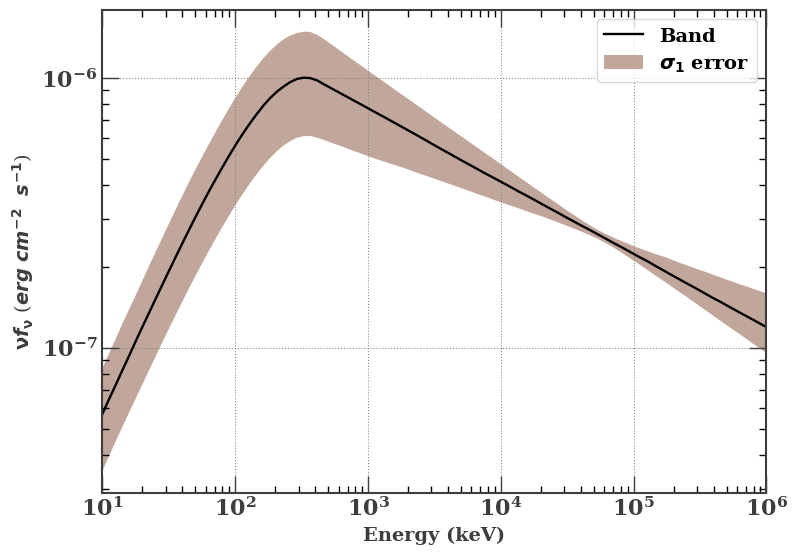}
          
    \end{subfigure}
        \hfill
    \begin{subfigure}[b]{0.40\textwidth}
        \centering
        \includegraphics[width=\textwidth]{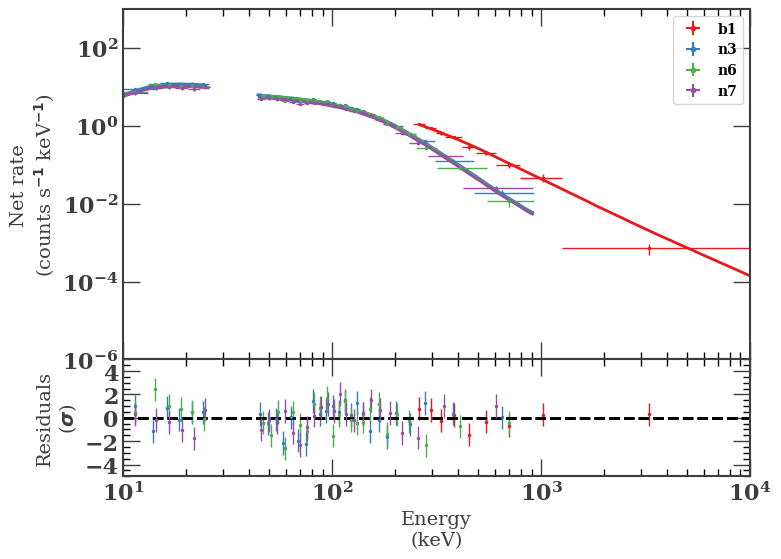}
          
    \end{subfigure}
        \hfill
    \begin{subfigure}[b]{0.40\textwidth}
        \centering
        \includegraphics[width=\textwidth]{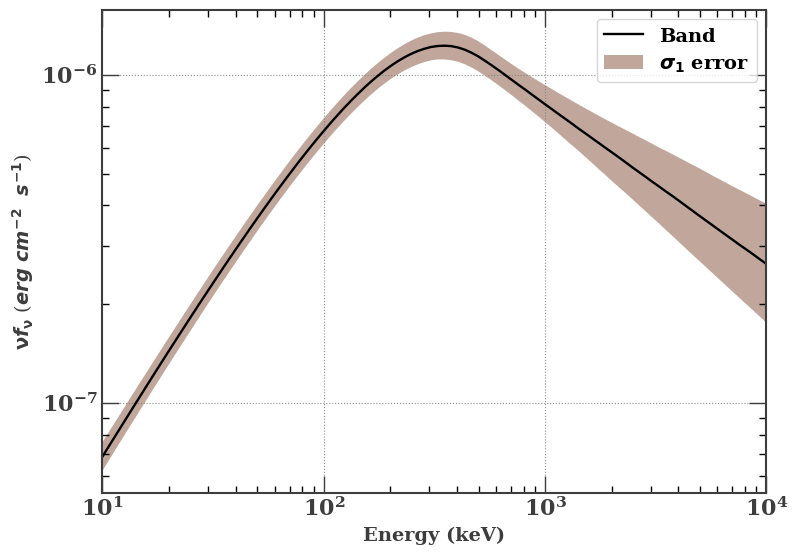}
          
    \end{subfigure}
    \caption{GRB 131108A: The count spectra (left panels) and ${\rm \nu}f_{\rm \nu}$ spectra (right panels).  The top (bottom) panels are for the joint (GBM-only) fits.}
    \label{fig_a16}
\end{figure*}


\begin{figure*}
    \centering
    \begin{subfigure}[b]{0.40\textwidth}
        \centering
        \includegraphics[width=\textwidth]{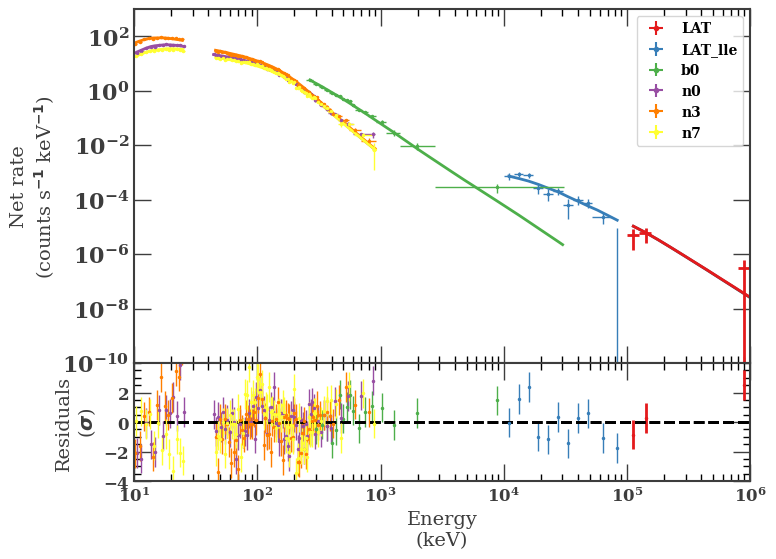}
            
    \end{subfigure}
    \hfill
    \begin{subfigure}[b]{0.40\textwidth}
        \centering
        \includegraphics[width=\textwidth]{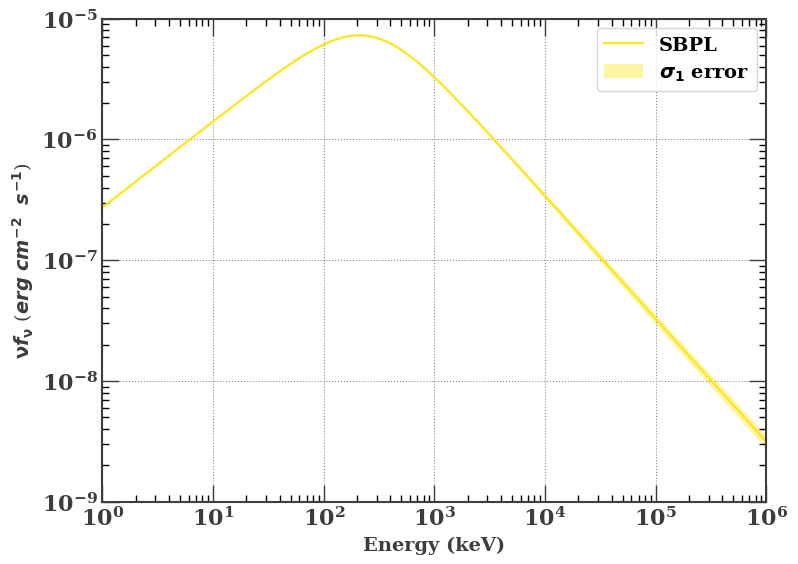}
          
    \end{subfigure}
        \hfill
    \begin{subfigure}[b]{0.40\textwidth}
        \centering
        \includegraphics[width=\textwidth]{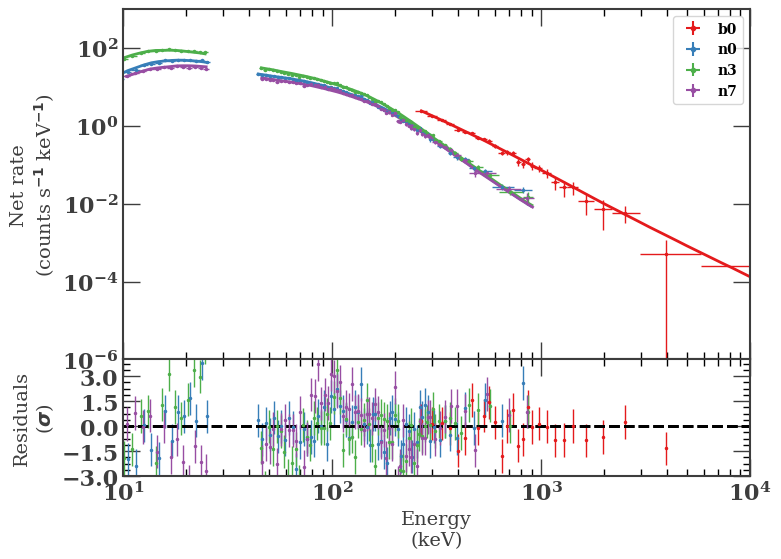}
          
    \end{subfigure}
        \hfill
    \begin{subfigure}[b]{0.40\textwidth}
        \centering
        \includegraphics[width=\textwidth]{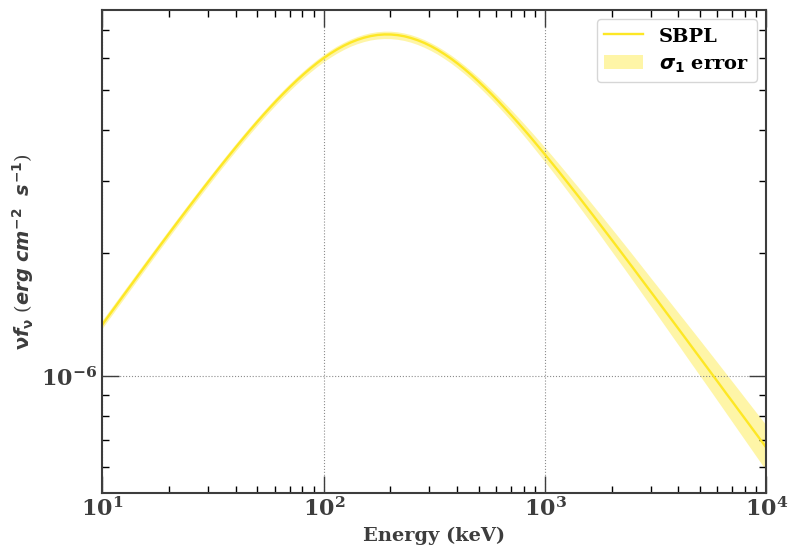}
          
    \end{subfigure}
    \caption{GRB 131231A: The count spectra (left panels) and ${\rm \nu}f_{\rm \nu}$ spectra (right panels).  The top (bottom) panels are for the joint (GBM-only) fits.}
    \label{fig_a17}
\end{figure*}


\begin{figure*}
    \centering
    \begin{subfigure}[b]{0.40\textwidth}
        \centering
        \includegraphics[width=\textwidth]{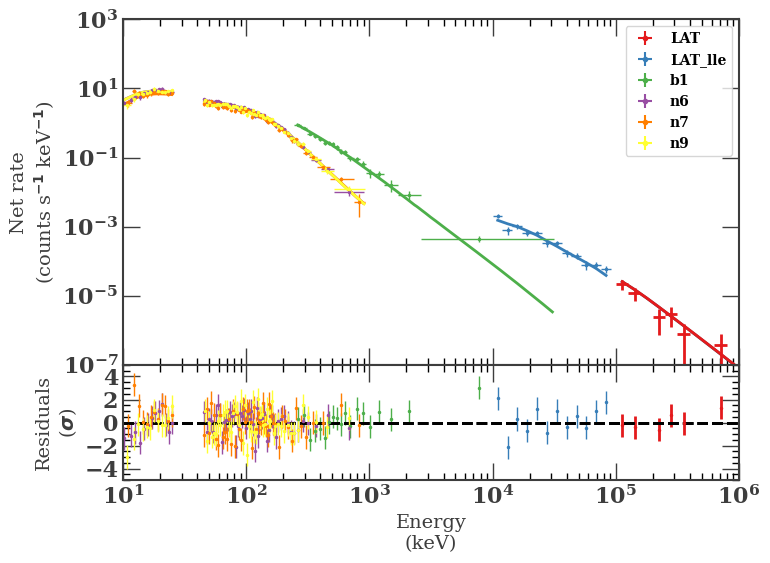}
            
    \end{subfigure}
    \hfill
    \begin{subfigure}[b]{0.40\textwidth}
        \centering
        \includegraphics[width=\textwidth]{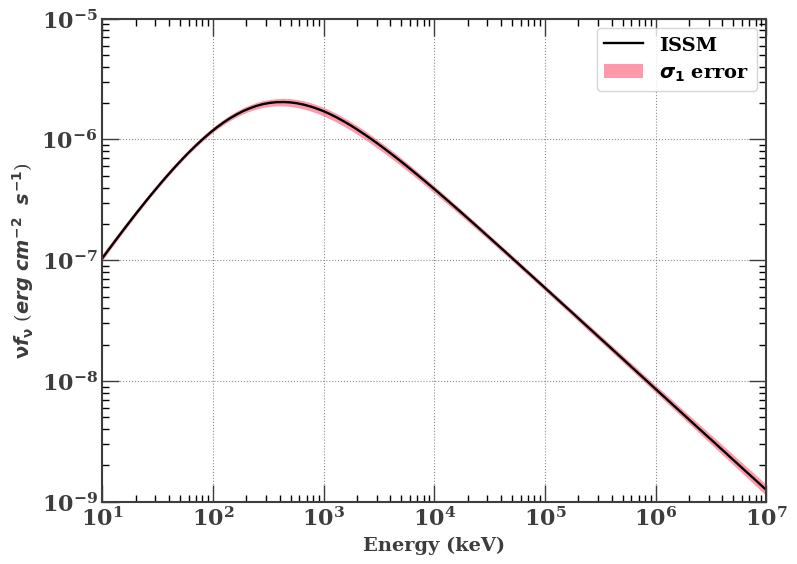}
          
    \end{subfigure}
        \hfill
    \begin{subfigure}[b]{0.40\textwidth}
        \centering
        \includegraphics[width=\textwidth]{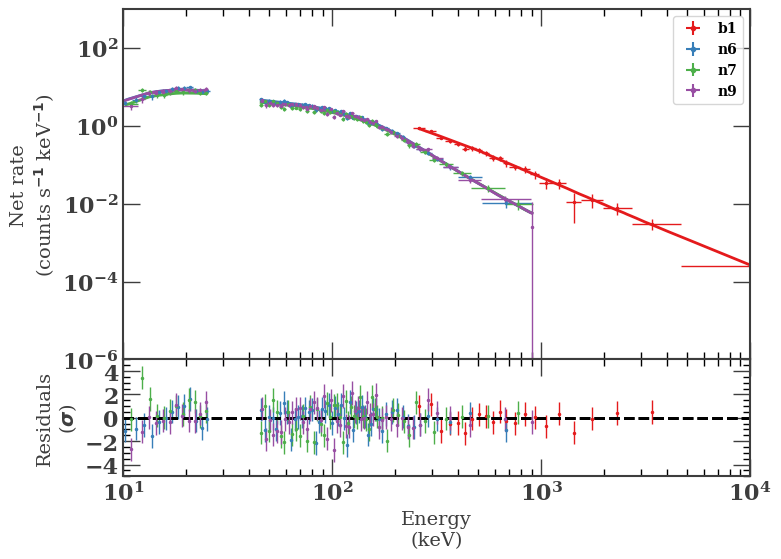}
          
    \end{subfigure}
        \hfill
    \begin{subfigure}[b]{0.40\textwidth}
        \centering
        \includegraphics[width=\textwidth]{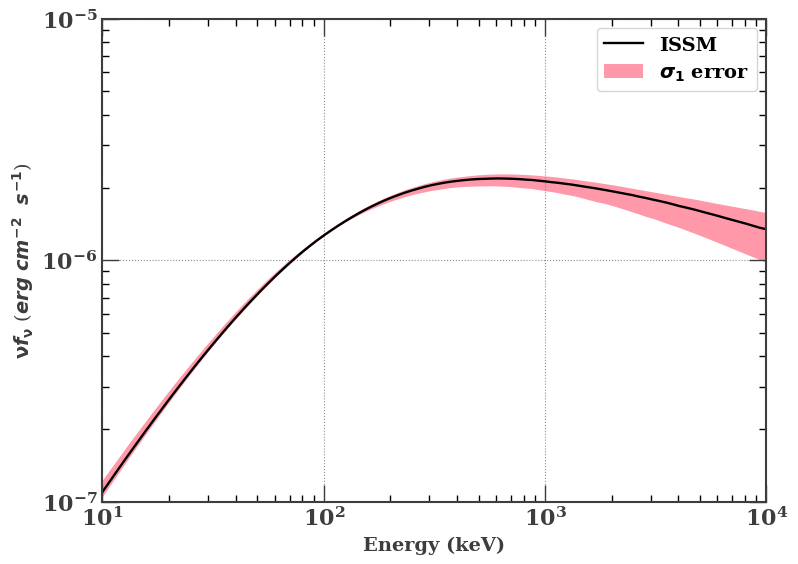}
          
    \end{subfigure}
    \caption{GRB 141028A: The count spectra (left panels) and ${\rm \nu}f_{\rm \nu}$ spectra (right panels).  The top (bottom) panels are for the joint (GBM-only) fits.}
    \label{fig_a18}
\end{figure*}


\begin{figure*}
    \centering
    \begin{subfigure}[b]{0.40\textwidth}
        \centering
        \includegraphics[width=\textwidth]{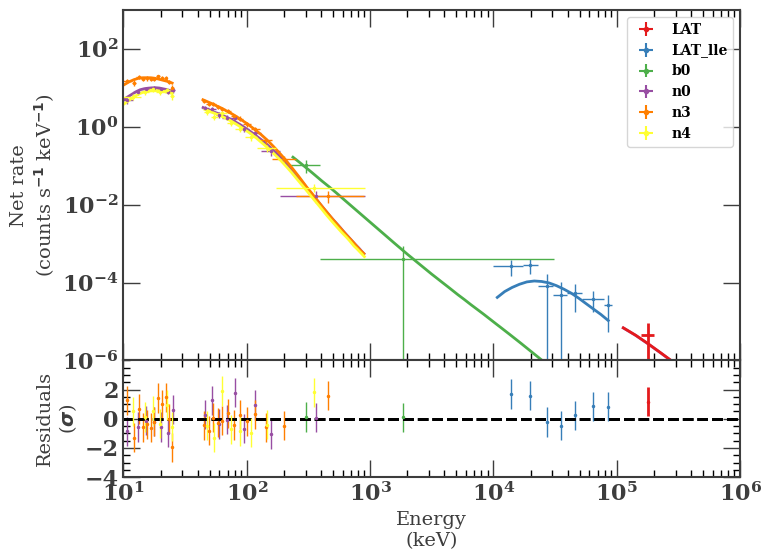}
            
    \end{subfigure}
    \hfill
    \begin{subfigure}[b]{0.40\textwidth}
        \centering
        \includegraphics[width=\textwidth]{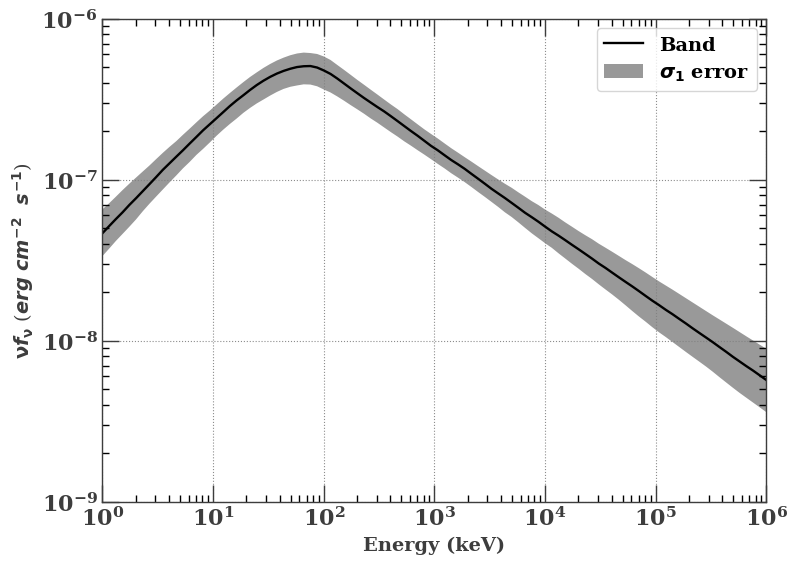}
          
    \end{subfigure}
        \hfill
    \begin{subfigure}[b]{0.40\textwidth}
        \centering
        \includegraphics[width=\textwidth]{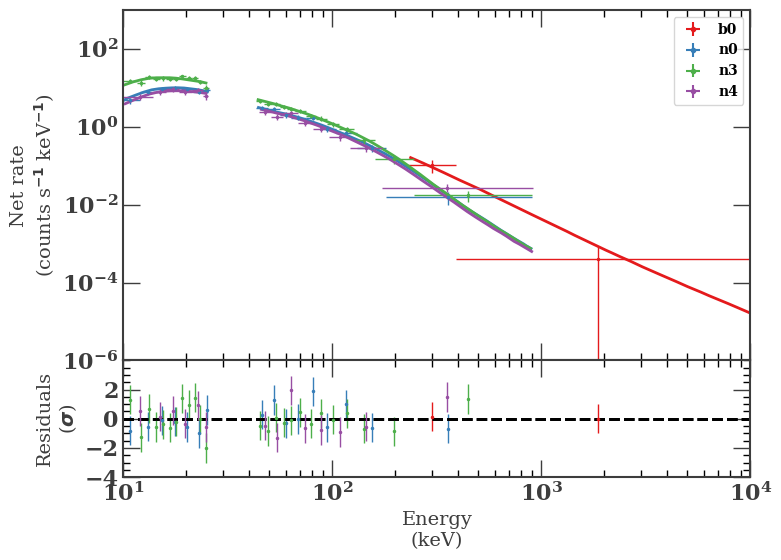}
          
    \end{subfigure}
        \hfill
    \begin{subfigure}[b]{0.40\textwidth}
        \centering
        \includegraphics[width=\textwidth]{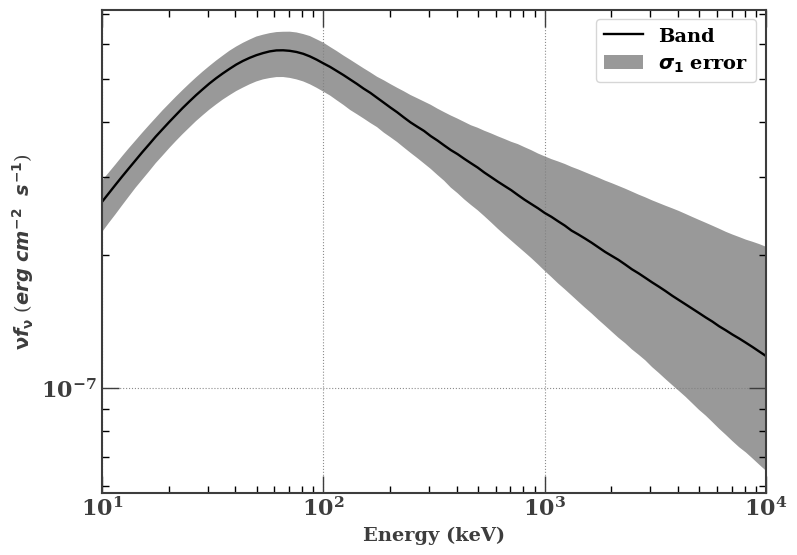}
          
    \end{subfigure}
    \caption{GRB 150314A: The count spectra (left panels) and ${\rm \nu}f_{\rm \nu}$ spectra (right panels).  The top (bottom) panels are for the joint (GBM-only) fits.}
    \label{fig_a19}
\end{figure*}


\begin{figure*}
    \centering
    \begin{subfigure}[b]{0.40\textwidth}
        \centering
        \includegraphics[width=\textwidth]{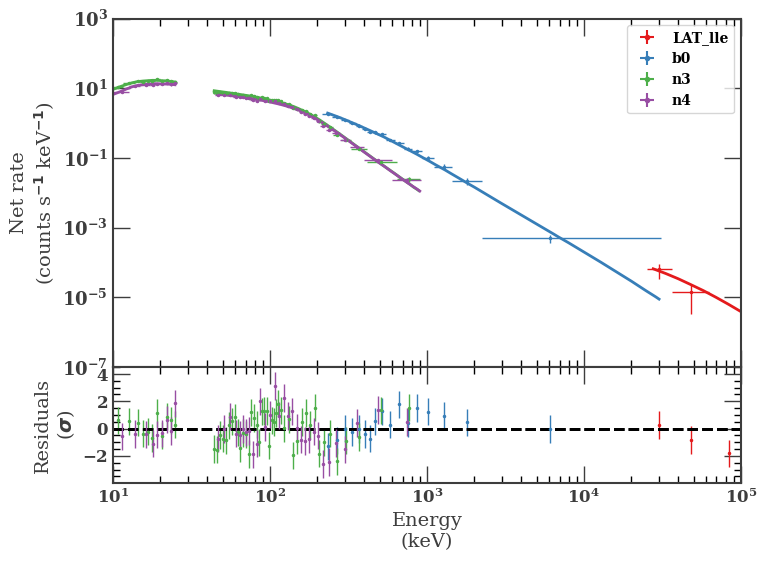}
            
    \end{subfigure}
    \hfill
    \begin{subfigure}[b]{0.40\textwidth}
        \centering
        \includegraphics[width=\textwidth]{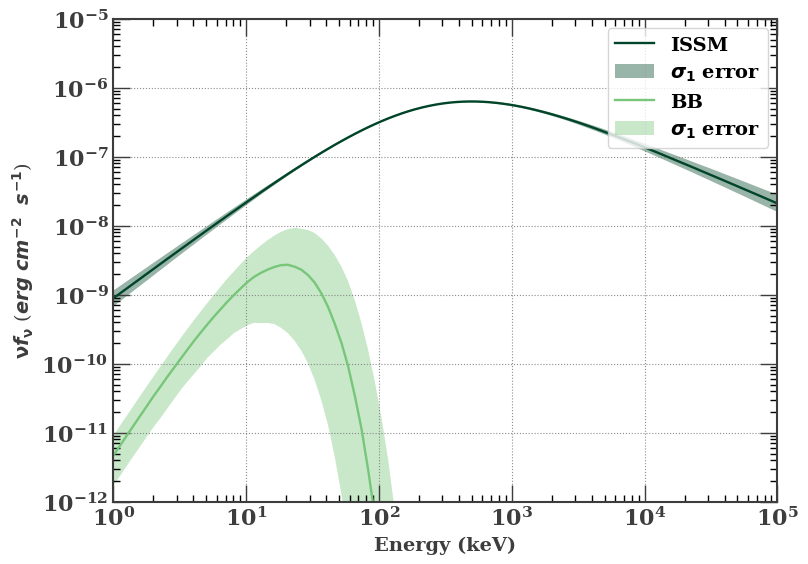}
          
    \end{subfigure}
        \hfill
    \begin{subfigure}[b]{0.40\textwidth}
        \centering
        \includegraphics[width=\textwidth]{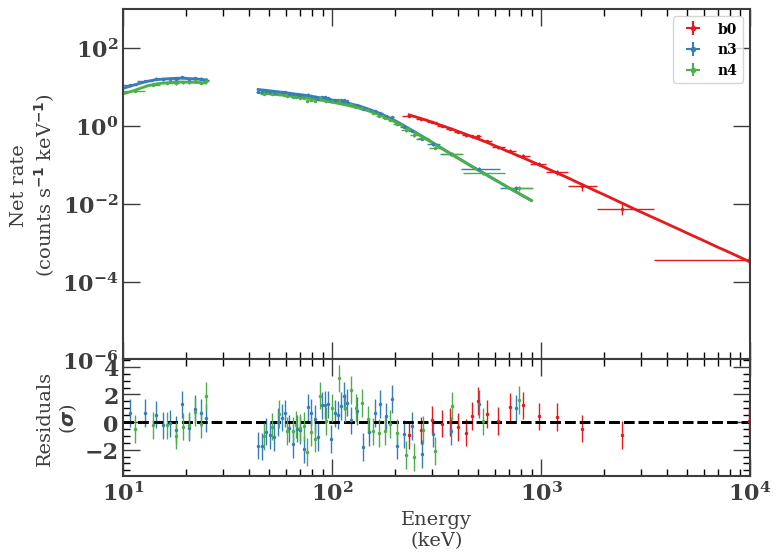}
          
    \end{subfigure}
        \hfill
    \begin{subfigure}[b]{0.40\textwidth}
        \centering
        \includegraphics[width=\textwidth]{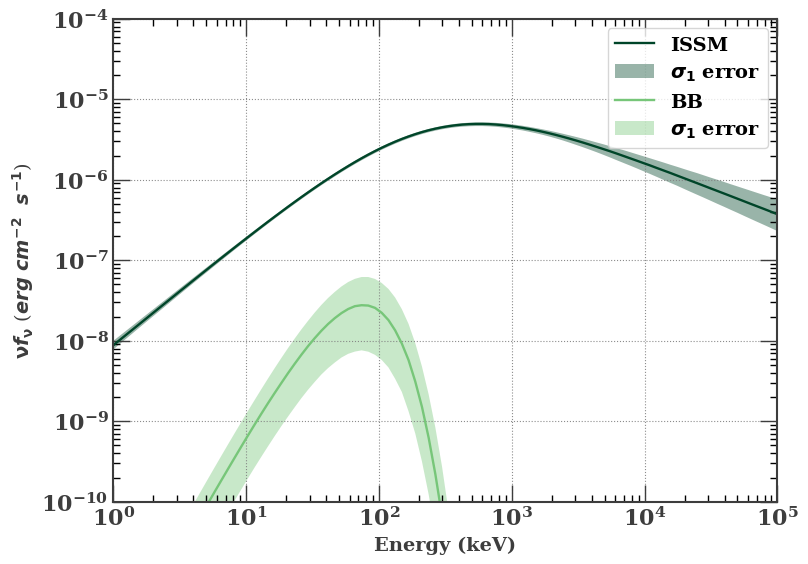}
          
    \end{subfigure}
    \caption{GRB 150403A: The count spectra (left panels) and ${\rm \nu}f_{\rm \nu}$ spectra (right panels).  The top (bottom) panels are for the joint (GBM-only) fits.}
    \label{fig_a20}
\end{figure*}


\begin{figure*}
    \centering
    \begin{subfigure}[b]{0.40\textwidth}
        \centering
        \includegraphics[width=\textwidth]{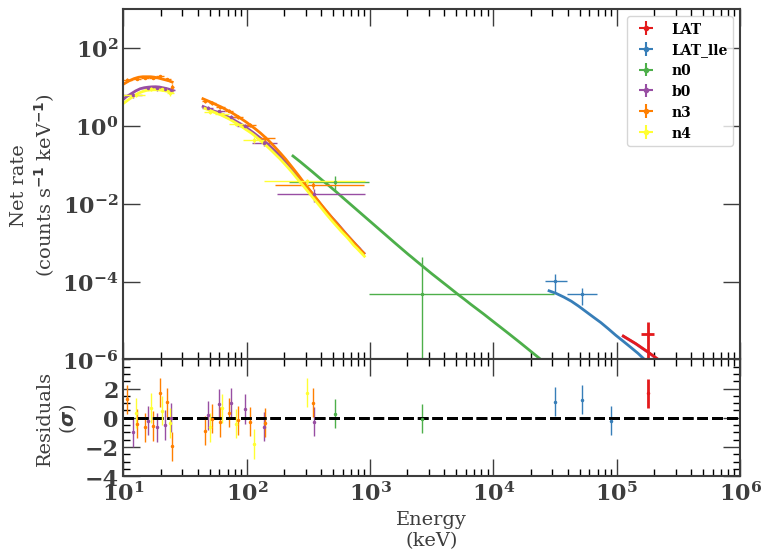}
            
    \end{subfigure}
    \hfill
    \begin{subfigure}[b]{0.40\textwidth}
        \centering
        \includegraphics[width=\textwidth]{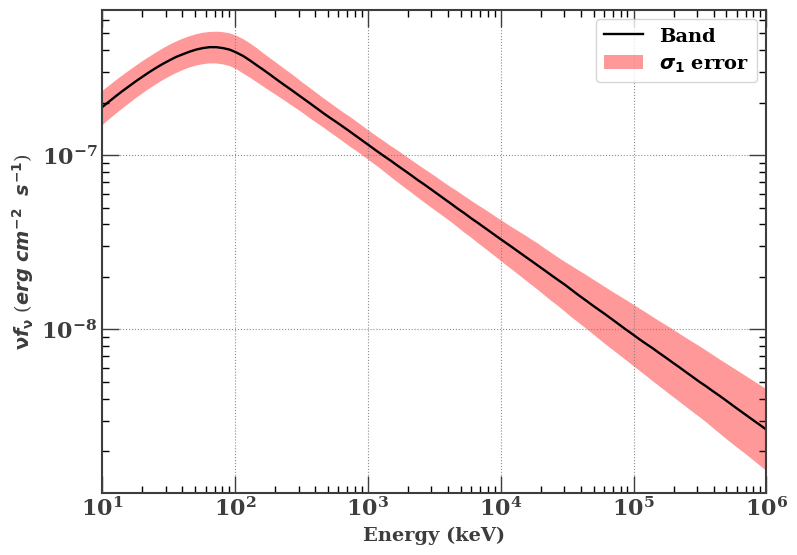}
          
    \end{subfigure}
        \hfill
    \begin{subfigure}[b]{0.40\textwidth}
        \centering
        \includegraphics[width=\textwidth]{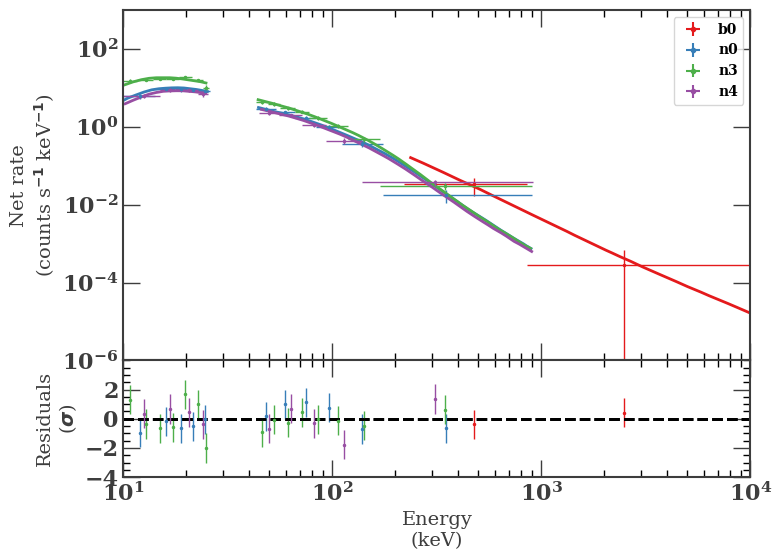}
          
    \end{subfigure}
        \hfill
    \begin{subfigure}[b]{0.40\textwidth}
        \centering
        \includegraphics[width=\textwidth]{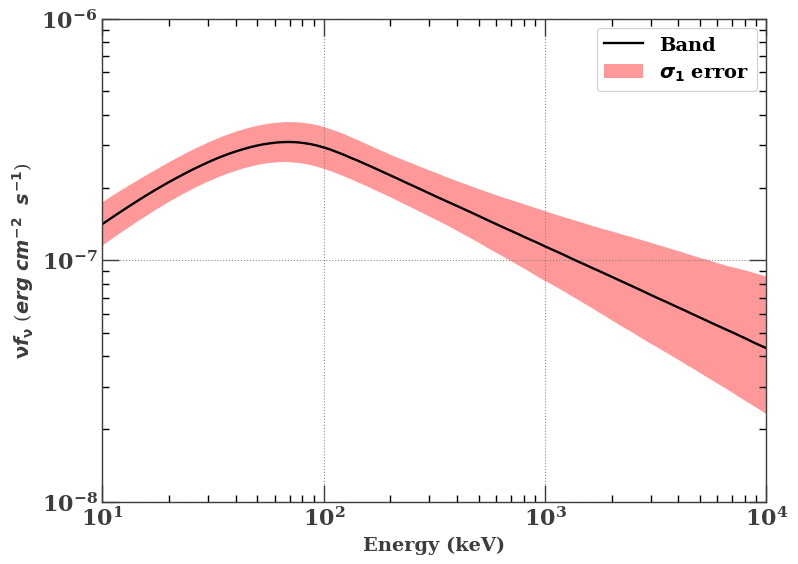}
          
    \end{subfigure}
    \caption{GRB 150514A: The count spectra (left panels) and ${\rm \nu}f_{\rm \nu}$ spectra (right panels).  The top (bottom) panels are for the joint (GBM-only) fits.}
    \label{fig_a_a20}
\end{figure*}


\begin{figure*}
    \centering
    \begin{subfigure}[b]{0.40\textwidth}
        \centering
        \includegraphics[width=\textwidth]{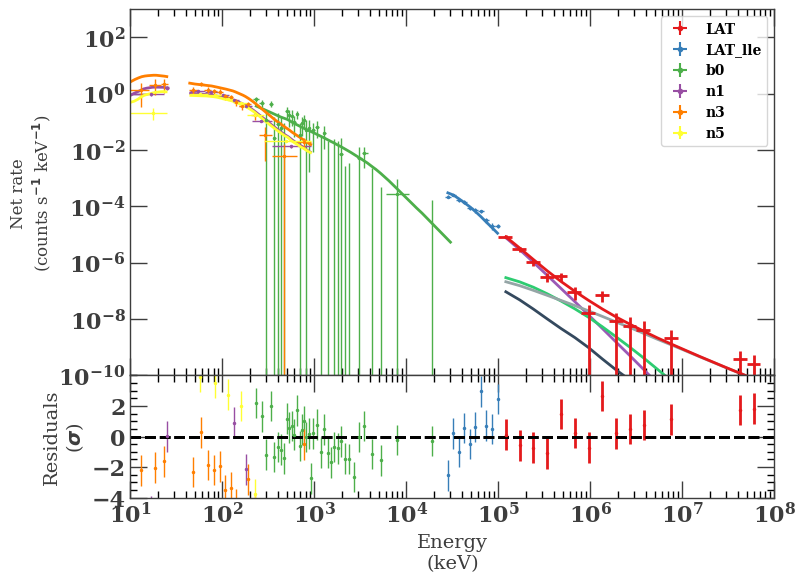}
            
    \end{subfigure}
    \hfill
    \begin{subfigure}[b]{0.40\textwidth}
        \centering
        \includegraphics[width=\textwidth]{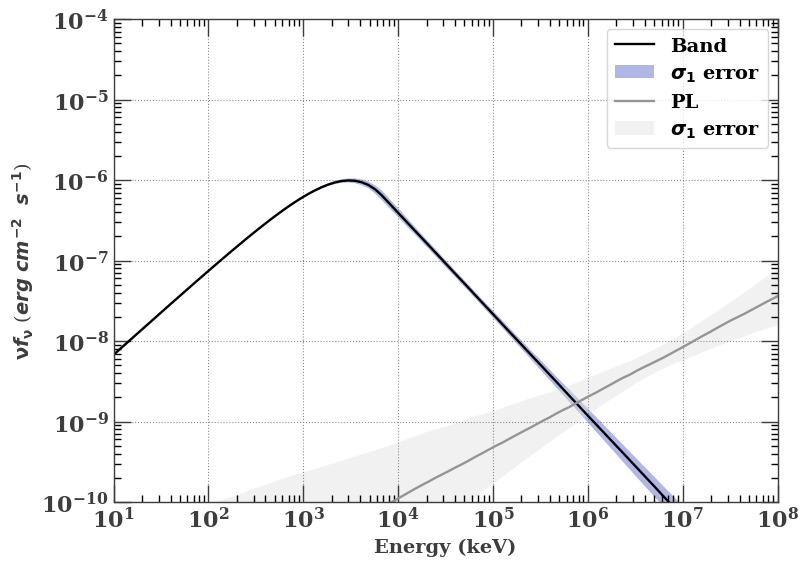}
          
    \end{subfigure}
        \hfill
    \begin{subfigure}[b]{0.40\textwidth}
        \centering
        \includegraphics[width=\textwidth]{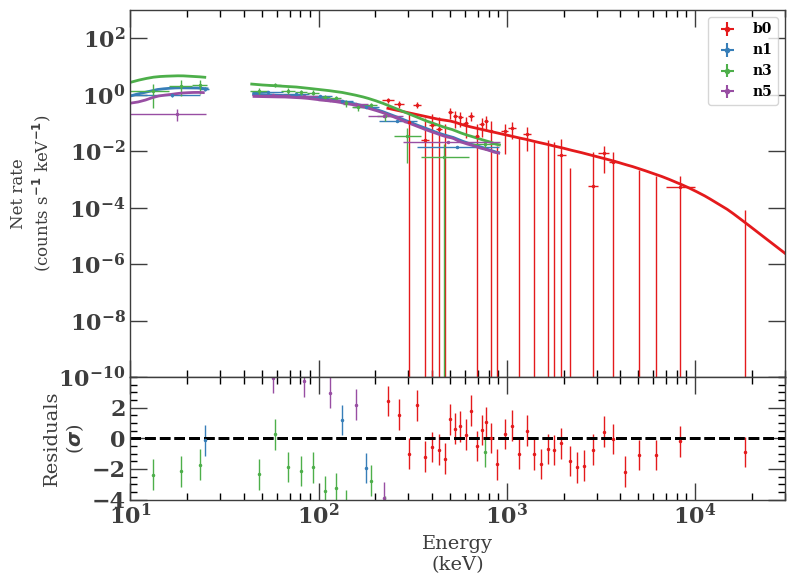}
          
    \end{subfigure}
        \hfill
    \begin{subfigure}[b]{0.40\textwidth}
        \centering
        \includegraphics[width=\textwidth]{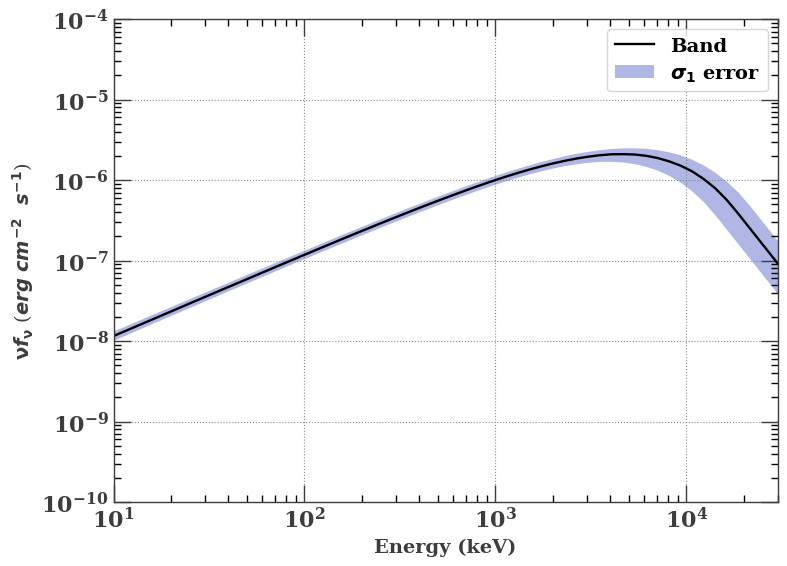}
          
    \end{subfigure}
    \caption{GRB 160509A: The count spectra (left panels) and ${\rm \nu}f_{\rm \nu}$ spectra (right panels).  The top (bottom) panels are for the joint (GBM-only) fits.}
    \label{fig_a21}
\end{figure*}


\begin{figure*}
    \centering
    \begin{subfigure}[b]{0.40\textwidth}
        \centering
        \includegraphics[width=\textwidth]{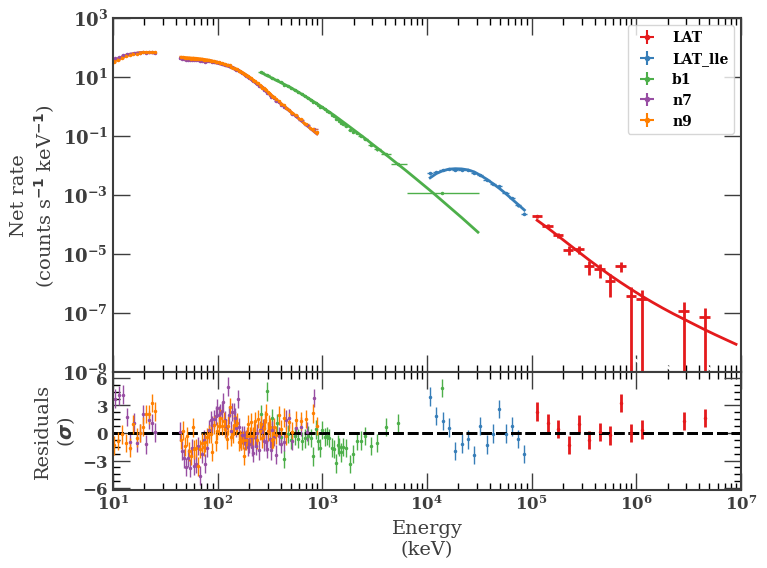}
            
    \end{subfigure}
    \hfill
    \begin{subfigure}[b]{0.40\textwidth}
        \centering
        \includegraphics[width=\textwidth]{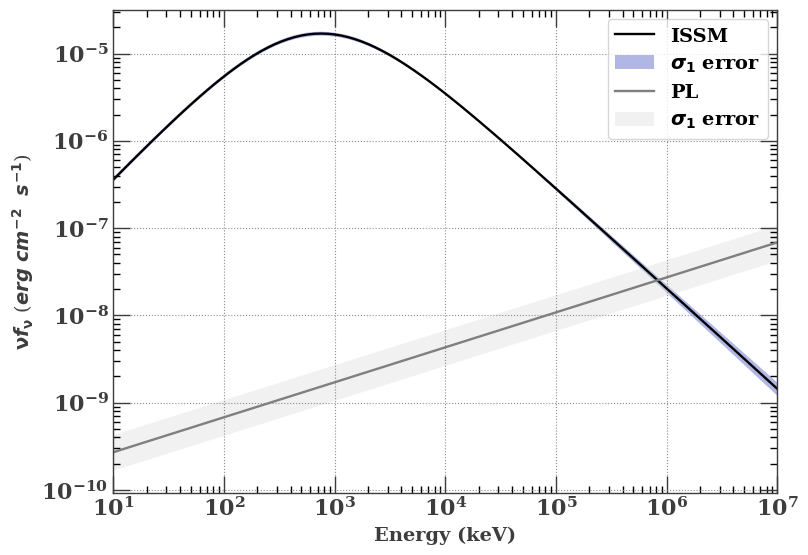}
          
    \end{subfigure}
        \hfill
    \begin{subfigure}[b]{0.40\textwidth}
        \centering
        \includegraphics[width=\textwidth]{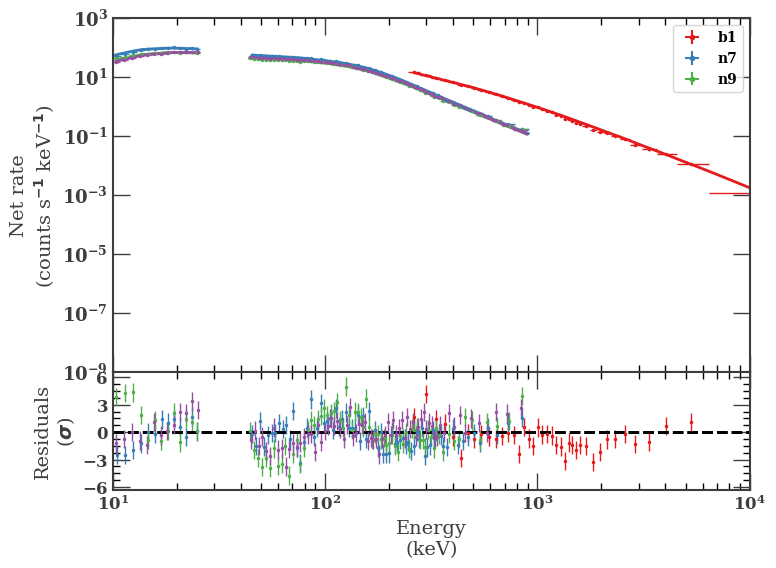}
          
    \end{subfigure}
        \hfill
    \begin{subfigure}[b]{0.40\textwidth}
        \centering
        \includegraphics[width=\textwidth]{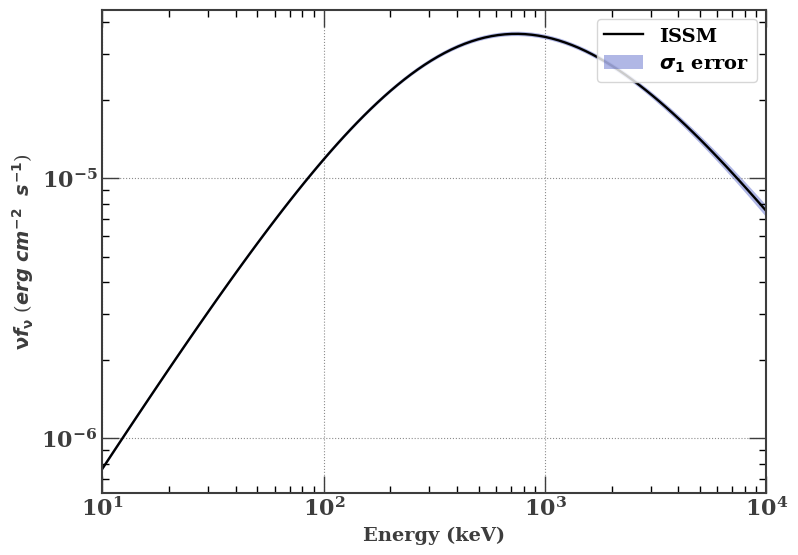}
          
    \end{subfigure}
    \caption{GRB 160625B: The count spectra (left panels) and ${\rm \nu}f_{\rm \nu}$ spectra (right panels).  The top (bottom) panels are for the joint (GBM-only) fits.}
    \label{fig_a22}
\end{figure*}


\begin{figure*}
    \centering
    \begin{subfigure}[b]{0.40\textwidth}
        \centering
        \includegraphics[width=\textwidth]{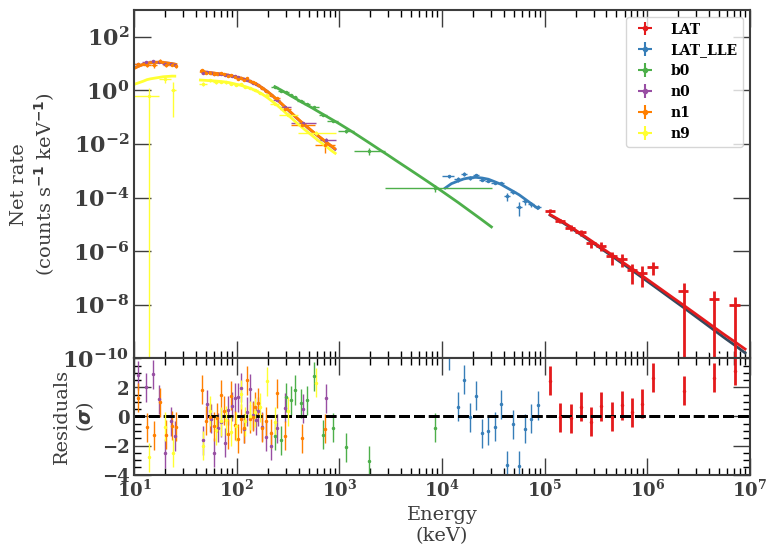}
            
    \end{subfigure}
    \hfill
    \begin{subfigure}[b]{0.40\textwidth}
        \centering
        \includegraphics[width=\textwidth]{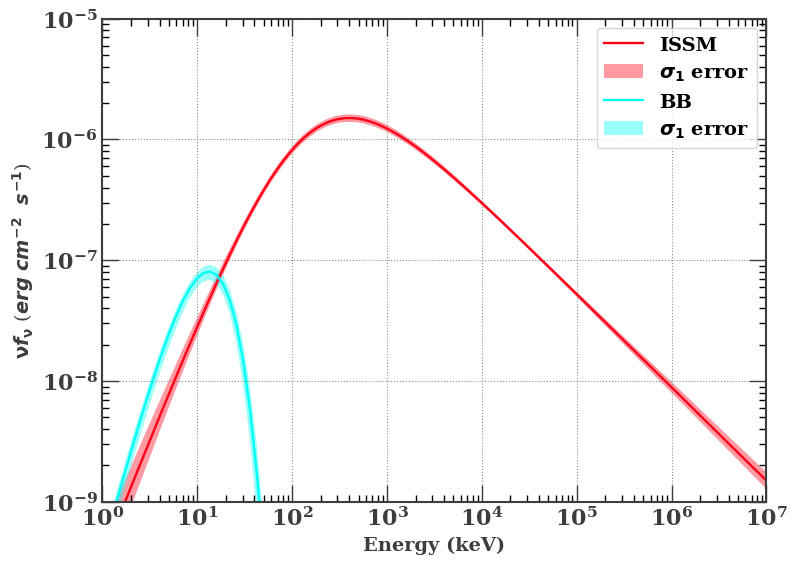}
          
    \end{subfigure}
        \hfill
    \begin{subfigure}[b]{0.40\textwidth}
        \centering
        \includegraphics[width=\textwidth]{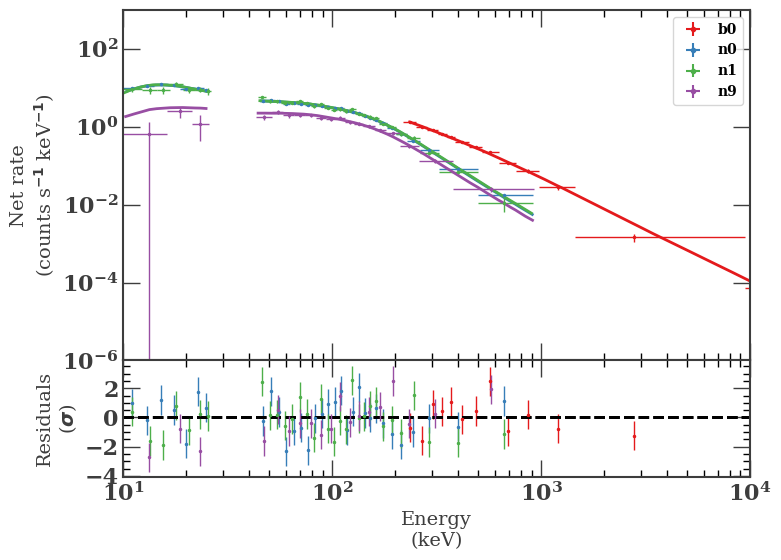}
          
    \end{subfigure}
        \hfill
    \begin{subfigure}[b]{0.40\textwidth}
        \centering
        \includegraphics[width=\textwidth]{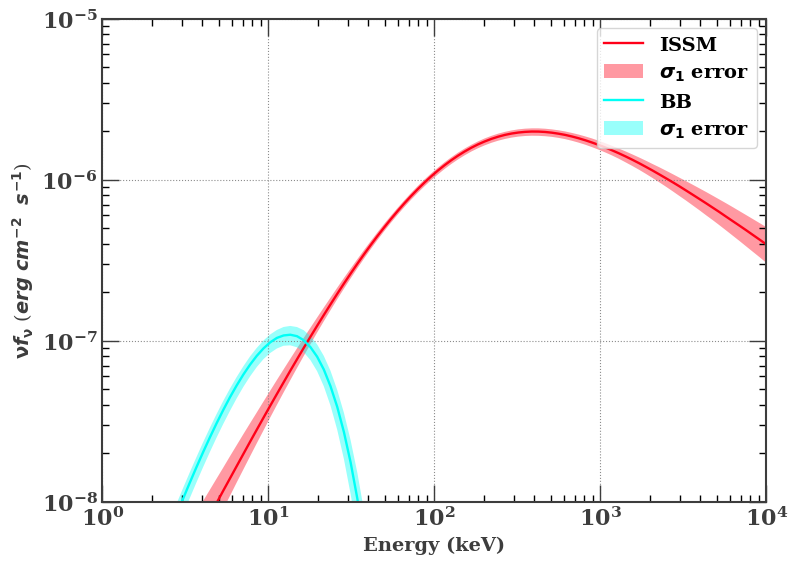}
          
    \end{subfigure}
    \caption{GRB 170214A: The count spectra (left panels) and ${\rm \nu}f_{\rm \nu}$ spectra (right panels).  The top (bottom) panels are for the joint (GBM-only) fits.}
    \label{fig_a23}
\end{figure*}


\begin{figure*}
    \centering
    \begin{subfigure}[b]{0.40\textwidth}
        \centering
        \includegraphics[width=\textwidth]{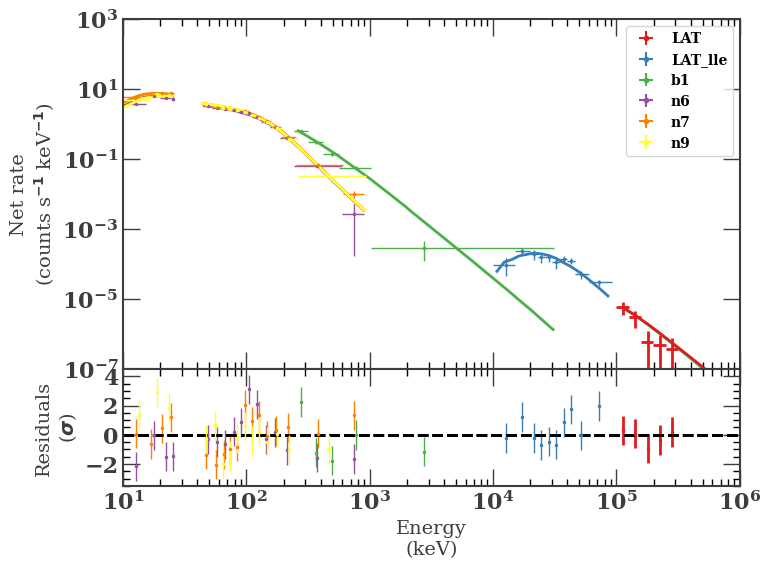}
            
    \end{subfigure}
    \hfill
    \begin{subfigure}[b]{0.40\textwidth}
        \centering
        \includegraphics[width=\textwidth]{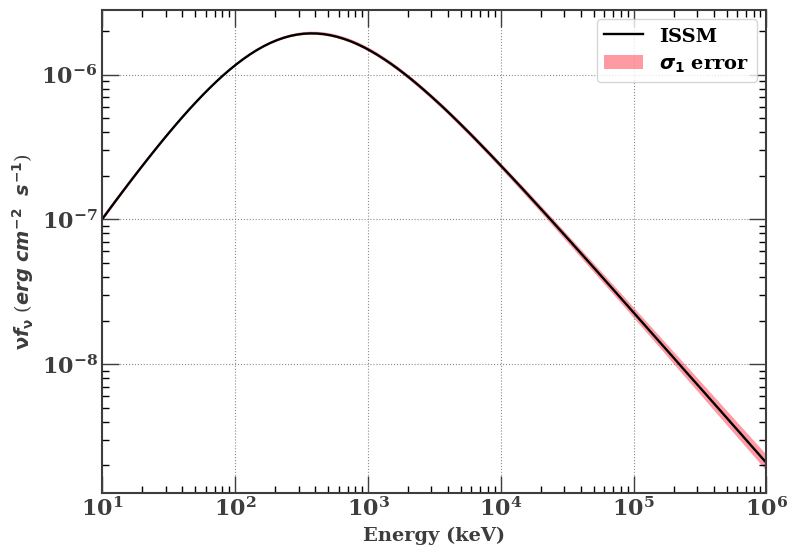}
          
    \end{subfigure}
        \hfill
    \begin{subfigure}[b]{0.40\textwidth}
        \centering
        \includegraphics[width=\textwidth]{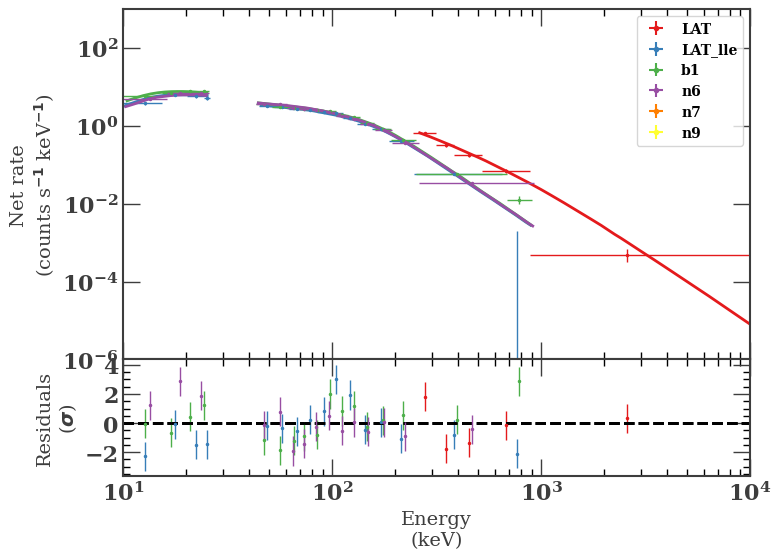}
          
    \end{subfigure}
        \hfill
    \begin{subfigure}[b]{0.40\textwidth}
        \centering
        \includegraphics[width=\textwidth]{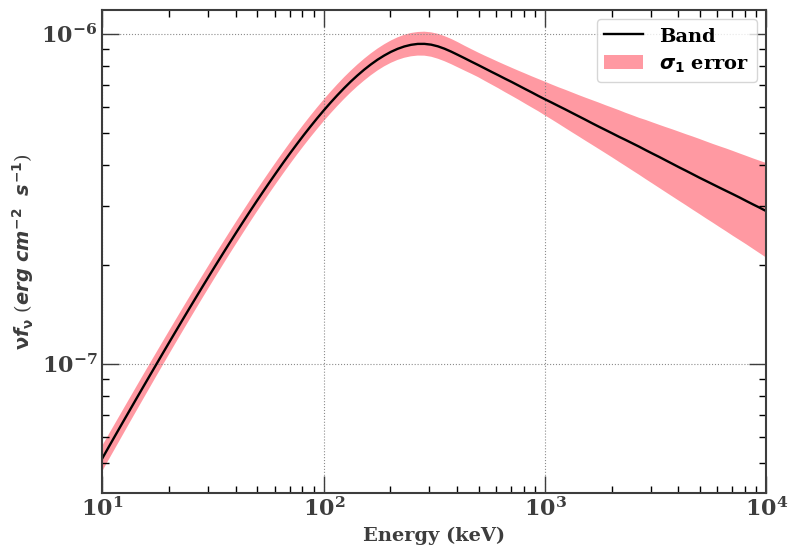}
          
    \end{subfigure}
    \caption{GRB 170405A: The count spectra (left panels) and ${\rm \nu}f_{\rm \nu}$ spectra (right panels).  The top (bottom) panels are for the joint (GBM-only) fits.}
    \label{fig_a24}
\end{figure*}


\begin{figure*}
    \centering
    \begin{subfigure}[b]{0.40\textwidth}
        \centering
        \includegraphics[width=\textwidth]{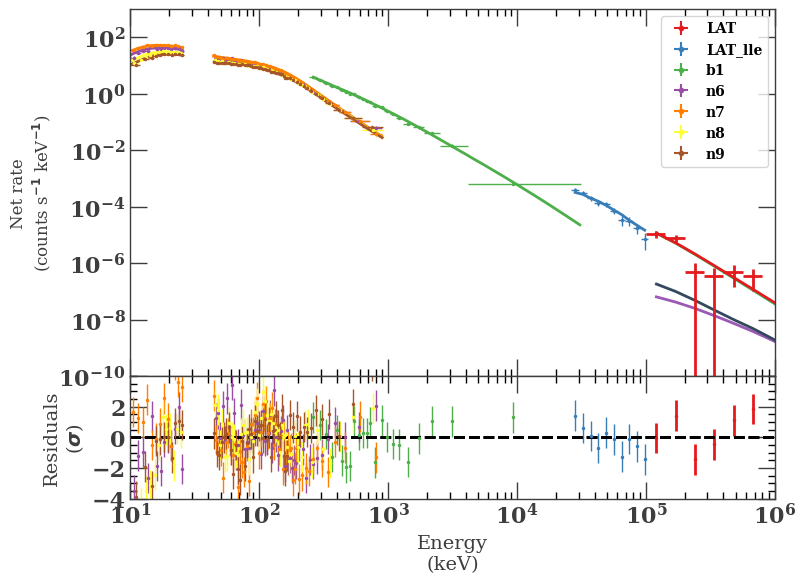}
            
    \end{subfigure}
    \hfill
    \begin{subfigure}[b]{0.40\textwidth}
        \centering
        \includegraphics[width=\textwidth]{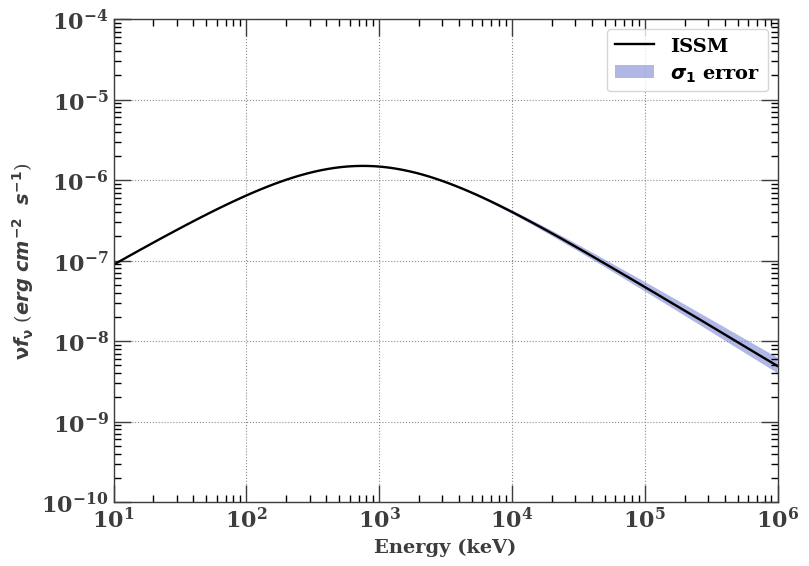}
          
    \end{subfigure}
        \hfill
    \begin{subfigure}[b]{0.40\textwidth}
        \centering
        \includegraphics[width=\textwidth]{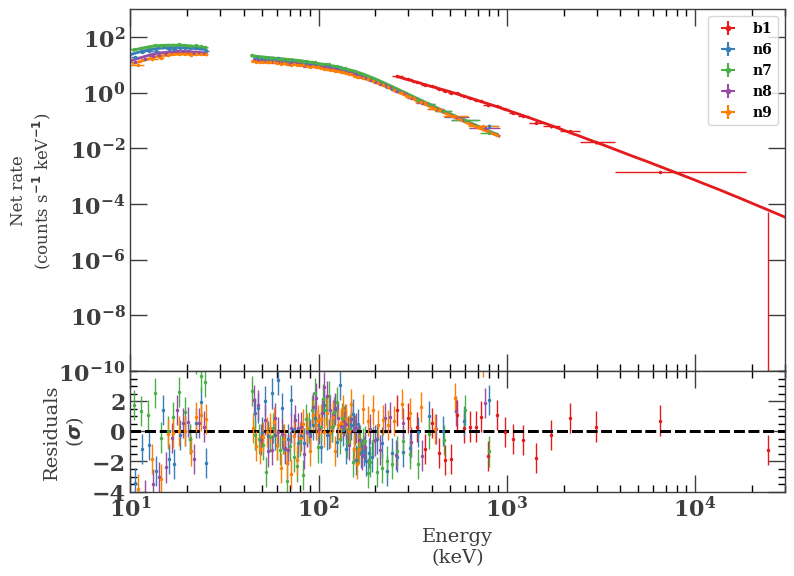}
          
    \end{subfigure}
        \hfill
    \begin{subfigure}[b]{0.40\textwidth}
        \centering
        \includegraphics[width=\textwidth]{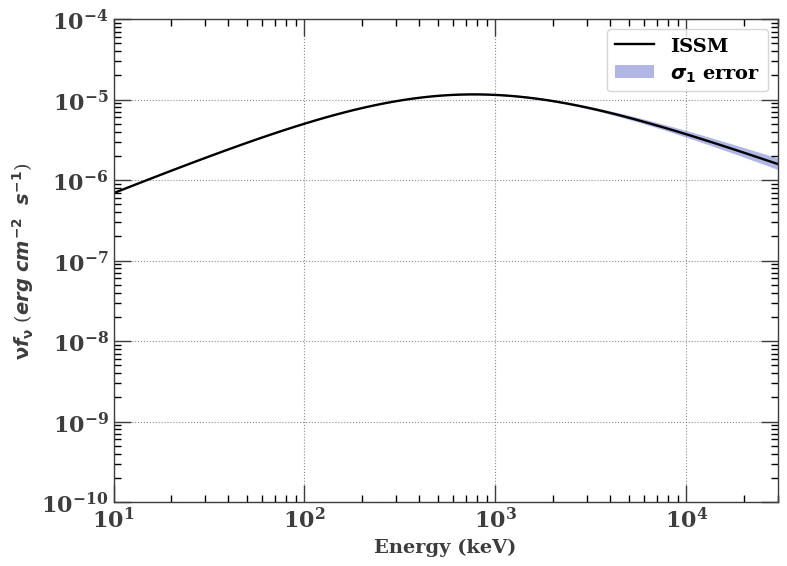}
        \caption{${\rm \nu}f_{\rm \nu}$ plot: GBM}
          
    \end{subfigure}
    \caption{GRB 180720B: The count spectra (left panels) and ${\rm \nu}f_{\rm \nu}$ spectra (right panels).  The top (bottom) panels are for the joint (GBM-only) fits.}
    \label{fig_a25}
\end{figure*}


\begin{figure*}
    \centering
    \begin{subfigure}[b]{0.40\textwidth}
        \centering
        \includegraphics[width=\textwidth]{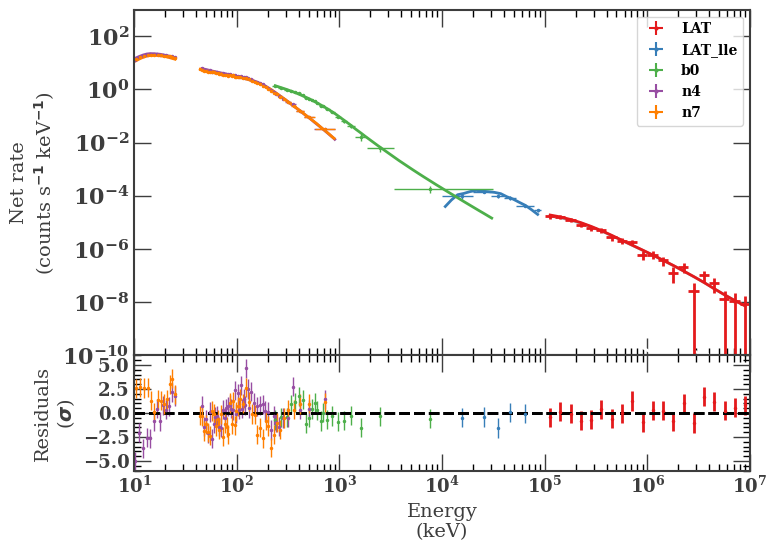}
            
    \end{subfigure}
    \hfill
    \begin{subfigure}[b]{0.40\textwidth}
        \centering
        \includegraphics[width=\textwidth]{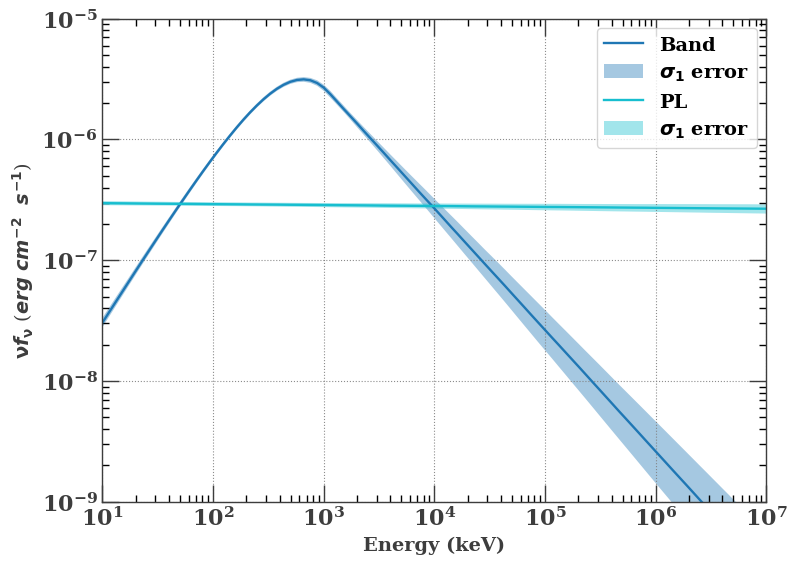}
          
    \end{subfigure}
        \hfill
    \begin{subfigure}[b]{0.40\textwidth}
        \centering
        \includegraphics[width=\textwidth]{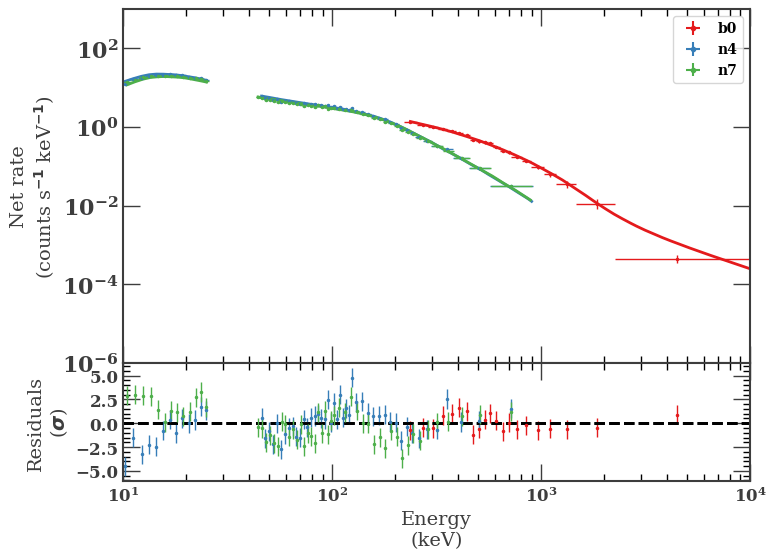}
          
    \end{subfigure}
        \hfill
    \begin{subfigure}[b]{0.40\textwidth}
        \centering
        \includegraphics[width=\textwidth]{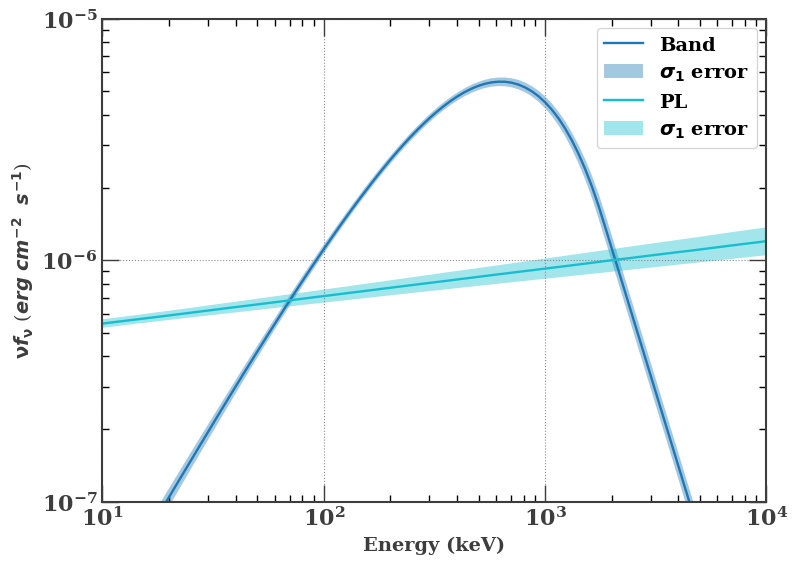}
          
    \end{subfigure}
    \caption{GRB 190114C: The count spectra (left panels) and ${\rm \nu}f_{\rm \nu}$ spectra (right panels).  The top (bottom) panels are for the joint (GBM-only) fits.}
    \label{fig_a26}
\end{figure*}


\begin{figure*}
    \centering
    \begin{subfigure}[b]{0.40\textwidth}
        \centering
        \includegraphics[width=\textwidth]{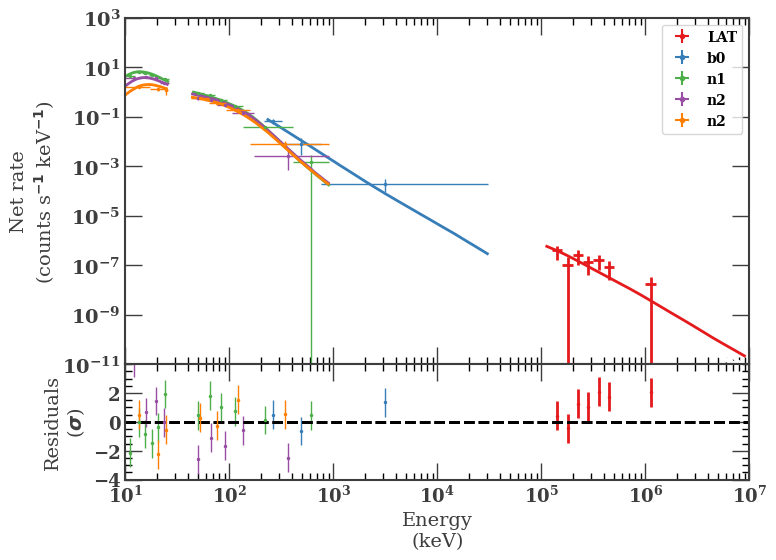}
            
    \end{subfigure}
    \hfill
    \begin{subfigure}[b]{0.40\textwidth}
        \centering
        \includegraphics[width=\textwidth]{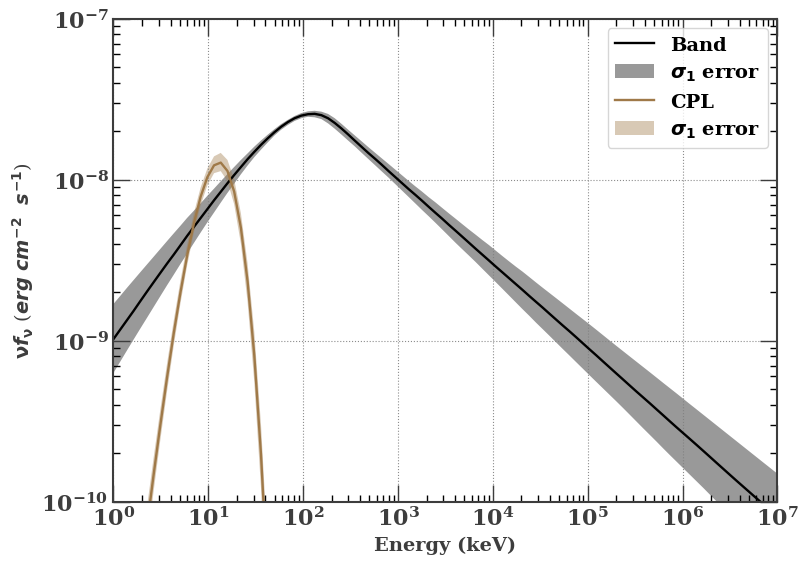}
          
    \end{subfigure}
        \hfill
    \begin{subfigure}[b]{0.40\textwidth}
        \centering
        \includegraphics[width=\textwidth]{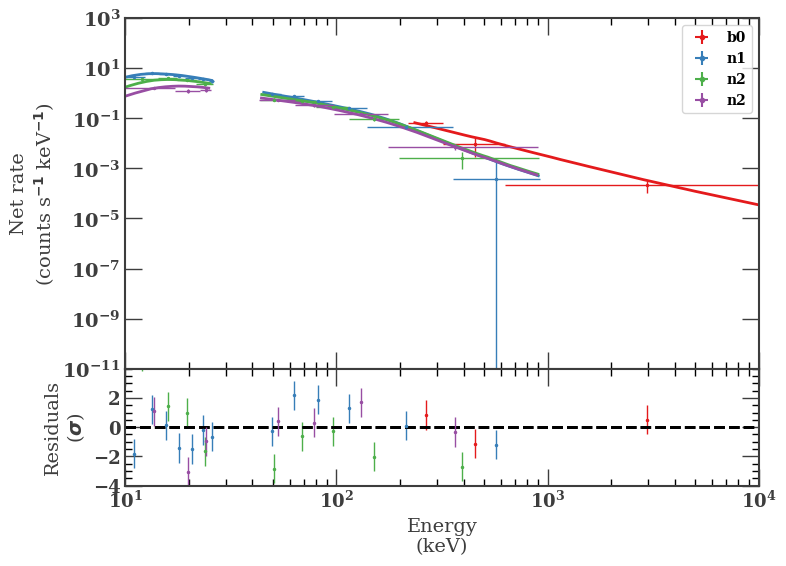}
          
    \end{subfigure}
        \hfill
    \begin{subfigure}[b]{0.40\textwidth}
        \centering
        \includegraphics[width=\textwidth]{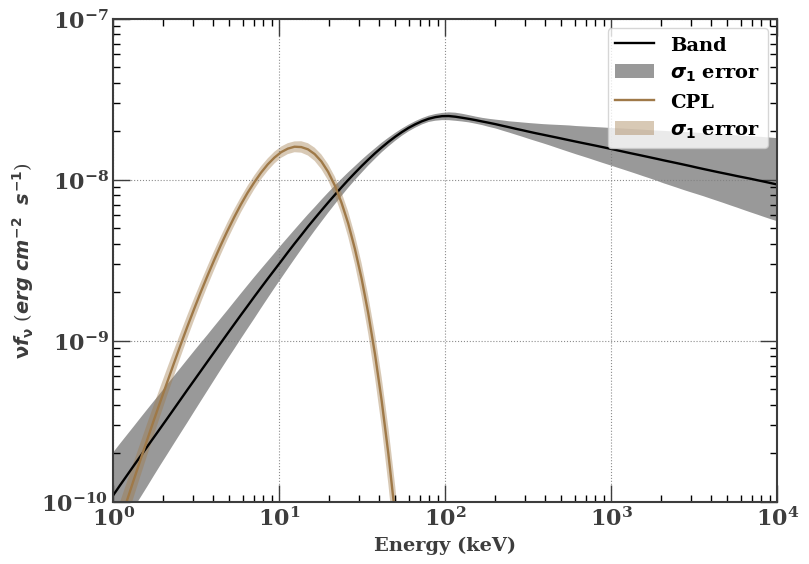}
          
    \end{subfigure}
    \caption{GRB 200613A: The count spectra (left panels) and ${\rm \nu}f_{\rm \nu}$ spectra (right panels).  The top (bottom) panels are for the joint (GBM-only) fits.}
    \label{fig_a27}
\end{figure*}


\begin{figure*}
    \centering
    \begin{subfigure}[b]{0.40\textwidth}
        \centering
        \includegraphics[width=\textwidth]{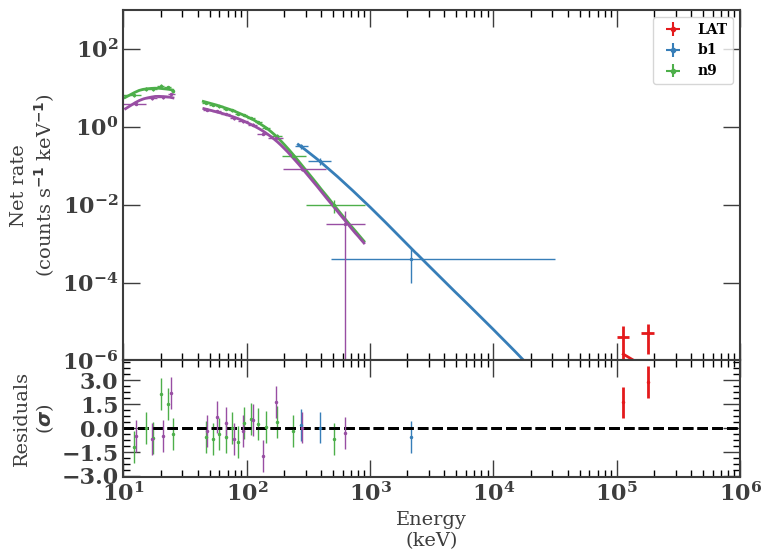}
            
    \end{subfigure}
    \hfill
    \begin{subfigure}[b]{0.40\textwidth}
        \centering
        \includegraphics[width=\textwidth]{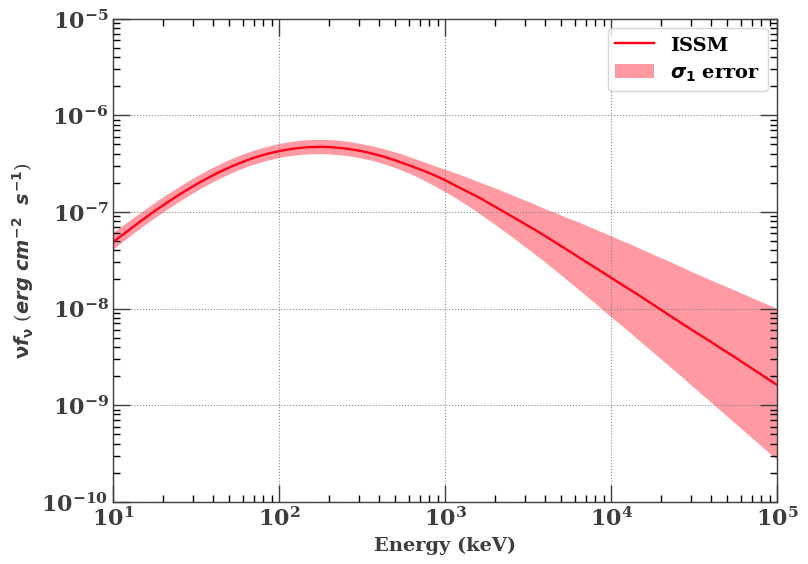}
          
    \end{subfigure}
        \hfill
    \begin{subfigure}[b]{0.40\textwidth}
        \centering
        \includegraphics[width=\textwidth]{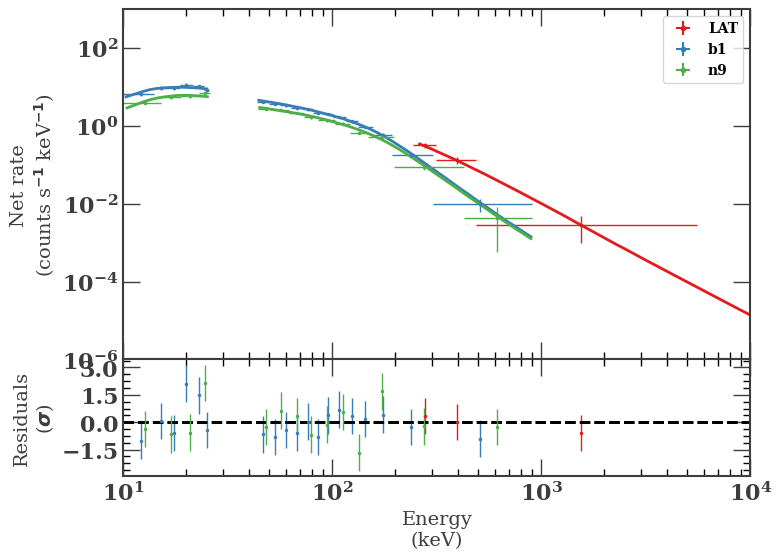}
          
    \end{subfigure}
        \hfill
    \begin{subfigure}[b]{0.40\textwidth}
        \centering
        \includegraphics[width=\textwidth]{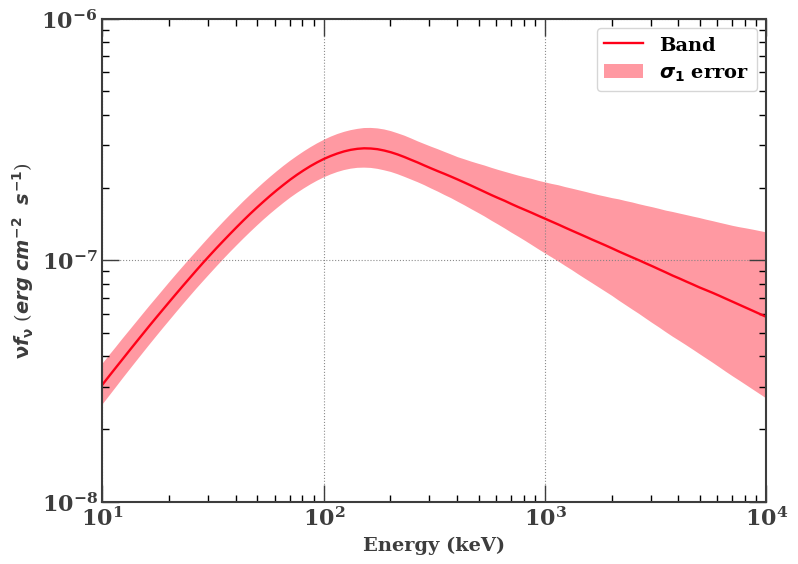}
          
    \end{subfigure}
    \caption{GRB 210826A: The count spectra (left panels) and ${\rm \nu}f_{\rm \nu}$ spectra (right panels).  The top (bottom) panels are for the joint (GBM-only) fits.}
    \label{fig_a28}
\end{figure*}


\begin{figure*}
    \centering
    \begin{subfigure}[b]{0.40\textwidth}
        \centering
        \includegraphics[width=\textwidth]{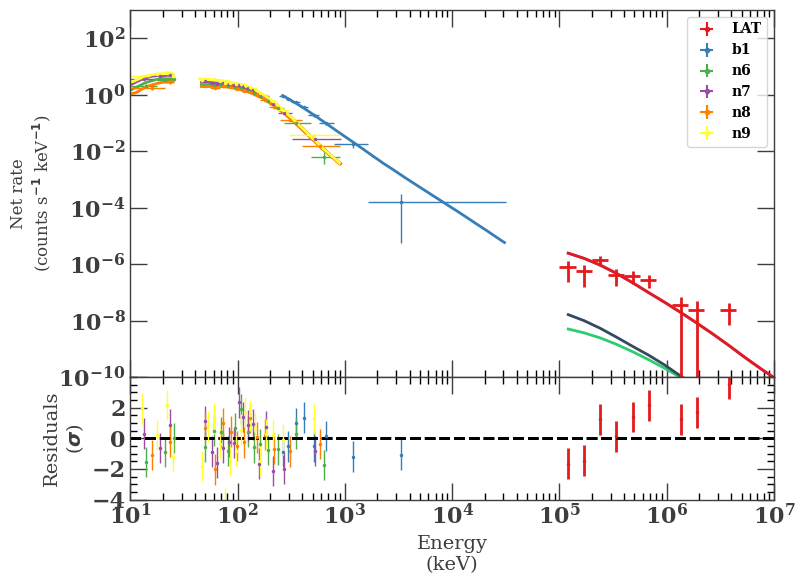}
            
    \end{subfigure}
    \hfill
    \begin{subfigure}[b]{0.40\textwidth}
        \centering
        \includegraphics[width=\textwidth]{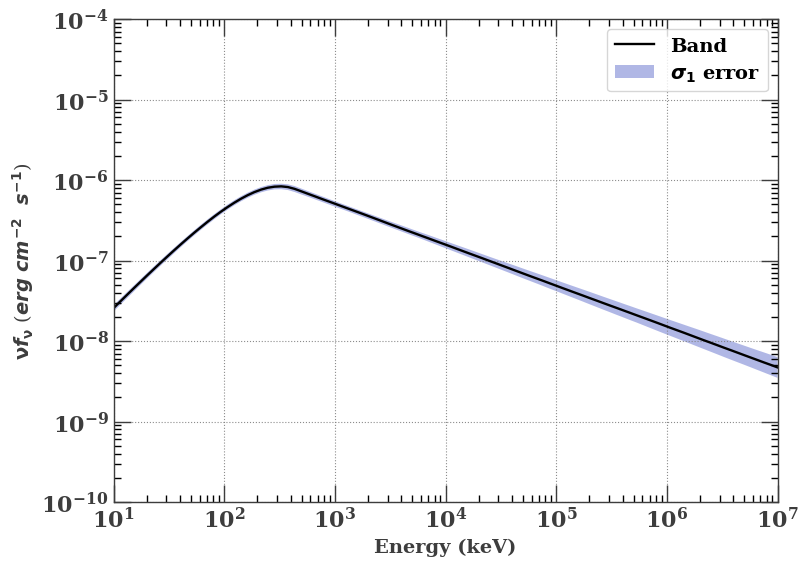}
          
    \end{subfigure}
        \hfill
    \begin{subfigure}[b]{0.40\textwidth}
        \centering
        \includegraphics[width=\textwidth]{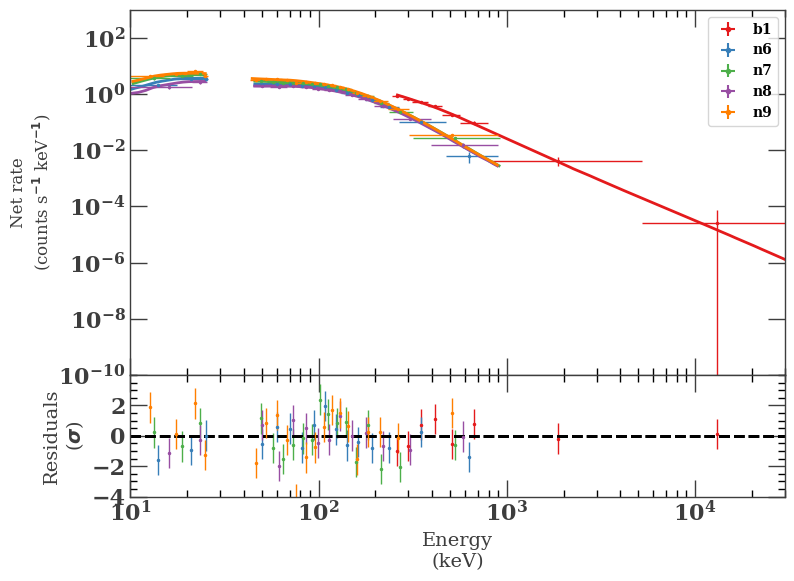}
          
    \end{subfigure}
        \hfill
    \begin{subfigure}[b]{0.40\textwidth}
        \centering
        \includegraphics[width=\textwidth]{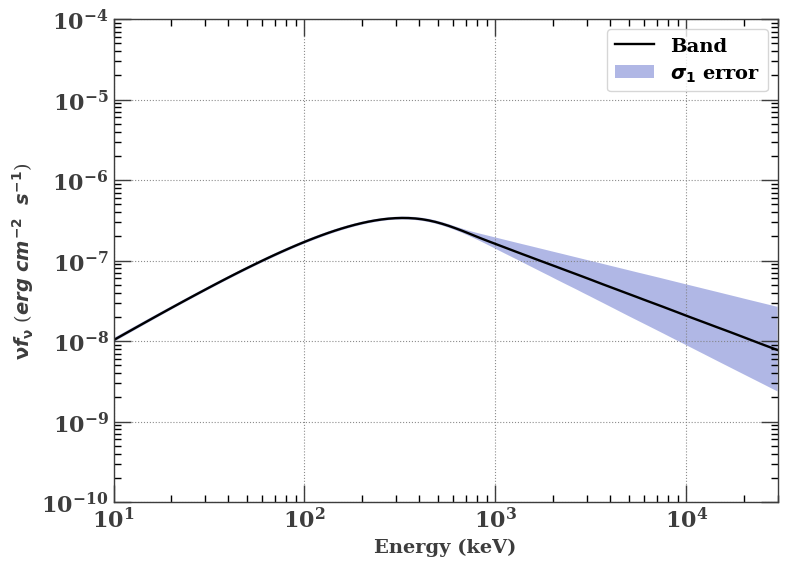}
          
    \end{subfigure}
    \caption{GRB 211018A: The count spectra (left panels) and ${\rm \nu}f_{\rm \nu}$ spectra (right panels).  The top (bottom) panels are for the joint (GBM-only) fits.}
    \label{fig_a29}
\end{figure*}


\begin{figure*}
    \centering
    \begin{subfigure}[b]{0.40\textwidth}
        \centering
        \includegraphics[width=\textwidth]{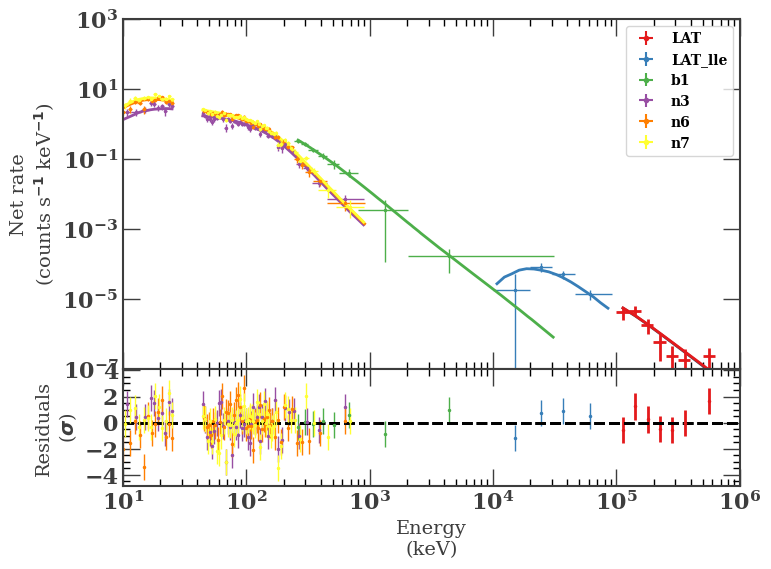}
            
    \end{subfigure}
    \hfill
    \begin{subfigure}[b]{0.40\textwidth}
        \centering
        \includegraphics[width=\textwidth]{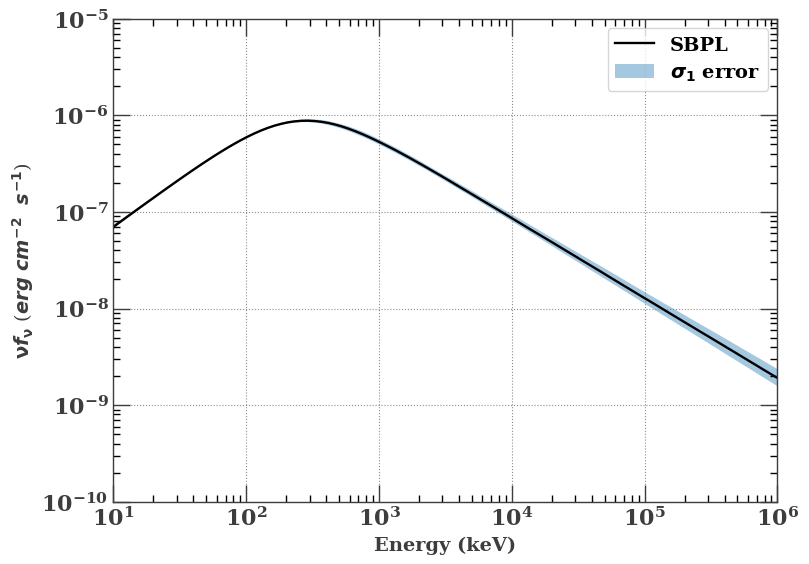}
          
    \end{subfigure}
        \hfill
    \begin{subfigure}[b]{0.40\textwidth}
        \centering
        \includegraphics[width=\textwidth]{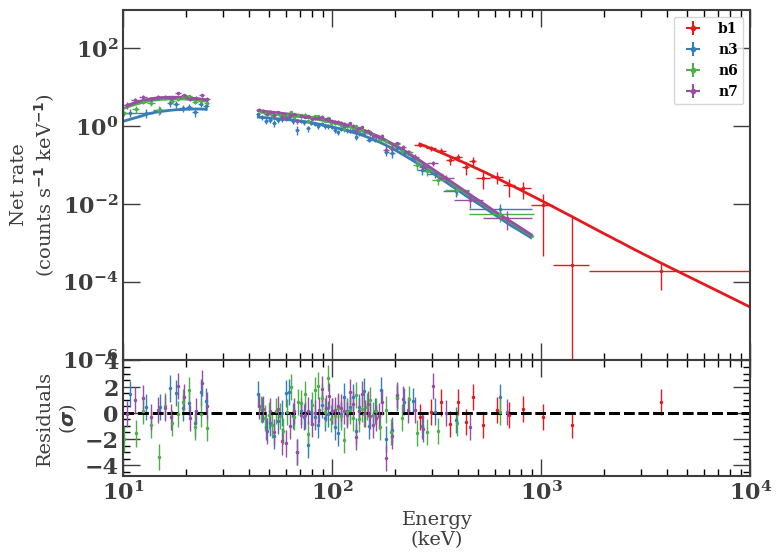}
          
    \end{subfigure}
        \hfill
    \begin{subfigure}[b]{0.40\textwidth}
        \centering
        \includegraphics[width=\textwidth]{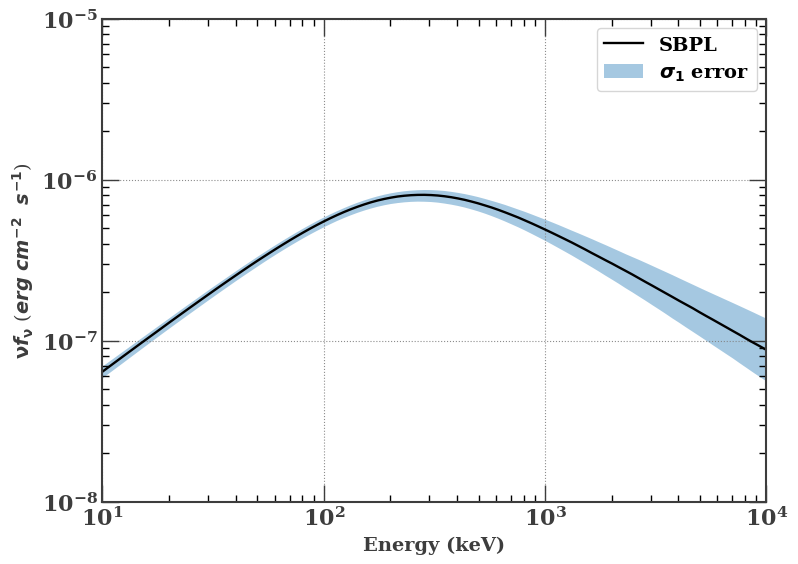}
          
    \end{subfigure}
    \caption{GRB 220101A: The count spectra (left panels) and ${\rm \nu}f_{\rm \nu}$ spectra (right panels).  The top (bottom) panels are for the joint (GBM-only) fits.}
    \label{fig_a_a30}
\end{figure*}


\begin{figure*}
    \centering
    \begin{subfigure}[b]{0.40\textwidth}
        \centering
        \includegraphics[width=\textwidth]{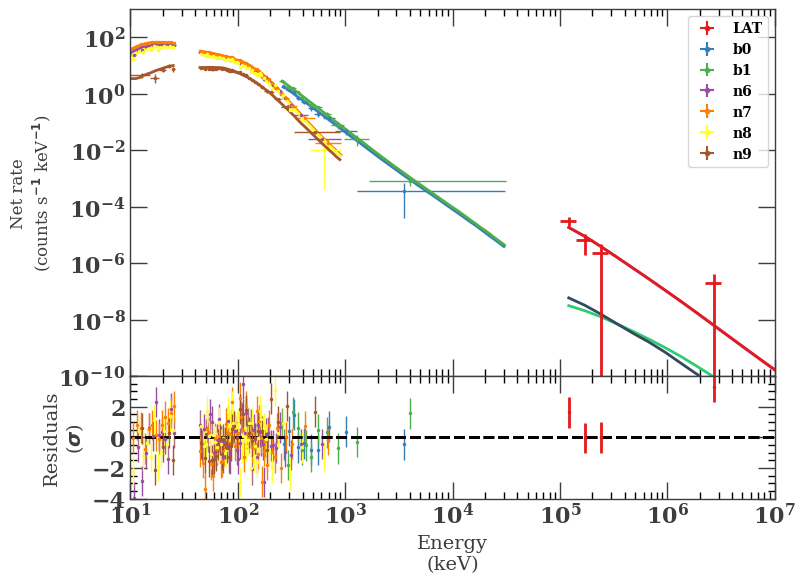}
            
    \end{subfigure}
    \hfill
    \begin{subfigure}[b]{0.40\textwidth}
        \centering
        \includegraphics[width=\textwidth]{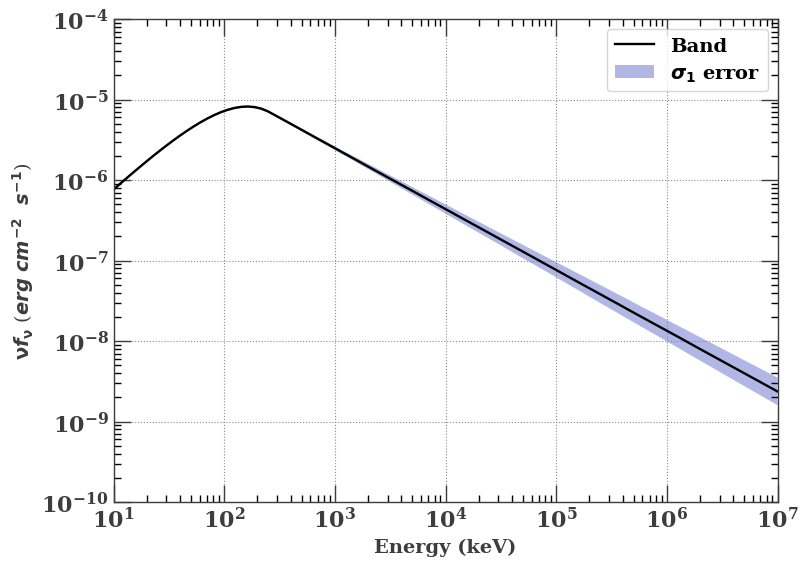}
          
    \end{subfigure}
        \hfill
    \begin{subfigure}[b]{0.40\textwidth}
        \centering
        \includegraphics[width=\textwidth]{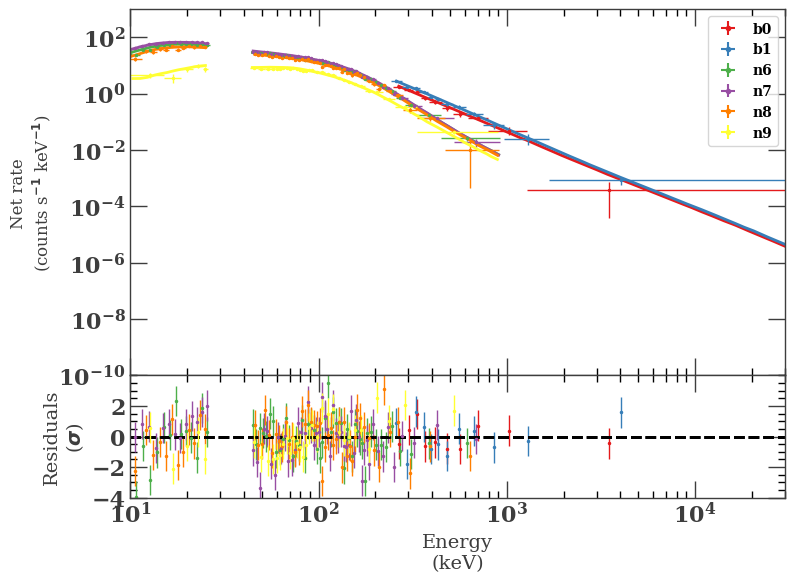}
          
    \end{subfigure}
        \hfill
    \begin{subfigure}[b]{0.40\textwidth}
        \centering
        \includegraphics[width=\textwidth]{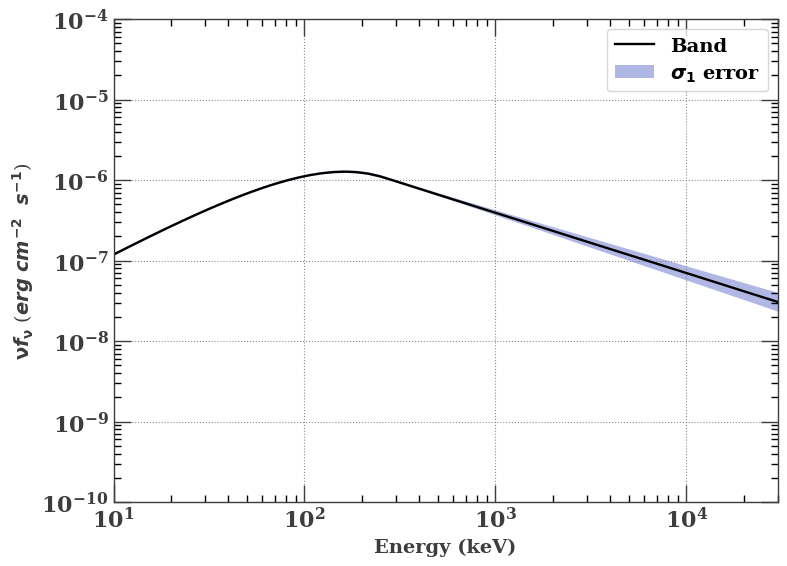}
          
    \end{subfigure}
    \caption{GRB 220527A: The count spectra (left panels) and ${\rm \nu}f_{\rm \nu}$ spectra (right panels).  The top (bottom) panels are for the joint (GBM-only) fits.}
    \label{fig_a31}
\end{figure*}

\begin{figure*}
    \centering
    \begin{subfigure}[b]{0.40\textwidth}
        \centering
        \includegraphics[width=\textwidth]{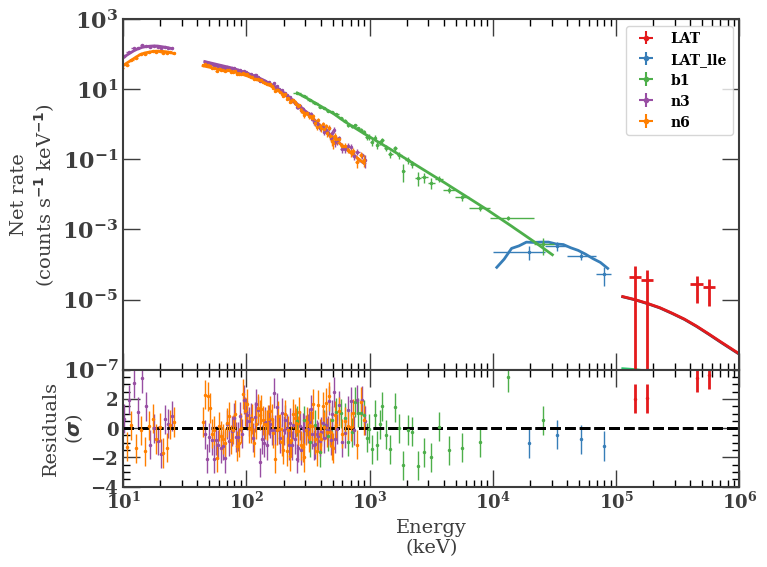}
            
    \end{subfigure}
    \hfill
    \begin{subfigure}[b]{0.40\textwidth}
        \centering
        \includegraphics[width=\textwidth]{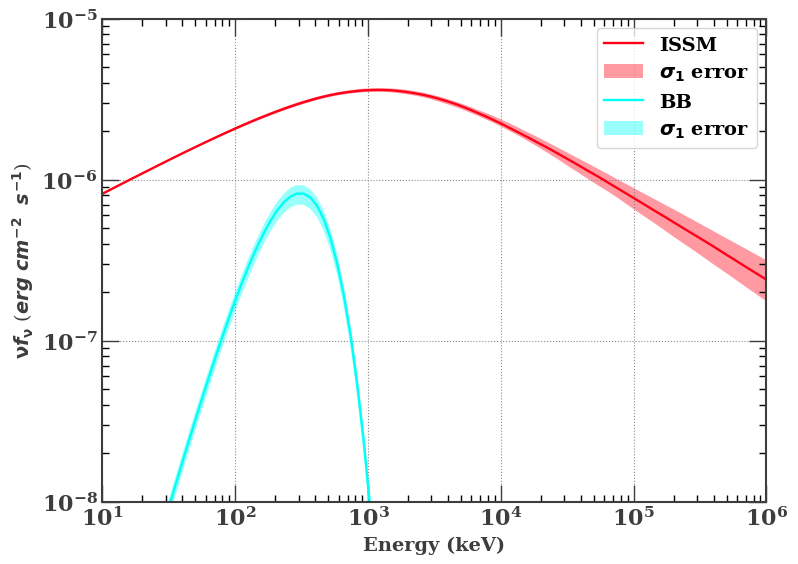}
          
    \end{subfigure}
        \hfill
    \begin{subfigure}[b]{0.40\textwidth}
        \centering
        \includegraphics[width=\textwidth]{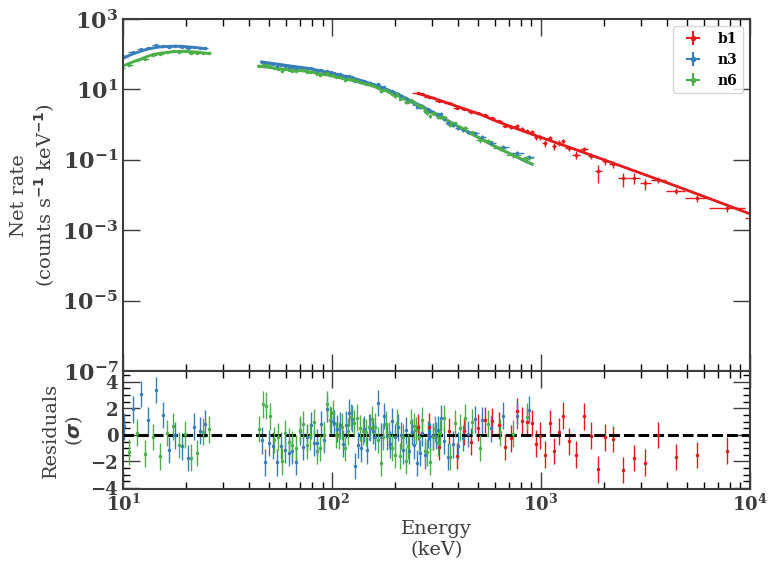}
          
    \end{subfigure}
        \hfill
    \begin{subfigure}[b]{0.40\textwidth}
        \centering
        \includegraphics[width=\textwidth]{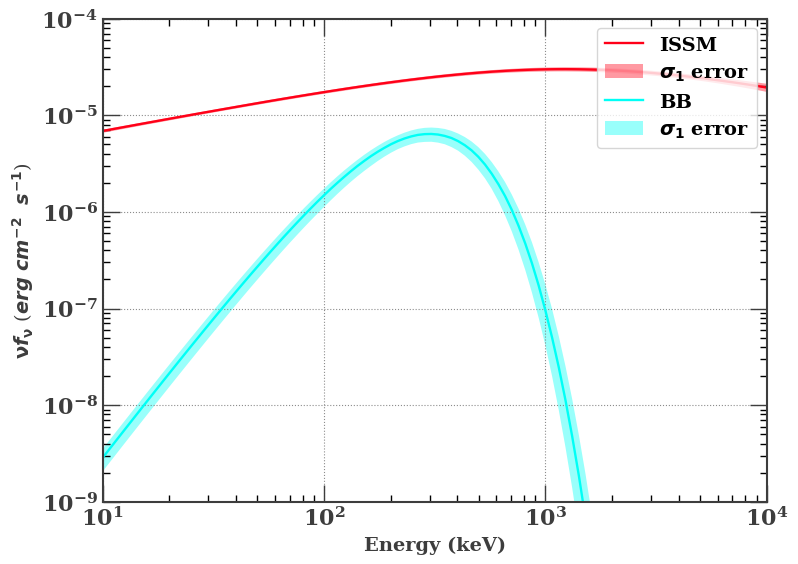}
          
    \end{subfigure}
    \caption{GRB 221009A: The count spectra (left panels) and ${\rm \nu}f_{\rm \nu}$ spectra (right panels).  The top (bottom) panels are for the joint (GBM-only) fits.}
    \label{fig24_2}
\end{figure*}

\begin{figure*}
    \centering
    \begin{subfigure}[b]{0.40\textwidth}
        \centering
        \includegraphics[width=\textwidth]{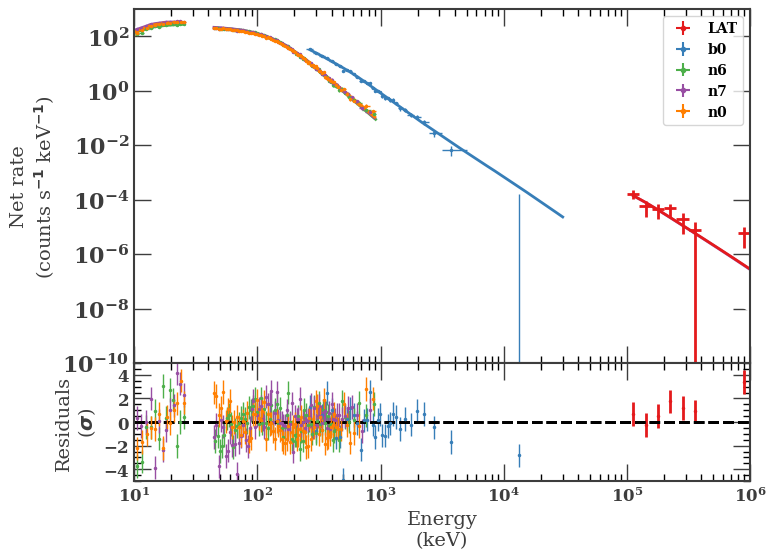}
            
    \end{subfigure}
    \hfill
    \begin{subfigure}[b]{0.40\textwidth}
        \centering
        \includegraphics[width=\textwidth]{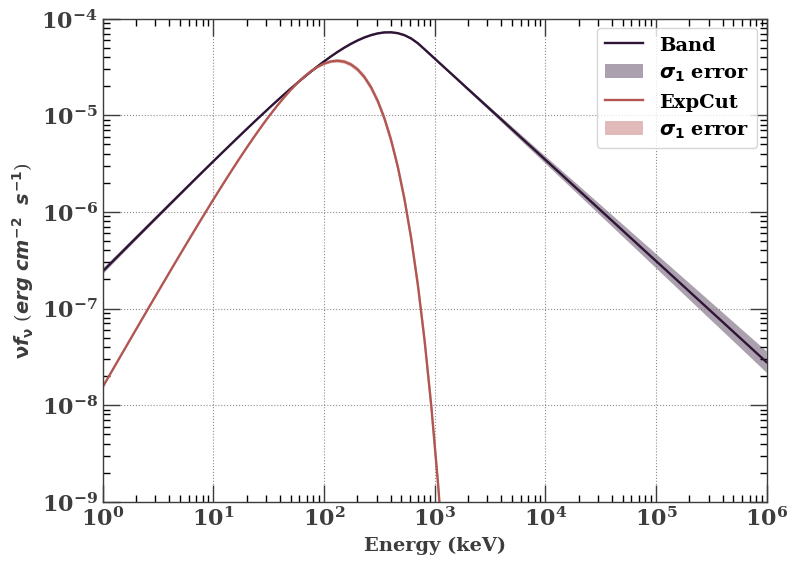}
          
    \end{subfigure}
        \hfill
    \begin{subfigure}[b]{0.40\textwidth}
        \centering
        \includegraphics[width=\textwidth]{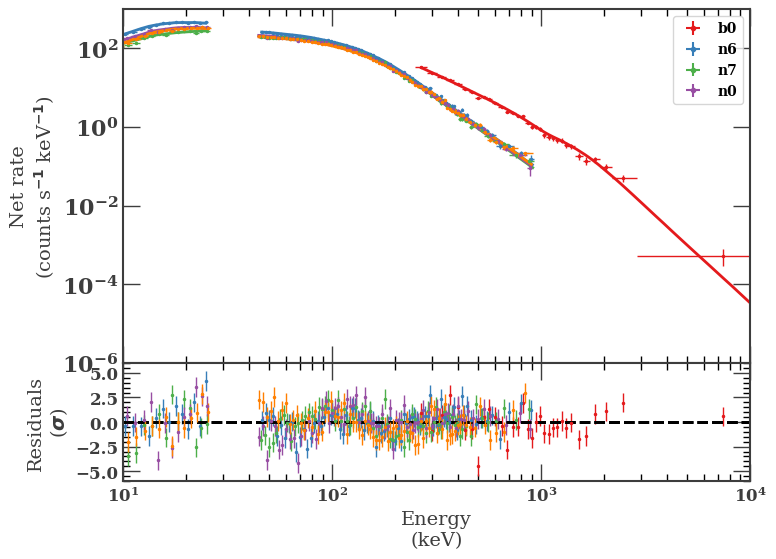}
          
    \end{subfigure}
        \hfill
    \begin{subfigure}[b]{0.40\textwidth}
        \centering
        \includegraphics[width=\textwidth]{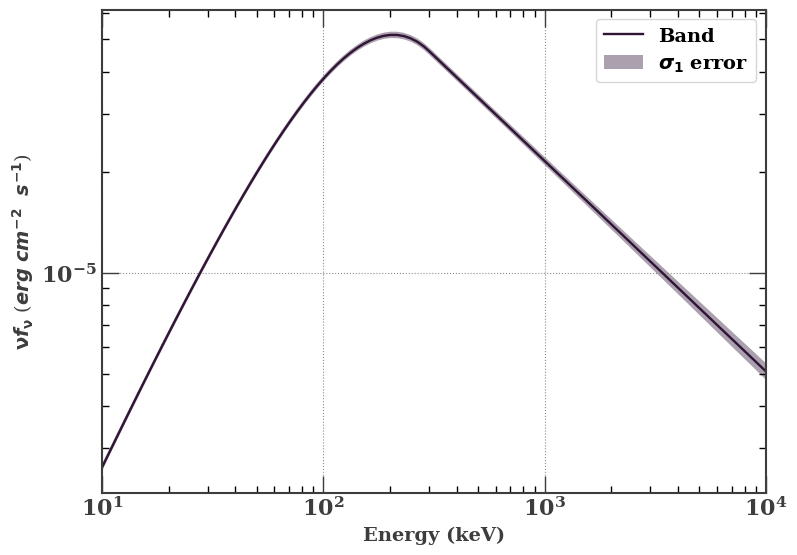}
          
    \end{subfigure}
    \caption{GRB 230812B: The count spectra (left panels) and ${\rm \nu}f_{\rm \nu}$ spectra (right panels).  The top (bottom) panels are for the joint (GBM-only) fits.}
    \label{fig_a32}
\end{figure*}

\begin{figure*}
    \centering
    \begin{subfigure}[b]{0.40\textwidth}
        \centering
        \includegraphics[width=\textwidth]{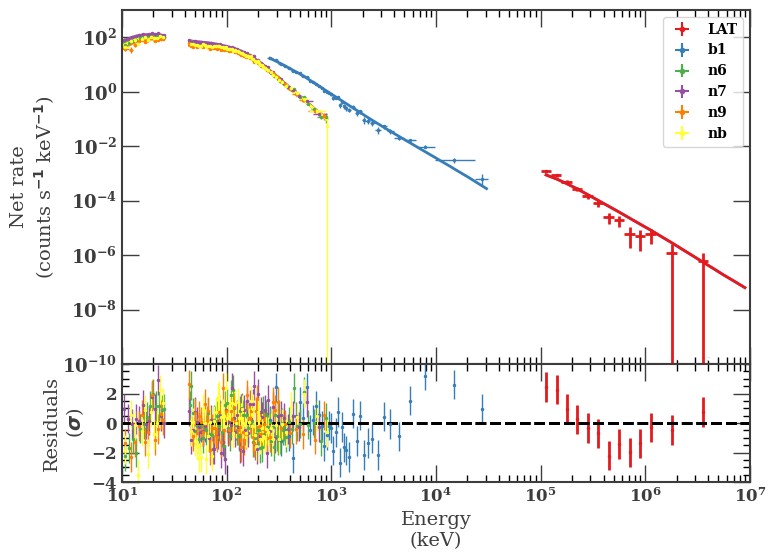}
            
    \end{subfigure}
    \hfill
    \begin{subfigure}[b]{0.40\textwidth}
        \centering
        \includegraphics[width=\textwidth]{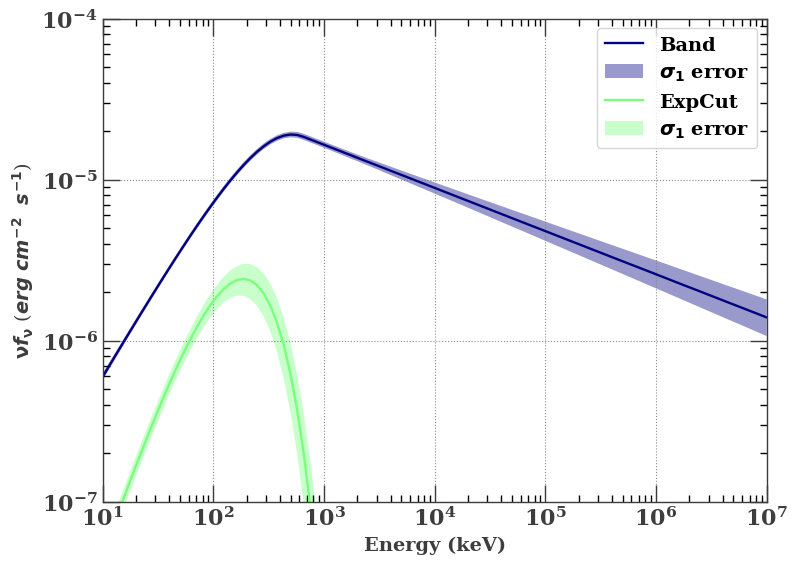}
          
    \end{subfigure}
        \hfill
    \begin{subfigure}[b]{0.40\textwidth}
        \centering
        \includegraphics[width=\textwidth]{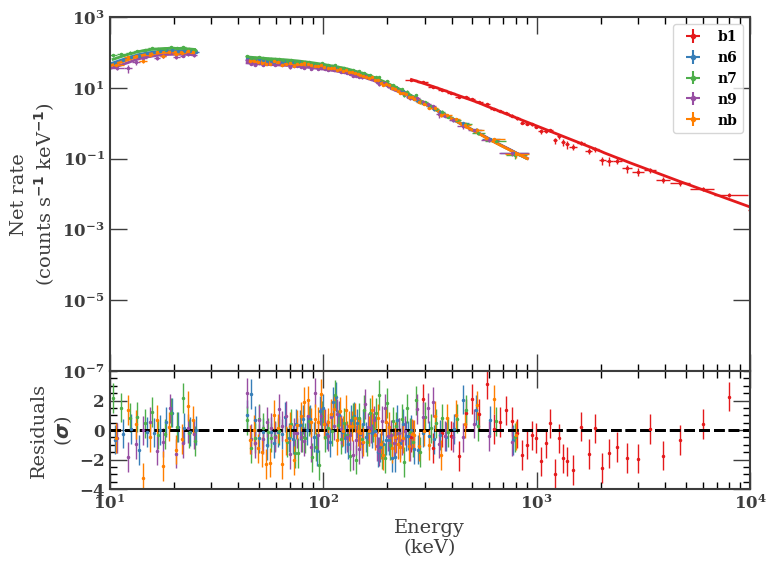}
          
    \end{subfigure}
        \hfill
    \begin{subfigure}[b]{0.40\textwidth}
        \centering
        \includegraphics[width=\textwidth]{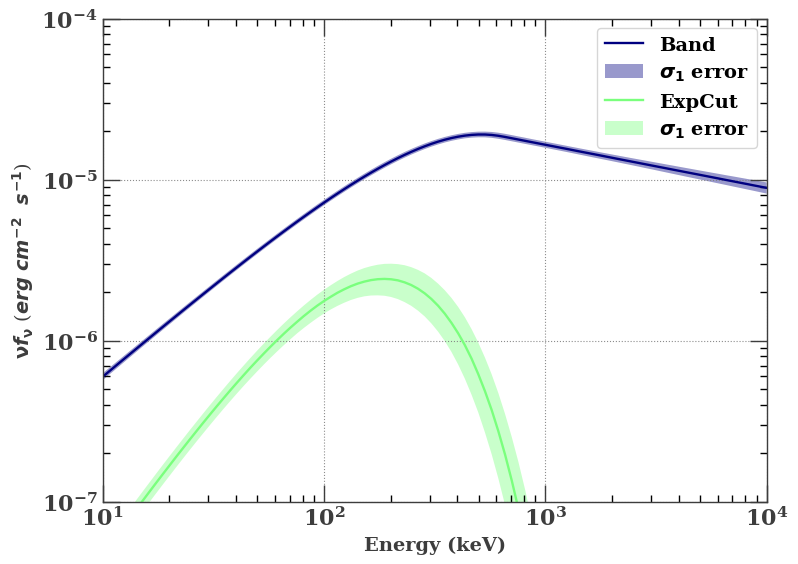}
          
    \end{subfigure}
    \caption{GRB\,240825A: The count spectra (left panels) and ${\rm \nu}f_{\rm \nu}$ spectra (right panels).  The top (bottom) panels are for the joint (GBM-only) fits.}
    \label{fig_a33}
\end{figure*}

\begin{figure*}
    \centering
    \begin{subfigure}[b]{0.40\textwidth}
        \centering
        \includegraphics[width=\textwidth]{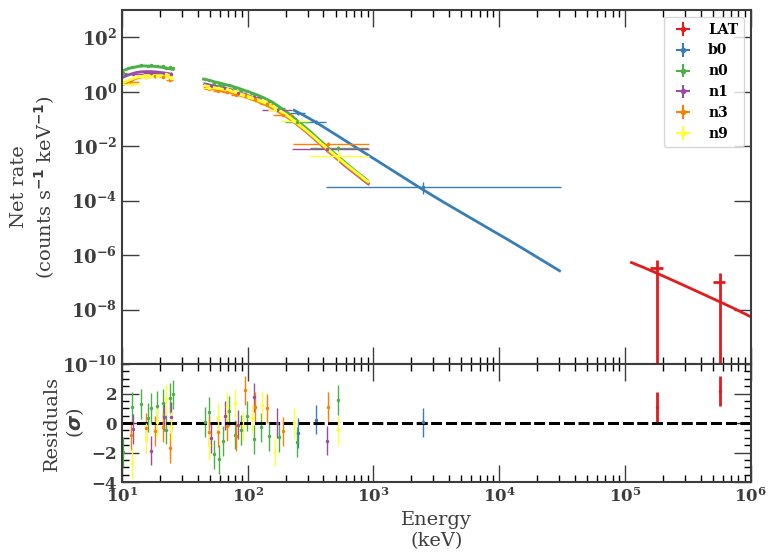}
            
    \end{subfigure}
    \hfill
    \begin{subfigure}[b]{0.40\textwidth}
        \centering
        \includegraphics[width=\textwidth]{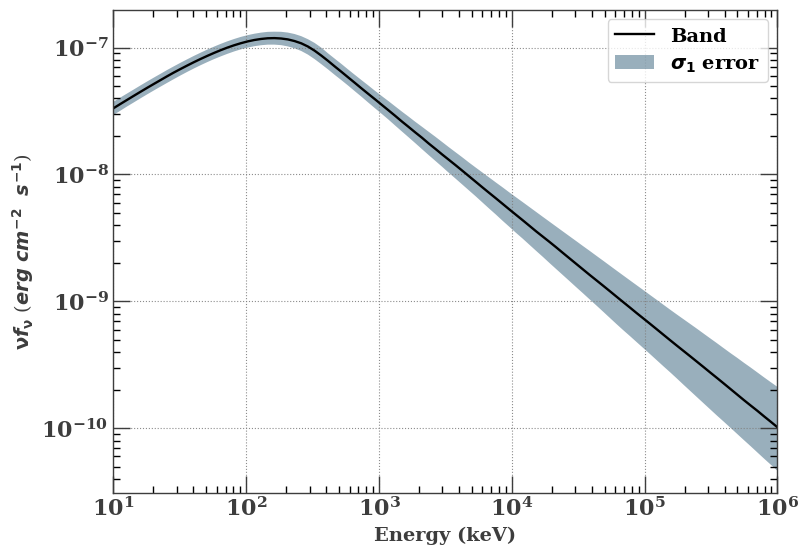}
          
    \end{subfigure}
        \hfill
    \begin{subfigure}[b]{0.40\textwidth}
        \centering
        \includegraphics[width=\textwidth]{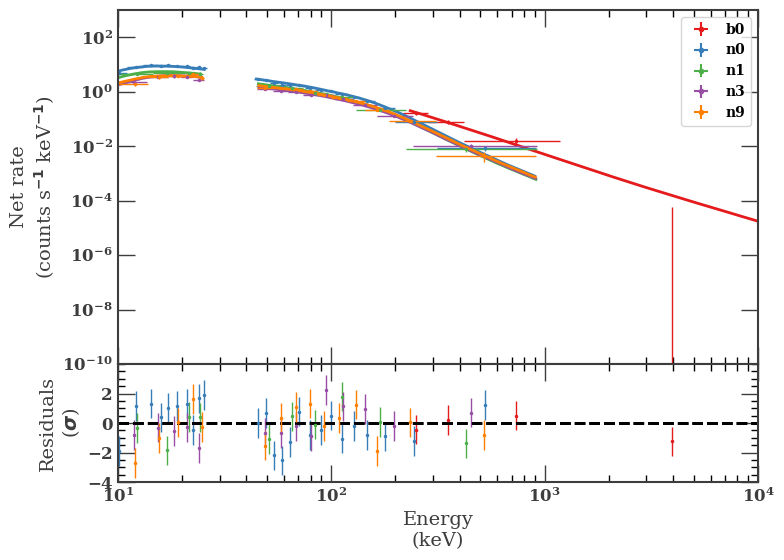}
          
    \end{subfigure}
        \hfill
    \begin{subfigure}[b]{0.40\textwidth}
        \centering
        \includegraphics[width=\textwidth]{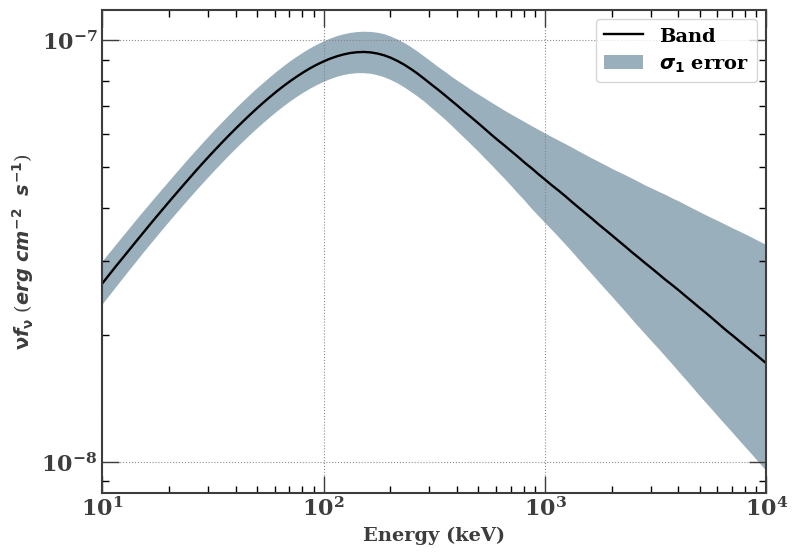}
          
    \end{subfigure}
    \caption{GRB\,241030A: The count spectra (left panels) and ${\rm \nu}f_{\rm \nu}$ spectra (right panels).  The top (bottom) panels are for the joint (GBM-only) fits.}
    \label{fig_a34}
\end{figure*}

\onecolumn
\begin{figure*}
    \centering
    \begin{subfigure}[b]{0.40\textwidth}
        \centering
        \includegraphics[width=\textwidth]{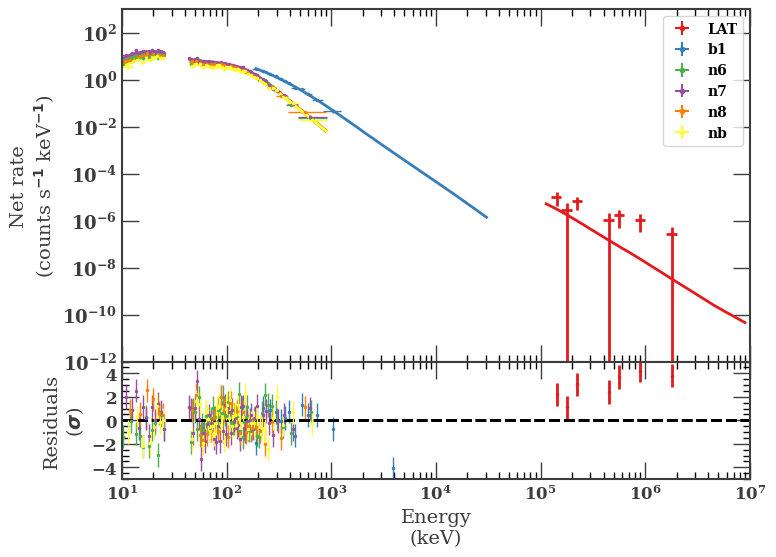}

    \end{subfigure}
    \hfill
    \begin{subfigure}[b]{0.40\textwidth}
        \centering
        \includegraphics[width=\textwidth]{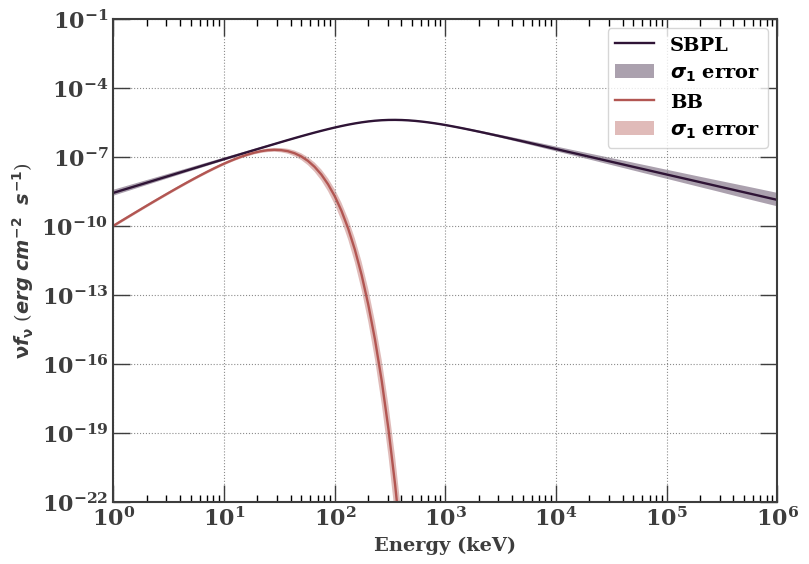}
          
    \end{subfigure}
        \hfill
    \begin{subfigure}[b]{0.40\textwidth}
        \centering
        \includegraphics[width=\textwidth]{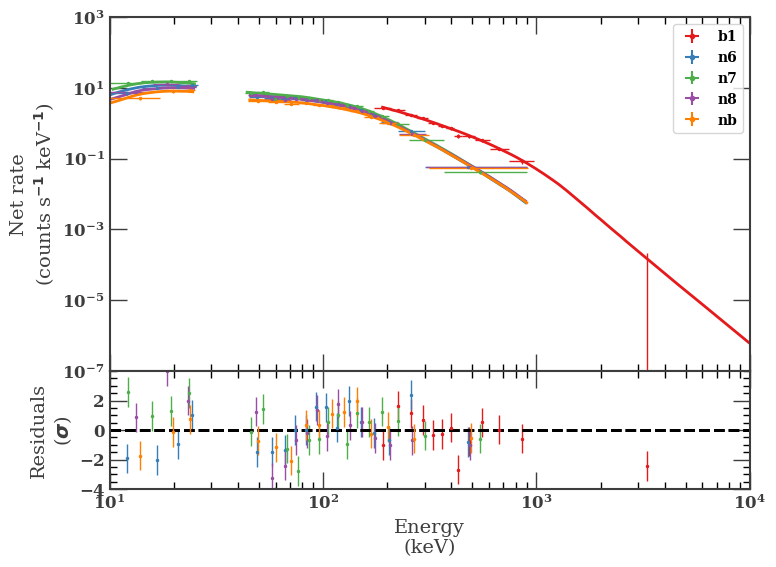}

    \end{subfigure}
        \hfill
    \begin{subfigure}[b]{0.40\textwidth}
        \centering
        \includegraphics[width=\textwidth]{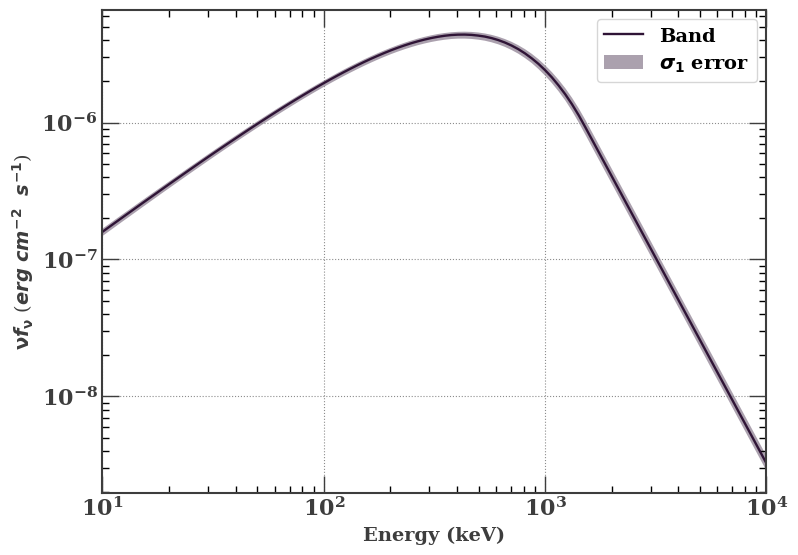}
          
    \end{subfigure}
    \caption{GRB\,241228B: The count spectra (left panels) and ${\rm \nu}f_{\rm \nu}$ spectra (right panels).  The top (bottom) panels are for the joint (GBM-only) fits.}
    \label{fig_a35}
\end{figure*}


\setcounter{section}{3}

\renewcommand{\thefigure}{\thesection\arabic{figure}}
\setcounter{figure}{0}

\clearpage
\begin{figure*}
    \centering
    \begin{subfigure}[b]{0.40\textwidth}
        \centering
        \includegraphics[width=\textwidth]{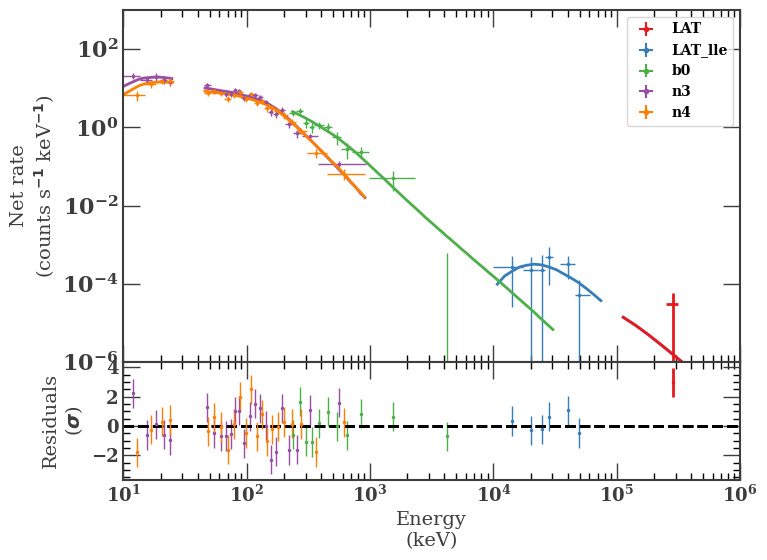}
            
    \end{subfigure}
    \hfill
    \begin{subfigure}[b]{0.40\textwidth}
        \centering
        \includegraphics[width=\textwidth]{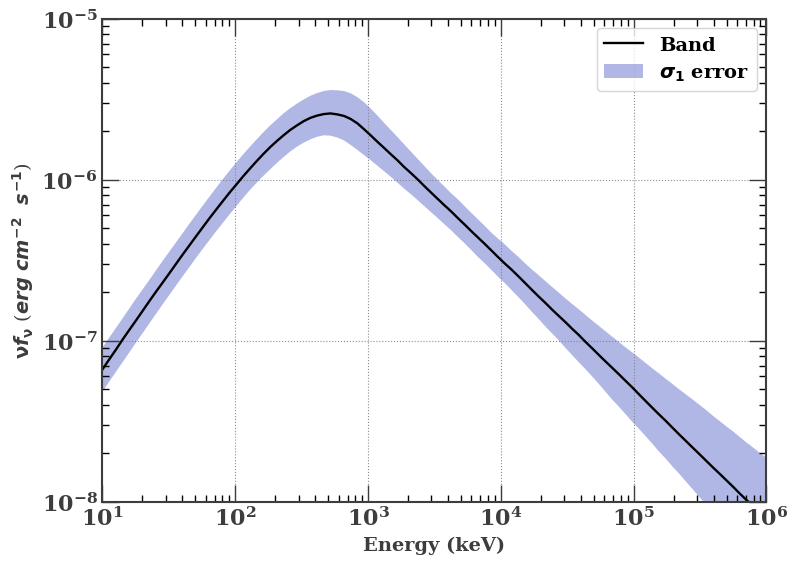}
          
    \end{subfigure}
        \hfill
    \begin{subfigure}[b]{0.40\textwidth}
        \centering
        \includegraphics[width=\textwidth]{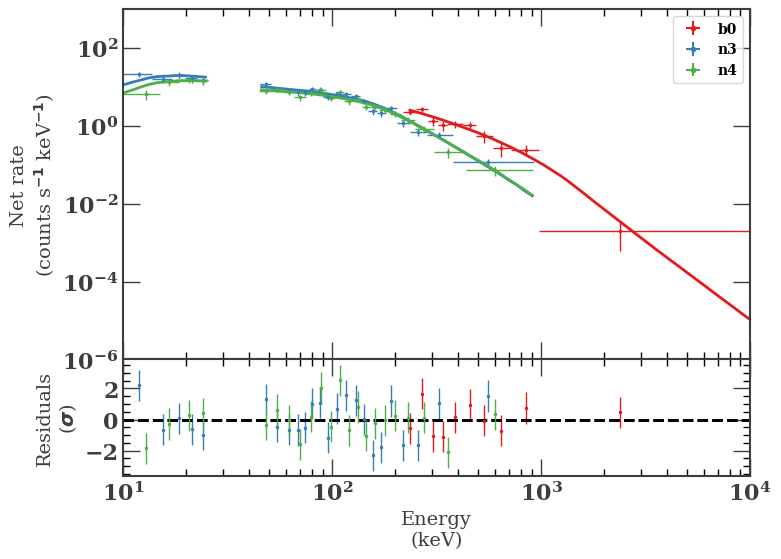}
          
    \end{subfigure}
        \hfill
    \begin{subfigure}[b]{0.40\textwidth}
        \centering
        \includegraphics[width=\textwidth]{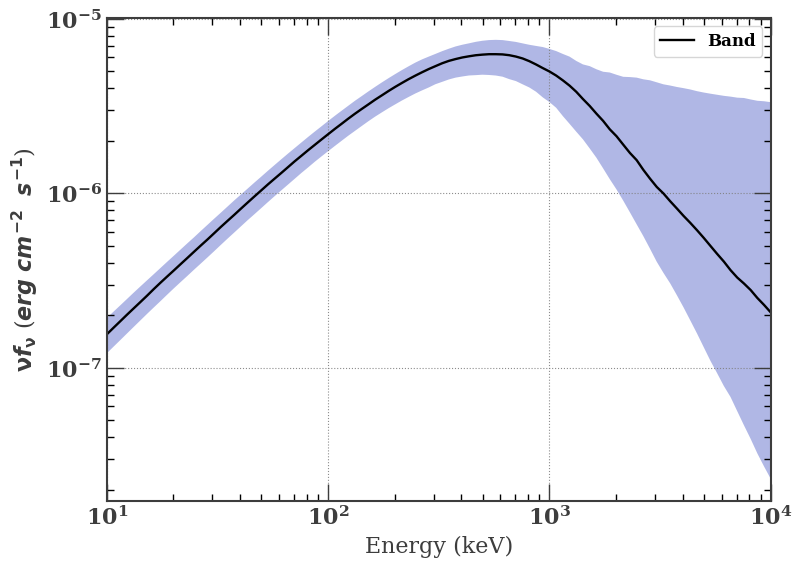}
          
    \end{subfigure}
    \caption{GRB 080916C: The count spectra (left panels) and ${\rm \nu}f_{\rm \nu}$ spectra (right panels).  The top (bottom) panels are for the joint (GBM-only) fits.}
    \label{fig_b1}
\end{figure*}


\begin{figure*}
    \centering
    \begin{subfigure}[b]{0.40\textwidth}
        \centering
        \includegraphics[width=\textwidth]{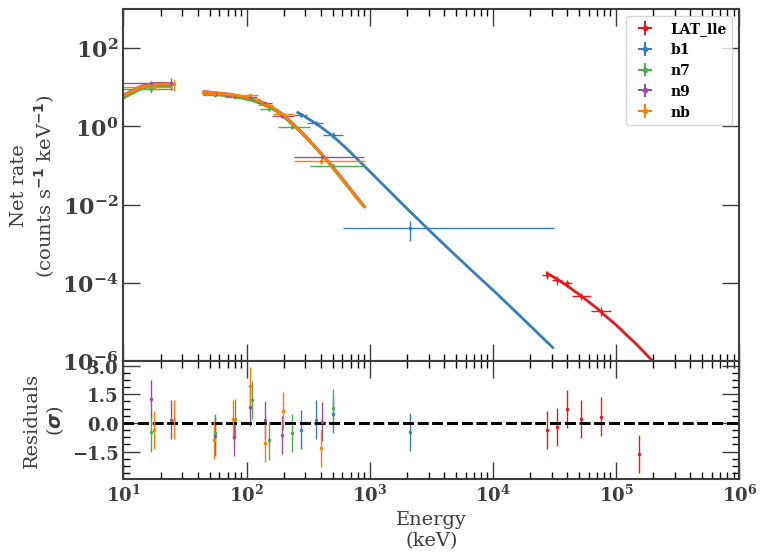}
            
    \end{subfigure}
    \hfill
    \begin{subfigure}[b]{0.40\textwidth}
        \centering
        \includegraphics[width=\textwidth]{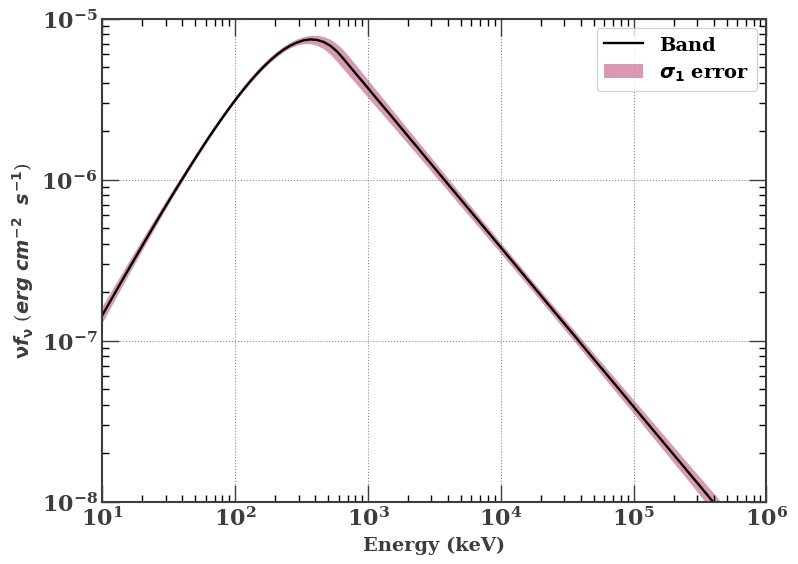}
          
    \end{subfigure}
        \hfill
    \begin{subfigure}[b]{0.40\textwidth}
        \centering
        \includegraphics[width=\textwidth]{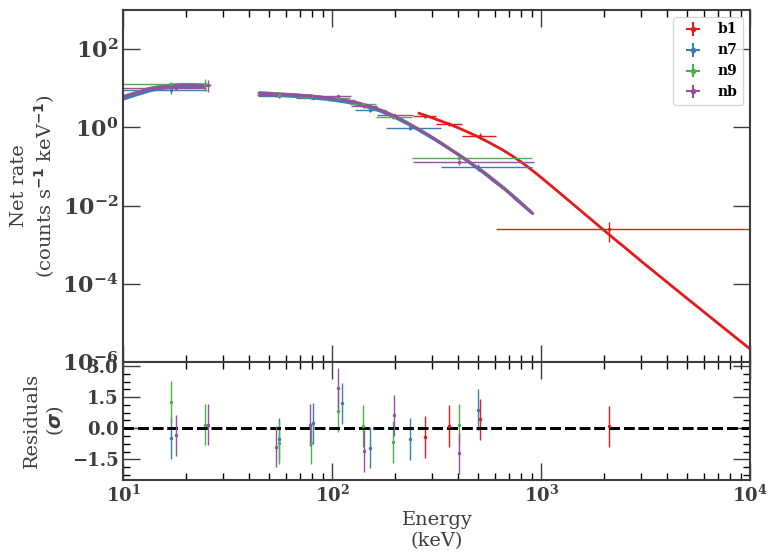}
          
    \end{subfigure}
        \hfill
    \begin{subfigure}[b]{0.40\textwidth}
        \centering
        \includegraphics[width=\textwidth]{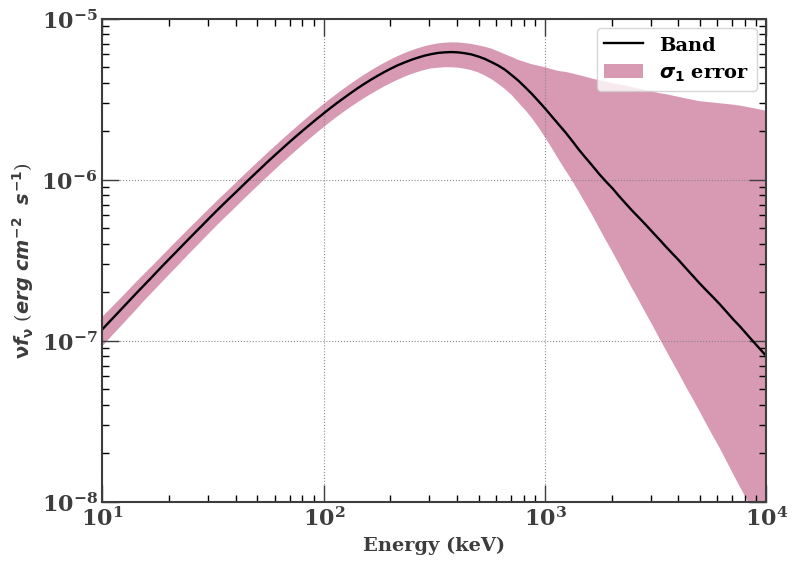}
          
    \end{subfigure}
    \caption{GRB 090323: The count spectra (left panels) and ${\rm \nu}f_{\rm \nu}$ spectra (right panels).  The top (bottom) panels are for the joint (GBM-only) fits.}
    \label{fig_b2}
\end{figure*}


\begin{figure*}
    \centering
    \begin{subfigure}[b]{0.40\textwidth}
        \centering
        \includegraphics[width=\textwidth]{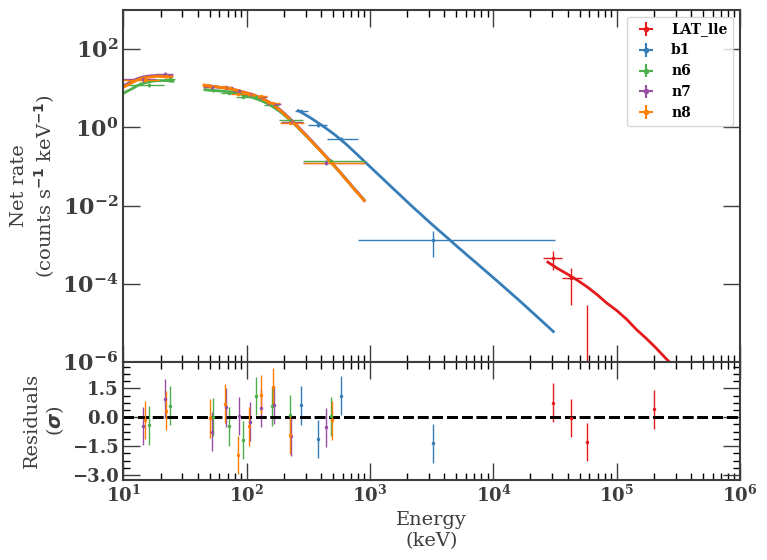}
            
    \end{subfigure}
    \hfill
    \begin{subfigure}[b]{0.40\textwidth}
        \centering
        \includegraphics[width=\textwidth]{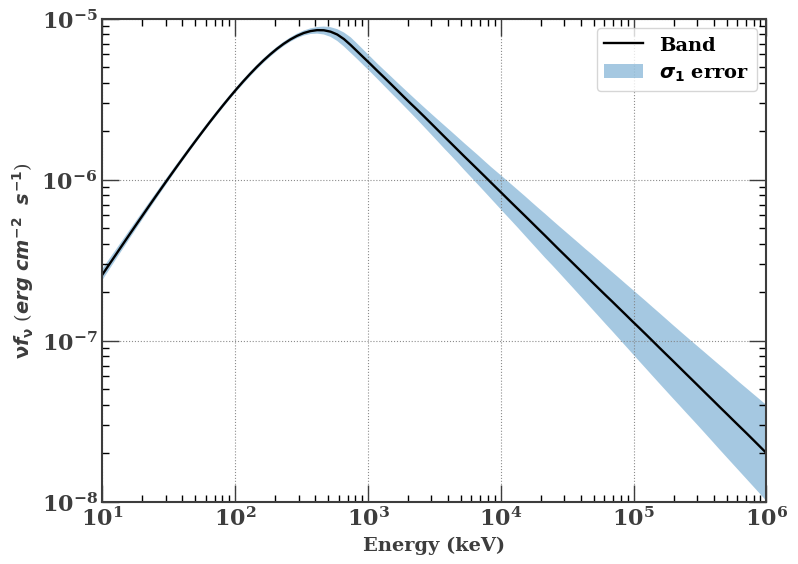}
          
    \end{subfigure}
        \hfill
    \begin{subfigure}[b]{0.40\textwidth}
        \centering
        \includegraphics[width=\textwidth]{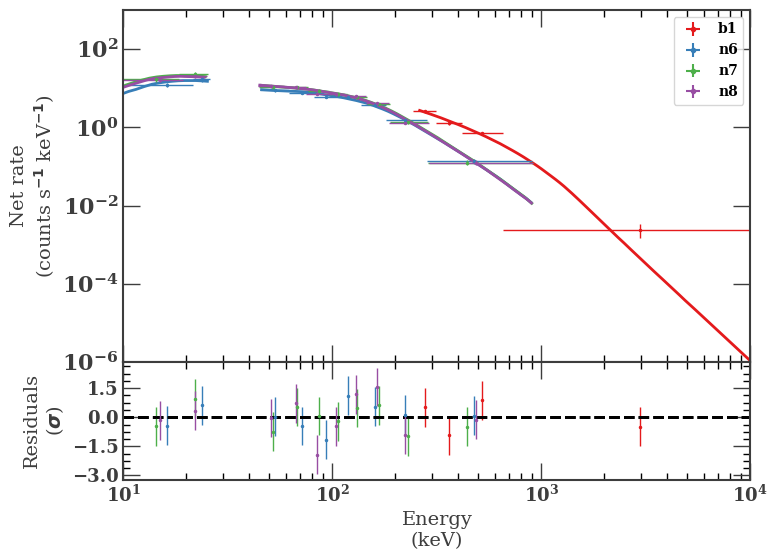}
          
    \end{subfigure}
        \hfill
    \begin{subfigure}[b]{0.40\textwidth}
        \centering
        \includegraphics[width=\textwidth]{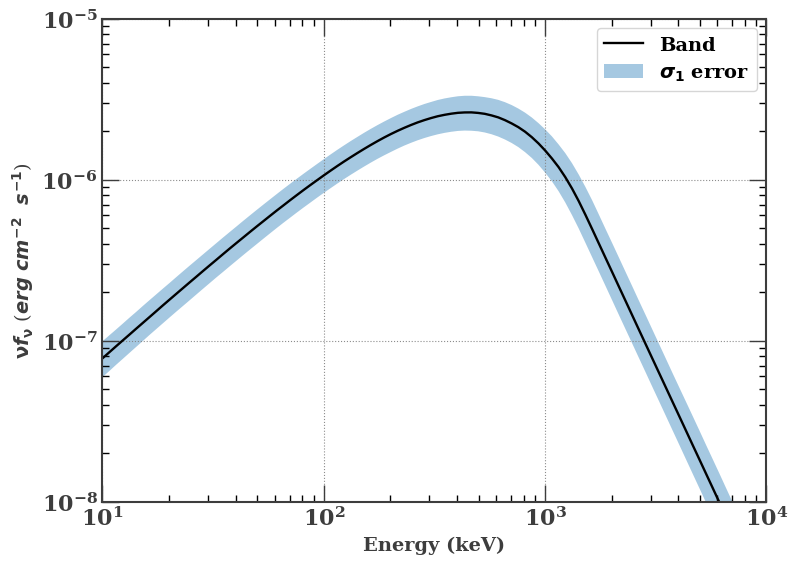}
          
    \end{subfigure}
    \caption{GRB 090328: The count spectra (left panels) and ${\rm \nu}f_{\rm \nu}$ spectra (right panels).  The top (bottom) panels are for the joint (GBM-only) fits.}
    \label{fig_b3}
\end{figure*}


\begin{figure*}
    \centering
    \begin{subfigure}[b]{0.40\textwidth}
        \centering
        \includegraphics[width=\textwidth]{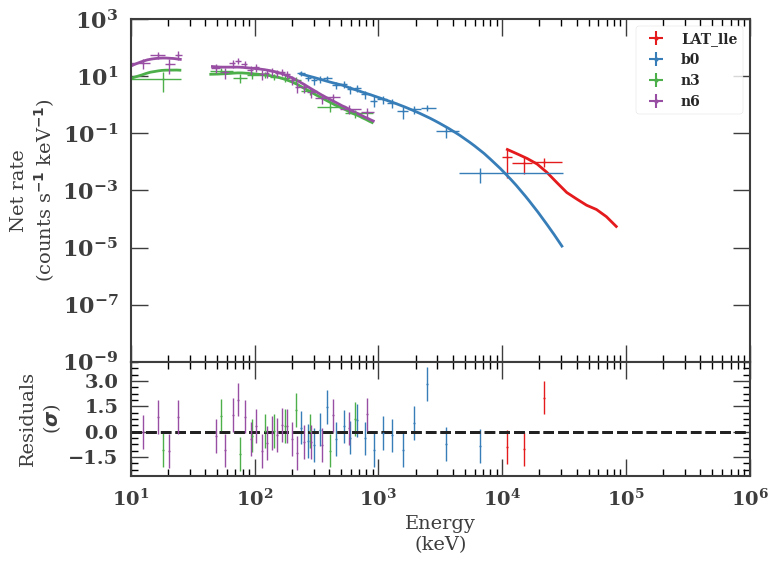}
            
    \end{subfigure}
    \hfill
    \begin{subfigure}[b]{0.40\textwidth}
        \centering
        \includegraphics[width=\textwidth]{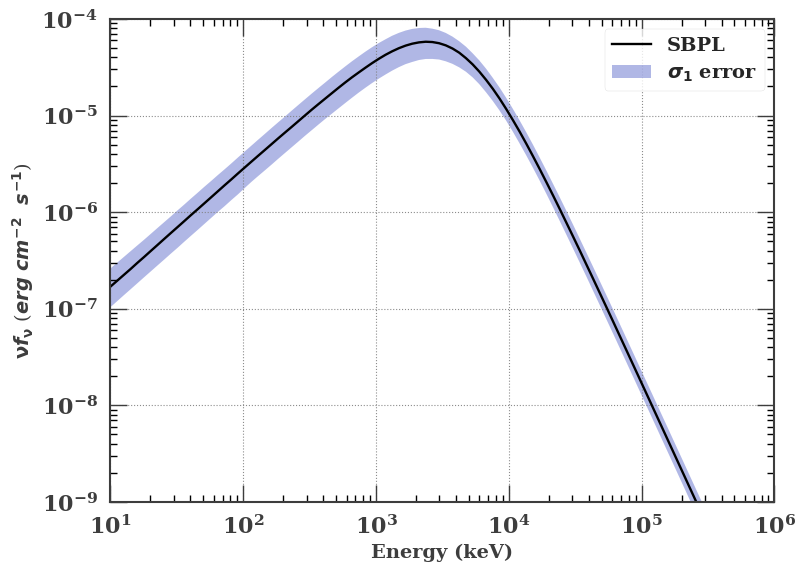}
          
    \end{subfigure}
        \hfill
    \begin{subfigure}[b]{0.40\textwidth}
        \centering
        \includegraphics[width=\textwidth]{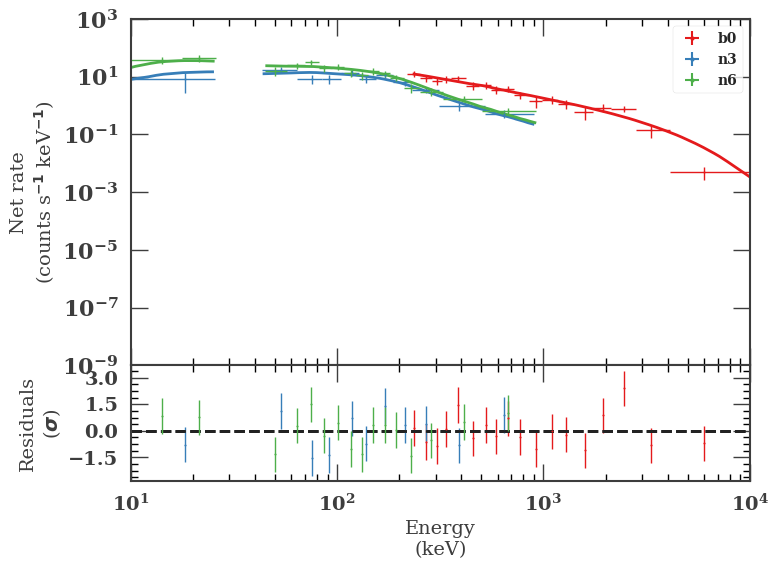}
          
    \end{subfigure}
        \hfill
    \begin{subfigure}[b]{0.40\textwidth}
        \centering
        \includegraphics[width=\textwidth]{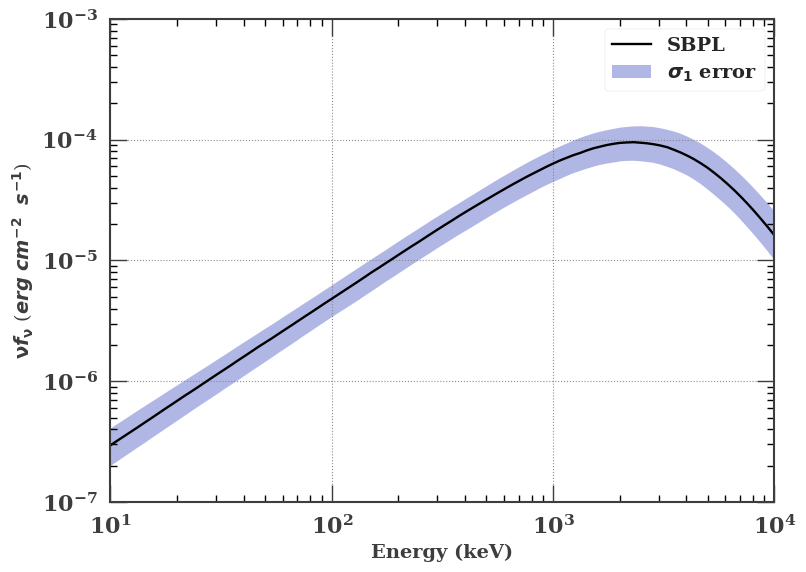}
          
    \end{subfigure}
    \caption{GRB 090510: \textbf{SGRB} The count spectra (left panels) and ${\rm \nu}f_{\rm \nu}$ spectra (right panels).  The top (bottom) panels are for the joint (GBM-only) fits.}
    \label{fig_b3_1}
\end{figure*}


\begin{figure*}
    \centering
    \begin{subfigure}[b]{0.40\textwidth}
        \centering
        \includegraphics[width=\textwidth]{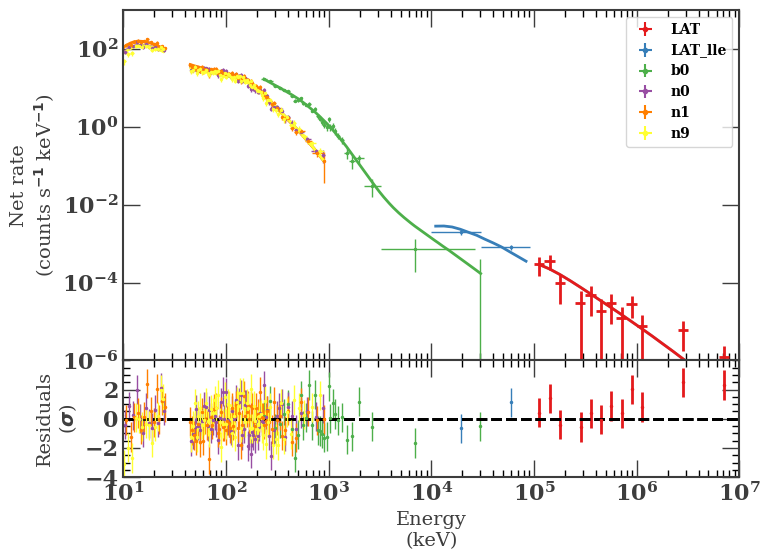}
            
    \end{subfigure}
    \hfill
    \begin{subfigure}[b]{0.40\textwidth}
        \centering
        \includegraphics[width=\textwidth]{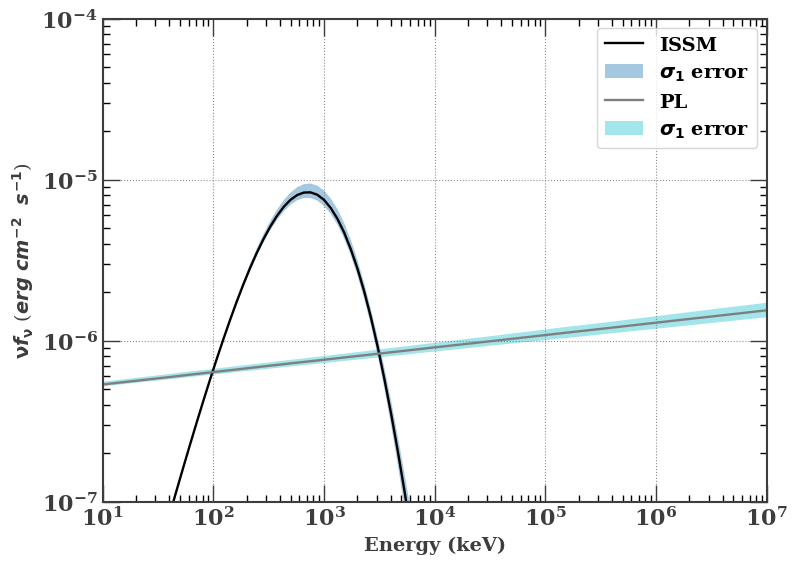}
          
    \end{subfigure}
        \hfill
    \begin{subfigure}[b]{0.40\textwidth}
        \centering
        \includegraphics[width=\textwidth]{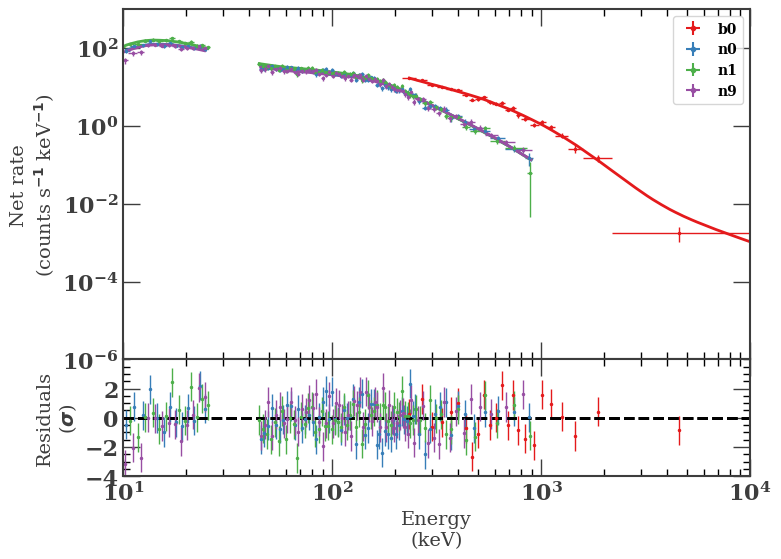}
          
    \end{subfigure}
        \hfill
    \begin{subfigure}[b]{0.40\textwidth}
        \centering
        \includegraphics[width=\textwidth]{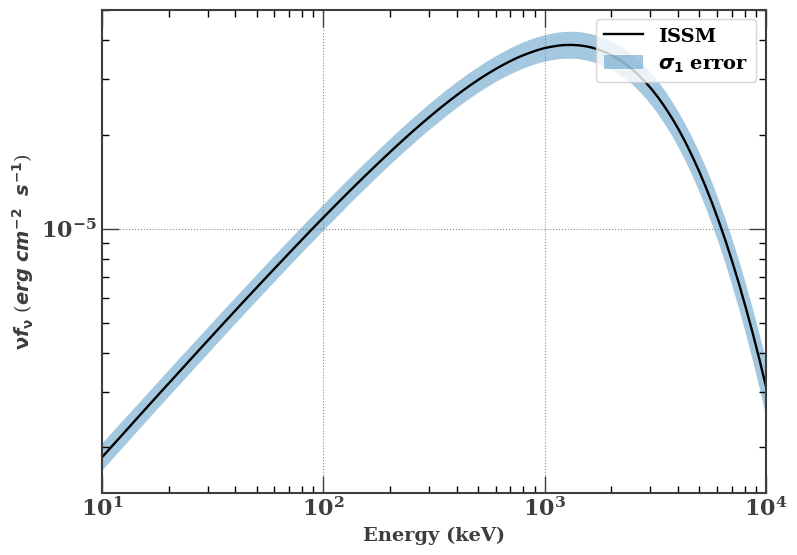}
          
    \end{subfigure}
    \caption{GRB 090902B: The count spectra (left panels) and ${\rm \nu}f_{\rm \nu}$ spectra (right panels).  The top (bottom) panels are for the joint (GBM-only) fits.}
    \label{fig_b4}
\end{figure*}


\begin{figure*}
    \centering
    \begin{subfigure}[b]{0.40\textwidth}
        \centering
        \includegraphics[width=\textwidth]{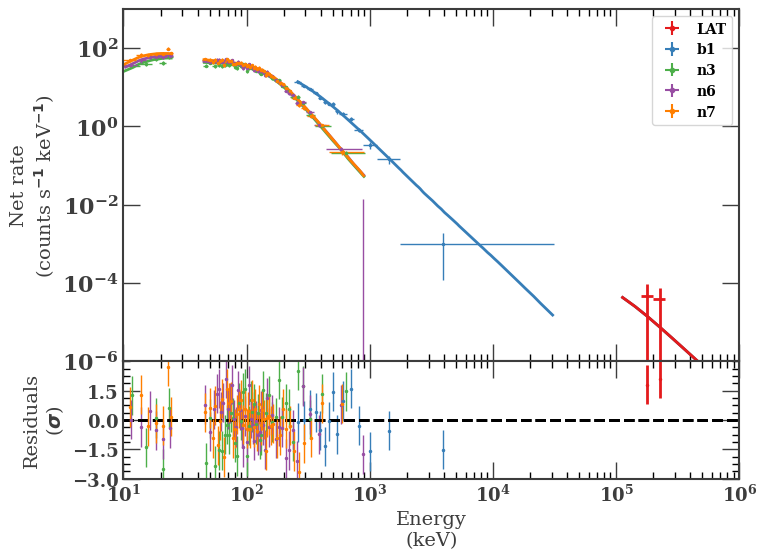}
            
    \end{subfigure}
    \hfill
    \begin{subfigure}[b]{0.40\textwidth}
        \centering
        \includegraphics[width=\textwidth]{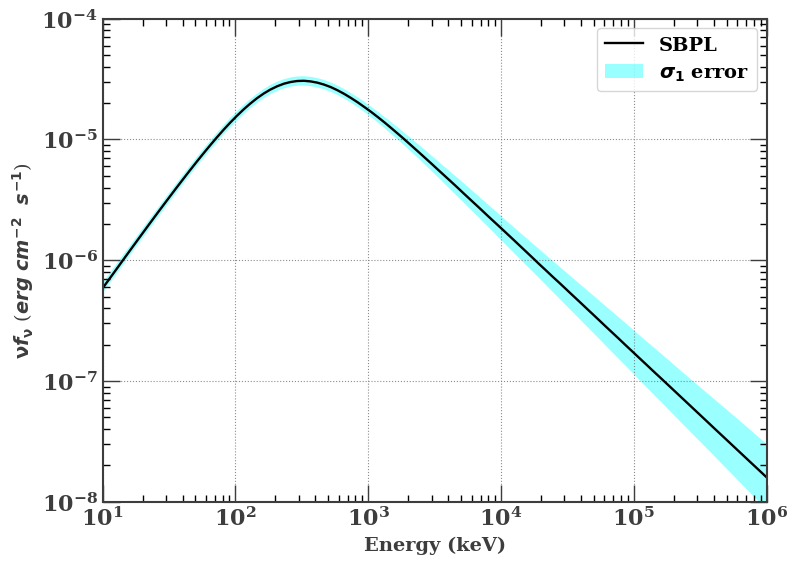}
          
    \end{subfigure}
        \hfill
    \begin{subfigure}[b]{0.40\textwidth}
        \centering
        \includegraphics[width=\textwidth]{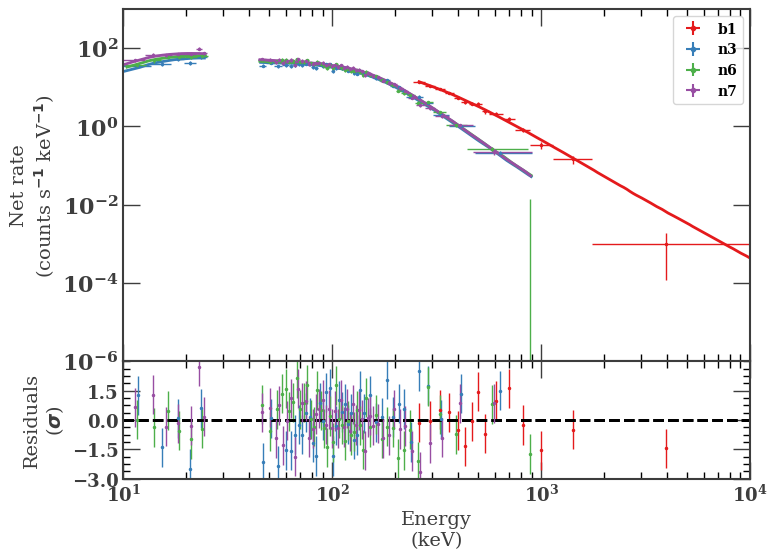}
          
    \end{subfigure}
        \hfill
    \begin{subfigure}[b]{0.40\textwidth}
        \centering
        \includegraphics[width=\textwidth]{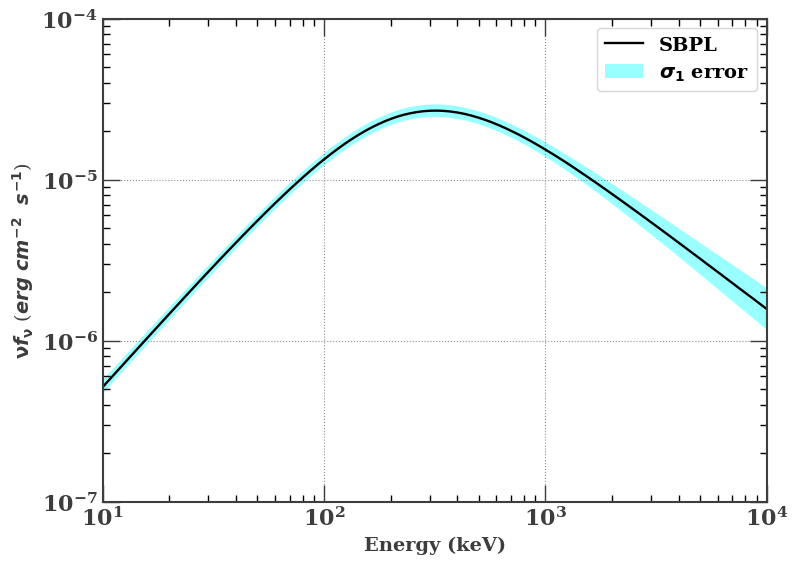}
          
    \end{subfigure}
    \caption{GRB 090926A: The count spectra (left panels) and ${\rm \nu}f_{\rm \nu}$ spectra (right panels).  The top (bottom) panels are for the joint (GBM-only) fits.}
    \label{fig_b5}
\end{figure*}


\begin{figure*}
    \centering
    \begin{subfigure}[b]{0.40\textwidth}
        \centering
        \includegraphics[width=\textwidth]{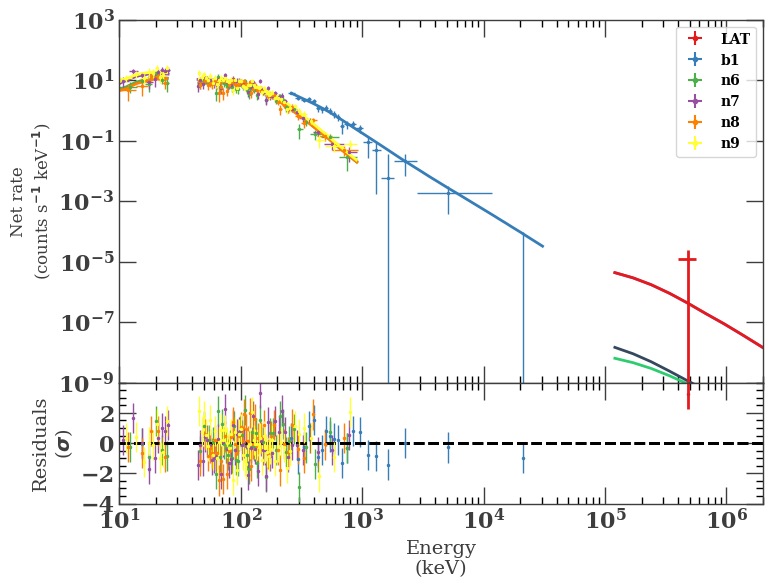}
            
    \end{subfigure}
    \hfill
    \begin{subfigure}[b]{0.40\textwidth}
        \centering
        \includegraphics[width=\textwidth]{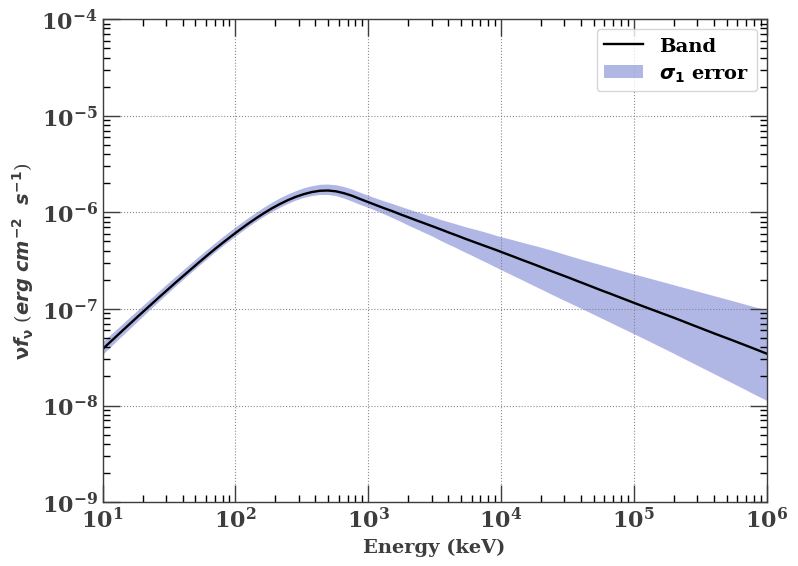}
          
    \end{subfigure}
        \hfill
    \begin{subfigure}[b]{0.40\textwidth}
        \centering
        \includegraphics[width=\textwidth]{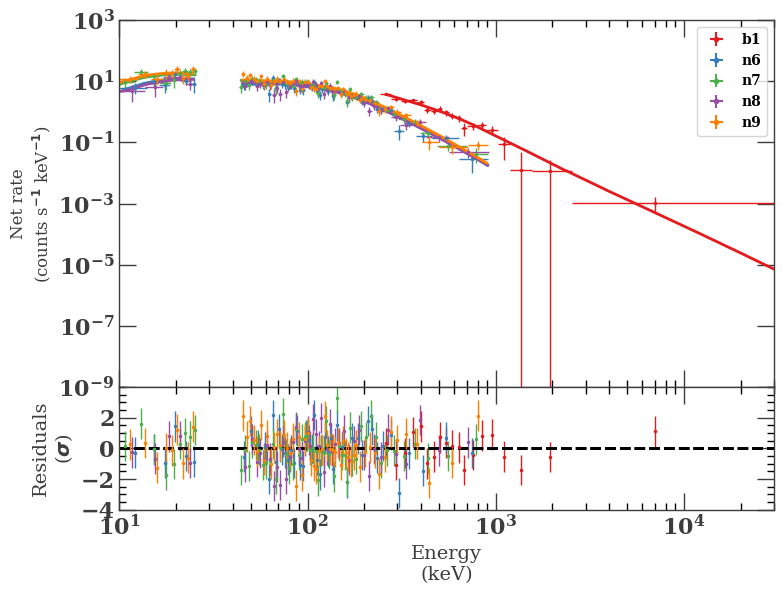}
          
    \end{subfigure}
        \hfill
    \begin{subfigure}[b]{0.40\textwidth}
        \centering
        \includegraphics[width=\textwidth]{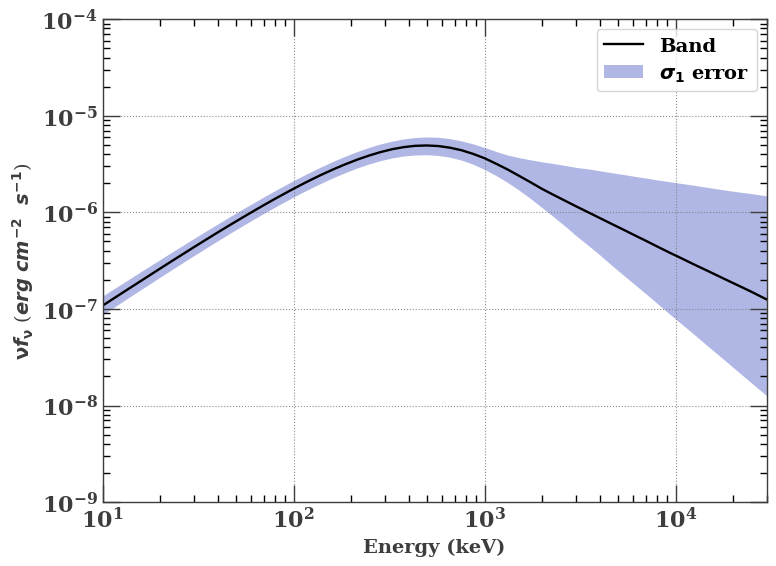}
          
    \end{subfigure}
    \caption{GRB 100414A: The count spectra (left panels) and ${\rm \nu}f_{\rm \nu}$ spectra (right panels).  The top (bottom) panels are for the joint (GBM-only) fits.}
    \label{fig_b6}
\end{figure*}


\begin{figure*}
    \centering
    \begin{subfigure}[b]{0.40\textwidth}
        \centering
        \includegraphics[width=\textwidth]{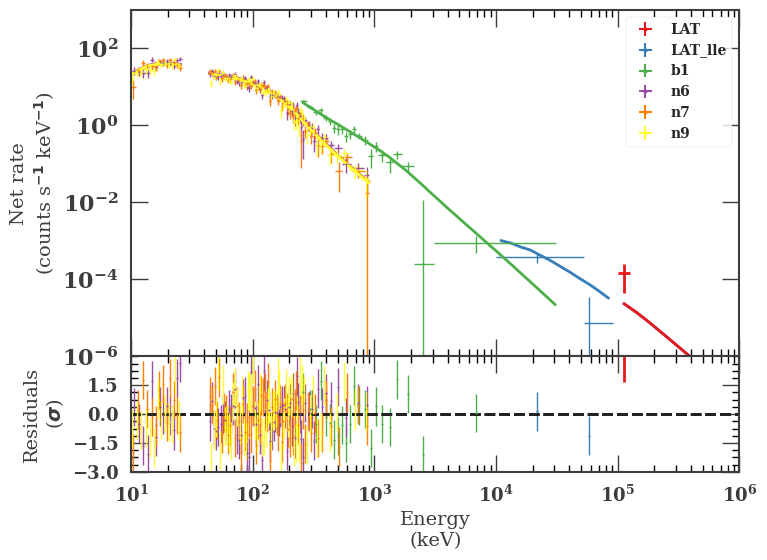}
            
    \end{subfigure}
    \hfill
    \begin{subfigure}[b]{0.40\textwidth}
        \centering
        \includegraphics[width=\textwidth]{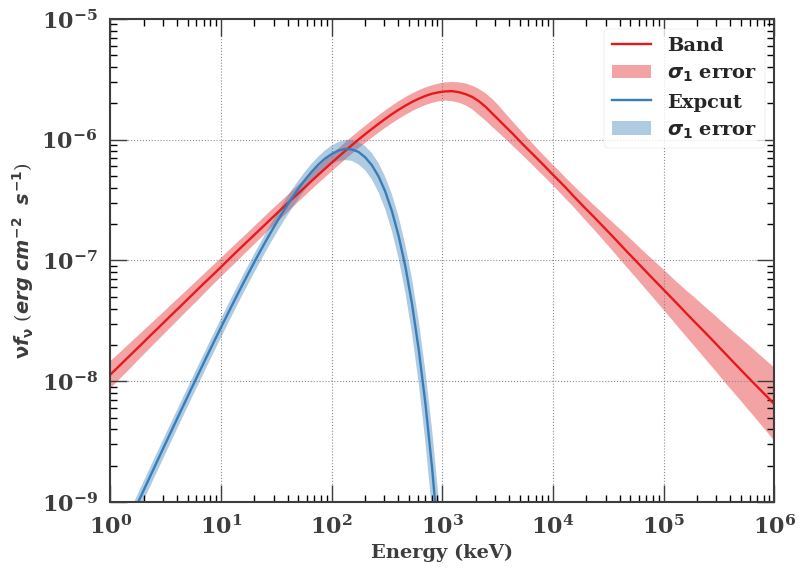}
          
    \end{subfigure}
        \hfill
    \begin{subfigure}[b]{0.40\textwidth}
        \centering
        \includegraphics[width=\textwidth]{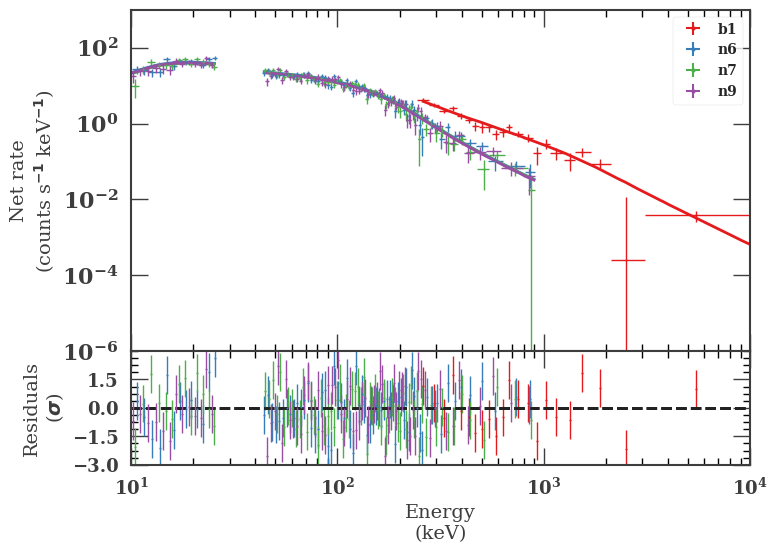}
          
    \end{subfigure}
        \hfill
    \begin{subfigure}[b]{0.40\textwidth}
        \centering
        \includegraphics[width=\textwidth]{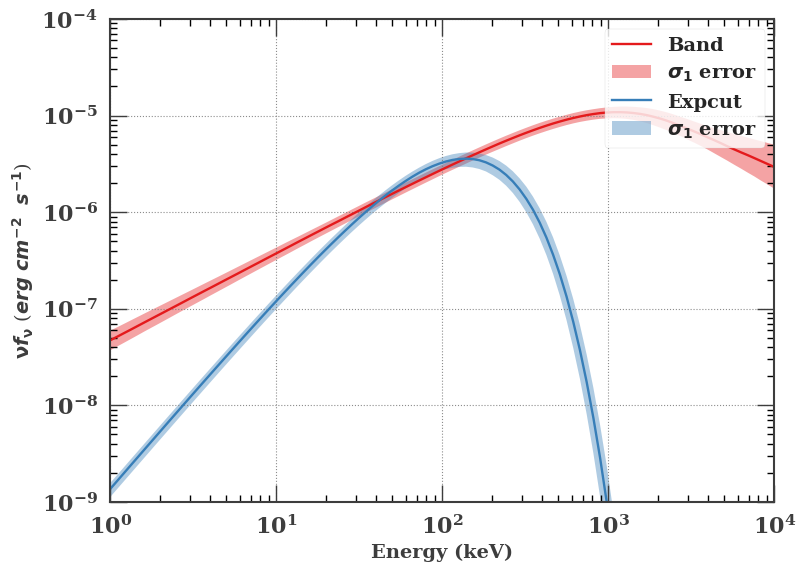}

    \end{subfigure}
    \caption{GRB 110721A: The count spectra (left panels) and ${\rm \nu}f_{\rm \nu}$ spectra (right panels).  The top (bottom) panels are for the joint (GBM-only) fits.}
    \label{fig_b7}
\end{figure*}


\begin{figure*}
    \centering
    \begin{subfigure}[b]{0.40\textwidth}
        \centering
        \includegraphics[width=\textwidth]{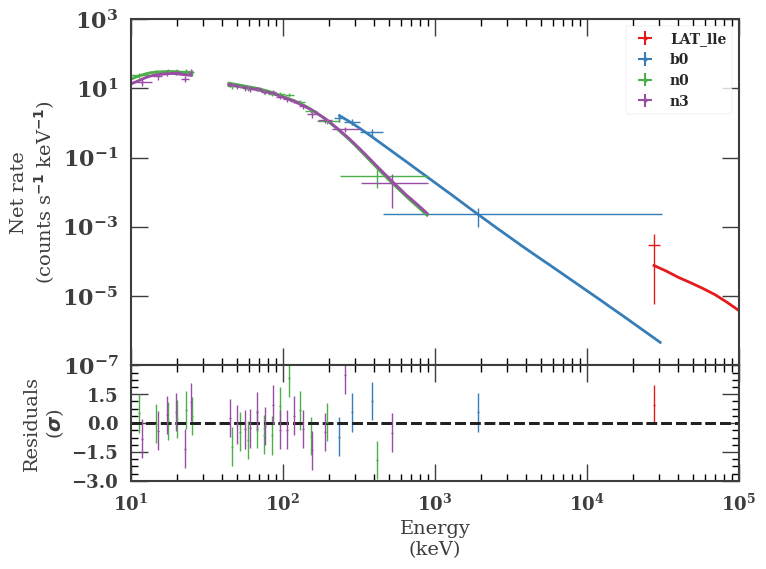}
            
    \end{subfigure}
    \hfill
    \begin{subfigure}[b]{0.40\textwidth}
        \centering
        \includegraphics[width=\textwidth]{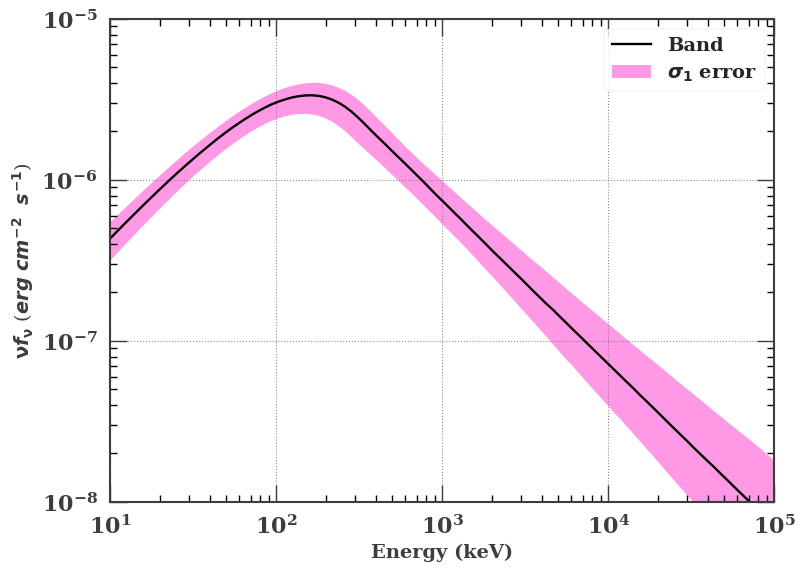}
          
    \end{subfigure}
        \hfill
    \begin{subfigure}[b]{0.40\textwidth}
        \centering
        \includegraphics[width=\textwidth]{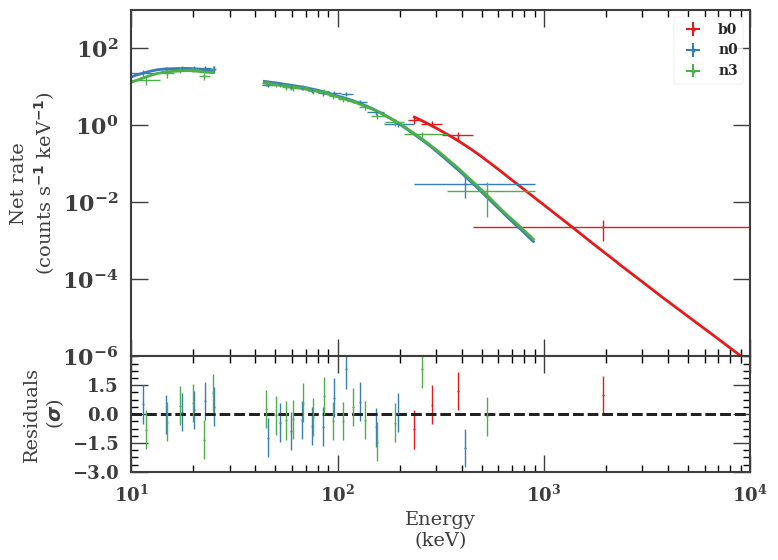}
          
    \end{subfigure}
        \hfill
    \begin{subfigure}[b]{0.40\textwidth}
        \centering
        \includegraphics[width=\textwidth]{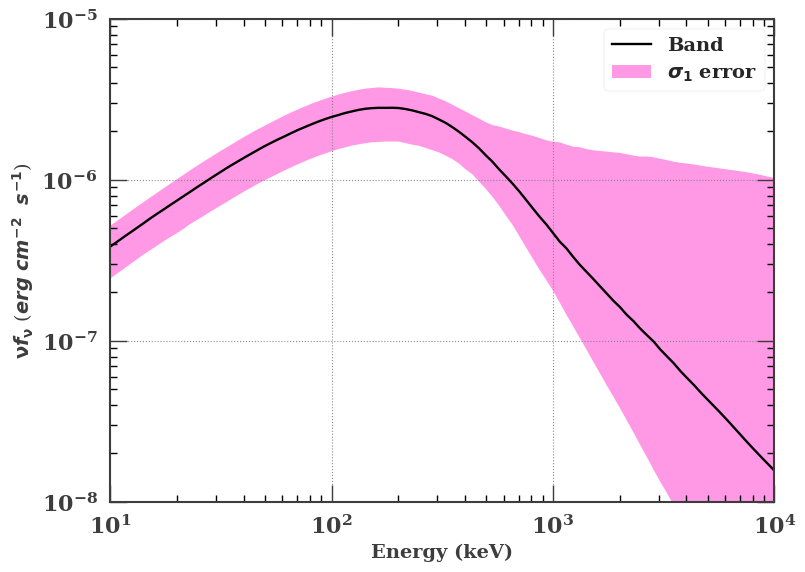}
          
    \end{subfigure}
    \caption{GRB 110731A: The count spectra (left panels) and ${\rm \nu}f_{\rm \nu}$ spectra (right panels).  The top (bottom) panels are for the joint (GBM-only) fits.}
    \label{fig_b8}
\end{figure*}


\begin{figure*}
    \centering
    \begin{subfigure}[b]{0.40\textwidth}
        \centering
        \includegraphics[width=\textwidth]{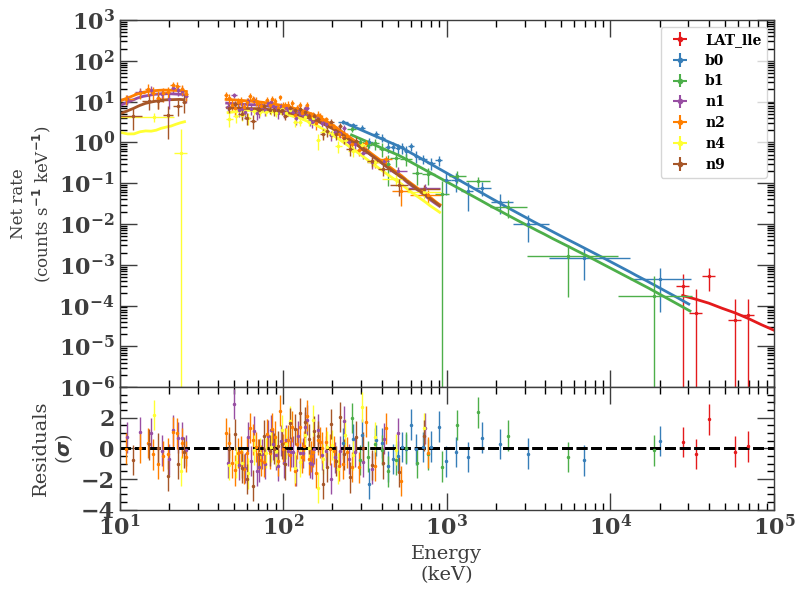}
            
    \end{subfigure}
    \hfill
    \begin{subfigure}[b]{0.40\textwidth}
        \centering
        \includegraphics[width=\textwidth]{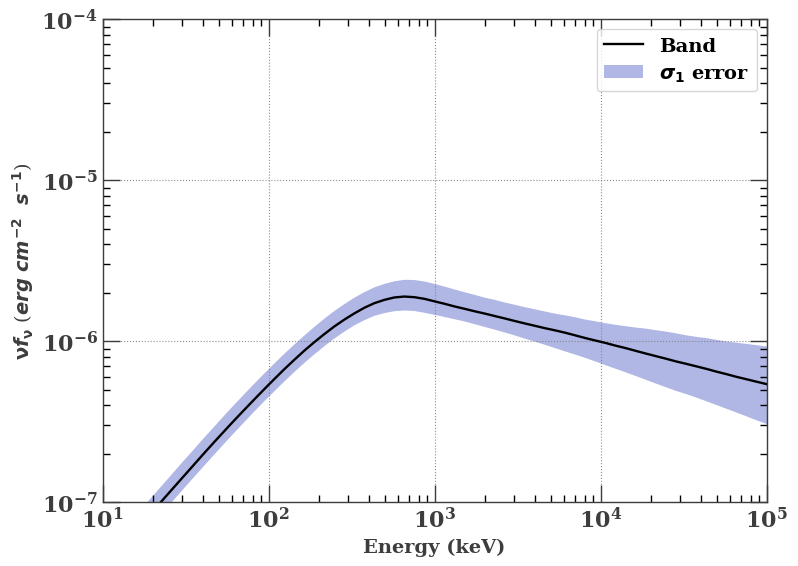}
          
    \end{subfigure}
        \hfill
    \begin{subfigure}[b]{0.40\textwidth}
        \centering
        \includegraphics[width=\textwidth]{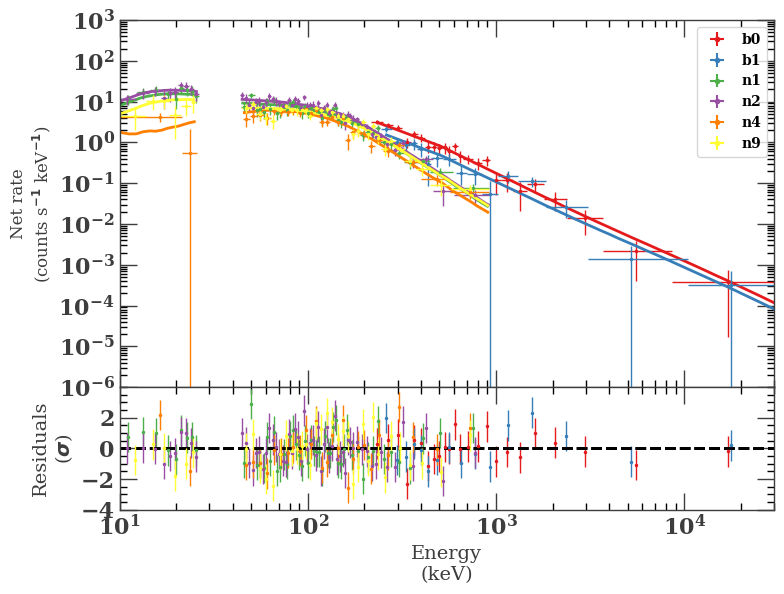}
          
    \end{subfigure}
        \hfill
    \begin{subfigure}[b]{0.40\textwidth}
        \centering
        \includegraphics[width=\textwidth]{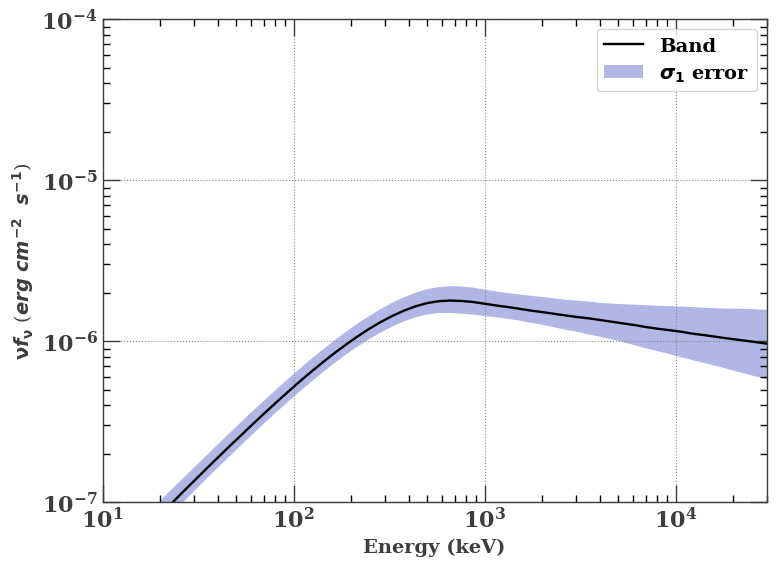} 
          
    \end{subfigure}
    \caption{GRB 120624B: The count spectra (left panels) and ${\rm \nu}f_{\rm \nu}$ spectra (right panels).  The top (bottom) panels are for the joint (GBM-only) fits.}
    \label{fig_b9}
\end{figure*}


\begin{figure*}
    \centering
    \begin{subfigure}[b]{0.40\textwidth}
        \centering
        \includegraphics[width=\textwidth]{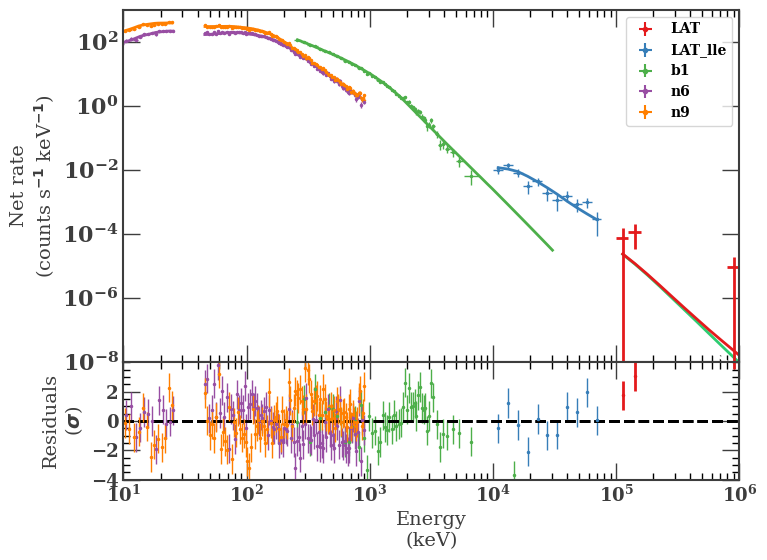}
            
    \end{subfigure}
    \hfill
    \begin{subfigure}[b]{0.40\textwidth}
        \centering
        \includegraphics[width=\textwidth]{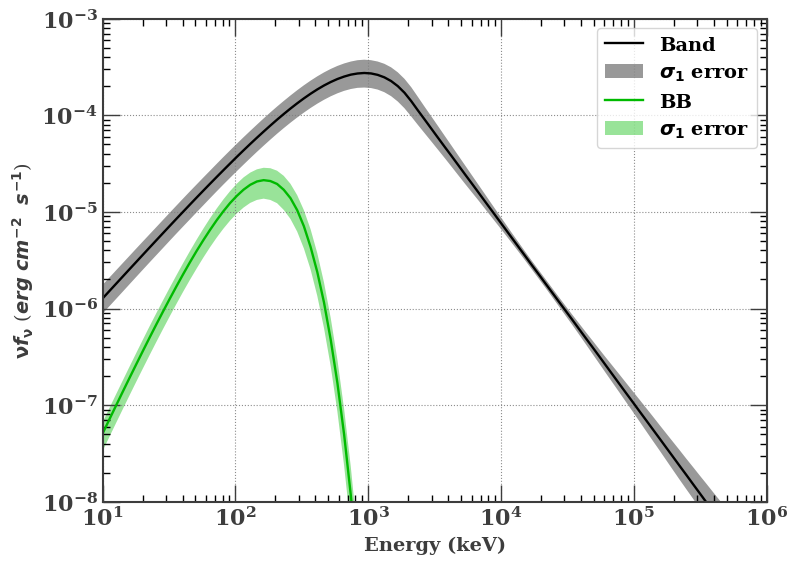}
          
    \end{subfigure}
        \hfill
    \begin{subfigure}[b]{0.40\textwidth}
        \centering
        \includegraphics[width=\textwidth]{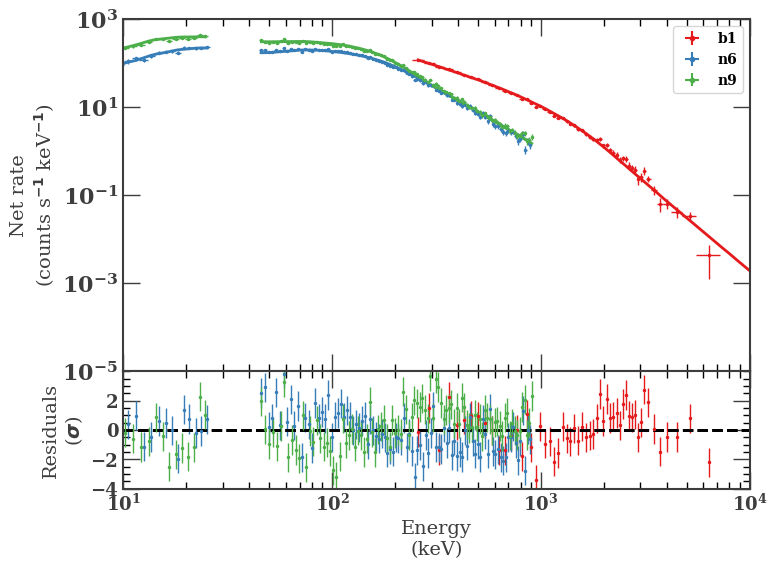}
          
    \end{subfigure}
        \hfill
    \begin{subfigure}[b]{0.40\textwidth}
        \centering
        \includegraphics[width=\textwidth]{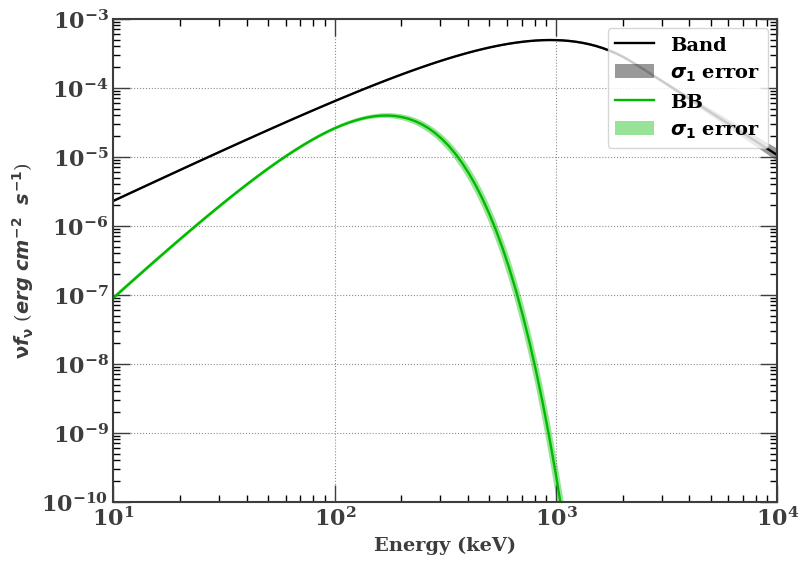}
          
    \end{subfigure}
    \caption{GRB 130427A: The count spectra (left panels) and ${\rm \nu}f_{\rm \nu}$ spectra (right panels).  The top (bottom) panels are for the joint (GBM-only) fits.}
    \label{fig_b10}
\end{figure*}


\begin{figure*}
    \centering
    \begin{subfigure}[b]{0.40\textwidth}
        \centering
        \includegraphics[width=\textwidth]{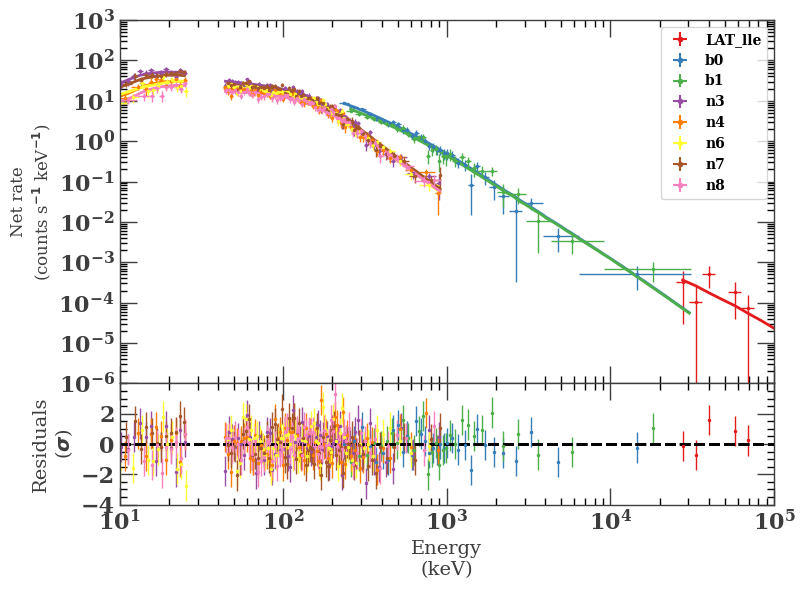}
            
    \end{subfigure}
    \hfill
    \begin{subfigure}[b]{0.40\textwidth}
        \centering
        \includegraphics[width=\textwidth]{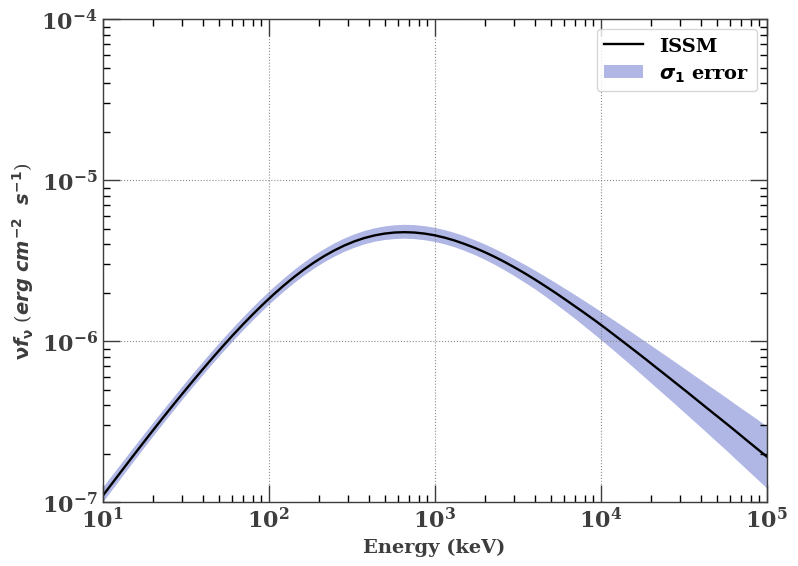}
          
    \end{subfigure}
        \hfill
    \begin{subfigure}[b]{0.40\textwidth}
        \centering
        \includegraphics[width=\textwidth]{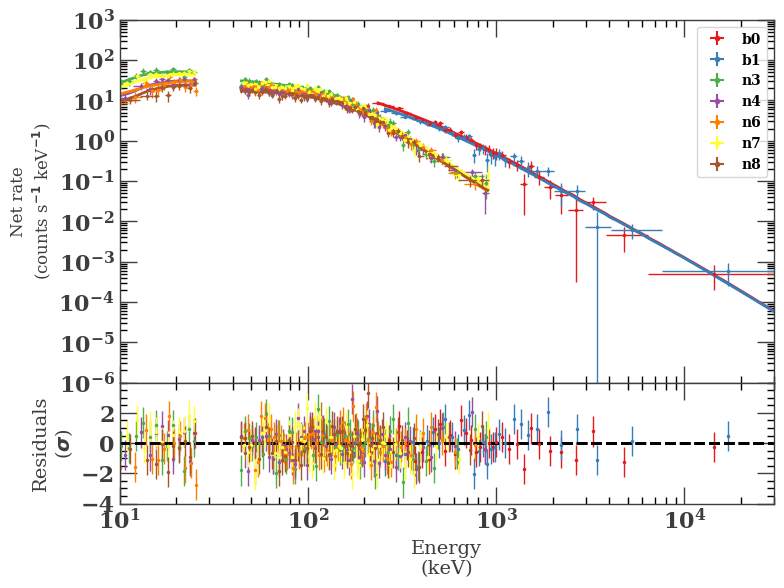}
          
    \end{subfigure}
        \hfill
    \begin{subfigure}[b]{0.40\textwidth}
        \centering
        \includegraphics[width=\textwidth]{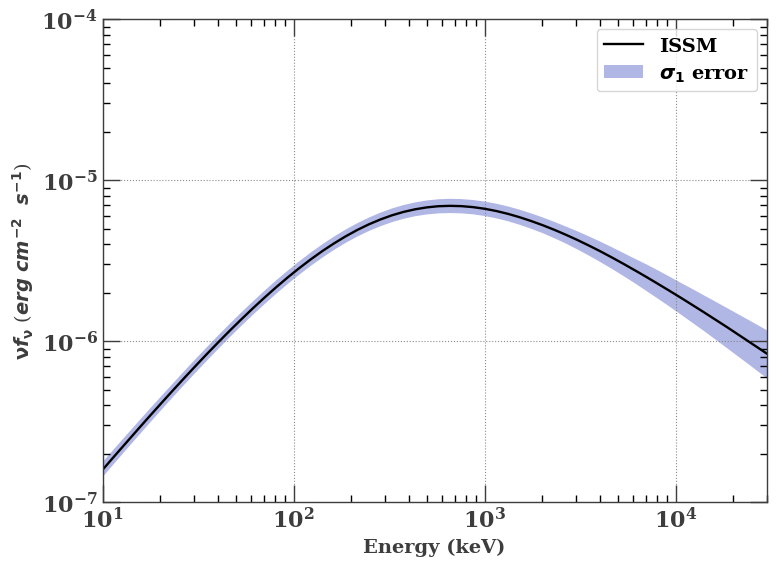}
          
    \end{subfigure}
    \caption{GRB 130518A: The count spectra (left panels) and ${\rm \nu}f_{\rm \nu}$ spectra (right panels).  The top (bottom) panels are for the joint (GBM-only) fits.}
    \label{fig_b11}
\end{figure*}


\begin{figure*}
    \centering
    \begin{subfigure}[b]{0.40\textwidth}
        \centering
        \includegraphics[width=\textwidth]{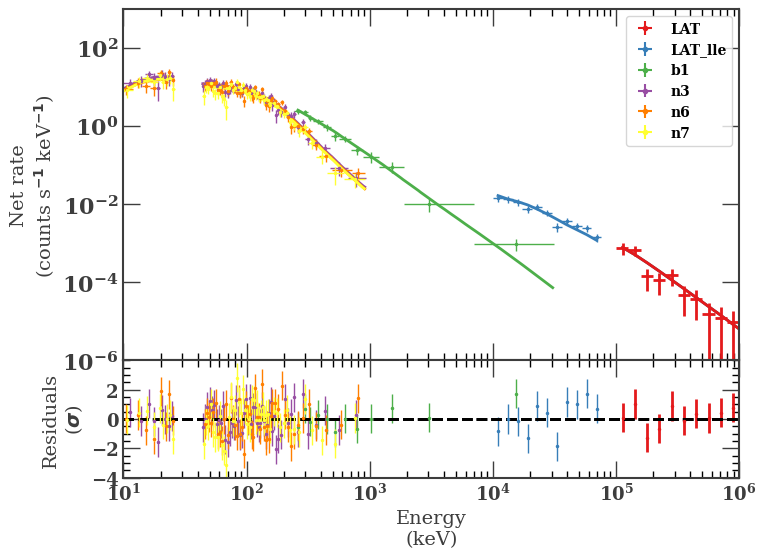}
            
    \end{subfigure}
    \hfill
    \begin{subfigure}[b]{0.40\textwidth}
        \centering
        \includegraphics[width=\textwidth]{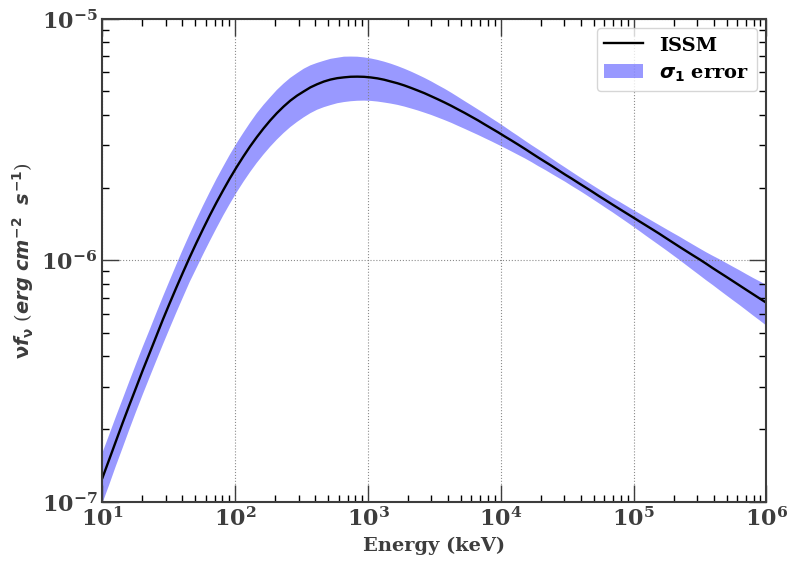}
          
    \end{subfigure}
        \hfill
    \begin{subfigure}[b]{0.40\textwidth}
        \centering
        \includegraphics[width=\textwidth]{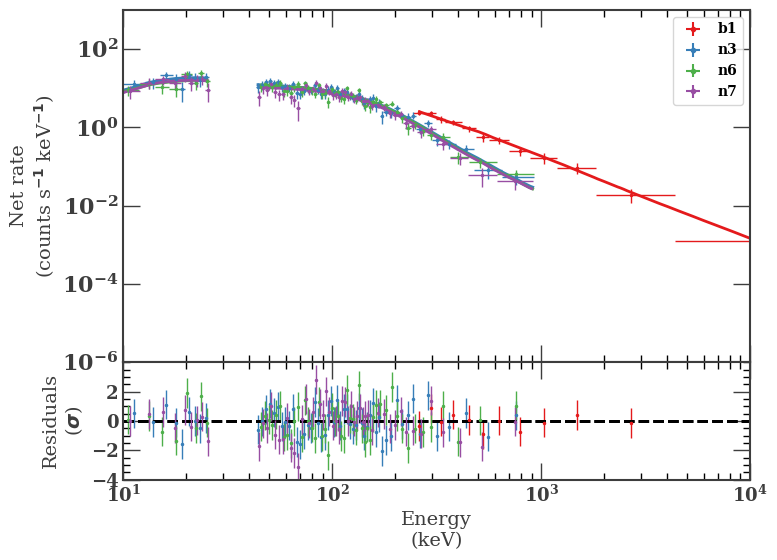}
          
    \end{subfigure}
        \hfill
    \begin{subfigure}[b]{0.40\textwidth}
        \centering
        \includegraphics[width=\textwidth]{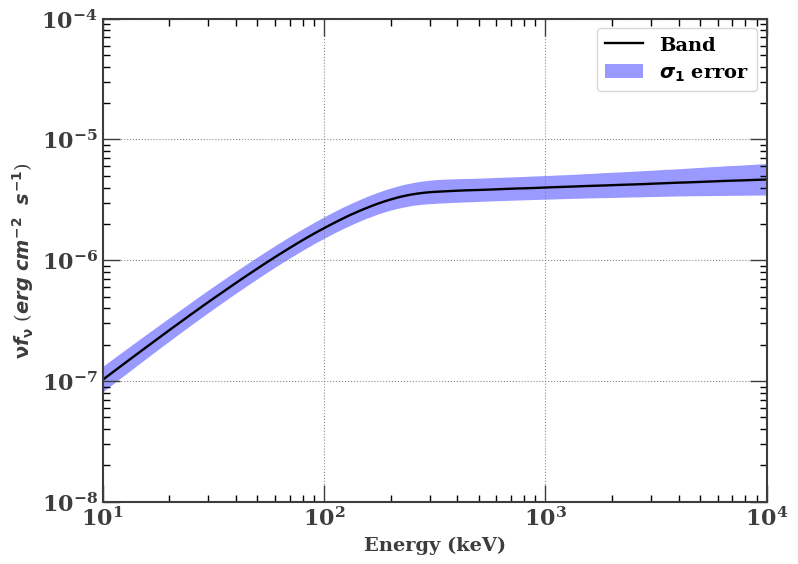}
          
    \end{subfigure}
    \caption{GRB 131108A: The count spectra (left panels) and ${\rm \nu}f_{\rm \nu}$ spectra (right panels).  The top (bottom) panels are for the joint (GBM-only) fits.}
    \label{fig_b12}
\end{figure*}


\begin{figure*}
    \centering
    \begin{subfigure}[b]{0.40\textwidth}
        \centering
        \includegraphics[width=\textwidth]{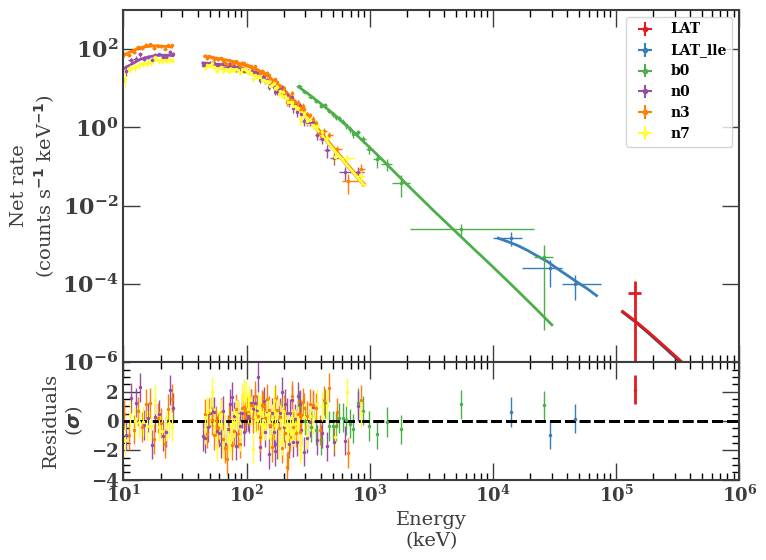}
            
    \end{subfigure}
    \hfill
    \begin{subfigure}[b]{0.40\textwidth}
        \centering
        \includegraphics[width=\textwidth]{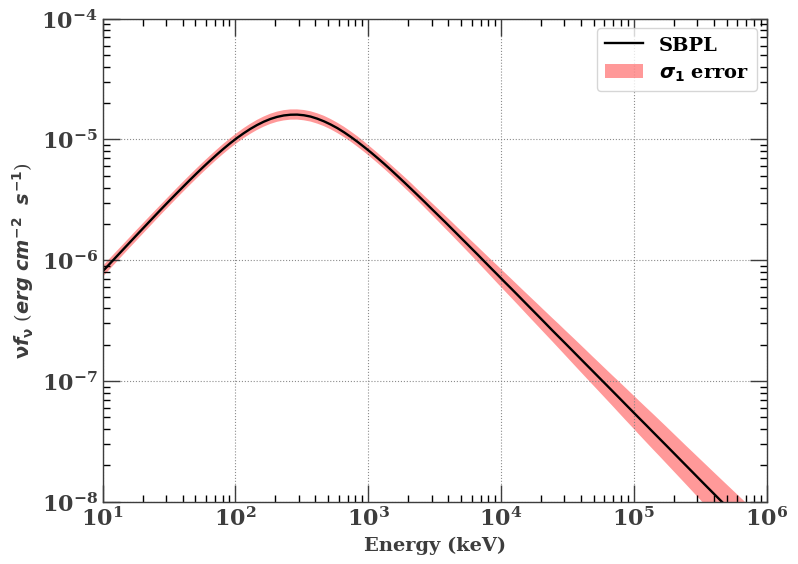}
          
    \end{subfigure}
        \hfill
    \begin{subfigure}[b]{0.40\textwidth}
        \centering
        \includegraphics[width=\textwidth]{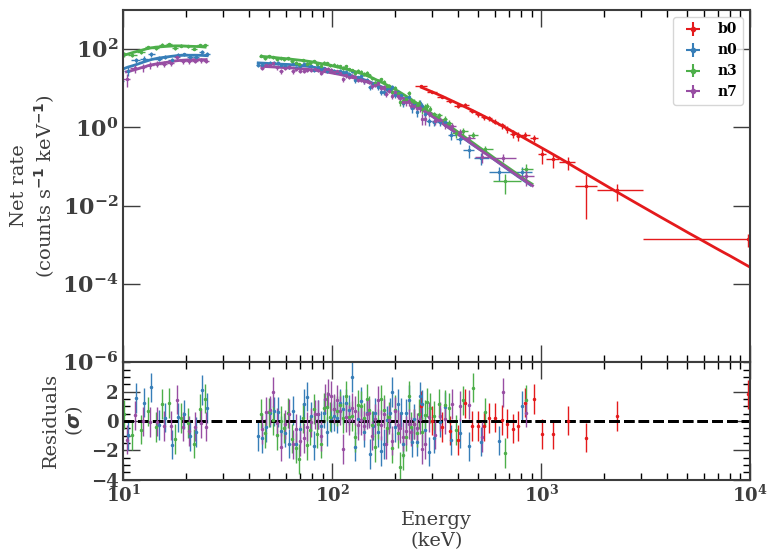}
          
    \end{subfigure}
        \hfill
    \begin{subfigure}[b]{0.40\textwidth}
        \centering
        \includegraphics[width=\textwidth]{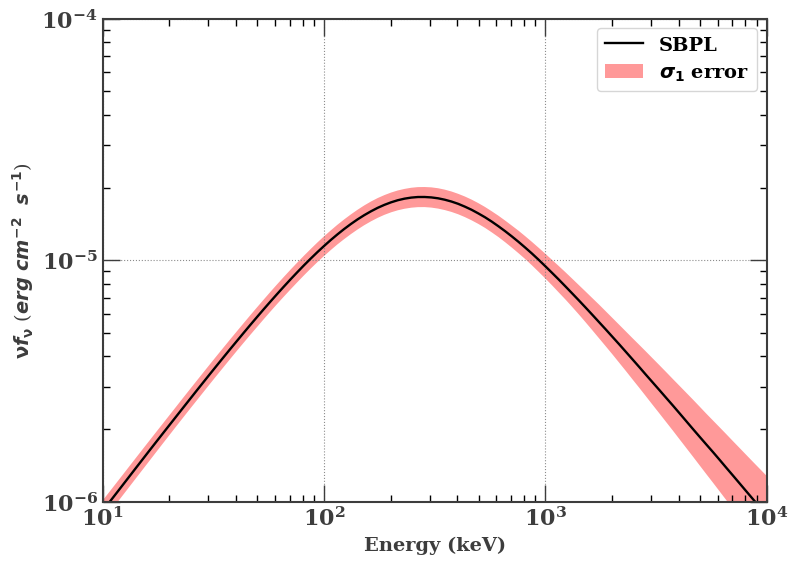}
          
    \end{subfigure}
    \caption{GRB 131231A: The count spectra (left panels) and ${\rm \nu}f_{\rm \nu}$ spectra (right panels).  The top (bottom) panels are for the joint (GBM-only) fits.}
    \label{fig_b13}
\end{figure*}


\begin{figure*}
    \centering
    \begin{subfigure}[b]{0.40\textwidth}
        \centering
        \includegraphics[width=\textwidth]{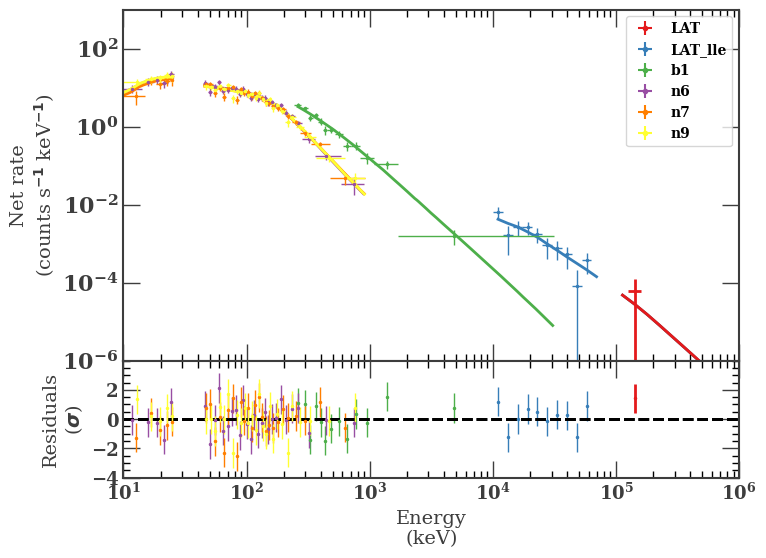}
            
    \end{subfigure}
    \hfill
    \begin{subfigure}[b]{0.40\textwidth}
        \centering
        \includegraphics[width=\textwidth]{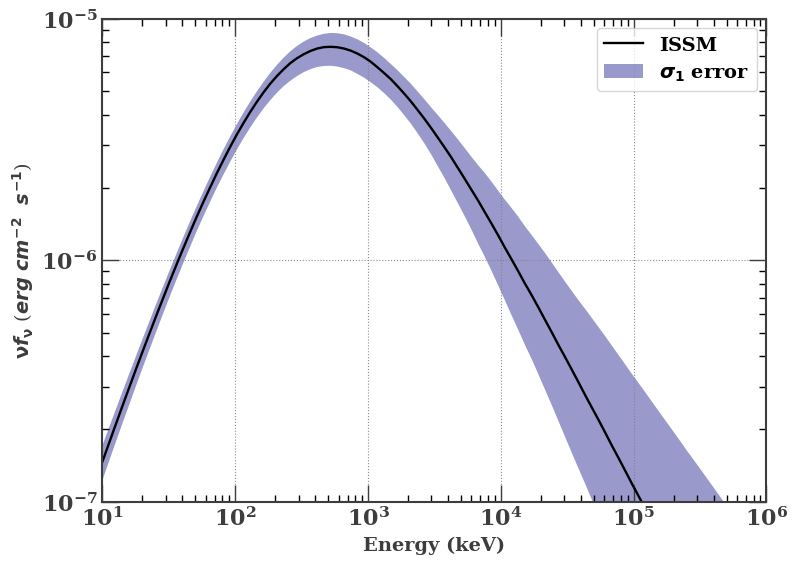}
          
    \end{subfigure}
        \hfill
    \begin{subfigure}[b]{0.40\textwidth}
        \centering
        \includegraphics[width=\textwidth]{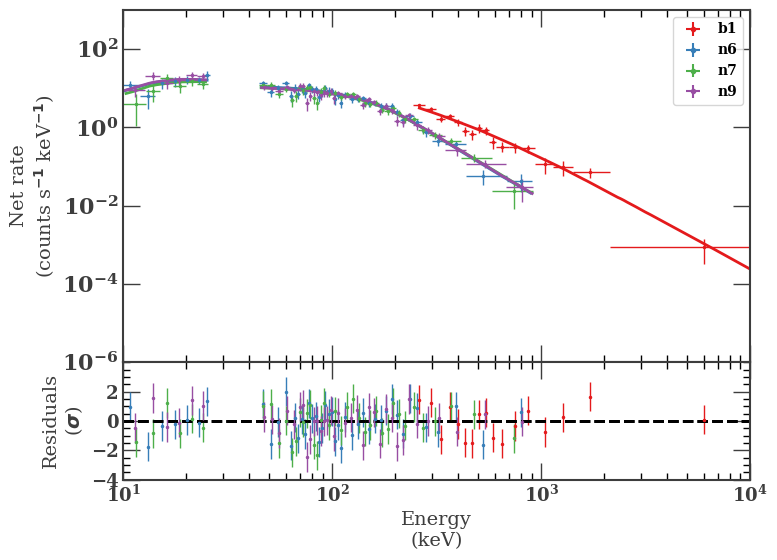}
          
    \end{subfigure}
        \hfill
    \begin{subfigure}[b]{0.40\textwidth}
        \centering
        \includegraphics[width=\textwidth]{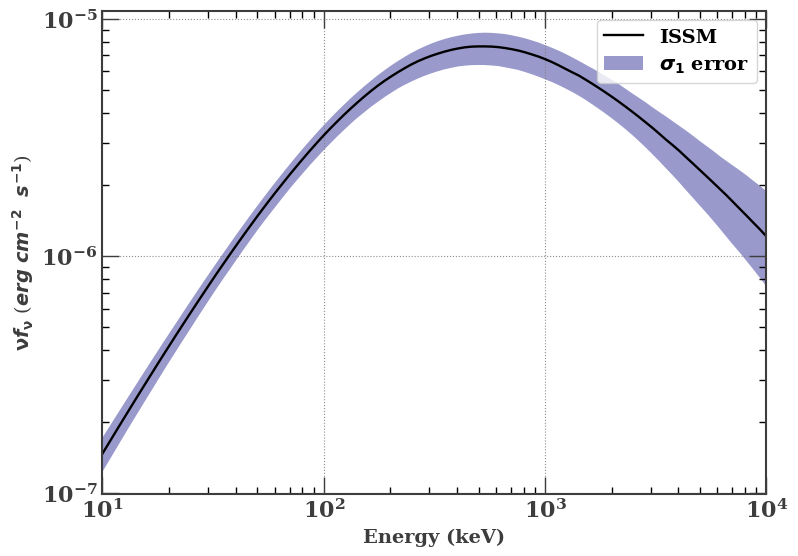}
          
    \end{subfigure}
    \caption{GRB 141028A: The count spectra (left panels) and ${\rm \nu}f_{\rm \nu}$ spectra (right panels).  The top (bottom) panels are for the joint (GBM-only) fits.}
    \label{fig_b14}
\end{figure*}


\begin{figure*}
    \centering
    \begin{subfigure}[b]{0.40\textwidth}
        \centering
        \includegraphics[width=\textwidth]{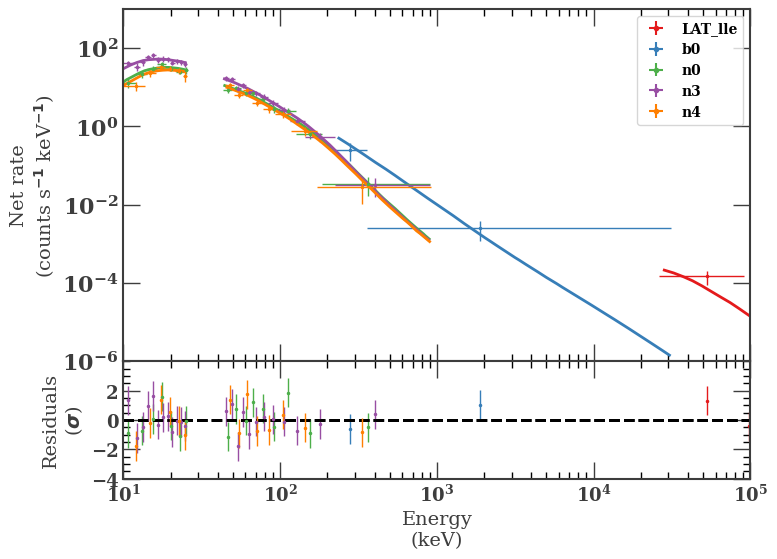}
            
    \end{subfigure}
    \hfill
    \begin{subfigure}[b]{0.40\textwidth}
        \centering
        \includegraphics[width=\textwidth]{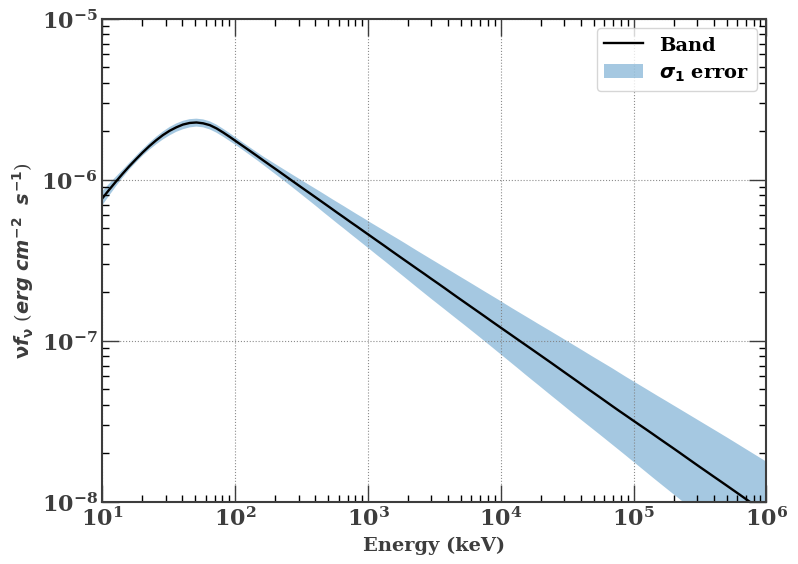}
          
    \end{subfigure}
        \hfill
    \begin{subfigure}[b]{0.40\textwidth}
        \centering
        \includegraphics[width=\textwidth]{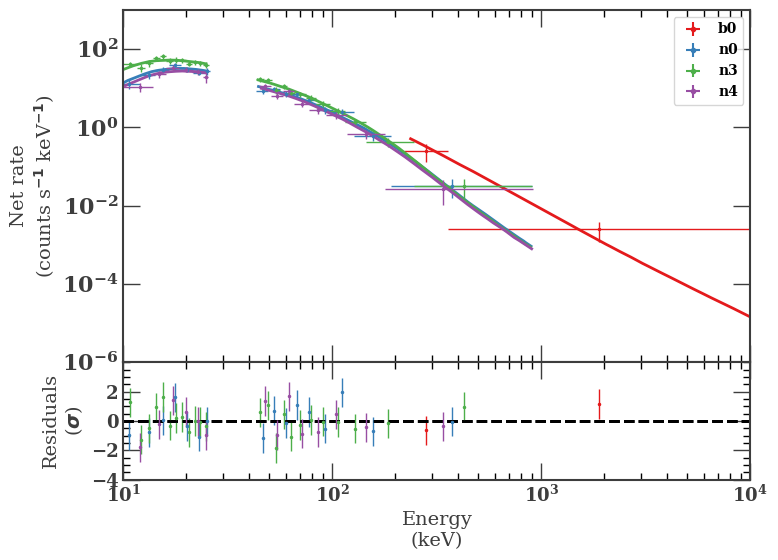}
          
    \end{subfigure}
        \hfill
    \begin{subfigure}[b]{0.40\textwidth}
        \centering
        \includegraphics[width=\textwidth]{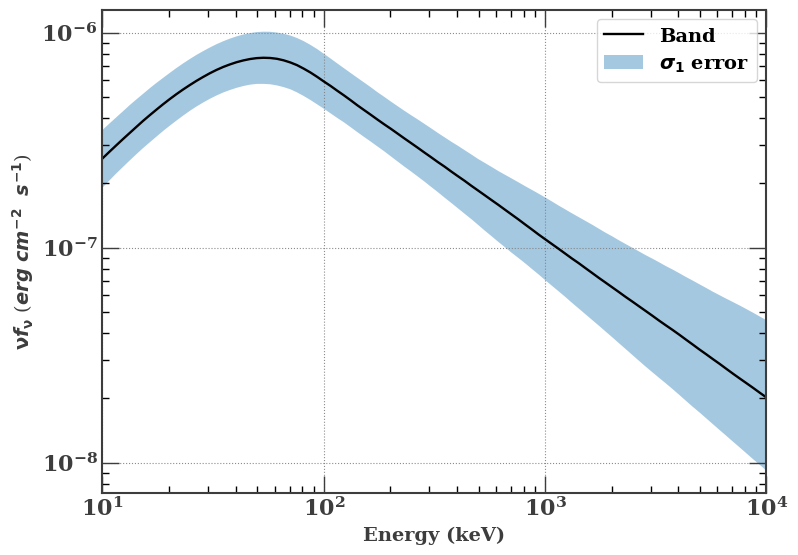}
          
    \end{subfigure}
    \caption{GRB 150314A: The count spectra (left panels) and ${\rm \nu}f_{\rm \nu}$ spectra (right panels).  The top (bottom) panels are for the joint (GBM-only) fits.}
    \label{fig_b15}
\end{figure*}


\begin{figure*}
    \centering
    \begin{subfigure}[b]{0.40\textwidth}
        \centering
        \includegraphics[width=\textwidth]{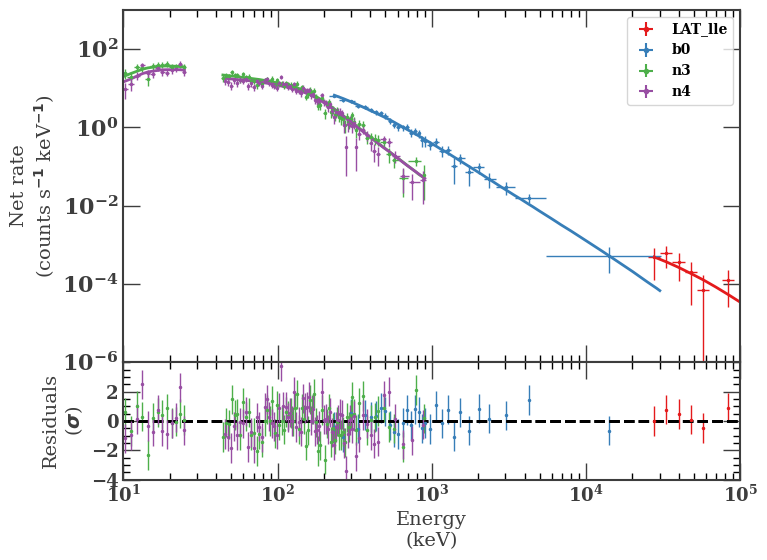}
            
    \end{subfigure}
    \hfill
    \begin{subfigure}[b]{0.40\textwidth}
        \centering
        \includegraphics[width=\textwidth]{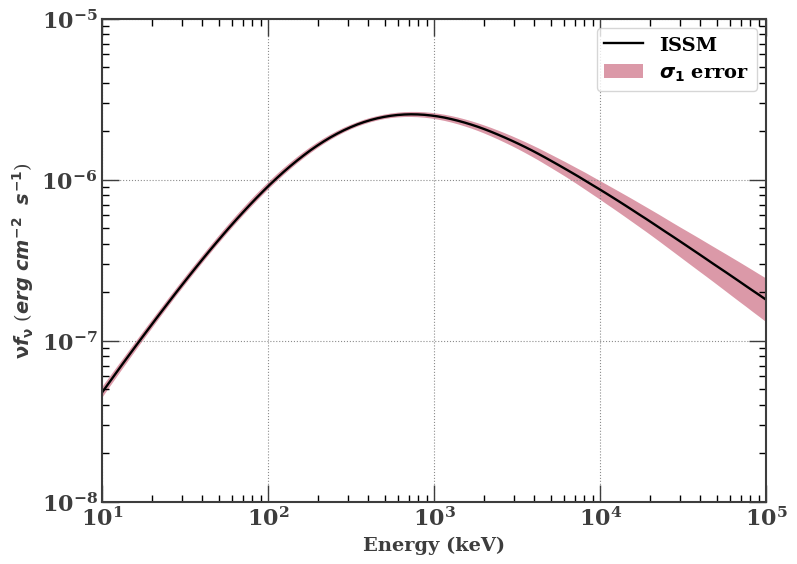}
          
    \end{subfigure}
        \hfill
    \begin{subfigure}[b]{0.40\textwidth}
        \centering
        \includegraphics[width=\textwidth]{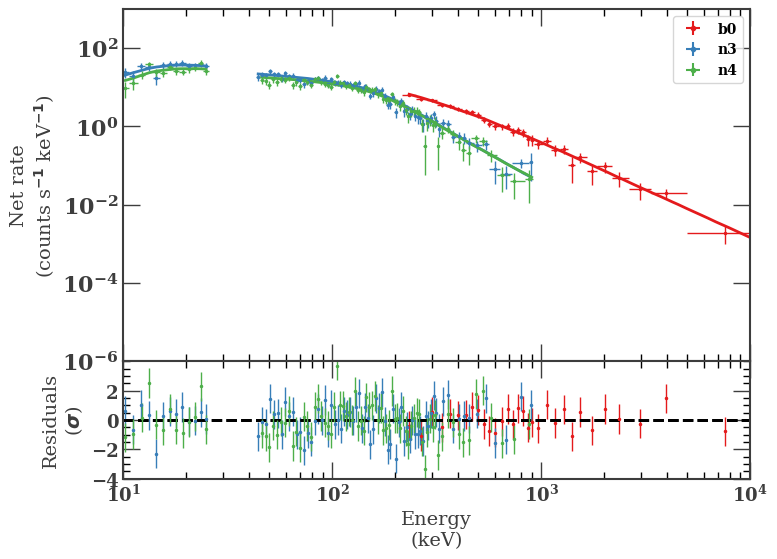}
          
    \end{subfigure}
        \hfill
    \begin{subfigure}[b]{0.40\textwidth}
        \centering
        \includegraphics[width=\textwidth]{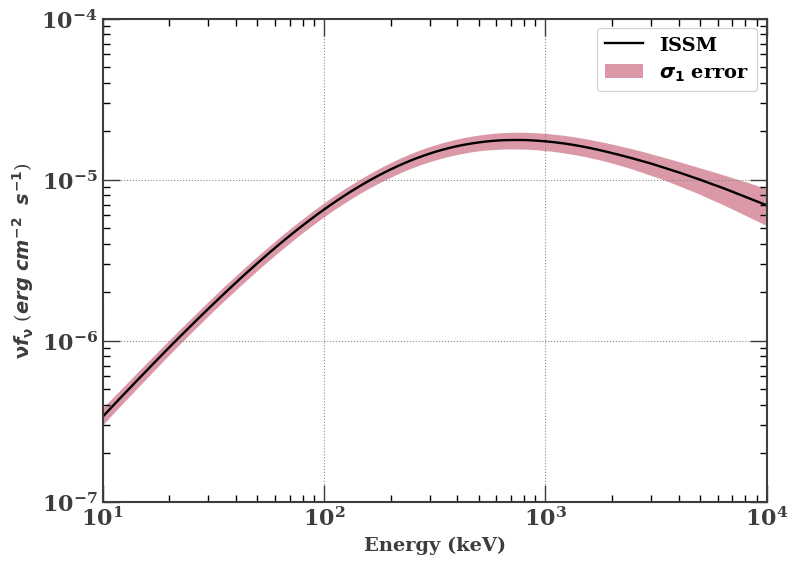}
          
    \end{subfigure}
    \caption{GRB 150403A: The count spectra (left panels) and ${\rm \nu}f_{\rm \nu}$ spectra (right panels).  The top (bottom) panels are for the joint (GBM-only) fits.}
    \label{fig_b16}
\end{figure*}


\begin{figure*}
    \centering
    \begin{subfigure}[b]{0.40\textwidth}
        \centering
        \includegraphics[width=\textwidth]{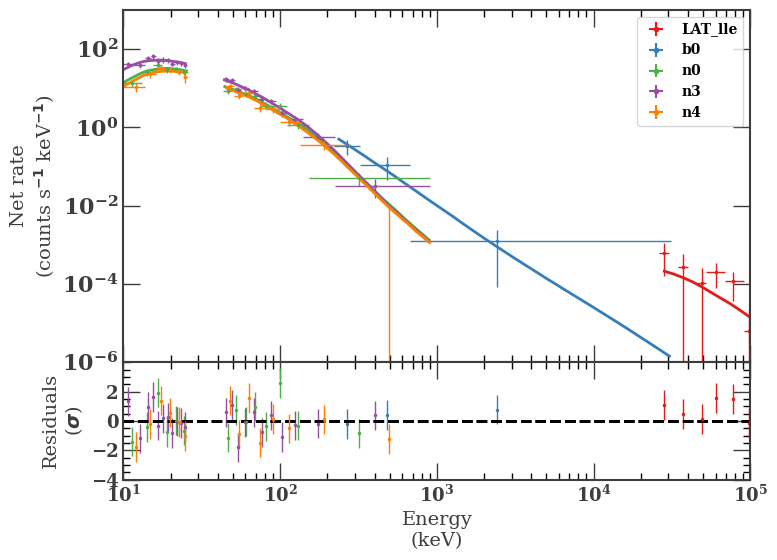}
            
    \end{subfigure}
    \hfill
    \begin{subfigure}[b]{0.40\textwidth}
        \centering
        \includegraphics[width=\textwidth]{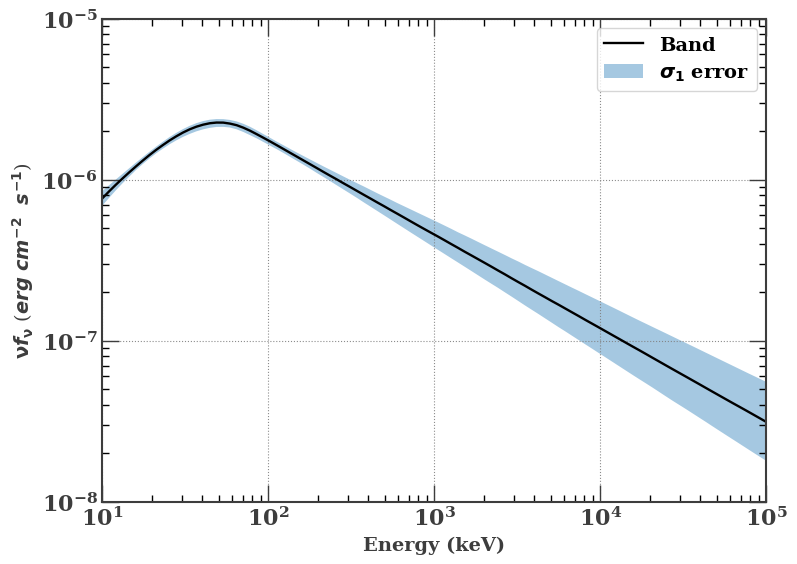}
          
    \end{subfigure}
        \hfill
    \begin{subfigure}[b]{0.40\textwidth}
        \centering
        \includegraphics[width=\textwidth]{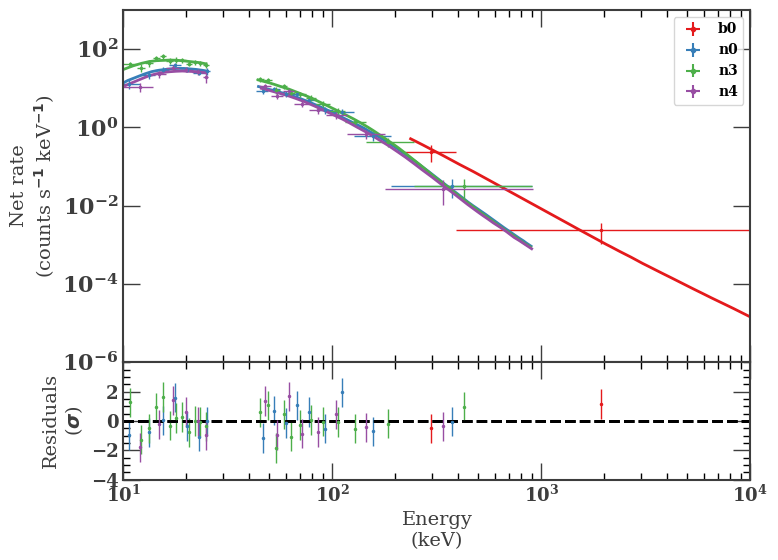}
          
    \end{subfigure}
        \hfill
    \begin{subfigure}[b]{0.40\textwidth}
        \centering
        \includegraphics[width=\textwidth]{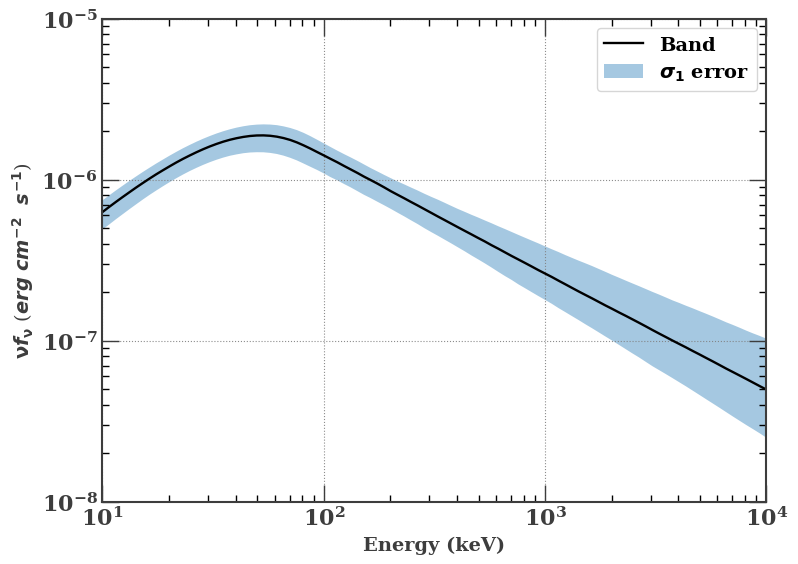}
          
    \end{subfigure}
    \caption{GRB 150514A: The count spectra (left panels) and ${\rm \nu}f_{\rm \nu}$ spectra (right panels).  The top (bottom) panels are for the joint (GBM-only) fits.}
    \label{fig_b17}
\end{figure*}


\begin{figure*}
    \centering
    \begin{subfigure}[b]{0.40\textwidth}
        \centering
        \includegraphics[width=\textwidth]{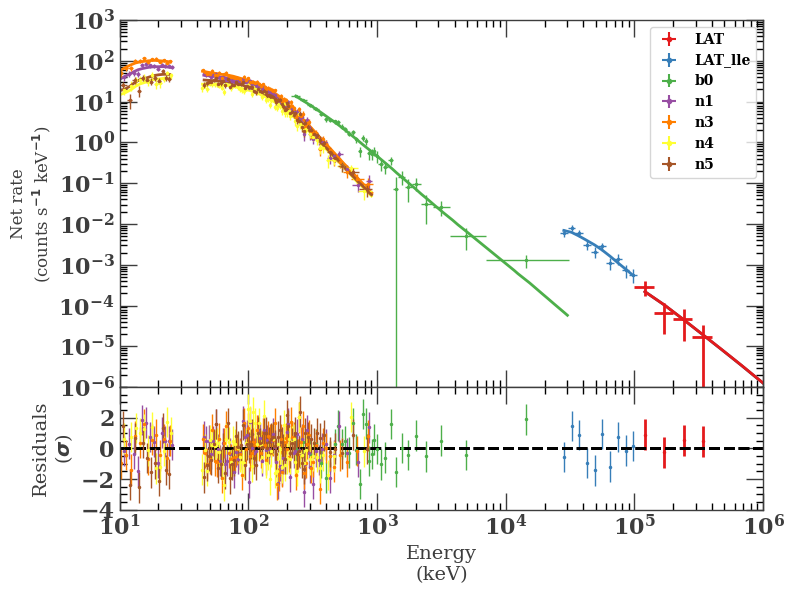}
            
    \end{subfigure}
    \hfill
    \begin{subfigure}[b]{0.40\textwidth}
        \centering
        \includegraphics[width=\textwidth]{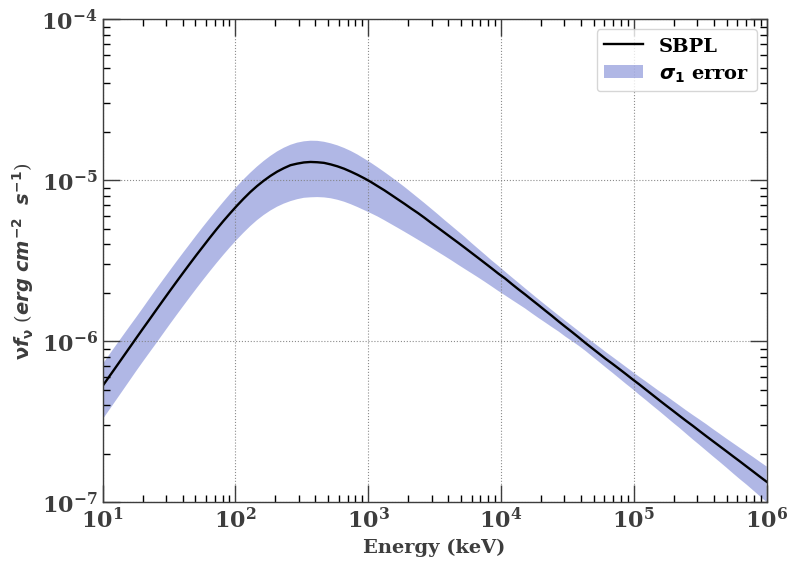}
          
    \end{subfigure}
        \hfill
    \begin{subfigure}[b]{0.40\textwidth}
        \centering
        \includegraphics[width=\textwidth]{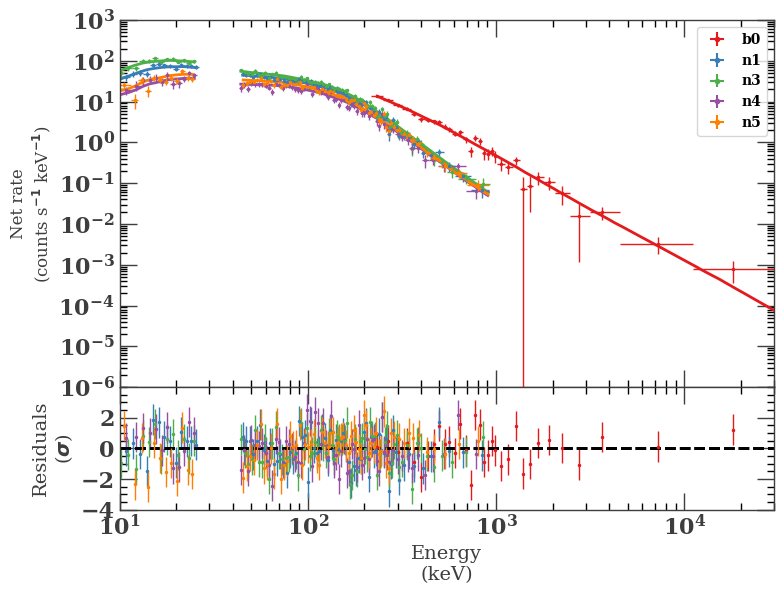}
          
    \end{subfigure}
        \hfill
    \begin{subfigure}[b]{0.40\textwidth}
        \centering
        \includegraphics[width=\textwidth]{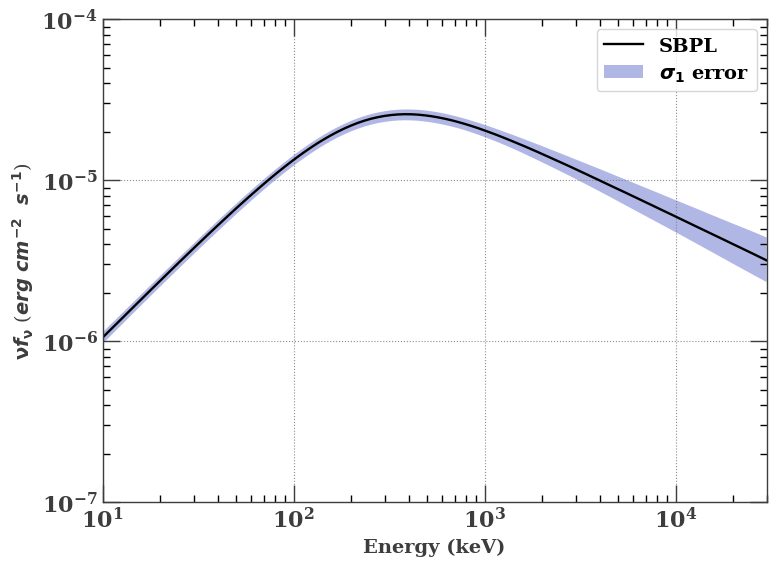}
          
    \end{subfigure}
    \caption{GRB 160509A: The count spectra (left panels) and ${\rm \nu}f_{\rm \nu}$ spectra (right panels).  The top (bottom) panels are for the joint (GBM-only) fits.}
    \label{fig_b18}
\end{figure*}


\begin{figure*}
    \centering
    \begin{subfigure}[b]{0.40\textwidth}
        \centering
        \includegraphics[width=\textwidth]{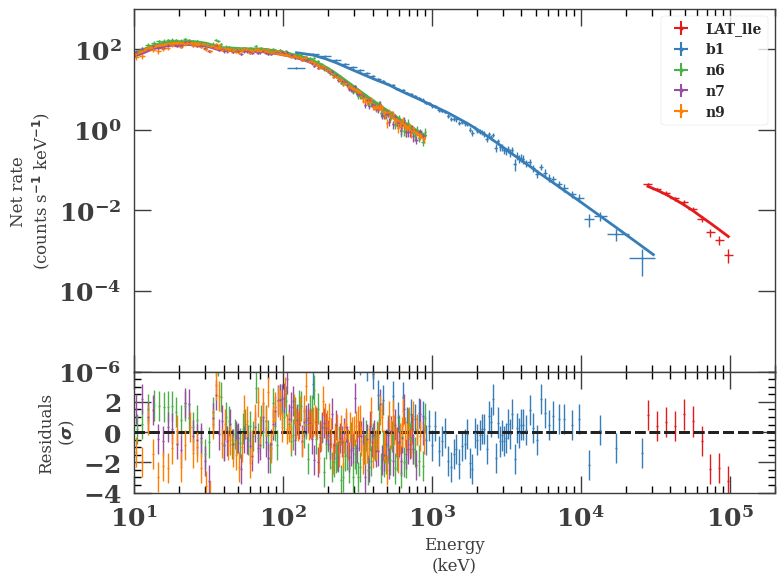}
            
    \end{subfigure}
    \hfill
    \begin{subfigure}[b]{0.40\textwidth}
        \centering
        \includegraphics[width=\textwidth]{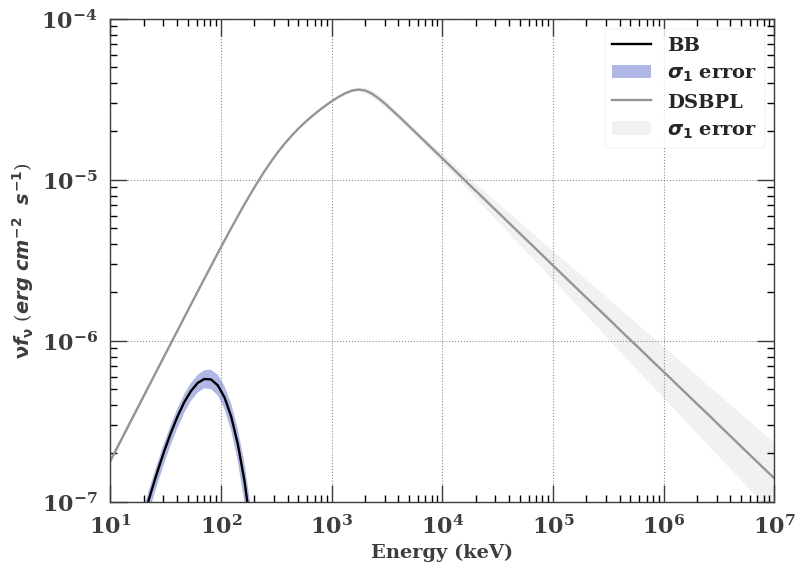}
          
    \end{subfigure}
        \hfill
    \begin{subfigure}[b]{0.40\textwidth}
        \centering
        \includegraphics[width=\textwidth]{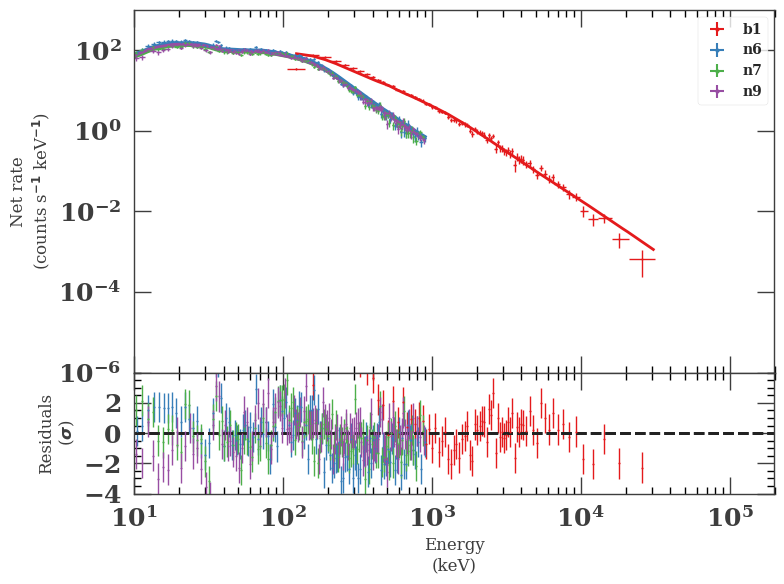}
          
    \end{subfigure}
        \hfill
    \begin{subfigure}[b]{0.40\textwidth}
        \centering
        \includegraphics[width=\textwidth]{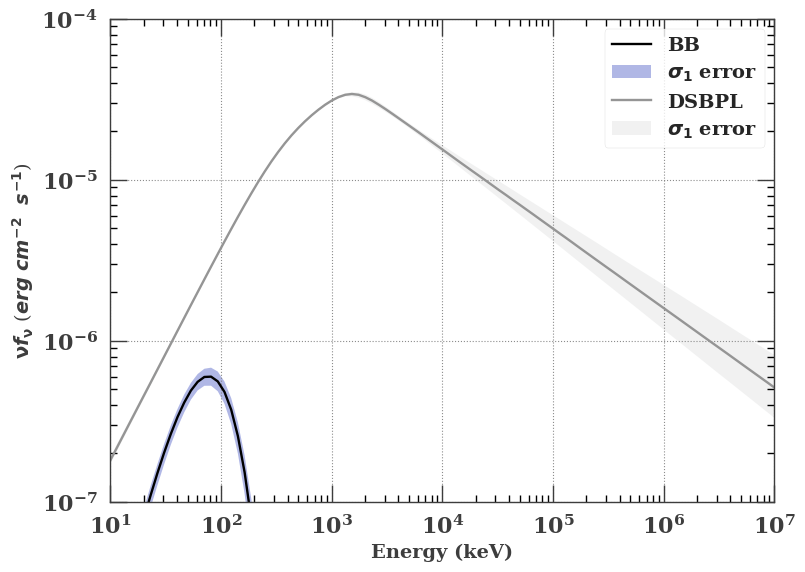}
          
    \end{subfigure}
    \caption{GRB 160625B: The count spectra (left panels) and ${\rm \nu}f_{\rm \nu}$ spectra (right panels).  The top (bottom) panels are for the joint (GBM-only) fits.}
    \label{fig_b19}
\end{figure*}


\begin{figure*}
    \centering
    \begin{subfigure}[b]{0.40\textwidth}
        \centering
        \includegraphics[width=\textwidth]{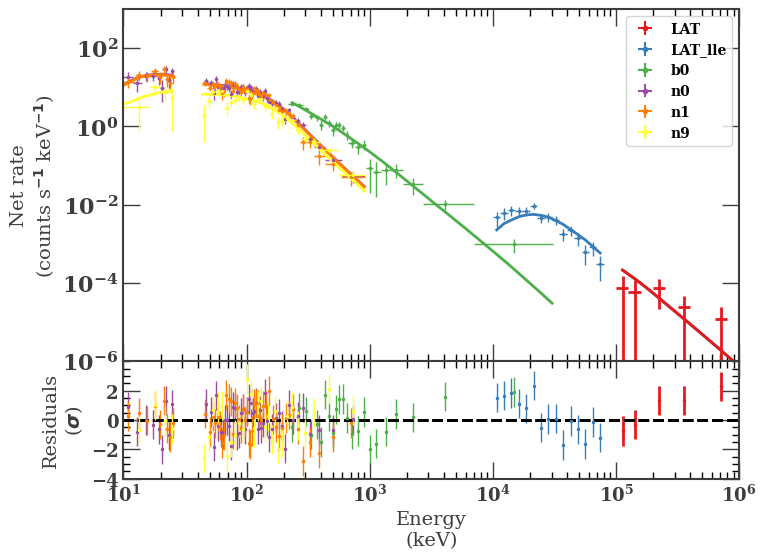}
            
    \end{subfigure}
    \hfill
    \begin{subfigure}[b]{0.40\textwidth}
        \centering
        \includegraphics[width=\textwidth]{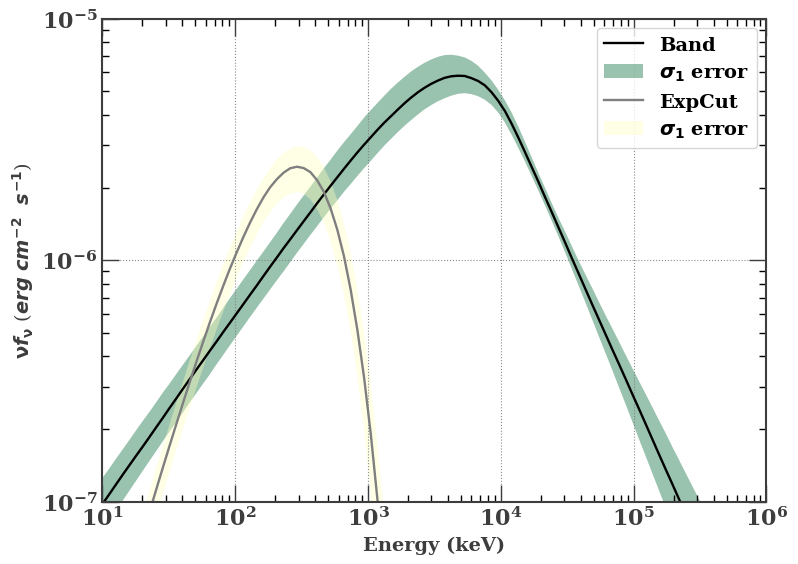}
          
    \end{subfigure}
        \hfill
    \begin{subfigure}[b]{0.40\textwidth}
        \centering
        \includegraphics[width=\textwidth]{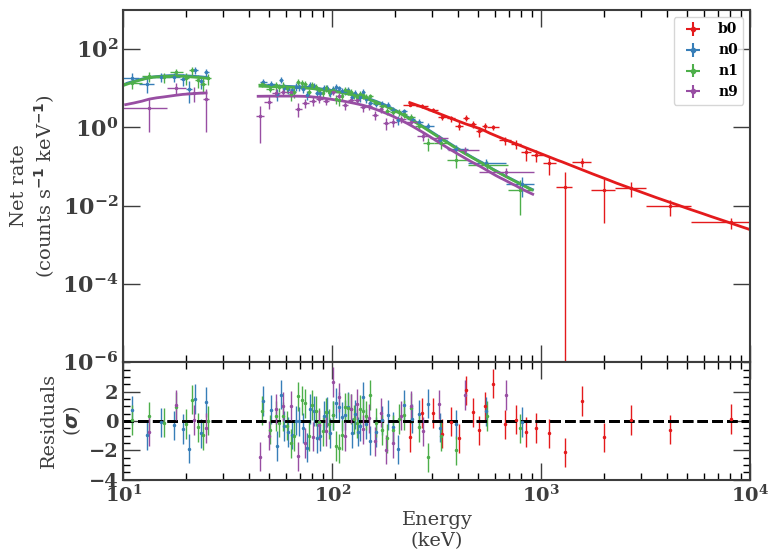}
          
    \end{subfigure}
        \hfill
    \begin{subfigure}[b]{0.40\textwidth}
        \centering
        \includegraphics[width=\textwidth]{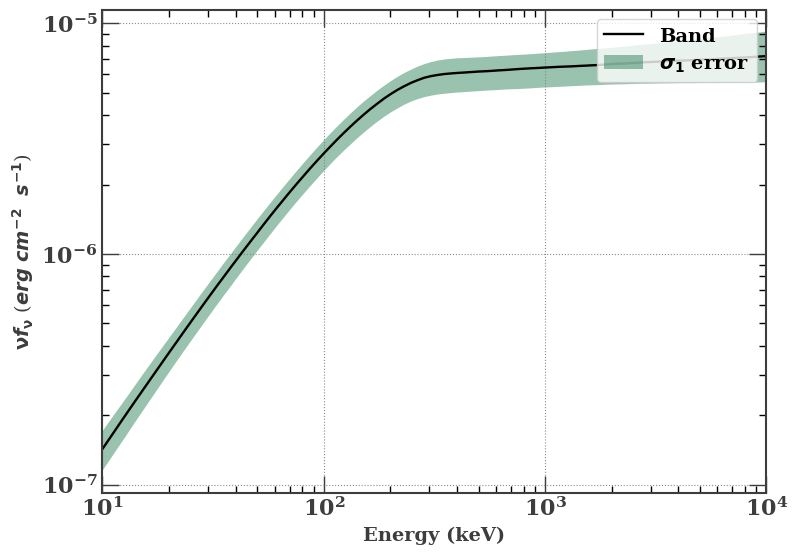}
          
    \end{subfigure}
    \caption{GRB 170214A: The count spectra (left panels) and ${\rm \nu}f_{\rm \nu}$ spectra (right panels).  The top (bottom) panels are for the joint (GBM-only) fits.
    }
    \label{fig_b20}
\end{figure*}


\begin{figure*}
    \centering
    \begin{subfigure}[b]{0.40\textwidth}
        \centering
        \includegraphics[width=\textwidth]{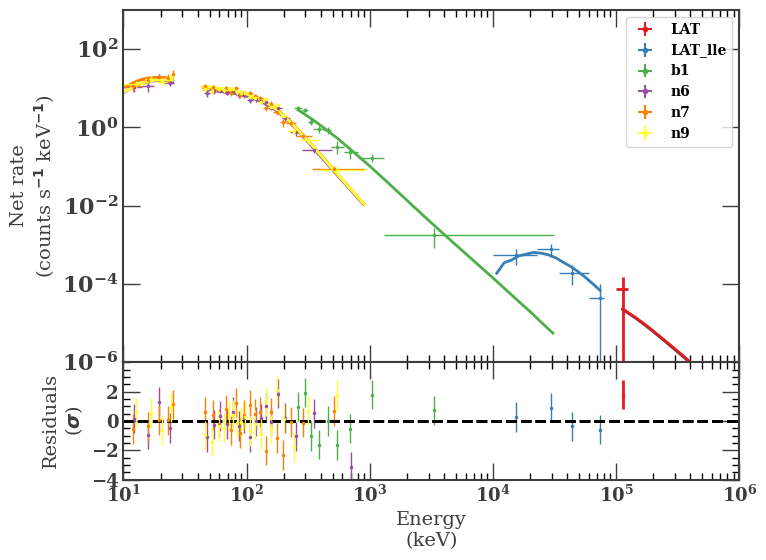}
            
    \end{subfigure}
    \hfill
    \begin{subfigure}[b]{0.40\textwidth}
        \centering
        \includegraphics[width=\textwidth]{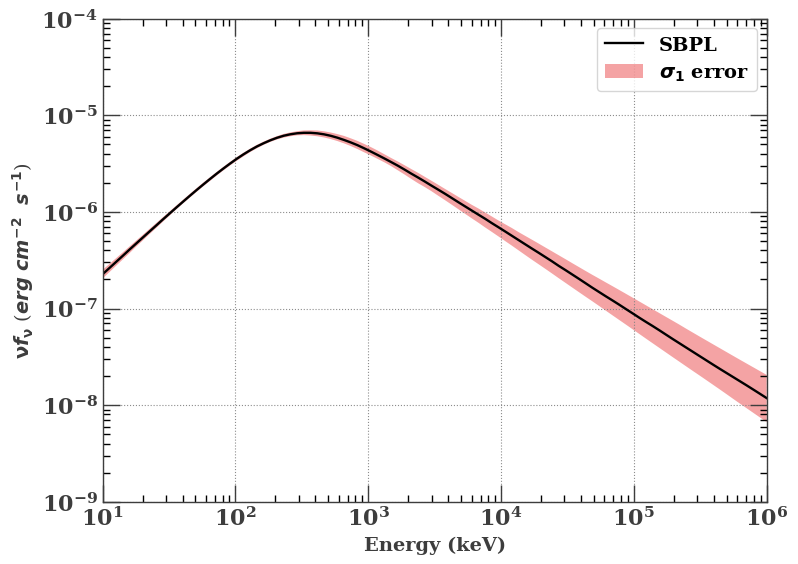}
          
    \end{subfigure}
        \hfill
    \begin{subfigure}[b]{0.40\textwidth}
        \centering
        \includegraphics[width=\textwidth]{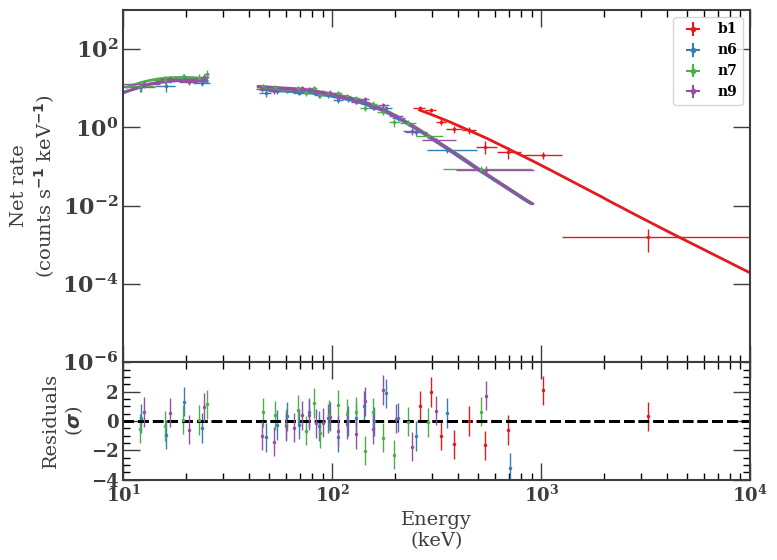}
          
    \end{subfigure}
        \hfill
    \begin{subfigure}[b]{0.40\textwidth}
        \centering
        \includegraphics[width=\textwidth]{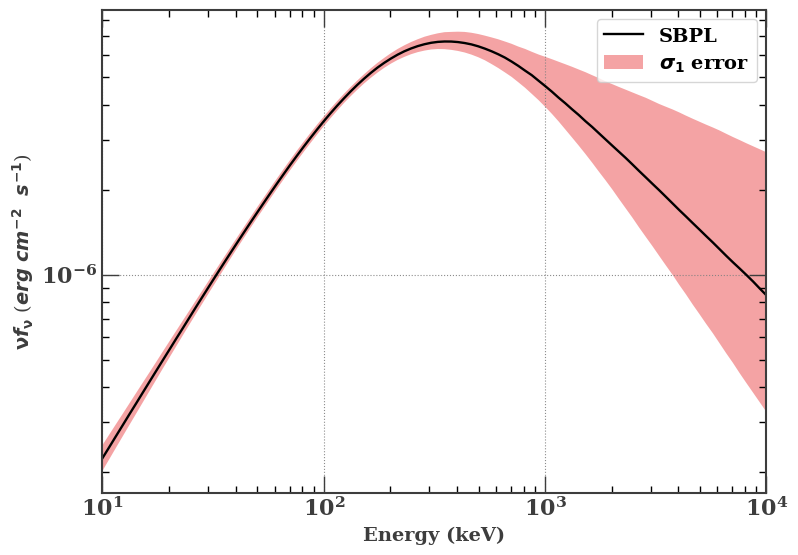}
          
    \end{subfigure}
    \caption{GRB 170405A: The count spectra (left panels) and ${\rm \nu}f_{\rm \nu}$ spectra (right panels).  The top (bottom) panels are for the joint (GBM-only) fits.}
    \label{fig_b21}
\end{figure*}


\begin{figure*}
    \centering
    \begin{subfigure}[b]{0.40\textwidth}
        \centering
        \includegraphics[width=\textwidth]{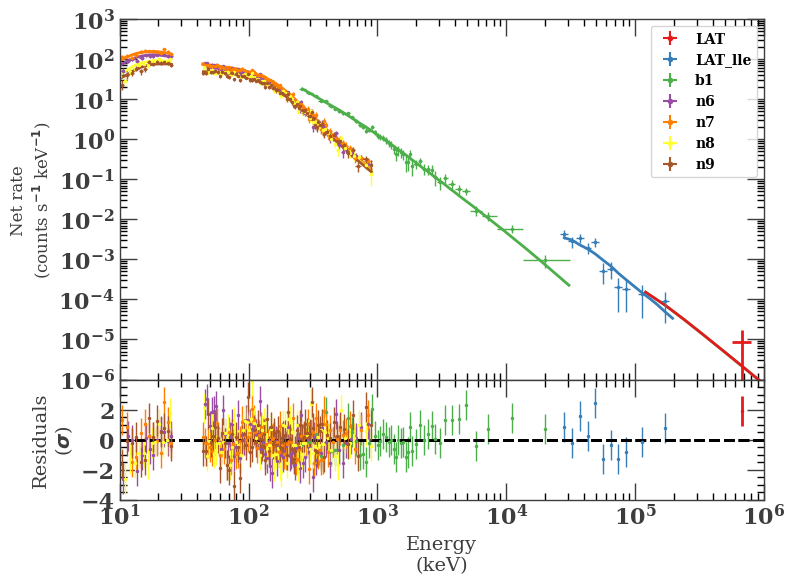}
            
    \end{subfigure}
    \hfill
    \begin{subfigure}[b]{0.40\textwidth}
        \centering
        \includegraphics[width=\textwidth]{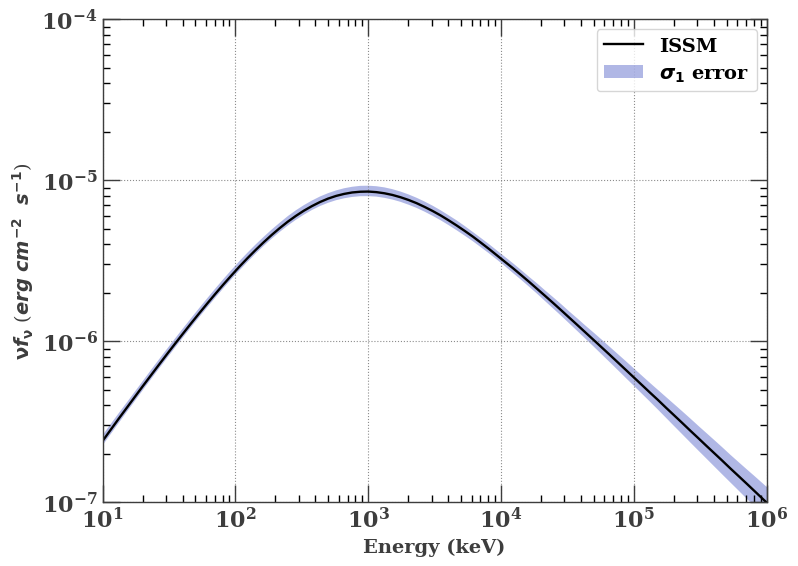}
          
    \end{subfigure}
        \hfill
    \begin{subfigure}[b]{0.40\textwidth}
        \centering
        \includegraphics[width=\textwidth]{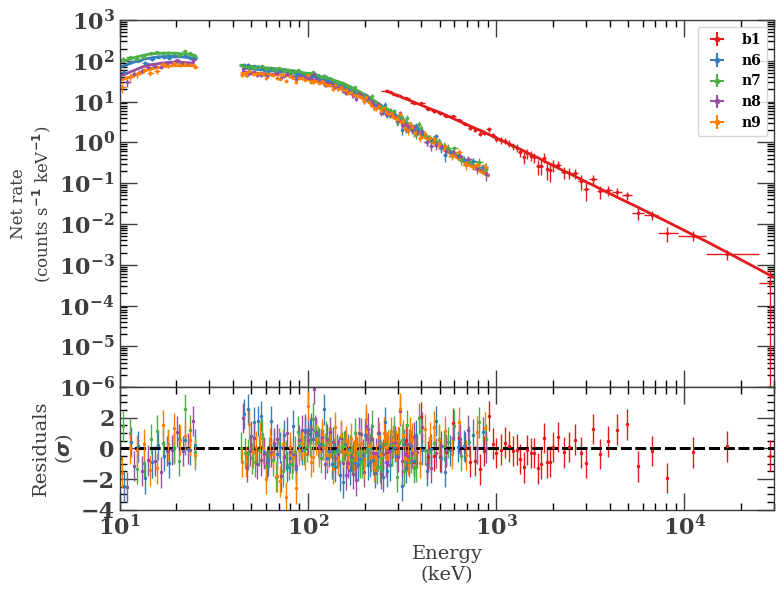}
          
    \end{subfigure}
        \hfill
    \begin{subfigure}[b]{0.40\textwidth}
        \centering
        \includegraphics[width=\textwidth]{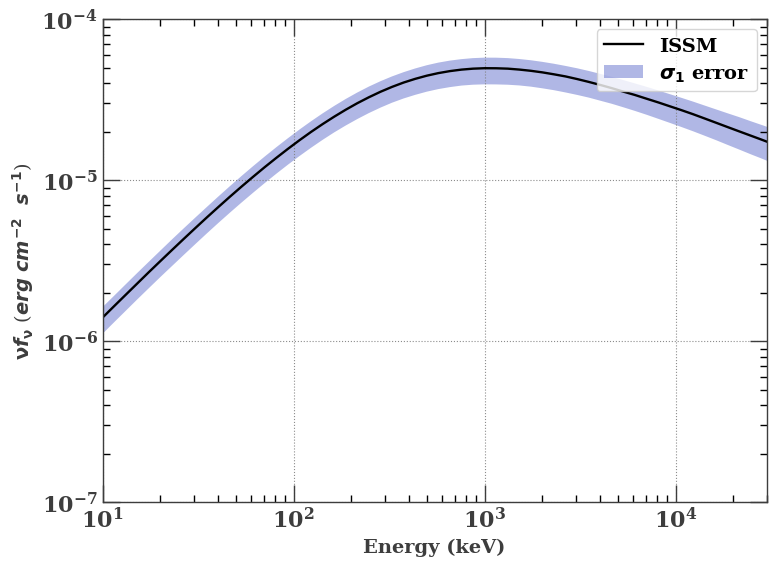}
          
    \end{subfigure}
    \caption{GRB 180720B: The count spectra (left panels) and ${\rm \nu}f_{\rm \nu}$ spectra (right panels).  The top (bottom) panels are for the joint (GBM-only) fits.}
    \label{fig_b22}
\end{figure*}


\begin{figure*}
    \centering
    \begin{subfigure}[b]{0.40\textwidth}
        \centering
        \includegraphics[width=\textwidth]{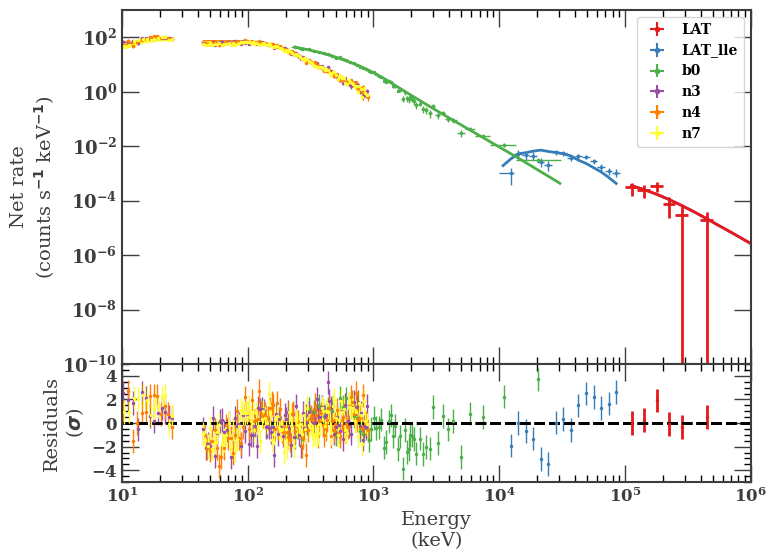}
            
    \end{subfigure}
    \hfill
    \begin{subfigure}[b]{0.40\textwidth}
        \centering
        \includegraphics[width=\textwidth]{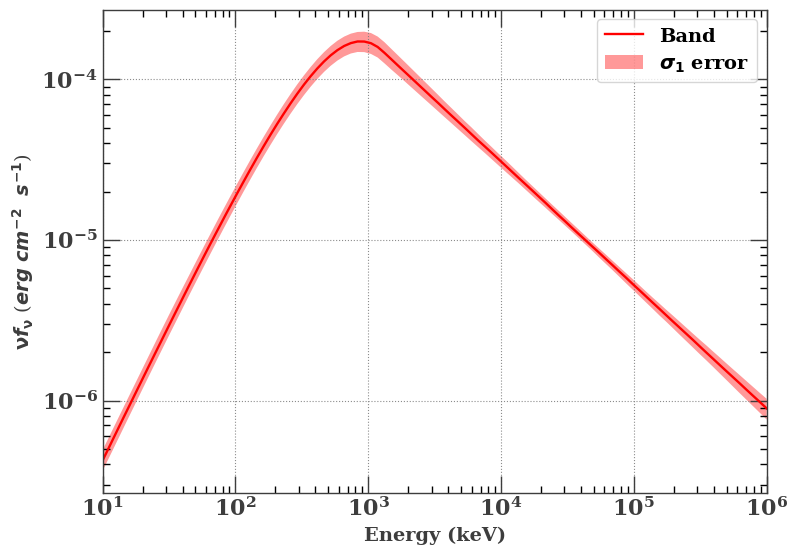}
          
    \end{subfigure}
        \hfill
    \begin{subfigure}[b]{0.40\textwidth}
        \centering
        \includegraphics[width=\textwidth]{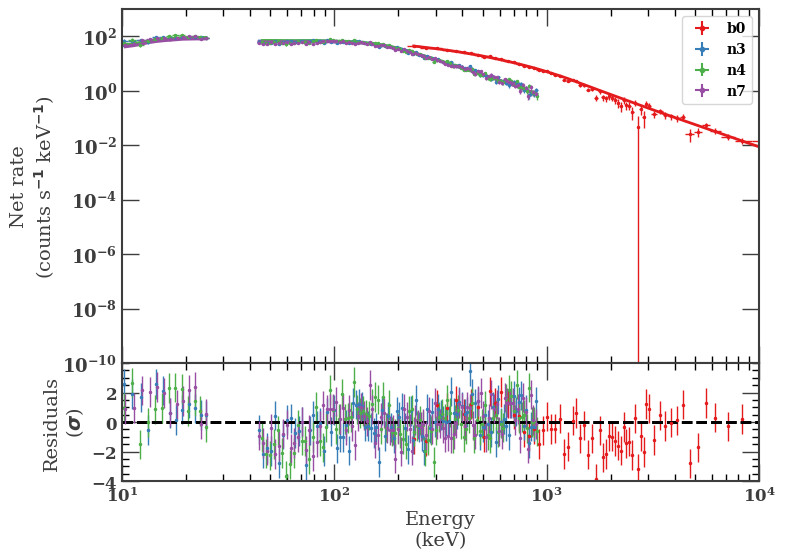}
          
    \end{subfigure}
        \hfill
    \begin{subfigure}[b]{0.40\textwidth}
        \centering
        \includegraphics[width=\textwidth]{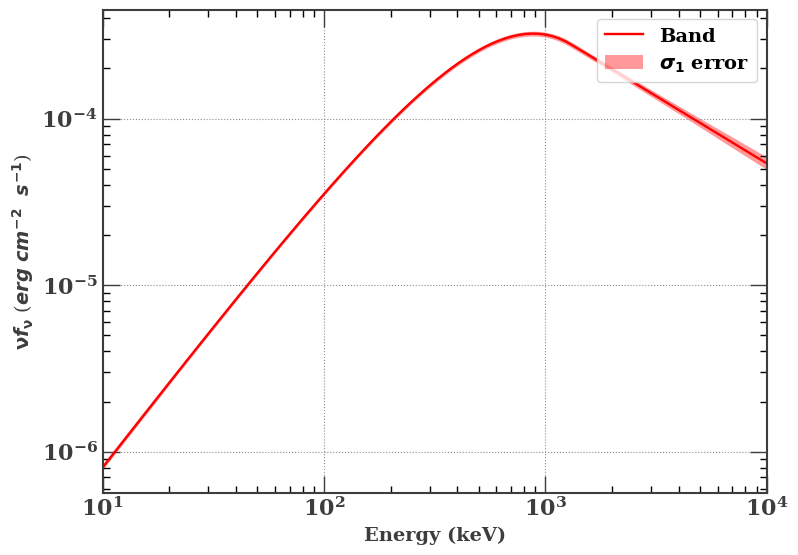}
          
    \end{subfigure}
    \caption{GRB 190114C: The count spectra (left panels) and ${\rm \nu}f_{\rm \nu}$ spectra (right panels).  The top (bottom) panels are for the joint (GBM-only) fits.}
    \label{fig_b23}
\end{figure*}


\begin{figure*}
    \centering
    \begin{subfigure}[b]{0.40\textwidth}
        \centering
        \includegraphics[width=\textwidth]{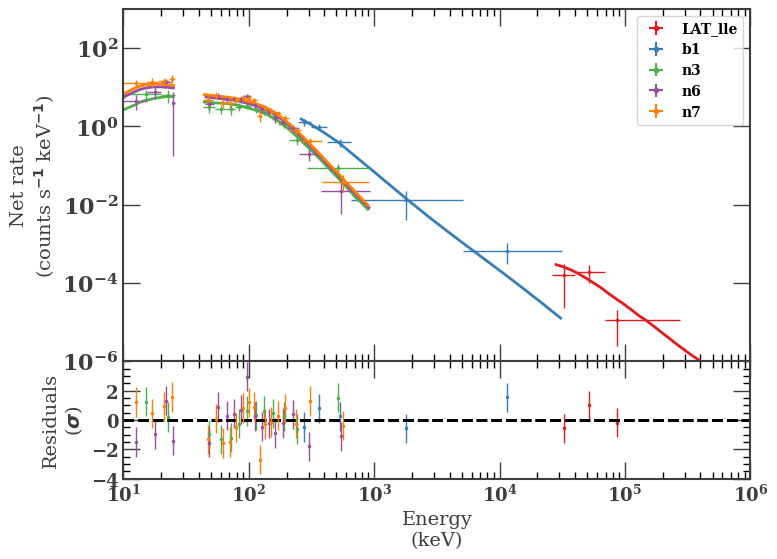}
            
    \end{subfigure}
    \hfill
    \begin{subfigure}[b]{0.40\textwidth}
        \centering
        \includegraphics[width=\textwidth]{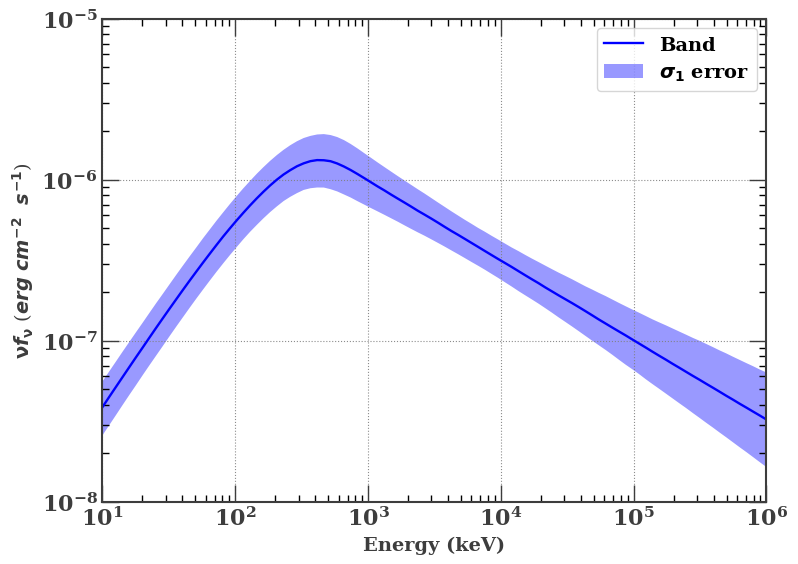}
          
    \end{subfigure}
        \hfill
    \begin{subfigure}[b]{0.40\textwidth}
        \centering
        \includegraphics[width=\textwidth]{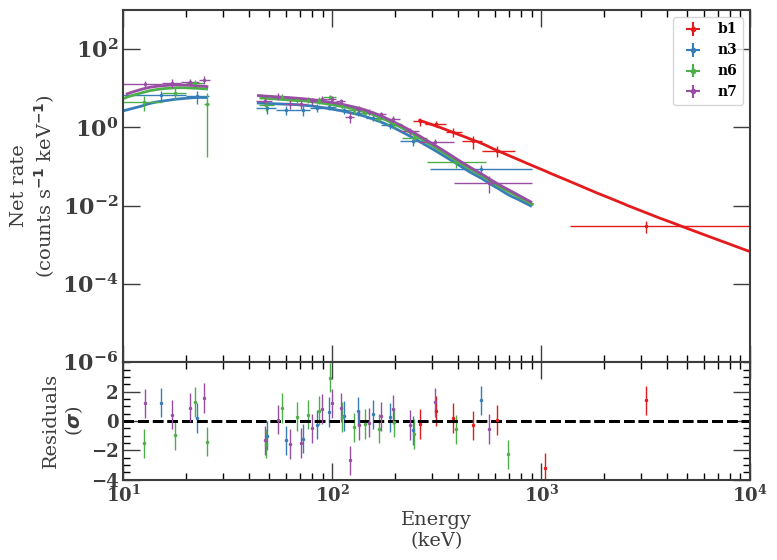}
          
    \end{subfigure}
        \hfill
    \begin{subfigure}[b]{0.40\textwidth}
        \centering
        \includegraphics[width=\textwidth]{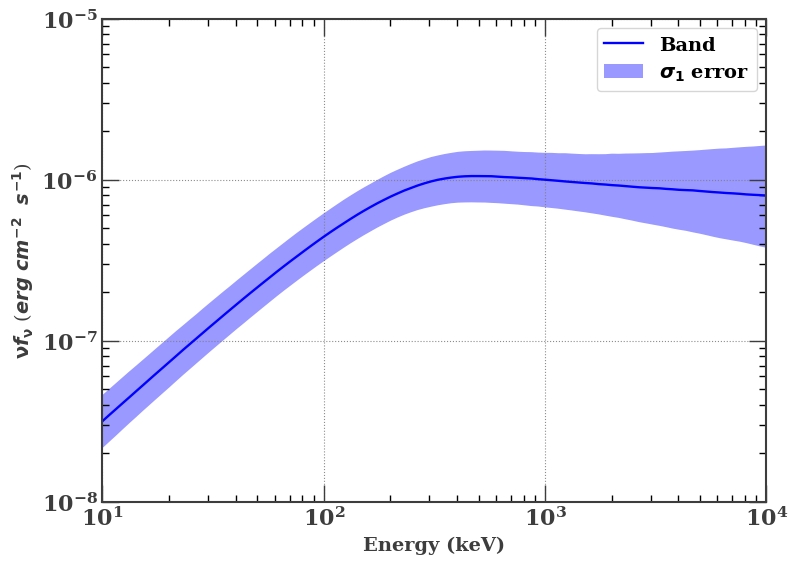}
          
    \end{subfigure}
    \caption{GRB 220101A: The count spectra (left panels) and ${\rm \nu}f_{\rm \nu}$ spectra (right panels).  The top (bottom) panels are for the joint (GBM-only) fits.}
    \label{fig_b24}
\end{figure*}
\begin{figure*}
    \centering
    \begin{subfigure}[b]{0.40\textwidth}
        \centering
        \includegraphics[width=\textwidth]{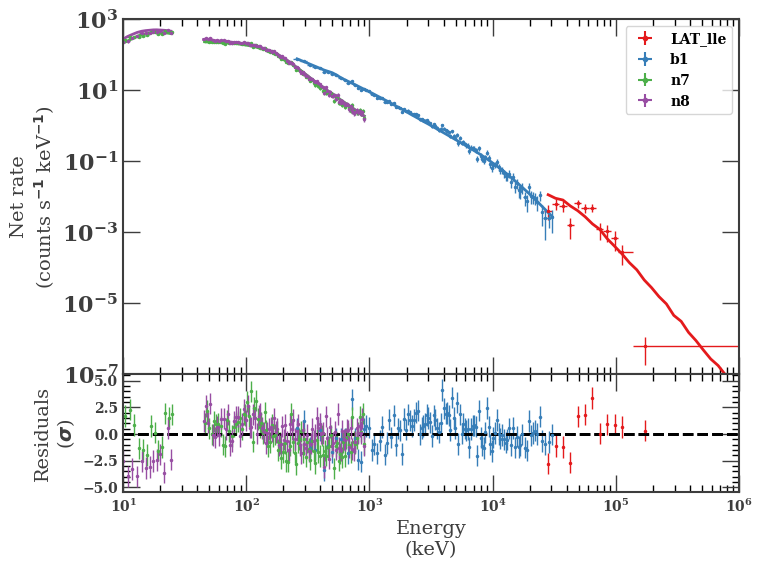}
            
    \end{subfigure}
    \hfill
    \begin{subfigure}[b]{0.40\textwidth}
        \centering
        \includegraphics[width=\textwidth]{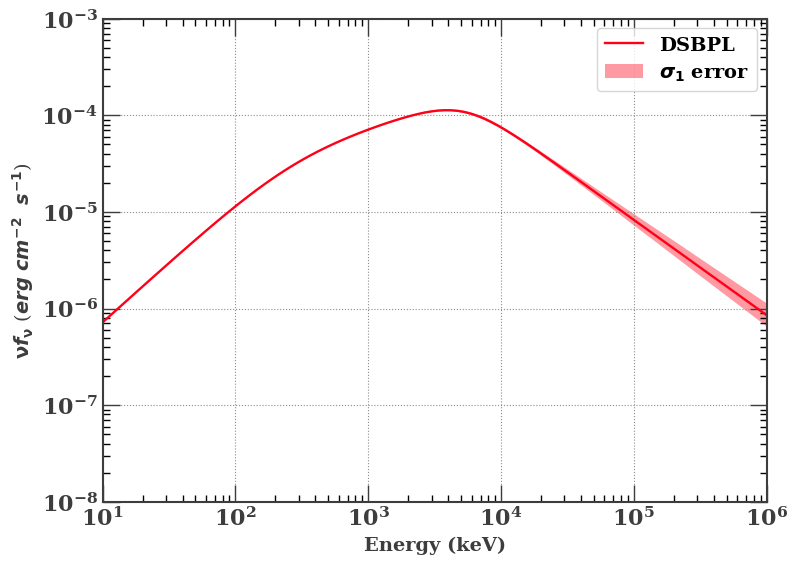}
          
    \end{subfigure}
        \hfill
    \begin{subfigure}[b]{0.40\textwidth}
        \centering
        \includegraphics[width=\textwidth]{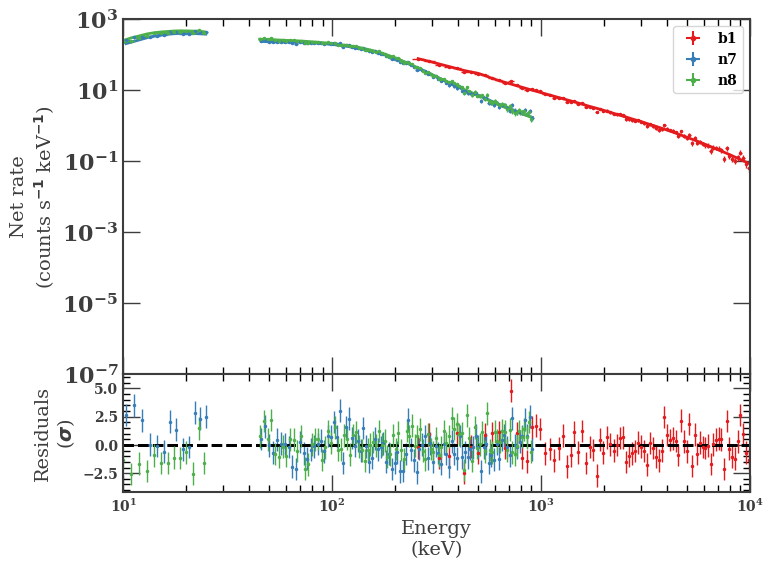}
          
    \end{subfigure}
        \hfill
    \begin{subfigure}[b]{0.40\textwidth}
        \centering
        \includegraphics[width=\textwidth]{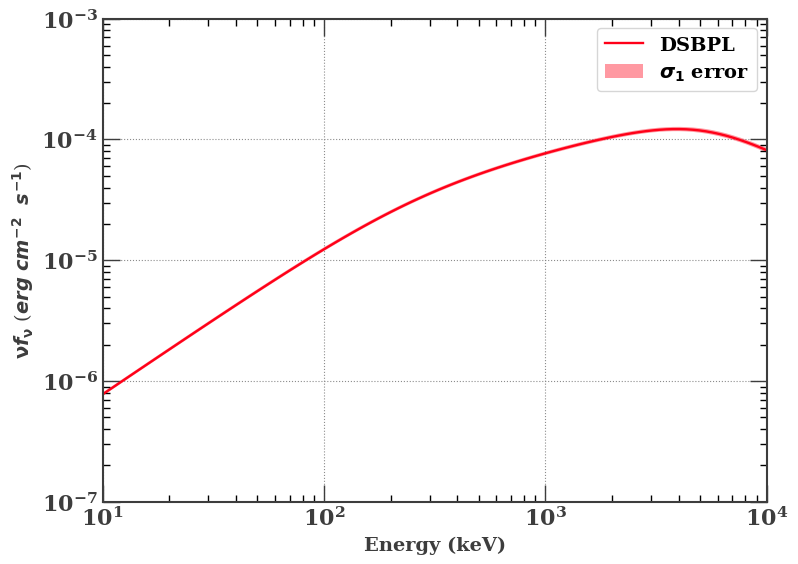}
          
    \end{subfigure}
    \caption{GRB 221009A: The count spectra (left panels) and ${\rm \nu}f_{\rm \nu}$ spectra (right panels).  The top (bottom) panels are for the joint (GBM-only) fits.}
    \label{fig44_1}
\end{figure*}

\begin{figure*}
    \centering
    \begin{subfigure}[b]{0.40\textwidth}
        \centering
        \includegraphics[width=\textwidth]{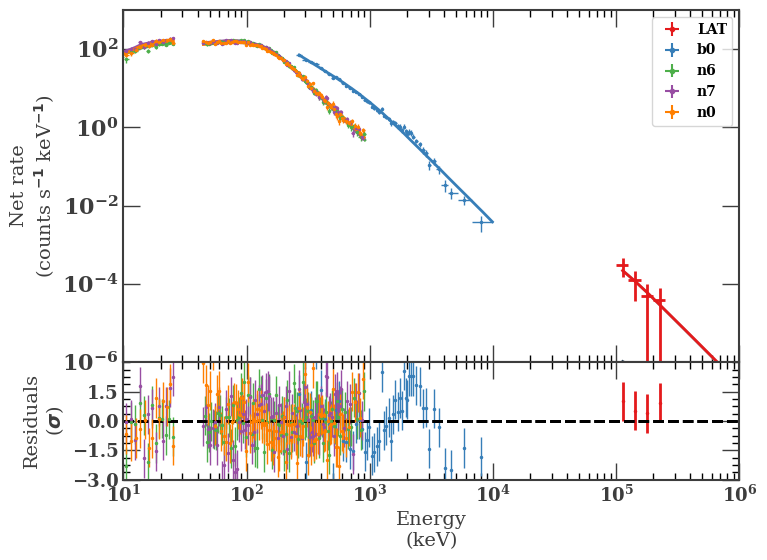}
            
    \end{subfigure}
    \hfill
    \begin{subfigure}[b]{0.40\textwidth}
        \centering
        \includegraphics[width=\textwidth]{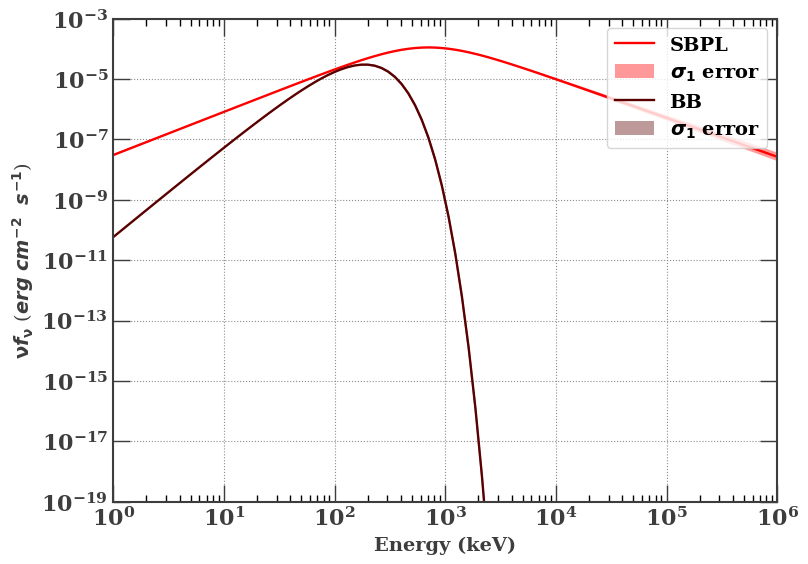}
          
    \end{subfigure}
        \hfill
    \begin{subfigure}[b]{0.40\textwidth}
        \centering
        \includegraphics[width=\textwidth]{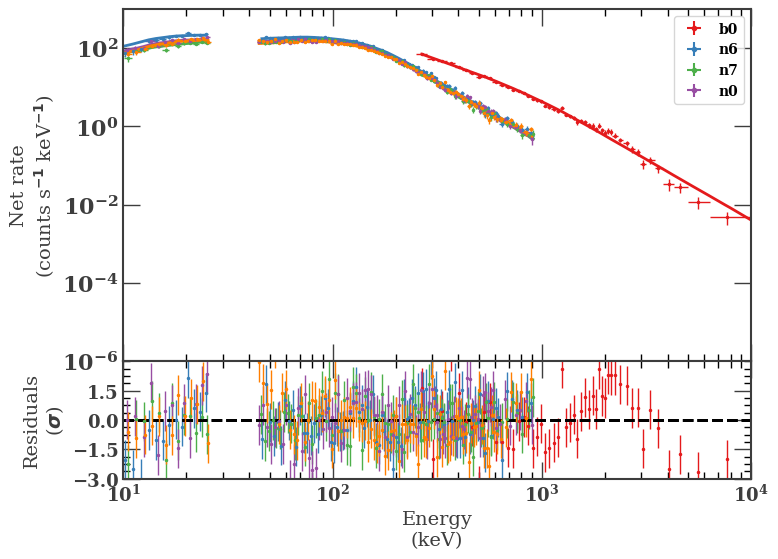}
          
    \end{subfigure}
        \hfill
    \begin{subfigure}[b]{0.40\textwidth}
        \centering
        \includegraphics[width=\textwidth]{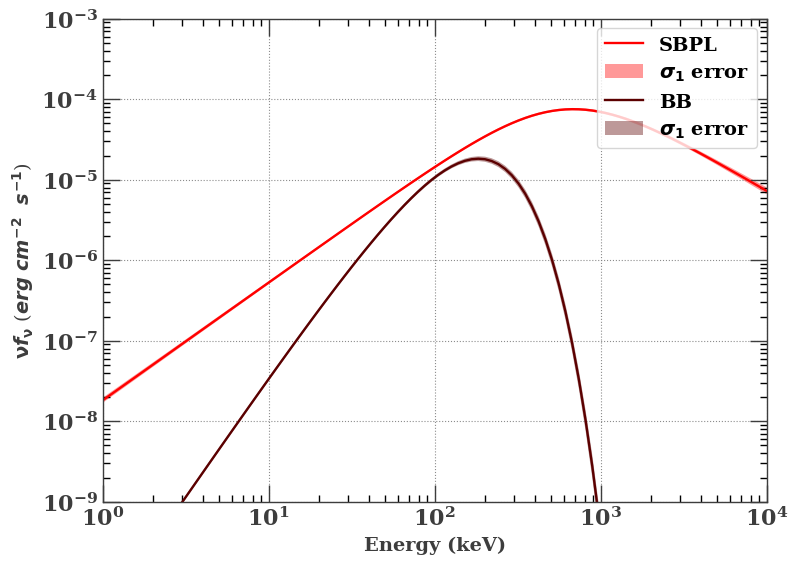}
          
    \end{subfigure}
    \caption{GRB 230812B: The count spectra (left panels) and ${\rm \nu}f_{\rm \nu}$ spectra (right panels).  The top (bottom) panels are for the joint (GBM-only) fits.}
    \label{fig_b25}
\end{figure*}


\begin{figure*}
    \centering
    \begin{subfigure}[b]{0.40\textwidth}
        \centering
        \includegraphics[width=\textwidth]{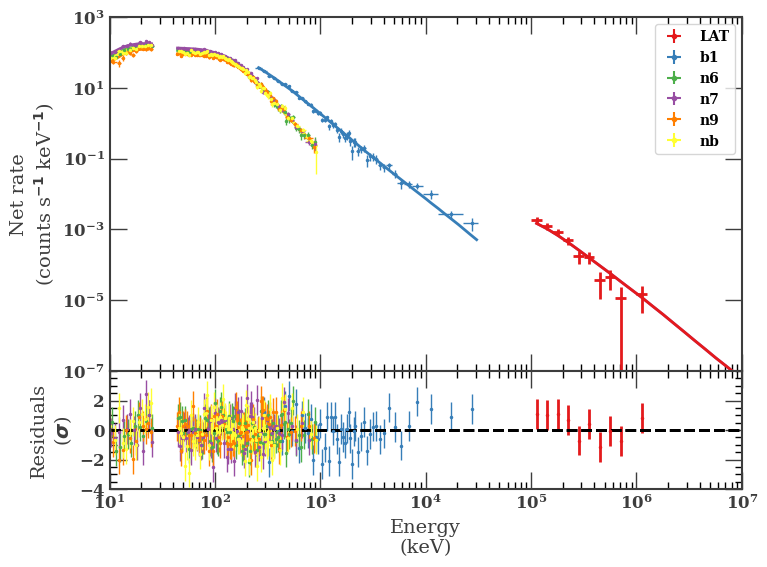}
            
    \end{subfigure}
    \hfill
    \begin{subfigure}[b]{0.40\textwidth}
        \centering
        \includegraphics[width=\textwidth]{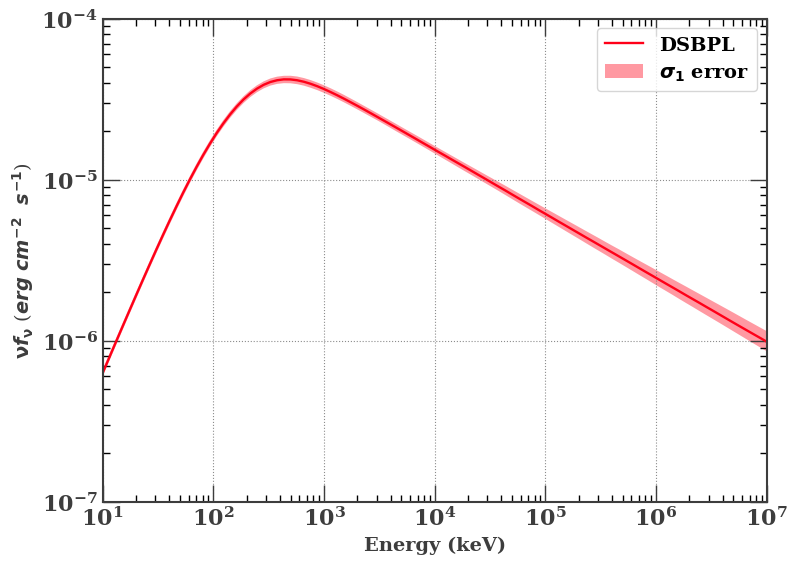}
          
    \end{subfigure}
        \hfill
    \begin{subfigure}[b]{0.40\textwidth}
        \centering
        \includegraphics[width=\textwidth]{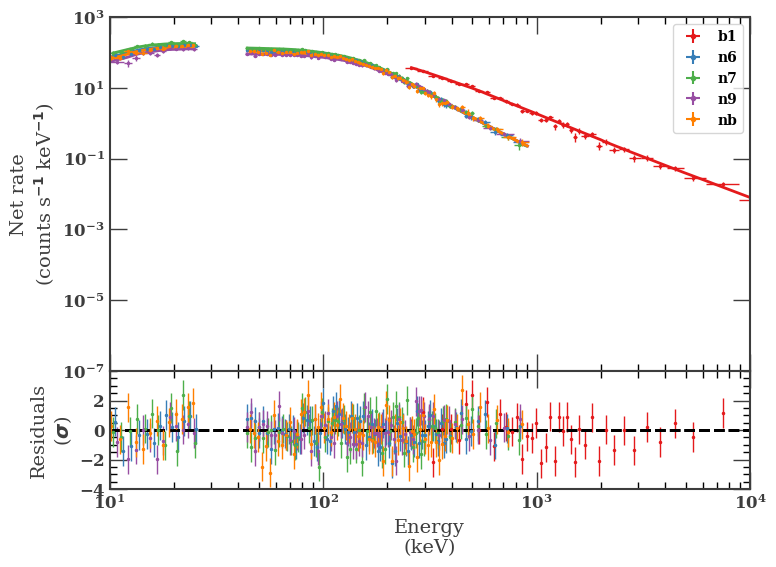}
          
    \end{subfigure}
        \hfill
    \begin{subfigure}[b]{0.40\textwidth}
        \centering
        \includegraphics[width=\textwidth]{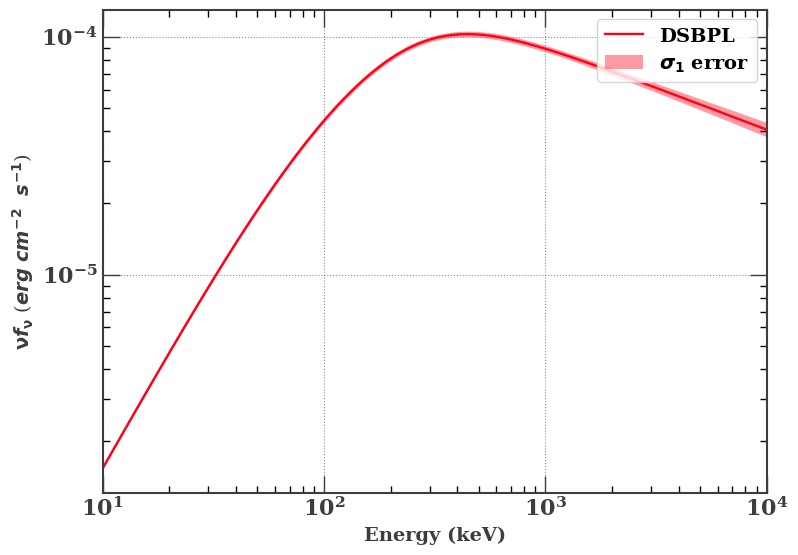}
          
    \end{subfigure}
    \caption{GRB\,240825A: The count spectra (left panels) and ${\rm \nu}f_{\rm \nu}$ spectra (right panels).  The top (bottom) panels are for the joint (GBM-only) fits.}
    \label{fig_b26}
\end{figure*}


\begin{figure*}
    \centering
    \begin{subfigure}[b]{0.40\textwidth}
        \centering
        \includegraphics[width=\textwidth]{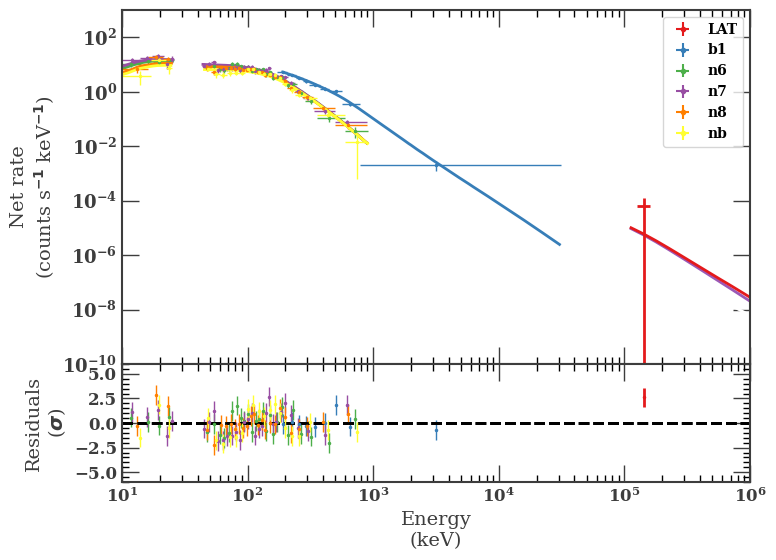}
            
    \end{subfigure}
    \hfill
    \begin{subfigure}[b]{0.40\textwidth}
        \centering
        \includegraphics[width=\textwidth]{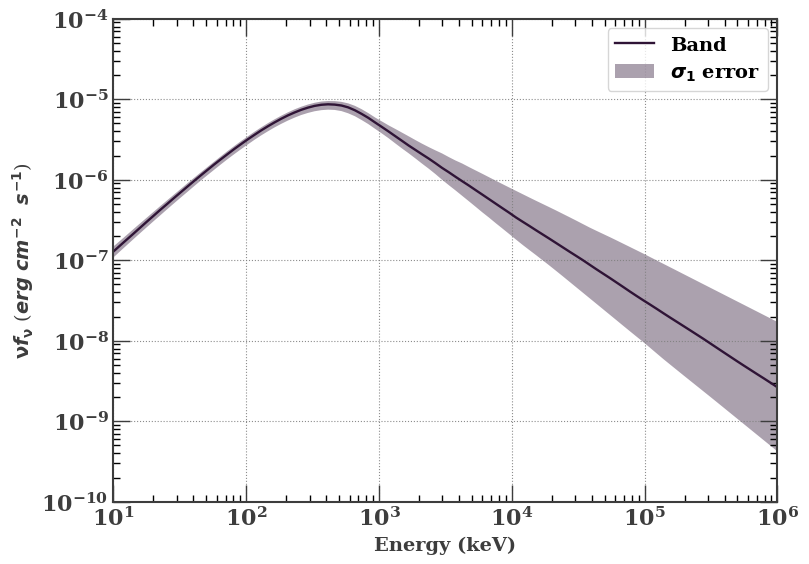}
          
    \end{subfigure}
        \hfill
    \begin{subfigure}[b]{0.40\textwidth}
        \centering
        \includegraphics[width=\textwidth]{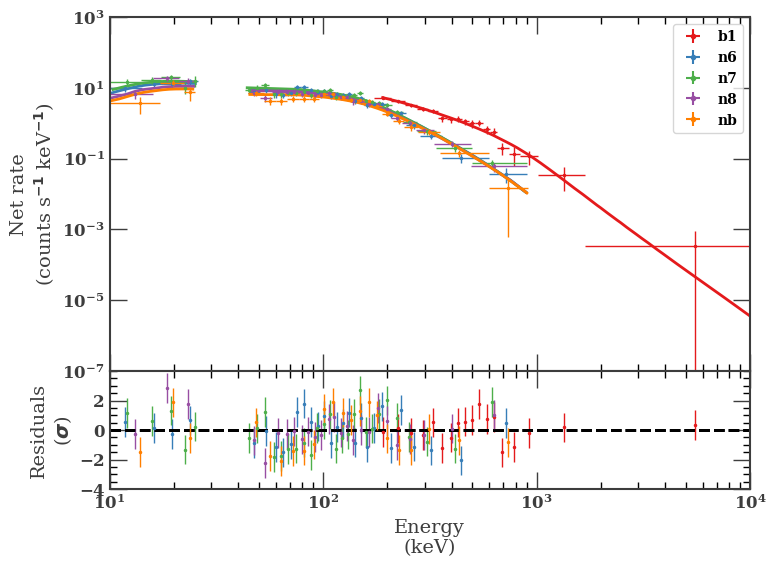}
          
    \end{subfigure}
        \hfill
    \begin{subfigure}[b]{0.40\textwidth}
        \centering
        \includegraphics[width=\textwidth]{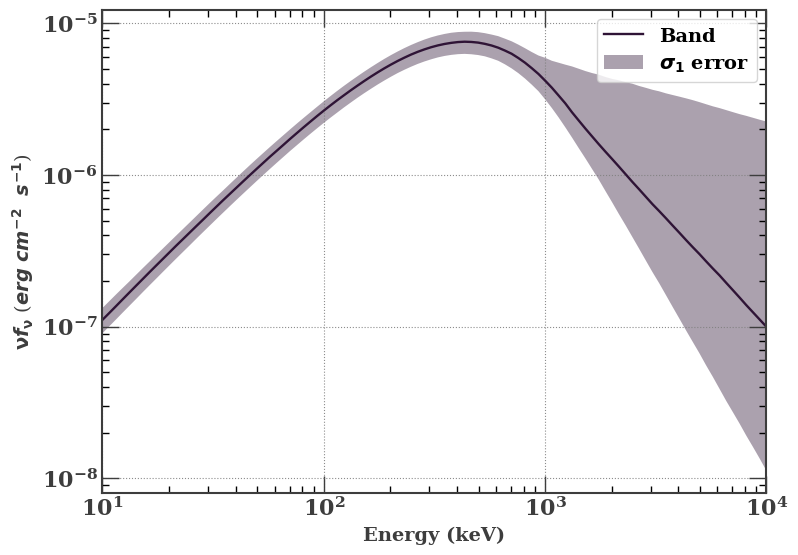}
          
    \end{subfigure}
    \caption{GRB\,241228B: The count spectra (left panels) and ${\rm \nu}f_{\rm \nu}$ spectra (right panels).  The top (bottom) panels are for the joint (GBM-only) fits.}
    \label{fig_b27}
\end{figure*}

\bsp	
\label{lastpage}
\end{document}